# A Search for Halo Axions

**by**

**Edward John Daw**

Bachelor of Arts, New College, Oxford University, England.

Submitted to the Department of Physics
in partial fulfillment of the requirements for the degree of

Doctor of Philosophy

at the

MASSACHUSETTS INSTITUTE OF TECHNOLOGY

February 1998

Author..........................................................................................

Department of Physics

January 3 1998

Certified by..........................................................................................

Leslie J Rosenberg

Professor of Physics

Thesis Supervisor

Accepted by..........................................................................................

George F. Koster

Chairman, Departmental Committee on Theses

# A Search for Halo Axions

**by**

**Edward John Daw**

Submitted to the Department of Physics
on January 4, 1998, in partial fulfillment of the
requirements for the degree of
Doctor of Philosophy

## Abstract

The axion is an unusually well motivated dark matter candidate which is the Goldstone Boson associated with spontaneous breaking of the PQ symmetry of QCD. I describe an experiment to search for axions in our Galactic halo using a high Q resonant cavity coupled to ultra-low-noise receiver electronics. I discuss the analysis of the first production data and present the first results of the experiment. At 90% confidence KSVZ axions of mass $2.9 \times 10^{-6}$eV to $3.3 \times 10^{-6}$eV are excluded as the dark matter in the halo of our Galaxy.

Thesis Supervisor: Leslie J Rosenberg
Title: Professor of Physics

# Acknowledgements

I would like to thank my advisor and friend Leslie Rosenberg. His intelligence, intellectual integrity, common sense and good humor have been invaluable throughout my career as a graduate student. I would also like to thank the other members of the PPC group at M.I.T. for many invigorating and enlightening discussions, particularly Larry Rosenson for his good advice during my early years at M.I.T., help with simulations later on and a recent careful reading of this thesis.

I would like to thank all my friends at Lawrence Livermore National Laboratory. In particular, thanks to Darin Kinion for his excellent data acquisition software and for companionship during the sometimes lonely months whilst I was sat at my computer doing analysis. Thanks also to Karl van Bibber for his encouragement and enthusiasm, to Wolfgang Stoeffl for invigorating discussions and debate and to Chris Hagmann and Hong Peng for good ideas and keeping me 'on my toes'.

I wish to thank my parents Stephen and Gillian for their love and support during my long stay in America.

Finally, my heartfelt thanks to my wife Anne for her for her CAD expertise, faith, help, encouragement and love during these eventful years.

# 1. Introduction

My thesis describes the commissioning, operation and analysis of first data for a large-scale axion search experiment. M.I.T. played a major role in proposing, commissioning and operating the detector and in developing the data acquisition system, and has spearheaded the analysis of data and the generation of the exclusion limits.

My experimental responsibilities have been threefold. Firstly, I designed, installed, tested, and commissioned the low noise microwave electronics for the axion detector. Secondly, I analyzed the production data, generating exclusion limits for axions as halo cold dark matter in the mass range 2.9 - 3.3μeV. Thirdly, I played a major role in the day-to-day running of the experiment.

The commissioning run of the detector commenced in early 1995. The production data for my analysis was taken between April 13th 1996 and October 22nd 1997.

# 2. Axion Theory and Phenomenology

## 2.1 Theory of Axions

### 2.1.1 QCD and CP violation

I begin with a brief introduction to CP violating effects in QCD. For a more detailed discussion see, e.g., reference [1]. The QCD Lagrangian can be written as follows:

$$\mathcal{L}_{QCD} = -\frac{1}{4}(F^{a}_{\mu\nu})^{2} + \overline{\psi}_{i}(i\gamma^{\mu}D_{\mu} - m)\psi_{i} \qquad (2.1)$$

where each $\psi$ is a Dirac field representing a massive quark and each $F_{\mu\nu}{}^{a}$ is the field strength tensor of a gluon. Both terms in this standard QCD Lagrangian are CP invariant. CP violation appears in QCD through the non-perturbative term:

$$\mathcal{L}_{CPV} = \frac{\overline{\theta}}{32\pi^{2}} F_{\mu\nu}{}^{a}\tilde{F}^{\mu\nu a} \text{ where } \tilde{F}^{\mu\nu a} = \varepsilon^{\mu\nu\rho\sigma}F_{\rho\sigma}{}^{a} \qquad (2.2)$$

where $\overline{\theta}$ is a number giving the magnitude of the CP violation. Introducing the QCD counterparts **E** and **B** of the electromagnetic fields for each gluon:

$$\mathcal{L}_{CPV} = \frac{\overline{\theta}}{16\pi^{2}} \mathbf{E}^{a} \cdot \mathbf{B}^{a} \qquad (2.3)$$

Equation 2.3 shows why $\tilde{a}_{CPV}$ violates CP. In electrodynamics **E** is a vector and **B** is an axial vector. Therefore $\mathbf{E} \cdot \mathbf{B}$ changes under CP. The same argument applies to QCD. Although $\mathcal{L}_{CPV}$ is often ignored, it is nonetheless there and part of 'standard model' QCD.

Physically, CP violation arises from the structure of the QCD vacuum. Briefly, the existence of tunneling solutions (instantons) between the degenerate but distinct QCD vacuum states imply that the physical vacuum is actually a superposition of these degenerate states. The effects of the physical vacuum, called the 'θ vacuum', on standard model physics can be written in to the theory in the form of an effective Lagrangian term of the same form as
$\mathcal{L}_{CPV}$. The coefficient of this Lagrangian term, θ, is called the vacuum angle.

A second and well known mechanism for CP violation in QCD arises as a consequence of electroweak unification. In the standard model, the quark masses are generated by spontaneously breaking the electroweak gauge symmetry. The resulting mass terms can be written $\bar{q}_{Ri} M_{ij} q_{Lj}$ where the quark spinors are eigenstates of helicity and the mass matrix M is neither hermitian nor diagonal. To diagonalize these mass terms such that the diagonal elements of M become the quark masses, chiral transformations are made on the quark fields. These chiral transformations also affect the θ vacuum, rotating the vacuum angle to a new value:

$$\bar{\theta} = \theta + \mathrm{Arg}(\det(M)) \qquad (2.4)$$

Since the two CP-violating effects arise from different sectors of the standard model it would be a remarkable coincidence if they combined to give $\bar{\theta} = 0$. Furthermore, numerical values of other strong-interaction parameters such as $\alpha_S$ lead us to expect $\bar{\theta}$ of order 1. Therefore we expect to observe CP-violating effects in QCD.

## 2.1.2 CP Violation and the Strong CP Problem

One sensitive way to look for CP violation in QCD by searching for an electric dipole moment (edm) of the neutron. The following argument explains the connection between the neutron edm and CP violation.

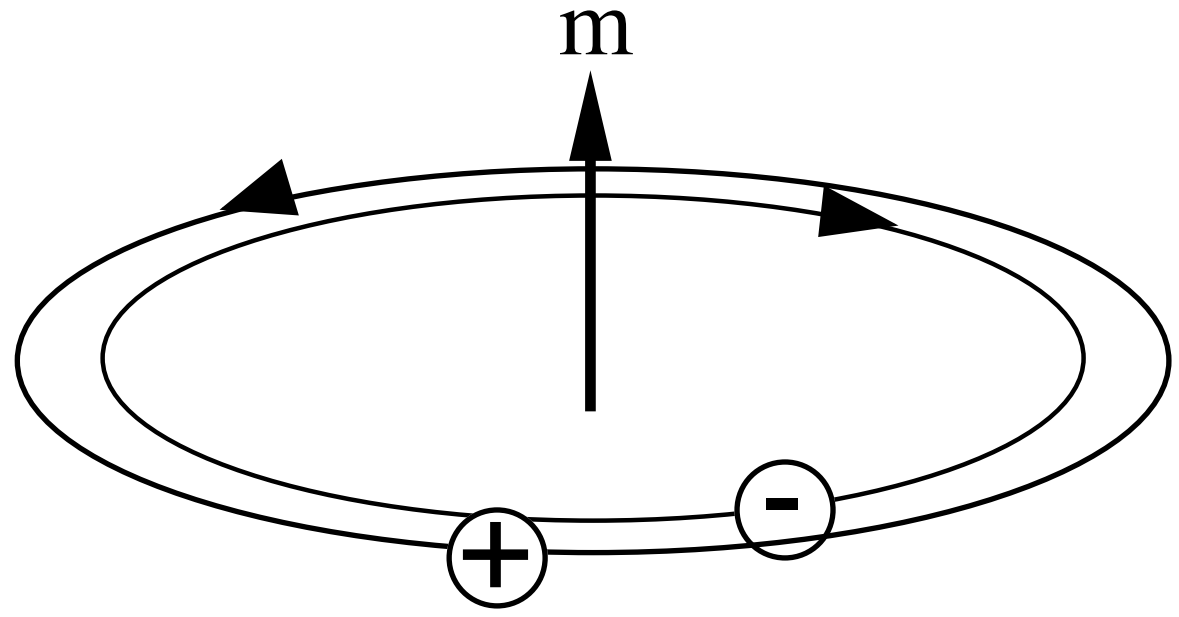

Figure 2.1 Model for a neutron with zero electric dipole moment

The neutron is an uncharged object with a non-zero magnetic dipole moment (mdm). Figure 2.1 shows a model neutron constructed out of electric charges that has these properties.

This model neutron has a magnetic dipole moment in the direction of the arrow and zero charge, but no electric dipole moment. To give the neutron a non-zero edm, we displace the plane of rotation of one of the charges along the symmetry axis and get the model neutron shown in Figure 2.2.

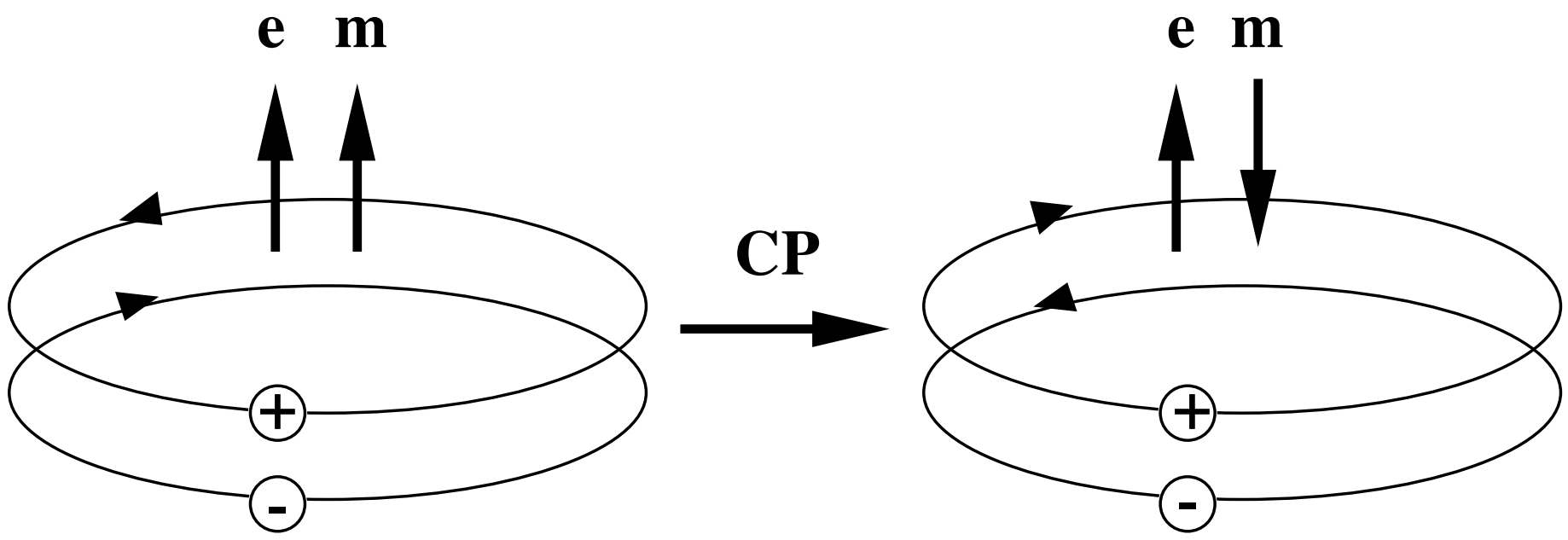

Figure 2.2 model neutron with edm ≠ 0 and its CP partner

CP operating on the neutron with a non-zero edm yields an object with different relative orientation of its electric and magnetic dipole moments, and hence different physical properties.

Many experiments have searched for a neutron edm. The present experimental bound is that a neutron electric dipole moment $d_n<1.1.10^{-25}$e cm [2]. This implies that $\bar{\theta} < 10^{-9}$. The question arises, why is there so little CP violation in the strong interactions? This is the strong CP problem.

## 2.1.3 The Peccei-Quinn Mechanism and The Axion

I will now describe briefly a solution to the strong CP problem proposed by Peccei and Quinn in 1977 [3], and outline how it implies the existence of the axion.

Peccei and Quinn proposed that QCD is invariant under an new global U(1) axial symmetry, called the PQ symmetry. The symmetry corresponds to conservation of a new charge carried by a complex pseudoscalar field $\phi$. Now $\mathrm{Arg}(\phi)$ is $\bar{\theta}$, the coefficient of the CP violating term in QCD. In other words, $\bar{\theta}$ has been promoted from a number to a field.

The classical QCD potential $V(\phi)$ has the 'wine bottle' shape necessary for spontaneous symmetry breaking. When the energy scale exceeds the symmetry breaking scale, the vacuum expectation value (VEV) of $\phi$ is zero and all values of $\bar{\theta}$ are equally probable. At these high energies, you would expect to see strong CP violation and a non-zero neutron electric dipole moment. Below $f_{PQ}$, $\phi$ has a non-zero VEV, and $\bar{\theta}$ is constrained to lie in the minimum of the potential. So far, the symmetry breaking scheme is similar to the Higgs mechanism. However, the vacuum expectation value of the field also has contributions from the $\theta$ vacuum which break the rotational symmetry of $V(\phi)$ about the origin. Peccei and Quinn showed that if at least one of the fermions acquires its mass by coupling to $\phi$, then the absolute minimum in $V(\phi)$ is at $\bar{\theta} = 0$. Thus the spontaneous breaking of the PQ symmetry forces the CP violating term in the QCD Lagrangian to vanish.

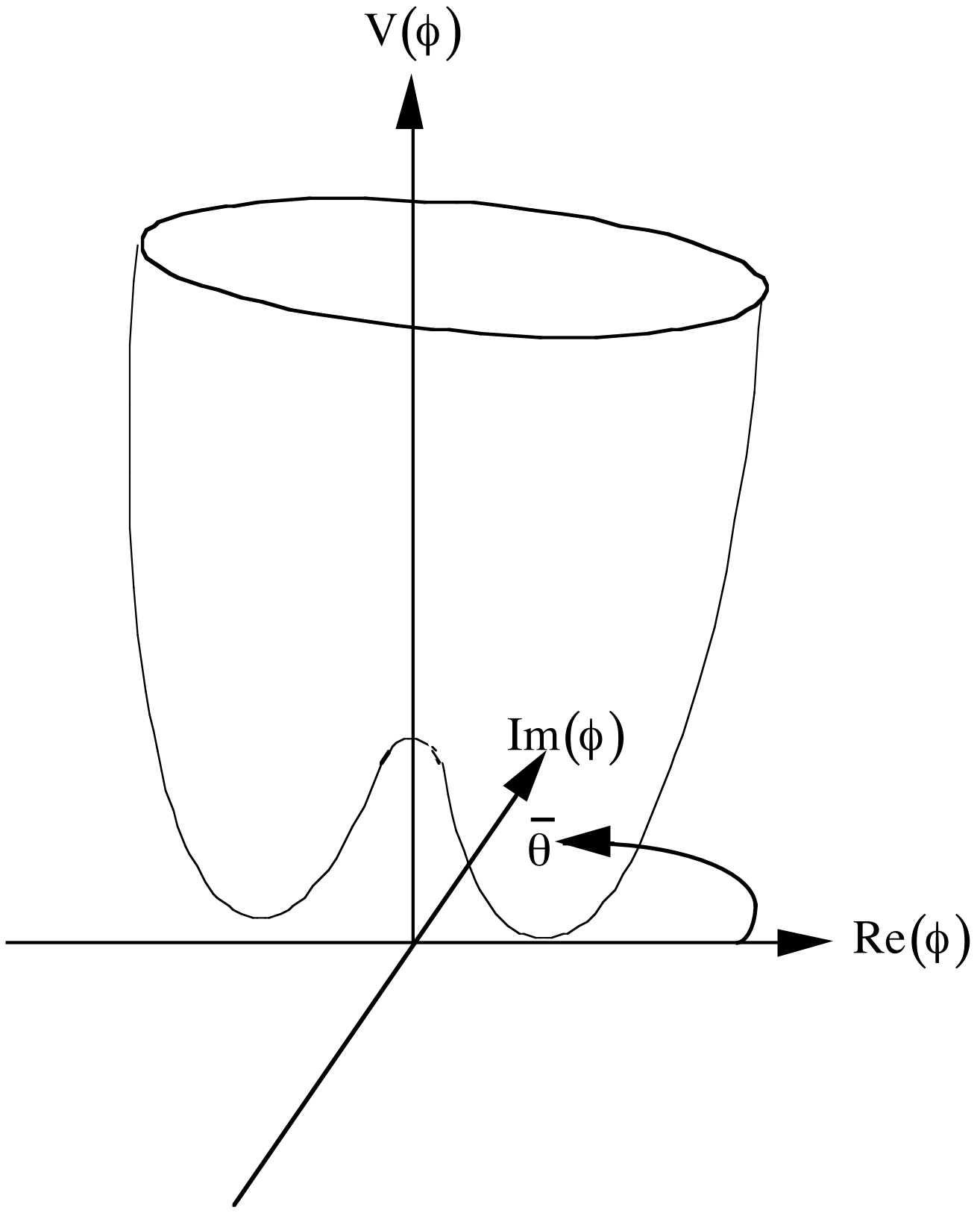

Figure 2.3 The classical QCD potential of the complex field $\phi$ introduced by Peccei and Quinn in the case where there is 1 quark flavor.

Figure 2.3 shows the classical potential including the perturbation due to the $\theta$ vacuum for the simplest case where there is only 1 quark flavor. In this case, the instanton perturbation causes a 'tipping' of the classical QCD potential towards $\bar{\theta} = 0$.

In 1978 Weinberg [4] and Wilczek [5] showed that the spontaneous breaking of the PQ symmetry implies the existence of a new pseudoscalar particle which was named the axion. Because the rotational symmetry of $V(\phi)$ about the origin is broken, the axion has a non-zero mass $m_a$ which is related to $f_{PQ}$ by:

$$m_a = 0.62\,\text{eV}\,\frac{10^7\,\text{GeV}}{\left(f_{PQ}/N\right)} \qquad (2.5)$$

where N is the number of quark flavors. Because of its non-zero mass, the axion is referred to a 'Pseudo-Goldstone boson'.

## 2.1.4 Some Axion Couplings

Figure 2.4 shows some of the axion couplings.

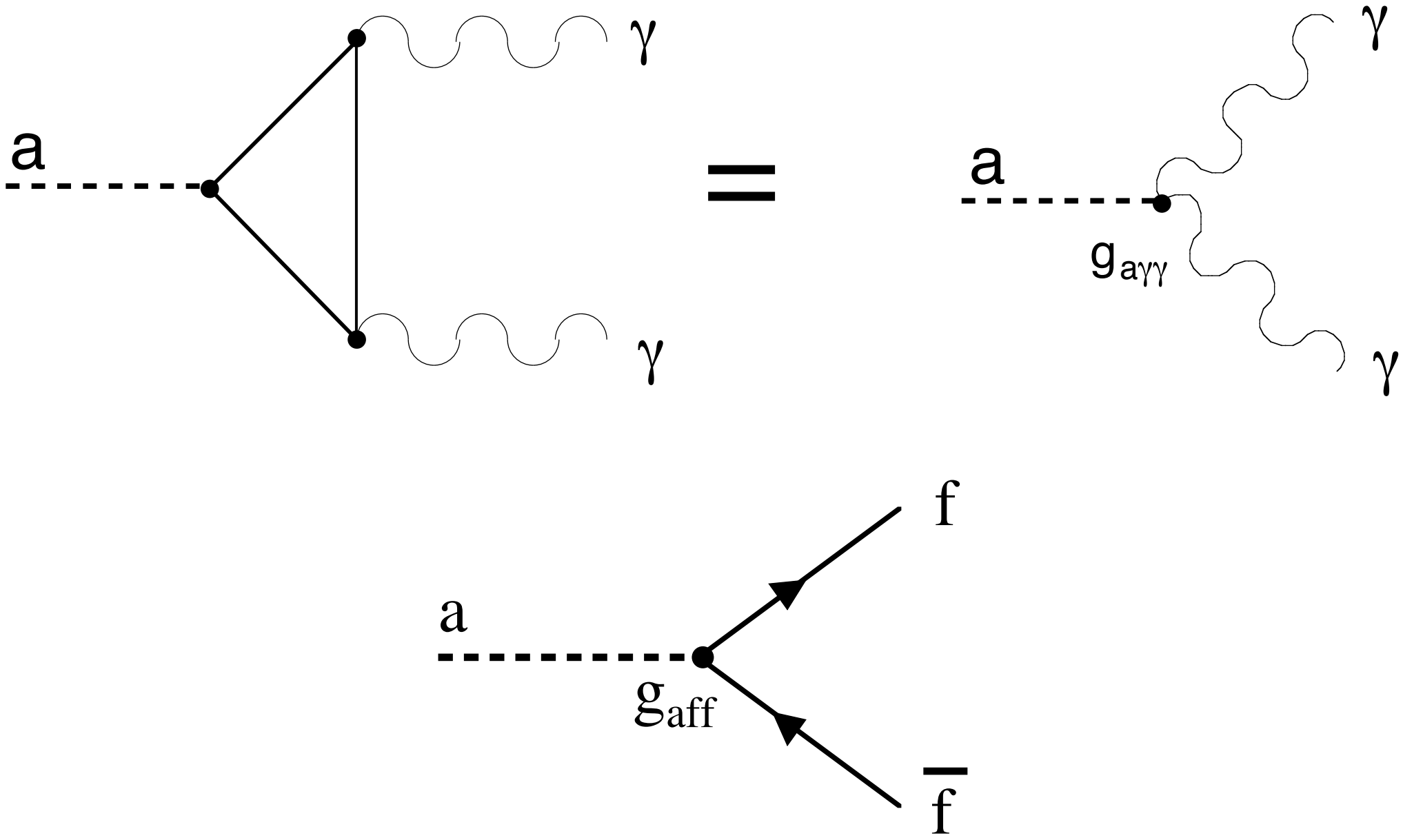

Figure 2.4 Diagrams of axion couplings to photons and fermions

The upper graph shows the axion coupling to two photons via an intermediate fermion loop. The lower graph is a tree-level coupling of the axion to fermions. Which fermions the axion couples to at tree level is model-dependent. In the KSVZ axion model, the axion couples to the quarks at tree level, but not to the charged leptons. In the DFSZ model, the axion couples to both charged leptons and quarks.

Like the axion mass, the couplings of axions to fermions and photons are related to the symmetry-breaking scale $f_{PQ}$. The coupling to photons which we will exploit in our axion detector is given by:

$$g_{a\gamma\gamma} = \frac{\alpha g_\gamma}{\pi\left(f_{PQ}/N\right)} \qquad (2.6)$$

where $g_\gamma$ is a model-dependent constant. For KSVZ (DFSZ) axions $g_\gamma$=-0.97 (+0.36). The axion couplings to fermions are complicated and model-dependent, and we will not need them. Axions and their interactions with photons are represented by the following Lagrangian terms:

$$\mathcal{L}_{axion} = \frac{1}{2}(\partial_\mu a)^2 + \frac{1}{2}m^2a^2 - \frac{g_{a\gamma\gamma}}{4\pi} a\mathbf{E}\cdot\mathbf{B} \qquad (2.7)$$

# 2.2 Axions and Cosmology

In section 2.2 I discuss axions as a candidate for the cold dark matter in galactic halos and experimental constraints on the axion mass.

## 2.2.1 Axions as Cold Dark Matter

Axions are a good cold dark matter candidate. One mechanism by which axions cold dark matter could have been produced, called misalignment production [6], is as follows: In the early universe when the energy density exceeds $f_{PQ}$, the PQ symmetry is unbroken and all values of $\bar{\theta}$ are equally probable. When the energy scale of the expanding universe reaches the symmetry breaking scale $f_{PQ,}$ the PQ symmetry is broken. $\bar{\theta}$ rolls down the classical QCD potential $V(\phi)$ and oscillates about $\bar{\theta} = 0$. The $\bar{\theta}$ oscillations correspond to a zero momentum condensate of axions, out of thermal equilibrium with the baryonic matter and therefore 'cold'. Although initially distributed with uniform density, the axions fall into the gravitational potential wells of clustered baryonic matter and could make up a substantial fraction of the dark matter in galactic halos today.

There are other scenarios for the formation of axion dark matter. In one scenario thermal axions produced around the time of nucleosynthesis form axion hot dark matter [7]. The important production mechanisms are $\gamma + q \rightarrow a + q$ and $N + \pi \rightarrow N + a$ where q is a quark and N is a nucleon. For $m_a < 10^{-3}$eV the cross sections for these processes are too small for thermal axions to account for a substantial fraction of the dark matter. As we shall see in Section 2.2.2, $m_a$ exceeding $10^{-3}$eV is ruled out experimentally. In another scenario, axion cold dark matter is produced through the decay of axion strings [8]. At present there is no reliable estimate of the density of axion dark matter from string decays.

The abundance of axions is expressed as $\Omega_a$, which is equal to 1 if axions provide closure density. $\Omega_a$ becomes larger with increasing $f_{PQ}$; axions produced earlier in the history of the universe are lighter and have a higher present-day number density.

### 2.2.2 Experimental Constraints on the Axion Mass

The mass and couplings of the axion are dependent on the unknown PQ symmetry breaking scale. Therefore the axion mass is not determined by theory. However, many observations and some direct searches constrain the axion mass. Figure 2.5 shows some of the experiments and observations and the corresponding excluded mass regions.

Initially it was thought that $f_{PQ}$ would be the weak symmetry breaking scale. This would mean axions would have masses and couplings similar to those of neutral pions and be detectable in accelerator experiments. The non-observation of axions in accelerators rules out axions at masses above ~10keV. The excluded regions due to red giants and SN1987A both use the same argument; if axions exist they should increase the rate of heat transport out of stars. The cross section for formation of axions at the center of the star is large whilst the interaction cross section for axions emitted from a star center is small. Hence axions can stream directly out of the cooling star once they are formed, acting as a rapid outlet for thermal energy. In particular, the neutrino burst from SN1987A would have been shortened if axions had provided extra heat transport.

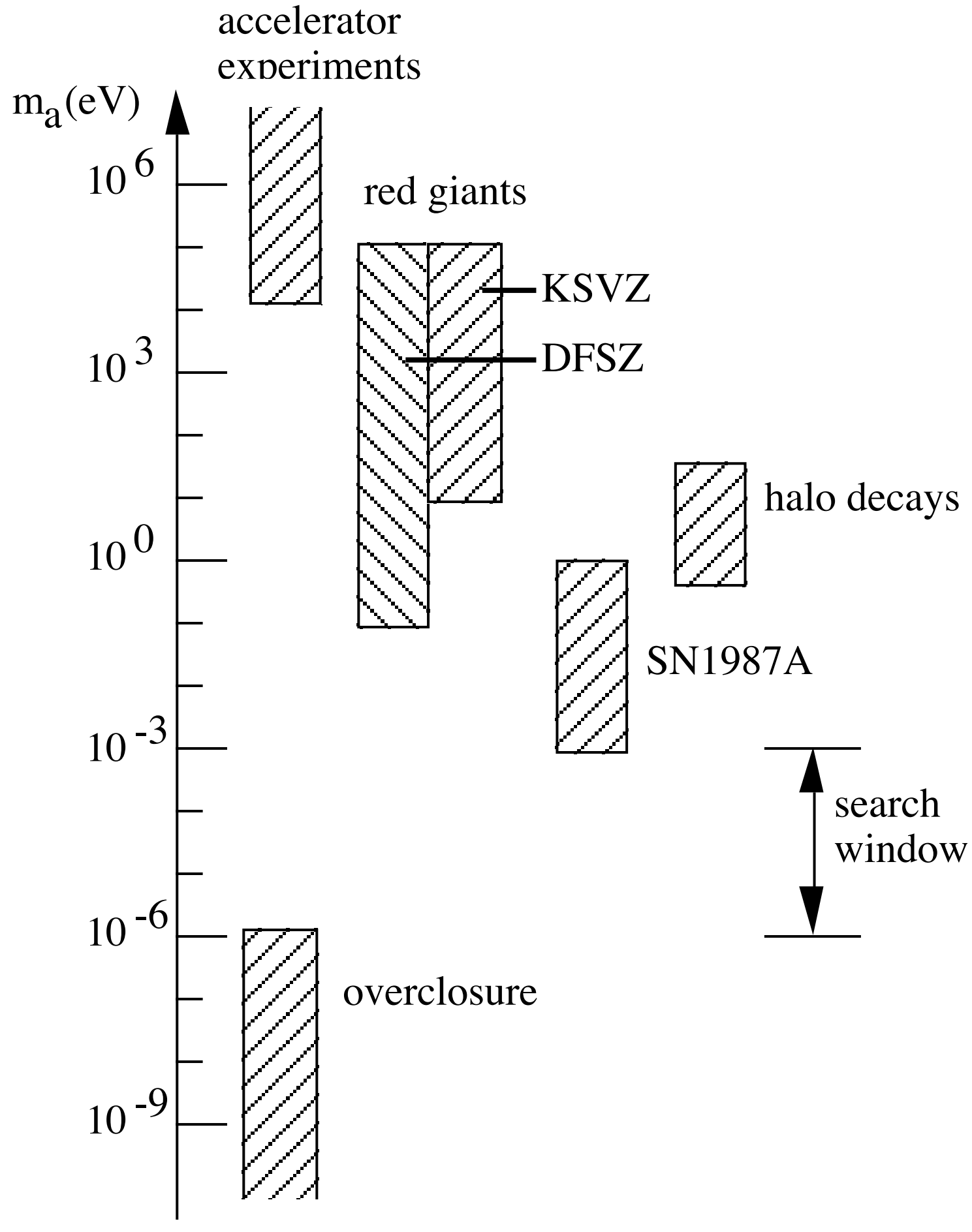

Figure 2.5 Axion masses excluded by terrestrial and astrophysical constraints

The region labeled halo decays was excluded by telescope searches using high resolution optical spectroscopy to look for narrow lines from axions in the halo decaying into photons. These experiments closed a window for KSVZ axions between 1 and 10eV. Finally, if axions are too light, they are abundant enough to overclose the universe. So axions below about $10^{-6}$eV are ruled out in most current cosmological models. Close to this lower limit, axions could make up the majority of the cold dark matter in our galactic halo.

For more detailed explanations of constraints on the axion mass and coupling, see [9].

### 2.2.3 The Axion Lineshape

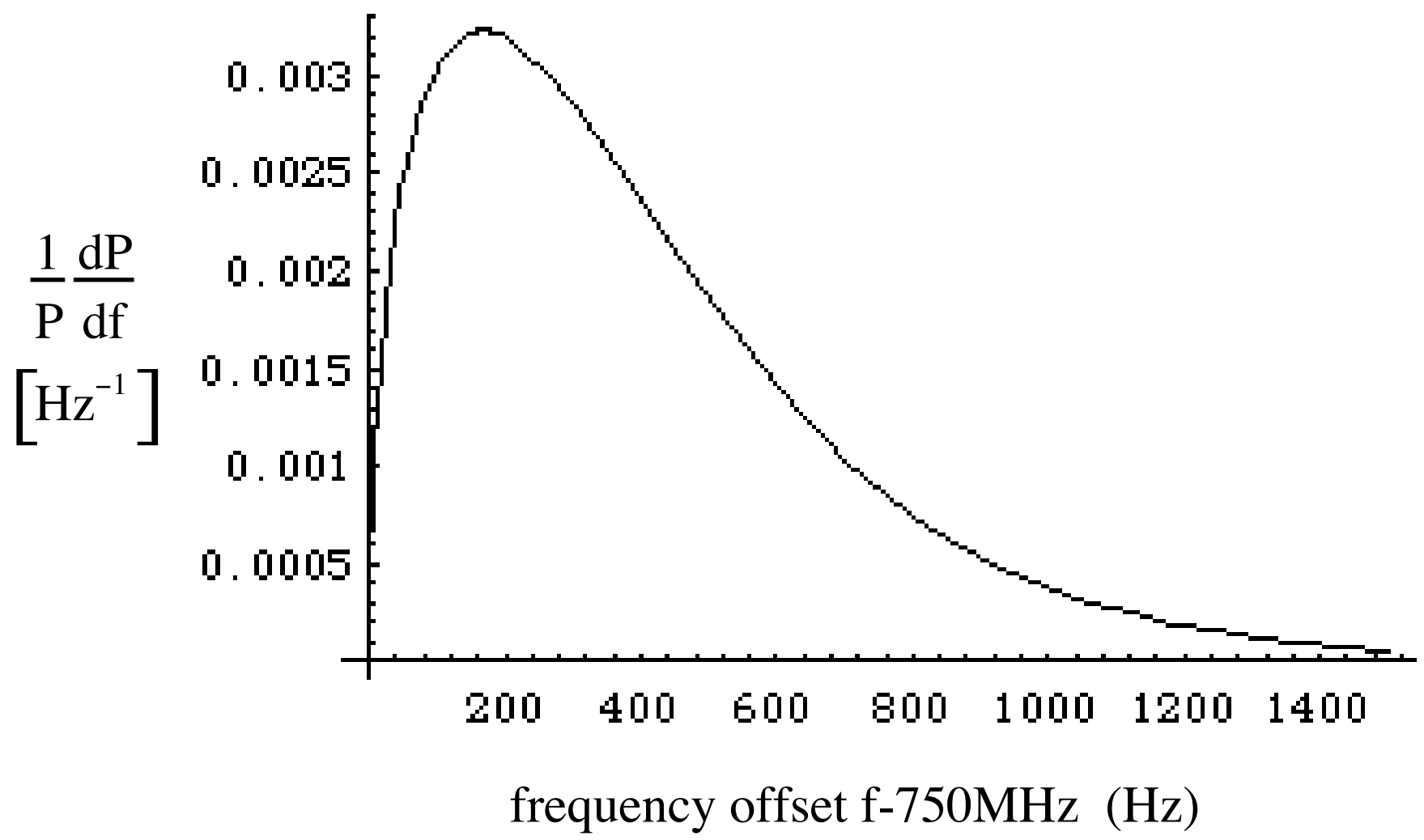

Figure 2.6 The axion lineshape for thermalized halo cold dark matter

What is the energy distribution of axions in a galactic halo? The axions pick up kinetic energy from random gravitational kicks as they fall into the galactic gravitational potential; they become thermalized through gravitational interactions so that the distribution of axion velocities becomes Bolzmann-like with the mean velocity equal to the virial velocity in the galactic gravitational potential. The local virial velocity near earth is about $10^{-3}c$. Hence, we expect dark matter axions to have a fractional linewidth of $\sim 10^{-6}$. Figure 2.6 shows the expected axion linewidth and shape for axions with a mass of 3.1μeV, corresponding to a frequency of 750MHz [18]. The Maxwellian is normalized to unit area. On the vertical axis is fraction of axion power per Hz and on the horizontal axis is the frequency of photons from axion conversion.

## 2.3 Resonant Cavity Detectors for Halo Axions

In section 2.3 I outline the principle of operation of a cavity axion detector and I derive the power appearing in it due to the conversion of halo axions into photons. I then consider what an axion signal would look like and estimate the dominant noise background.

### 2.3.1 Principle of Operation of Cavity Axion Detectors

For masses within the search window of $10^{-6}$-$10^{-3}$eV, the coupling of axions to photons is very weak. The mean life for an axion at rest to decay into 2 photons is ~$10^{42}$ years. It is therefore advantageous to search for axions that already exist in the galactic halo, rather than trying to create and detect them in a lab which requires two axion-photon coupling vertices.

Models in which the galactic halo is flattened predict that the local cold dark matter (CDM) density is ~0.45GeV/cc [10]. If the axion mass is $10^{-5}$eV then the number density is over $10^{13}$ per cc. Moreover, assuming a velocity of $10^{-3}$c, a $10^{-5}$eV axion has a De Broglie wavelength of ~100m. The wavelength of the photons from axion-photon conversion is about 12cm. So on the length scale of the wavelength of photons from axion conversion, a halo consisting of axions axions behaves like a classical field coherent over distances of order 100m.

The axion search is based on a resonant cavity detector proposed by Sikivie [11] and used previously on two considerably less sensitive pilot axion search experiments [12,13]. The detection device is a cylindrical resonant cavity of high quality factor Q threaded with a strong DC magnetic field. The $TM_{010}$ resonance is used which has the E field oriented along the cavity axis and parallel to the magnetic field. Axions from the galactic halo excite the cavity mode by resonant conversion to microwave photons in the cavity. One can imagine the photon scattering off a zero energy photon in the field and converting to a real photon with frequency corresponding to the axion mass. But a more realistic picture is that the axion field acts as a classical source of coherent radiation that excites the cavity resonance. The result of this excitation is a peak of fractional width ~$10^{-6}$ in the noise power spectrum of the cavity. The aim of the experiment is to detect this peak by monitoring the power spectrum of the cavity noise in the neighborhood of the resonant frequency using ultra-low-noise electronics. In Section 2.3.2 I derive the signal power from axion to photon conversion.

The dominant background is Johnson noise in the cavity mode due to its non-zero physical temperature and additional Johnson noise generated in the receiver electronics. To minimize the background level, the lowest noise electronics available are employed and the cavity is cooled to cryogenic temperatures. If the overall temperature of the thermal noise from the cavity and the receiver electronics is T, then this temperature is associated with a noise power $P_N$ given by Nyquist's theorem:

$$P_N = k_B TB \qquad (2.8)$$

where $k_B$ is Bolzmann's constant and B is the frequency bandwidth. Extensive use of Nyquist's theorem, which is derived in Appendix 2, is made throughout the remainder of this thesis. A detailed discussion of background noise requires knowledge of the nature of the receiver electronics and is deferred to Chapters 3 and 4.

Finally, there are several important practical details that govern how a cavity axion search is done. For example, because the axion mass is unknown, the cavity resonant frequency must be tuned across a range of frequencies corresponding to the mass range to be covered. Furthermore, the mass range we search corresponds to frequencies from several hundred MHz up to a few GHz. Great care must be taken in designing low noise receiver electronics in this difficult frequency range. These and other practicalities of the detector will be discussed in detail in Chapter 3.

### 2.3.2 Derivation of the Signal Power from Axion to Photon Conversion

A simple method for deriving the signal power due to axions decaying into photons in a Sikivie-type axion detector is to consider the equivalent circuit analogous to photons from axion conversion exciting a mode of a resonant cavity, shown in Figure 2.7.

The classical equivalent circuit Lagrangian [14] is:

$$L = \frac{1}{2}\dot{q}^2 - \frac{1}{2C}q^2 + qV(t) \qquad (2.9)$$

The dissipation in the resistor is included through a term:

$$F = \frac{1}{2}R\dot{q}^2 \qquad (2.10)$$

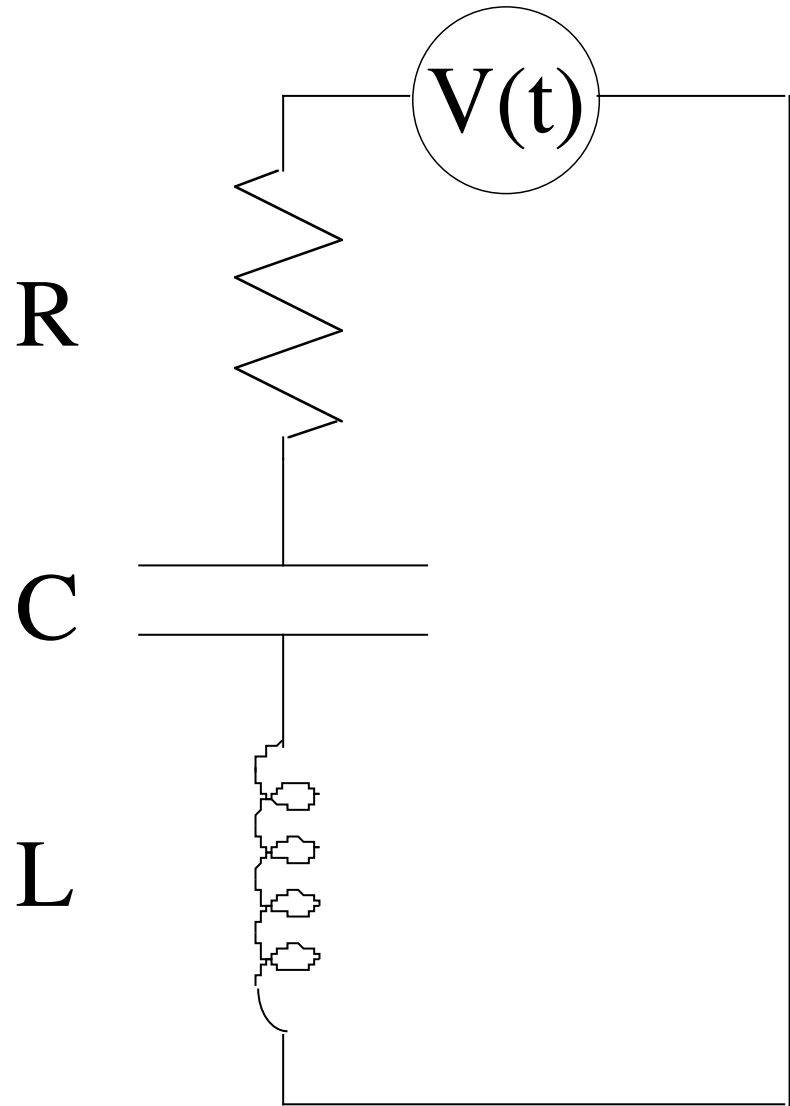

Figure 2.7 The equivalent circuit of a cavity mode
excited by a classical source (photons from axion conversion)

The Euler-Lagrange equation including the dissipative term is:

$$\frac{d}{dt}\left(\frac{\partial L}{\partial \dot{q}}\right) - \frac{\partial L}{\partial q} + \frac{\partial F}{\partial \dot{q}} = 0 \qquad (2.11)$$

so the equation of motion for the circuit is:

$$L\ddot{q} + R\dot{q} + \frac{q}{C} = V(t) \qquad (2.12)$$

The term qV(t) in the circuit Lagrangian represents a classical source driving the circuit. In a cavity axion detector the source is the axion field coupling to modes in the resonant cavity. In MKS units the Lagrangian term describing the coupling of the axion to the electric and magnetic field is (see appendix 1):

$$\mathcal{L} = -\varepsilon_0 g_{a\gamma\gamma} a\mathbf{E}.\mathbf{B} \qquad (2.13)$$

where **E** and **B** are the local values of the electromagnetic field, a is the axion field and $g_{a\gamma\gamma}$ is the axion photon coupling. The product $ag_{a\gamma\gamma}$ is dimensionless. The integral of this interaction over the cavity volume times c is equal to the classical source term in the Lagrangian for the equivalent circuit:

$$q(t)V(t) = -g_{a\gamma\gamma} a(t) c\varepsilon_0 \int_V dV\, \mathbf{E}(\mathbf{x}, t).\mathbf{B}(\mathbf{x}, t) \qquad (2.14)$$

In a Sikivie-type axion detector, the external DC magnetic field is axial, so:

$$q(t)V(t) = -g_{a\gamma\gamma} a(t) c\varepsilon_0 B_0 \int_V dV\, \mathbf{E}(\mathbf{x}, t).\hat{\mathbf{z}} \qquad (2.15)$$

At a frequency near one of the cavity resonant modes, the electric field will be dominated by the cavity fields for the mode. I therefore define a mode-dependent form factor f as follows:

$$f_{nlm} \equiv \frac{\left( \int_V dV\, \mathbf{E}(\mathbf{x}, t) \cdot \hat{\mathbf{z}} \right)^2}{V \int_V dV\, \varepsilon_r E^2} \qquad (2.16)$$

The subscripts (nlm) are indices labeling a mode of a cylindrical cavity; they may be replaced with whatever indices are required to specify modes in a cavity of any other geometry. For a cavity containing tuning rods, $f_{nlm}$ depends on the positions of the rods. The integral in the denominator is proportional to the energy stored in the cavity and $\varepsilon_r$ is the relative permittivity of any dielectric material filling the cavity volume. I equate the stored energy in the cavity mode with the stored energy in its equivalent circuit:

$$\frac{q^2}{2C} = \frac{\varepsilon_0}{2} \int_V dV\, \varepsilon_r E^2 \qquad (2.17)$$

Substituting this into the definition of the form factor defined in Equation 2.16 I obtain:

$$\int_V dV\, \mathbf{E}(\mathbf{x}, t) \cdot \hat{\mathbf{z}} = q \sqrt{\frac{f_{nlm} V}{\varepsilon_0 C}} \qquad (2.18)$$

Substituting Equation 2.18 into Equation 2.15 I obtain an expression for the voltage V(t) in the circuit equivalent to the effect of axion to photon conversion into a cavity mode:

$$V(t) = -g_{a\gamma\gamma} c B_0 \sqrt{\frac{f_{nlm} V \varepsilon_0}{C}}\, a(t) \qquad (2.19)$$

The power P deposited in the cavity mode due to axion-photon conversion is given by:

$$P = \frac{\langle V^2(t) \rangle}{R} = g_{a\gamma\gamma}{}^2 c^2 \varepsilon_0 B_0{}^2 V f_{nlm} \frac{1}{RC} \langle a^2(t) \rangle \qquad (2.20)$$

The remaining terms relating to the equivalent circuit may be re-written in terms of the usual angular frequency $\omega_0$ and the quality factor Q of the resonance (see Appendix 3):

$$Q = \frac{1}{R}\sqrt{\frac{L}{C}} \text{ and } \omega_0 = \frac{1}{\sqrt{LC}} \text{ so } \frac{1}{RC} = \omega_0 Q \qquad (2.21)$$

yielding:

$$P = g_{a\gamma\gamma}{}^2 c^2 \varepsilon_0 B_0{}^2 V f_{nlm} \omega_o Q \langle a^2(t) \rangle \qquad (2.22)$$

In MKS units the axion field a(t) has dimensions $\sqrt{kg/s}$. Its mean square amplitude is related to the mass density of axions $\rho_a$ in kg $m^{-3}$ and the axion mass in kg by:

$$\langle a^2(t) \rangle = \frac{\rho_a \hbar^2}{m_a{}^2 c} \qquad (2.23)$$

Using this definition of the halo density would require conversion of the axion photon coupling to MKS units. Instead I exploit the fact that the product $a(t)g_{a\gamma\gamma}$ is dimensionless. In natural units $(\hbar = c = 1)$ it can be rewritten in terms of the halo density and the axion mass:

$$g_{a\gamma\gamma}{}^2 \langle a^2(t) \rangle = \frac{g_{a\gamma\gamma}{}^2 \rho_a}{m_a{}^2} \qquad (2.24)$$

I assume a local density of halo dark matter of 0.45GeV/cc. In natural units $1cm^3 = 1.31 \times 10^{14} eV^{-3}$ so the halo density is $3.44 \times 10^{-6}$ $eV^4$. Combining Equations 2.5 and 2.6 I obtain an equation relating the axion mass and its coupling to photons:

$$g_{a\gamma\gamma}(GeV^{-1}) = 10^{-7} GeV^{-1} \left(\frac{m_a}{0.62 eV}\right) \frac{\alpha g_\gamma}{\pi} \qquad (2.25)$$

where $g_\gamma$ is 0.36 for DFSZ axions and -0.97 for the KSVZ axions. Hence assuming KSVZ axions and converting all energies into eV I get:

$$\frac{g_{a\gamma\gamma}{}^2 \rho_a}{m_a{}^2} = (3.64 \times 10^{-19})^2 \mathrm{eV}^{-4} .(3.44 \times 10^{-6}) \mathrm{eV}^4 = 4.56 \times 10^{-43} \qquad (2.26)$$

Substituting this expression into the equation for the axion photon conversion power I obtain:

$$P = 2.86 \times 10^{-42} c^2 \varepsilon_0 B_0{}^2 V f_{nlm} f_a Q \qquad (2.27)$$

where $f_a$ is the cavity resonant frequency. Finally the expression may be re-written in terms of physical parameters comparable with those of our axion receiver. Additional factors have been inserted to allow for axion models other than KSVZ and different values of the local halo density than that assumed above.

$$P_{a\to\gamma} = 1.52 \times 10^{-21} \mathrm{W} \left(\frac{V}{220\,\ell}\right)\left(\frac{B}{7.6\mathrm{T}}\right)^2 f_{nlm} \left(\frac{g_\gamma}{0.97}\right)^2 \left(\frac{\rho_a}{0.45\mathrm{GeV}/\mathrm{cc}}\right)\left(\frac{f_a}{750\,\mathrm{MHz}}\right)\left(\frac{Q}{70,000}\right) \qquad (2.28)$$

This expression is used to compute the expected signal power in my analysis.

# 3. The Axion Detector

## 3.1 Introduction

I will first give a brief description of the experiment and the data acquisition procedure.

The axion detector is a circular cylindrical copper plated stainless steel resonant cavity containing 2 moveable copper tuning rods. The cavity is coupled through a small electric field probe to ultra-low-noise receiver electronics. To search for axions, a cavity resonance to which axions would couple is swept by moving the tuning rods in discrete steps. At each tuning rod position the cavity resonant frequency is measured, then the receiver electronics is used to acquire power spectra of the electromagnetic field in the cavity about the resonant frequency $f_0$. The receiver electronics downconverts power from a 50kHz bandwidth about the cavity resonant frequency to the same bandwidth centered at 35kHz. Each power spectrum consists of 400 125Hz frequency bins covering the bandwidth of 50kHz about the cavity resonant frequency. The set of power spectra for each cavity resonant frequency together with other experimental parameters constitutes the raw data. As discussed in Section 2.3, the experimental signature of axions is excess power in a bandwidth of $\sim 10^{-6} f_0$ a peak about 6 bins wide in our power spectra. The dominant background is broadband Johnson noise from the cavity and the receiver electronics.

Figure 3.1 is a diagram of the apparatus. The cavity is cooled to 1.3K to reduce its Johnson noise. To stimulate axion to photon conversion, a 7.6T static magnetic field parallel to the cylinder axis threads the cavity volume. We use the $TM_{010}$ cavity mode since it has the largest form factor $f_{nlm}$ for axion to photon coupling. The quality factor Q of the $TM_{010}$ resonance is typically $7\times10^4$. The frequency step between adjacent power spectra is typically 2kHz. This is about 1/15th of the width of the receiver passband, so many power spectra overlap with any cavity frequency in the search range. The frequency range covered in the data set discussed in this thesis is 701-800MHz, corresponding to the axion mass range $2.9 - 3.3\mu eV$ .

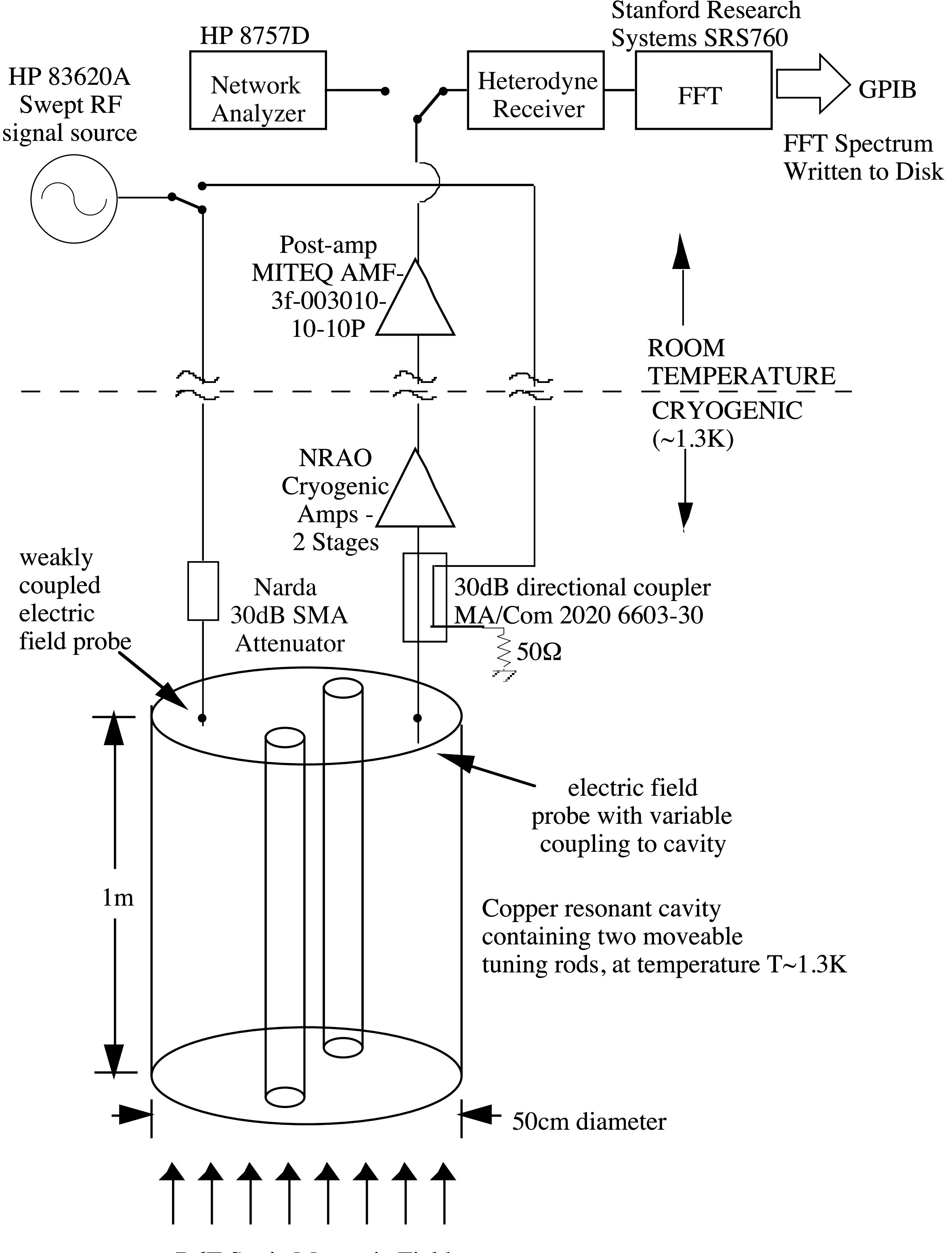

Figure 3.1 Schematic of the apparatus

The cavity is coupled to the front-end of the receiver chain by an electric field probe with an adjustable insersion depth. During normal running, the insertion depth is adjusted to maintain critical coupling. This means that the noise power deposited in the front-end of the electronics chain is equal to that dissipated in the cavity walls. A more detailed explanation of critical coupling is given in Appendix 3. The directional coupler between the cavity and the cryogenic amplifier is for the injection of test signals.

The signal from the electric field probe is amplified in two cryogenic HEMT amplifiers built by the National Radio Astronomy Observatory (NRAO) [15]. The overall gain of the two cryogenic amplifiers is ~34dB* . Each amplifier is cooled to a physical temperature of ~3K, and has a noise temperature of ~4.3K. The output of the second cryogenic amplifier is further amplified by a low noise room temperature amplifier with ~38dB of gain mounted in an RF shielded enclosure directly on the top flange of the detection apparatus. After room temperature amplification is fed through a 25m length of RG213 flexible coaxial cable to the axion receiver.

The axion receiver is a double heterodyne receiver designed, constructed and tested at MIT. It first downconverts the amplified signal to an intermediate frequency (IF) of 10.7MHz. A crystal bandpass filter in the 10.7MHz IF rejects power outside a 35kHz frequency window centered at $f_0$. A second mixing stage downconverts the filtered signal to an audio frequency (AF) bandwidth of 50kHz centered at 35kHz. The downconverted signal is applied to a commercial fast Fourier transform box which computes a power spectrum. A linear average of a set of 10,000 power spectra is written to disk together

---

* Throughout this thesis I will quote power gains in dB and power levels in dBm. The gain of a device in dB is given in terms of the power $P_{IN}$ applied to the device input and power $P_{OUT}$ measured by a perfectly absorbing detector at the device output by:

$$\text{Gain(dB)} = 10\log_{10}\left(\frac{P_{OUT}}{P_{IN}}\right) \quad \text{so that unity gain} = 0\text{dB} \qquad (3.1)$$

A power level equal to P(W) watts is given in dBm by:

$$P(\text{dBm}) = 10\log_{10}\left(\frac{P(W)}{10^{-3}}\right) \quad \text{so that } 1\text{mW} = 0\text{dBm} \qquad (3.2)$$

with other parameters of the experiment including the measured cavity Q and resonant frequency where the power spectrum was measured. I will refer to the linear average of 10,000 power spectra as a trace. A total of $\sim 4.5 \times 10^5$ traces for the axion search covering the mass range $2.9 - 3.3\mu eV$ .

The following Sections 3.2-3.5 are more detailed descriptions of the detector components. I will emphasize parameters of the different components of the detector important to my analysis of the data. The reader wishing to pass quickly on to data analysis may wish to concentrate on sections 3.3.5 where I discuss the cavity form factor, 3.4.6 where I discuss the noise temperature of the 1st cryogenic amplifier, 3.5.2 where I discuss the crystal filter transfer function and 3.5.4 where I discuss the raw data.

## 3.2 The Expected Signal to Noise Ratio

I now estimate signal and noise power levels for our experiment. Equation 2.28 is an expression for the axion to photon conversion power scaled to cavity parameters typical for our apparatus. Assuming a typical value for the form factor of $f_{010}$=0.4, the power from photons converting to axions is $6\times10^{-22}$W. With the cavity critically coupled to the receiver electronics, half of this power goes to the amplifier. The signal power entering the amplifier input is therefore $3\times10^{-22}$W.

The noise temperature of the receiver electronics is discussed in detail in Section 3.4. For this estimate I will assume that the noise temperature of the electronics is 4.3K. The cavity temperature is typically 1.3K, so the overall temperature of the Johnson noise background is 5.6K. If the axion frequency is 750MHz then the axion line width (see Section 2.2.3) is ~750Hz. Hence using Nyquist's theorem (Equation 2.8), the noise power in the bandwidth of the axion signal is $5.8\times10^{-20}$W. The ratio of the signal power to the noise power in the expected signal bandwidth is therefore ~0.5%.

How is it possible to see the signal given that the signal power is so small compared to the noise power? Remember that power due to thermal noise is broadband – every 750Hz bandwidth contains the same average noise power. The axion signal only contributes excess noise power in a single 750Hz bandwidth. Therefore if the noise power in many neighboring 750Hz bins is monitored for a time period long enough for the rms noise fluctuations in the background thermal noise to become less than the power excess due to the axion signal, the signal could be detected as a persistent non Gaussian power excess above the bin-to-bin statistical fluctuations in the thermal noise background. The signal to noise ratio SNR, defined as the ratio of the power excess $P_S$ due to axions divided by the rms of the noise fluctuations $\sigma_{P_N}$ is given in terms of the integration time t and the signal bandwidth B by Dicke's radiometer equation, (derived in Appendix 2):

$$\mathrm{SNR} = \frac{P_S}{\sigma_{P_N}} = \frac{P_S}{P_N}\sqrt{Bt} \qquad (3.3)$$

I assume for now that I can detect axions with reasonable confidence if the signal power is 4 times the rms fluctuation in the noise background, i.e., SNR=4. Substituting this and the raw signal and noise powers calculated above into the ratiometer equation I obtain an integration time of t=850s. This is a rough estimate of the integration time required in every 750Hz bandwidth within the range of masses covered in the axion search.

## 3.3 The Resonant Cavity

### 3.3.1 Construction

The resonant cavity consists of a stainless steel cylinder 1m long and 50cm in diameter with two endplates. The lower one is welded to the barrel and the upper is removable. The upper endplate is affixed to a knife edged lip and secured in place with 96 bolts. Currents of the $TM_{010}$ mode run across the joint so it is important that the joint is solid. The components of the cavity are plated with high purity oxygen free copper which has been annealed to increase the grain size of the copper crystals. Its volume is 220 liters. The tuning rods are copper plated with oxygen free copper, and are mounted at the end of alumina arms. Axles mounted on the other ends of these arms protrude through the top and bottom plates of the cavity; by rotating these axles the tuning rods may be swung in circular arcs from close to the cavity side wall to close to the center. Figure 3.2 is a photograph of two tuning rods sitting in the cavity with the lid removed. The axles used to move the tuning rods are clearly visible.

In this picture the upper right tuning rod is made of alumina, a dielectric. For our first data-taking run, both rods are copper, similar to the lower left rod in the picture. The tuning rods are driven by stepper motors mounted on top of the cryostat and a two stage gear reduction to allow for small steps in rod position and cavity frequency. There are 6 million stepper motor steps per full revolution, so a single stepper motor step corresponds to a change in the angle of the tuning rod by 0.22 arc seconds.

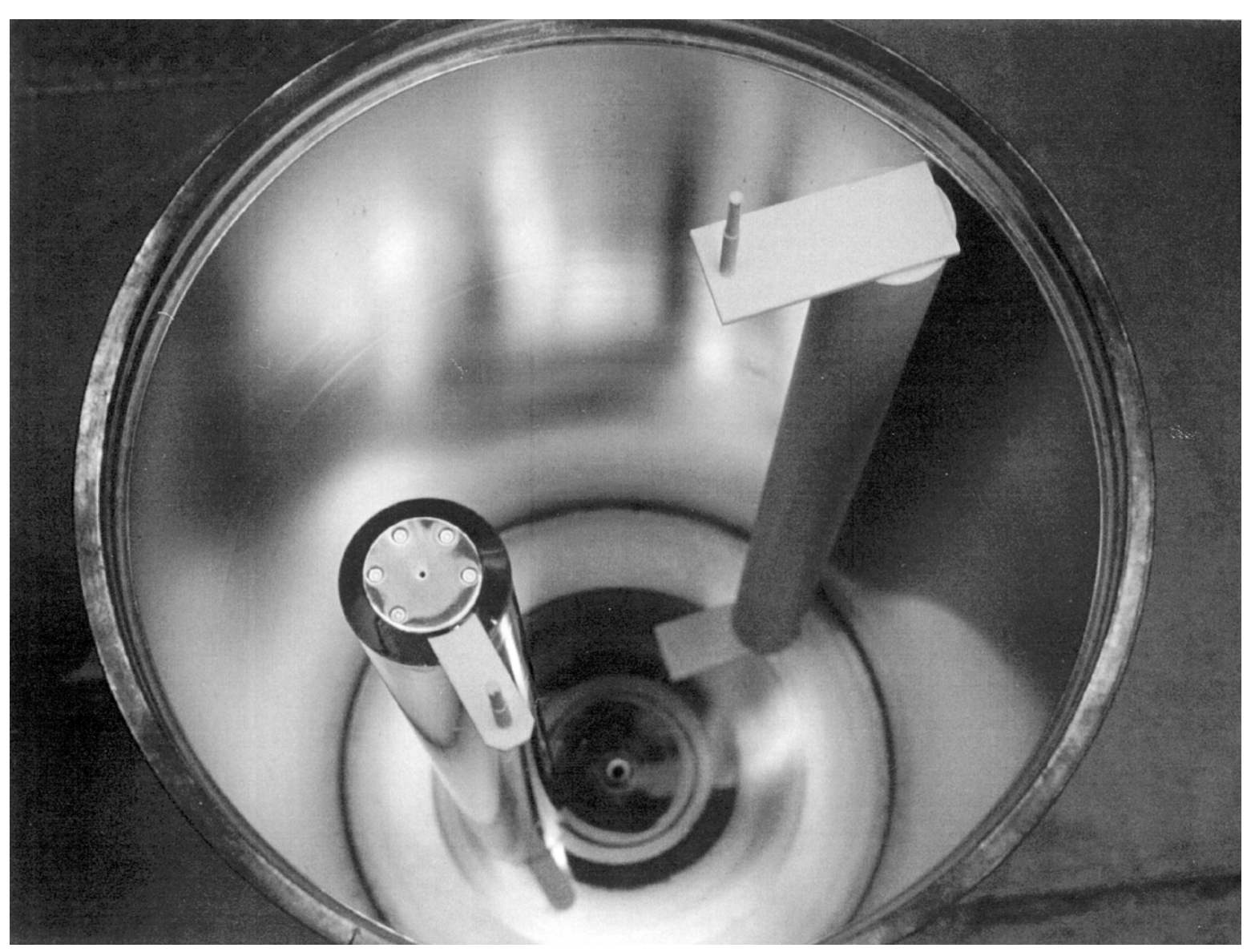

Figure 3.2. The resonant cavity viewed from above with the top flange removed. The diameter is 50cm and the depth is 1m.

## 3.3.2 Measuring the Cavity Resonant Frequency and Q

The Q of the $TM_{010}$ mode was measured by a transmission measurement. Refer to figure 3.1. An RF signal swept through a bandwidth several times the width of the cavity resonance was applied to the weakly coupled RF port. The power transmitted to the critically coupled port as a function of frequency was measured by connecting the output of the room temperature post amplifier to a scalar network analyzer. Figure 3.3 is a typical plot of the transmitted power as a function of frequency taken from the on-line data acquisition software. On the horizontal axis is frequency in MHz, on the vertical axis is power detected at the network analyzer in dBm.

During our first run the width of the resonance was typically 7 or 8 kHz with the cavity critically coupled to the receiver electronics. The width of the resonance is the sum of two equal contributions. Firstly the resonance has an intrinsic width because power is absorbed in the lossy walls of the cavity. Second the electronics chain extracts power from the cavity. The second RF probe is so weakly coupled that it makes no significant contribution to the resonance width.

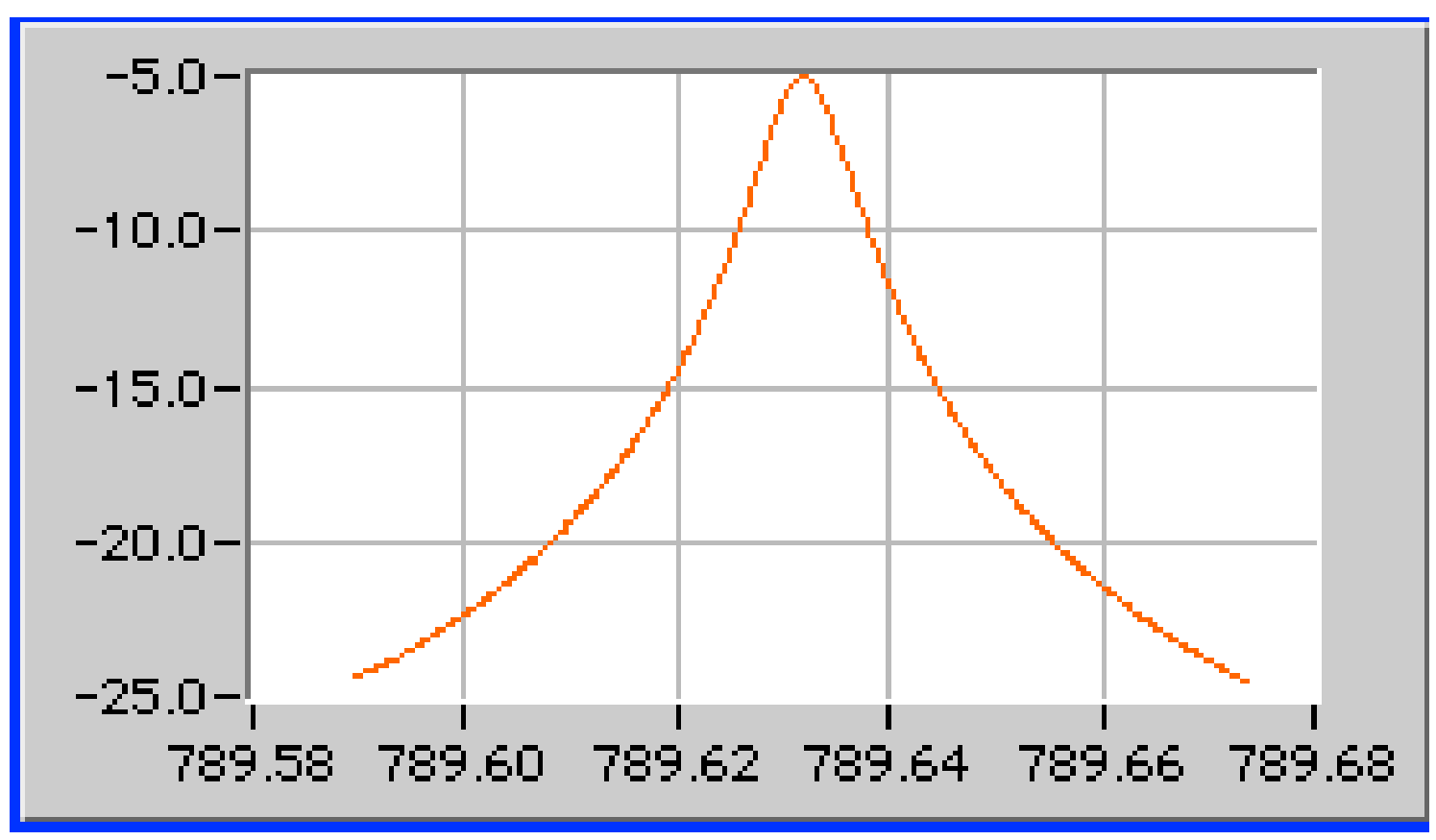

Figure 3.3 Display of the cavity resonance measured in transmission taken from on-line data acquisition screen display. On the vertical axis is power in dBm measured by a network analyzer connected to the output of the receiver electronics. On the horizontal axis is the frequency of the injected signal in MHz.

### 3.3.3 Critical Coupling The Cavity to The Receiver Electronics

During normal data taking the insertion depth of the electric field probe coupling the cavity to the receiver was periodically adjusted to maintain critical coupling between the cavity $TM_{010}$ mode and the receiver electronics. As discussed in detail in Appendix 3, critical coupling is achieved when the impedance looking towards the cavity down the electric field probe is equal to the probe characteristic impedance of 50Ω. Therefore an RF signal incident on the cavity from the field probe will be absorbed in the cavity with no reflection. To critically couple the cavity to the electronics a swept RF signal from the sweeper was injected into the field probe through the directional coupler (see figure 1). All but 0.1% of the signal power is absorbed in the cold 50Ω terminator. The remaining power is incident on the cavity. Far off resonance the cavity impedance is dominated by an imaginary component and all the power is reflected back towards the receiver. However at the cavity resonance the imaginary component of its impedance vanishes and the reflected power is reduced. If the probe is critically coupled then essentially no power is reflected. To achieve critical coupling during our experiment the probe insertion depth was adjusted continuously with the RF source on. Figure 3.4 is a plot of the reflected power in the neighborhood of the resonance after critical coupling vs. the frequency of the injected swept signal. The reflected power is measured through the amplifier chain using the scalar network analyzer (Figure 3.1). The on-resonance reflected power is a factor of about $10^3$ less than the off-resonance reflected power though actually the bottom of the dip represents the noise floor of the electronics and the reflected power is probably even less. We considered a 30dB dip to indicate that we had achieved critical coupling.

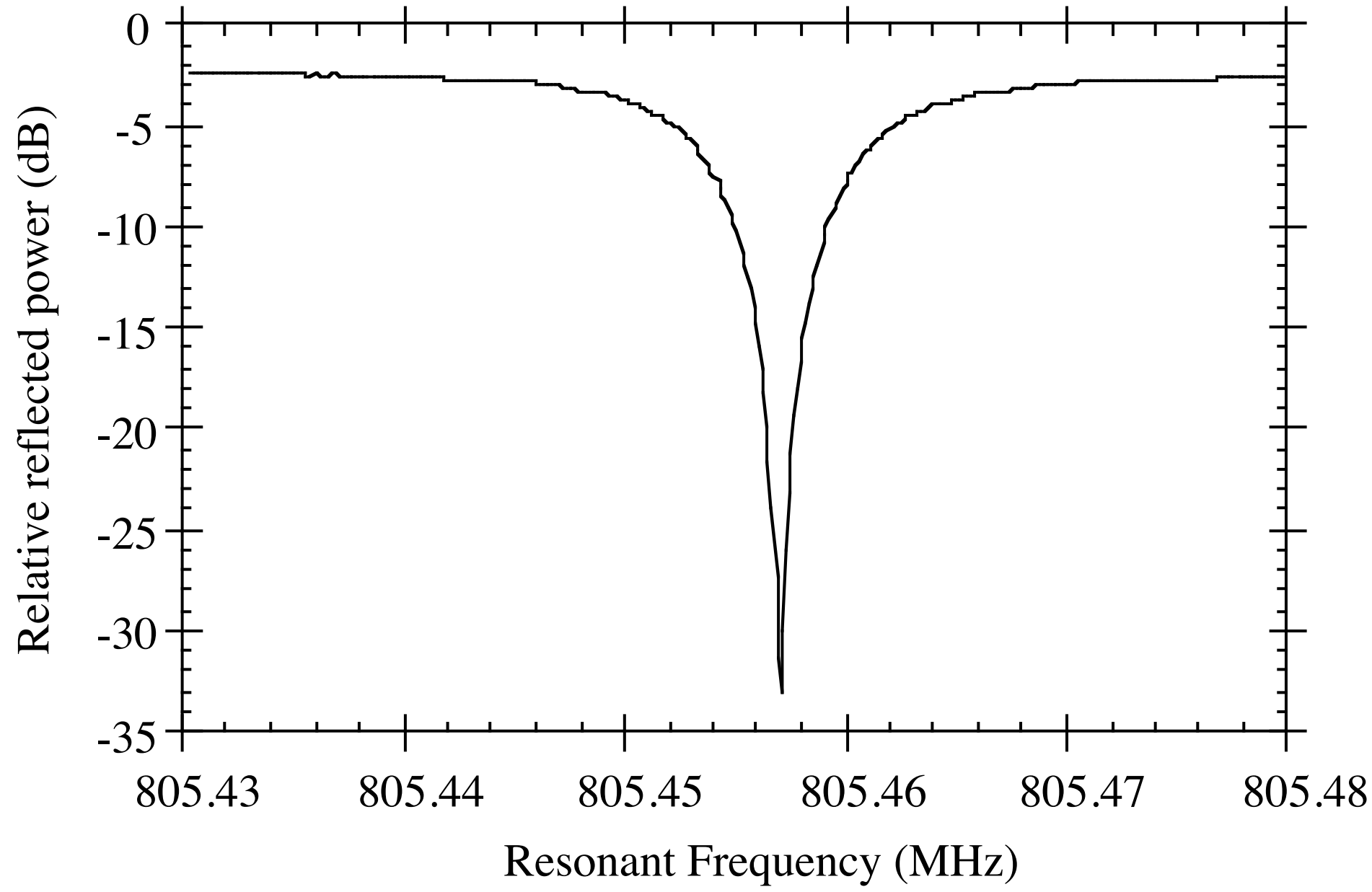

Figure 3.4 Reflected power off the variable insertion depth RF probe as a function of frequency around the cavity $TM_{010}$ resonance with the probe at critical coupling.

### 3.3.4 The Cavity Modes

In an empty cylindrical cavity the resonant modes fall into two categories. The TM modes have the **B** field perpendicular to the cavity axis everywhere and the TE modes have the **E** field perpendicular to the cavity axis everywhere. In the $TM_{010}$ mode the E field lines run along and the **B** field lines form closed circles around the cavity axis. With the tuning rods inserted no analytic solution is available for the mode frequencies and field pattern. The resonant frequencies can be measured and the form factor $f_{010}$ must be determined by numerical simulation. The presence of a tuning rod also introduces TEM modes.

With 2 metal tuning rods in the cavity the lowest mode frequency is achieved with both copper rods near the wall. To raise the frequency, one rod is initially left near the wall and the other rod moved through a half revolution to its closest approach to the cavity center. This tuning rod is then left near the center and the other tuning rod is moved from near the wall to near the center. Figures 3.5a and 3.5b are plots of the frequencies of various cavity modes measured as a function of the tuning rod positions.

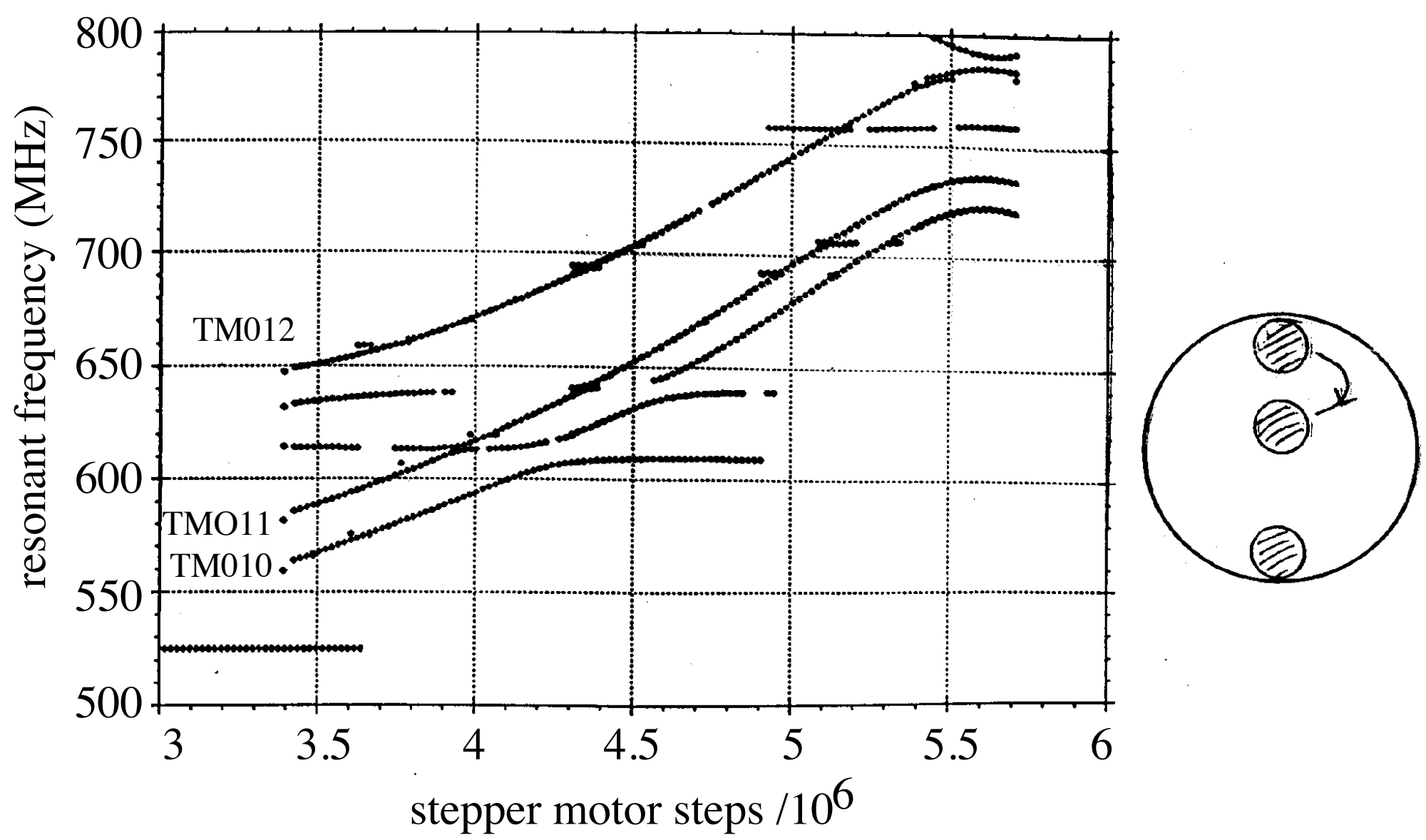

Figure 3.5a The cavity modes with one rod fixed near the wall and the other moved towards the center as shown in the sketch to the right.

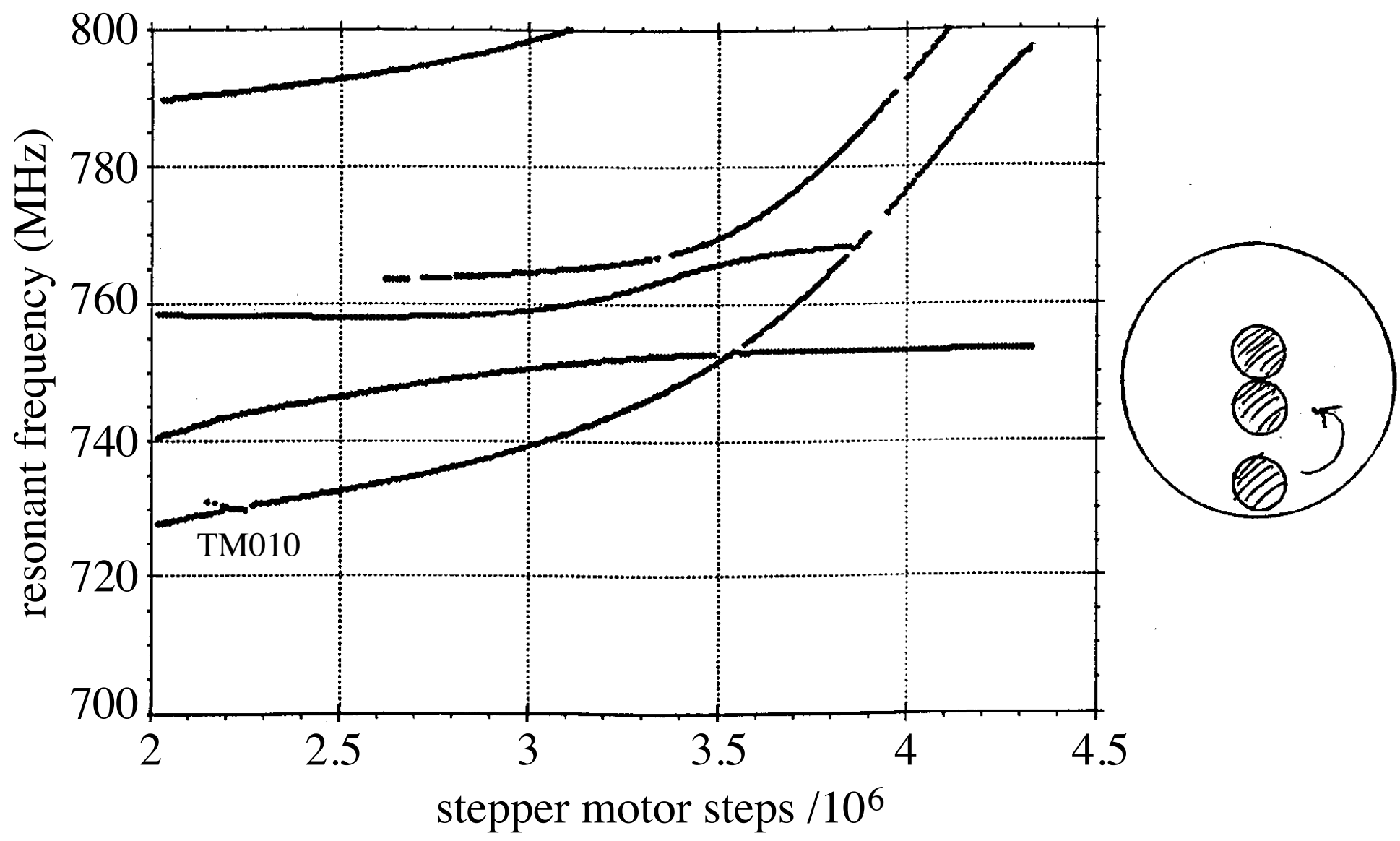

Figure 3.5b The cavity modes with one rod fixed near the center and the other moved from the edge to the center as shown in the sketch to the right.In each figure the vertical axis is cavity frequency $f_0$ and the horizontal axis is stepper motor steps.

The mode maps were obtained by feeding a swept RF signal in to the cavity through the weakly coupled RF port (see Figure 3.1) and measuring the transmitted power through the cavity using the network analyzer.

Notice that there are discontinuities in frequency of the various TM modes as the tuning rods are moved. These discontinuities are due to mode crossings. The TE and TEM modes of the cavity are only weakly excited by the RF probes since the RF probes are wires oriented along the symmetry axis of the cavity which is perpendicular to the electric fields of these modes. Hence the modes do not appear directly on the mode map. Moreover, the frequencies of the TE modes are independent of the tuning rod positions so the TM and mode frequencies sometimes coincide with the TE or TEM mode frequencies. In the mode mixing region, the field pattern of the TM mode and the field pattern of the TE or TEM mode it mixes with become similar. If our cavity had exact cylindrical symmetry then at the frequency where the frequencies of the two modes coincided the field patterns for the two modes would be the same. The degeneracy is broken for a real cavity because cylindrical symmetry is not exact. Figure 3.6 is a schematic of mode mixing.

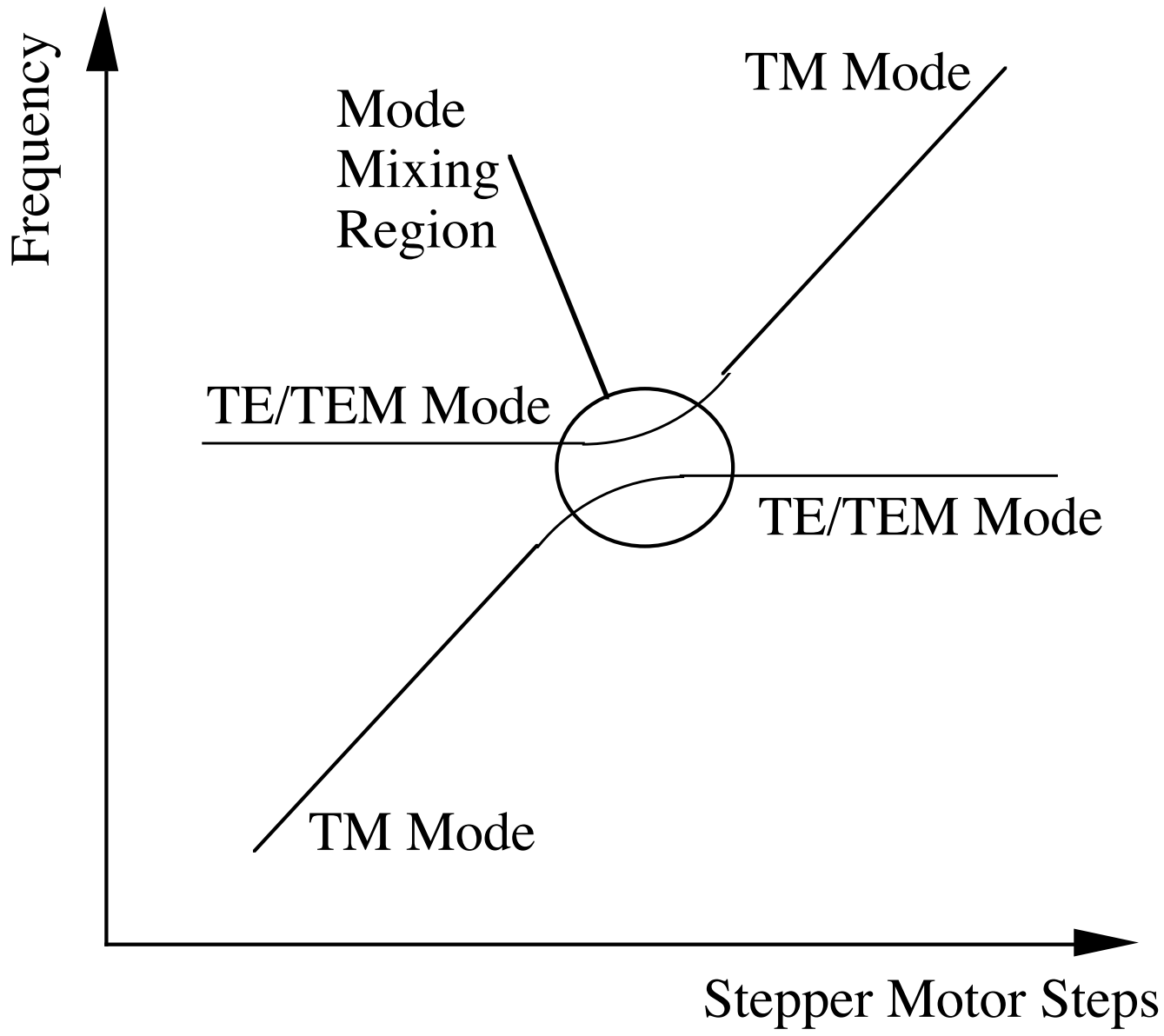

Figure 3.6 Behavior of a TM mode in the region of a mode crossing with a TE or TEM mode

Mode mixing means that there are frequency ranges at mode crossings which the pure $TM_{010}$ mode cannot reach. To solve this problem, frequency ranges at mode crossings were scanned with the cavity filled with liquid helium at 1.3K which has a relative dielectric permittivity of 1.055 and hence alters the speed of light in the cavity by a factor of 1.027. Thus the frequencies of all the modes of the cavity and the mode crossings are altered by 2.7% which at 700MHz is a shift of 19MHz. By combining data taken with the cavity filled with low pressure gas with data taken when filled with helium, we may cover the whole frequency range without holes due to mode crossings.

### 3.3.5 Form Factor of the $TM_{010}$ Mode

The form factor $f_{010}$ defined in Equation 2.16 is calculated from numerical solution of the wave equation for $E_z$ in the cavity with the tuning rods at different positions. A relaxation code using the Gauss Seidel method was used; this method is described in reference [19]. The numerical solution was computed for a two dimensional cross section of the cavity and tuning rods assuming cylindrical symmetry using a 100×100 grid. The effect of the ~1cm gap between the tuning rod ends and the cavity end plates were assumed to be negligible. Figure 3.7 is a plot of the form factor as a function of cavity resonant frequency. The two lines represent the form factor with the cavity filled with low density gas and the cavity filled with liquid helium.

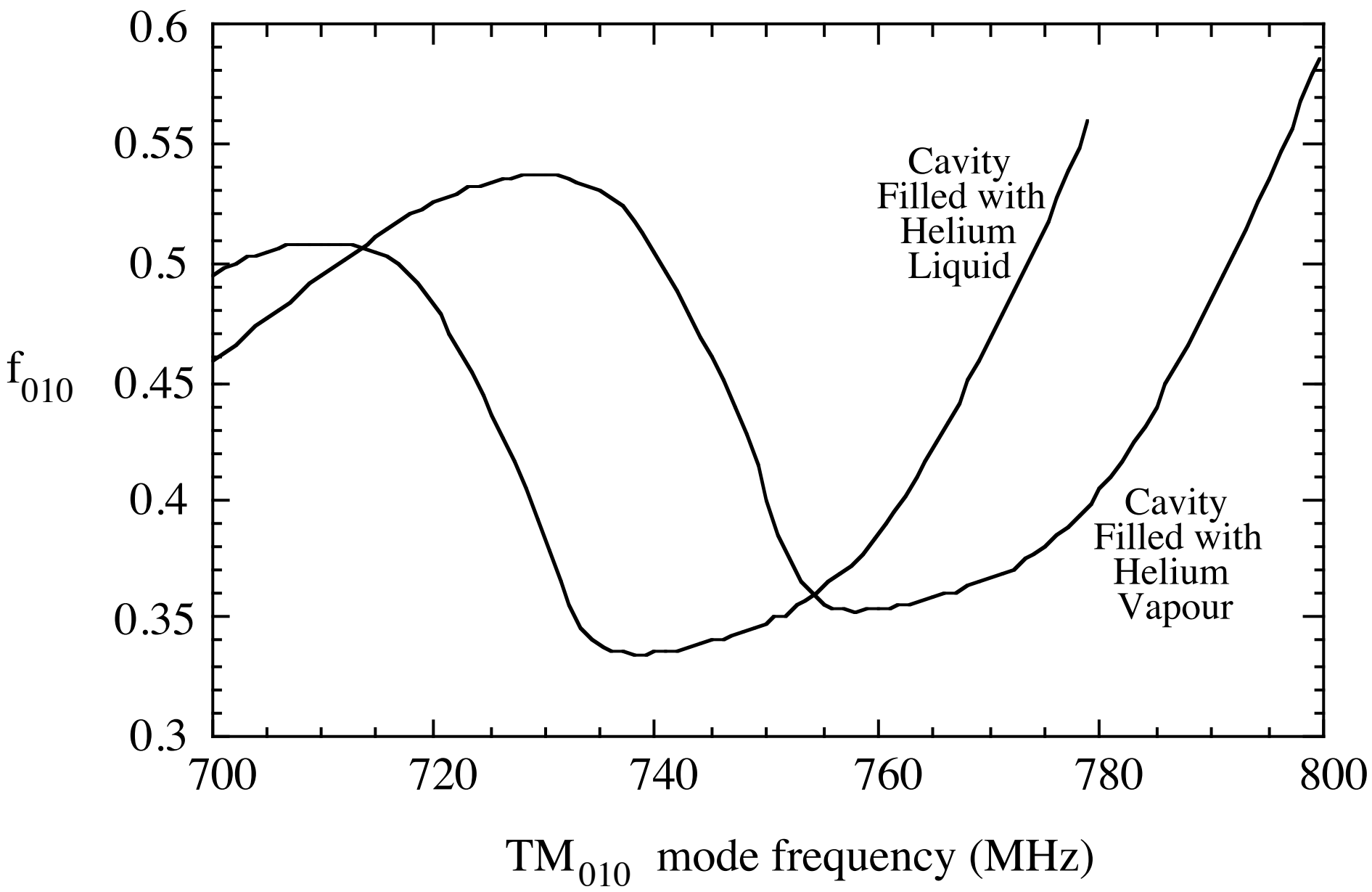

Figure 3.7 Form factor $f_{010}$ as a function of frequency calculated numerically as a function of the cavity resonant frequency with the rods moved as described in Section 3.3.4

There are two modifications made to the plot of form factor vs. frequency with the cavity empty to arrive at the form factor curve with the cavity filled with liquid helium. Firstly, the definition of the form factor from Equation 2.16 contains a factor of $\varepsilon_r$ in the denominator. Hence the form factor is reduced by a factor of 5.5% over the value for an empty cavity at all frequencies. Secondly, with the tuning rods in a fixed position, filling the cavity with helium shifts the mode frequencies down by 2.7%. If the cavity resonant frequency is $f_0$ before filling with helium, a tuning rod position must be adjusted after filling to re-tune the resonant frequency back to $f_0$. Combining these two effects, the form factor at frequency $f_0$ after filling with liquid helium is equal to the form factor at frequency $(1+0.027)f_0$ before filling, divided by 1.055.

# 3.4 The 1st-Stage Cryogenic Amplifier

Section 3.4 contains detailed descriptions of measurements on the 1st stage cryogenic amplifier that are critical to my analysis. This section may be skipped if the reader is not concerned with details of the noise characteristics of the cryogenic amplifier. However, it is necessary to read this section to understand Appendix 3 on the equivalent circuit of the cavity and amplifier.

### 3.4.1 The Importance of the First Cryogenic Amplifier

The first cryogenic amplifier is particularly important because its Johnson noise is the overwhelmingly dominant contribution to the noise background from the receiver electronics. To understand why the first amplifier is so important, consider two amplifiers in series, the first amplifier having gain $G_1$ and noise temperature $T_1$ and the second amplifier having gain $G_2$ and noise temperature $T_2$. If the input of the first amplifier is terminated in a matched load at temperature $T_L$ then using Nyquist's theorem (Equation 2.8) the total noise power $P_{OUT}$ in a bandwidth B at the output of the second amplifier is:

$$P_{OUT} = k_B B G_1 G_2 (T_L + T_1) + k_B B G_2 T_2 \qquad (3.4)$$

Notice that the relative contribution from the noise in the 2nd amplifier is surpressed by a factor of $1/G_1$. If the gain of the 1st amplifier is sufficiently large the noise contribution from the 2nd amplifier is negligable compared to that of the first amplifier. In our axion receiver, the 2nd stage amplifier is a cryogenic HEMT device with a noise temperature similar to the 1st stage, about 4K. The gain of the 1st stage is ~17dB which is a factor of 50 in power gain, hence the contribution from the 2nd stage amplifier is 4/50~0.1K. This is negligible compared to the noise contribution from the 1st stage. The same argument applies to all subsequent components in the receiver chain. It is therefore important to have the lowest noise amplifier possible for the 1st stage.

### 3.4.2 The Balanced Amplifiers Design

Both cryogenic amplifiers used in our receiver electronics are balanced designs. In this section I describe the balanced design and its advantages over conventional single ended devices. Figure 3.8 is a schematic of the balanced amplifier used in our receiver electronics.

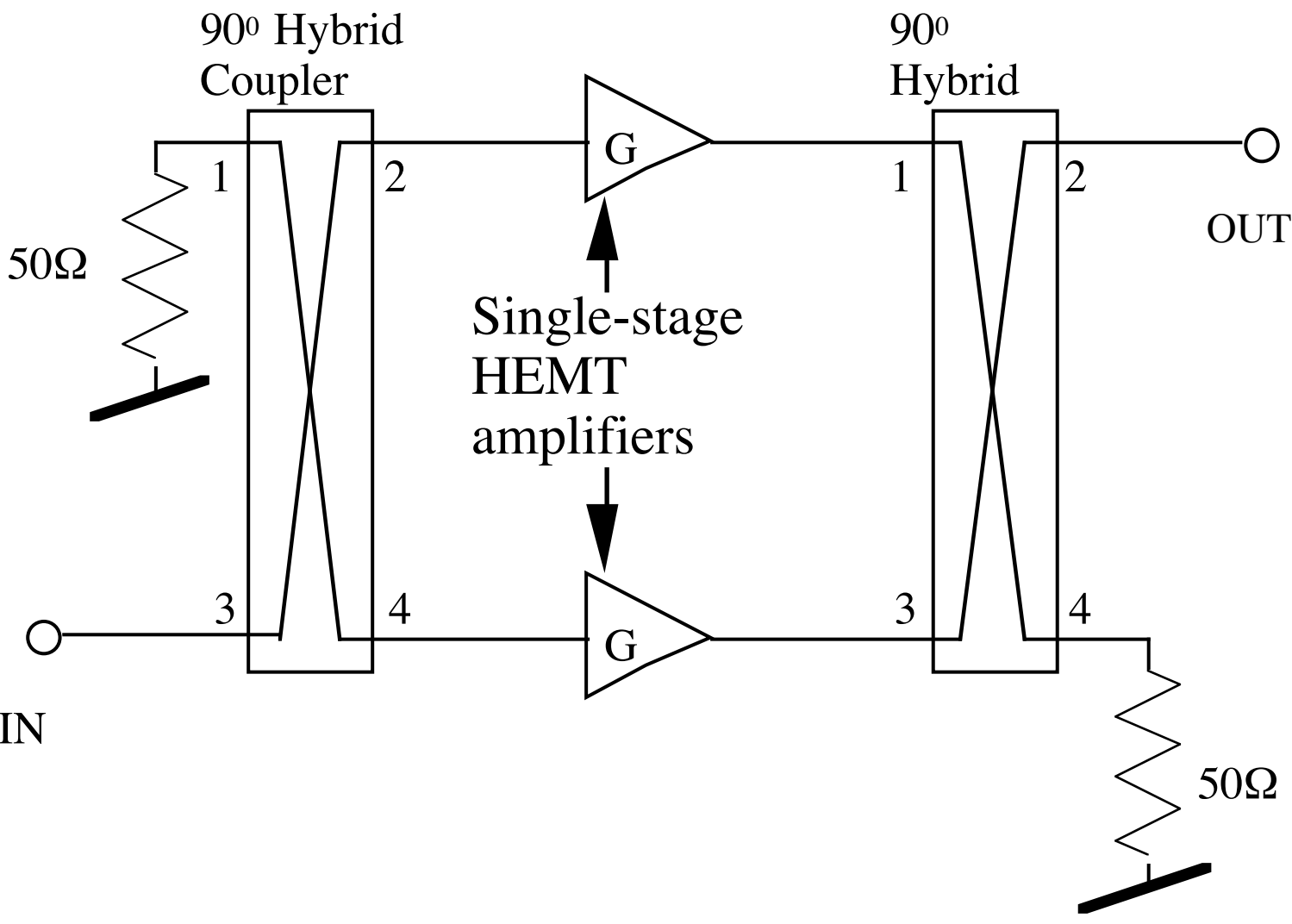

Figure 3.8 A schematic of the balanced HEMT amplifier used in each of the first two amplification stages in the axion receiver.

The balanced design consists of two single stage HEMT amplifiers, the two a matched pair with nearly identical characteristics. Each has power gain G and matched to the input and output transmission lines with simple single-pole matching networks. As a stand-alone amplifier, such a simple single-stage HEMT amplifier would only achieve good match over a bandwidth of a few 10s of MHz. However, the balanced design containing two such units is typically well-matched to a 50Ω source over bandwidths of around 500MHz. To understand the high bandwidth of this design, consider a plane wave incident at the amplifier input. The leftmost $90^0$ hybrid* splits the incoming wave into two components which are incident on the two HEMTs. Due to the similarity of the two single-ended amplifiers, the same fraction of the incident power from each of the two HEMTs is reflected back towards the left-hand $90^0$ hybrid. For clarity I have numbered the four ports of each $90^0$ hybrid. The reflected power from the upper HEMT is incident on port 2 of the hybrid. It is split into two components. The component sent to port 3 picks up an additional $90^0$ phase lag, so a total phase lag of $180^0$ has been introduced by the coupler. The component sent to port 1 picks up no additional phase lag, so the overall phase lag introduced remains at $90^0$. The reflected power from the lower HEMT is incident on port 4 of the hybrid. The component sent to port 1 picks up a $90^0$ phase lag, so it interferes constructively with the component of the reflected power from the upper HEMT, and the combined signal is absorbed in the upper-left termination. The component sent to port 3 is $180^0$ out of phase with the component reflected by the upper HEMT so the two components interfere destructively, and there is no power reflected back down the input line. To a signal incident on the input port, the amplifier is perfectly matched to 50Ω since all the incident power is absorbed by the amplifier with no reflection.

---

* A $90^0$ hybrid is a 4 port device. Two of them are drawn schematically in Figure 3.8. A signal incident on a port of the hybrid is split into two. Half of the power is emmitted at port diagonally opposite the input port, and half at the port directly opposite the input port. The component emmitted at the diagonally opposite port has a phase lag of $90^0$ relative to the component emmitted at the directly opposite port. No power is emmitted at the fourth port.

What happens to the fraction of the signal that is absorbed in the HEMTs? The component in each HEMT is amplified and the amplified signals are incident on ports 1 and 3 of the right-hand $90^0$ hybrid. The phase of the component amplified by the upper HEMT lags that amplified by the lower HEMT by $90^0$. So by the same argument as used for reflected power, the components combine constructively at the (upper right) output port, and destructively at the (lower right) termination. The same argument that we made when discussing the input impedance can be made for power incident on the amplifier output, so the balanced amplifier output is essentially perfectly matched to 50Ω as well.

### 3.4.3 Input Match of the First Cryogenic Amplifier

To measure the input match of the 1st cryogenic amplifier with the 50Ω transmission line, the power reflection coefficient at room temperature was determined according to the apparatus of Figure 3.9.

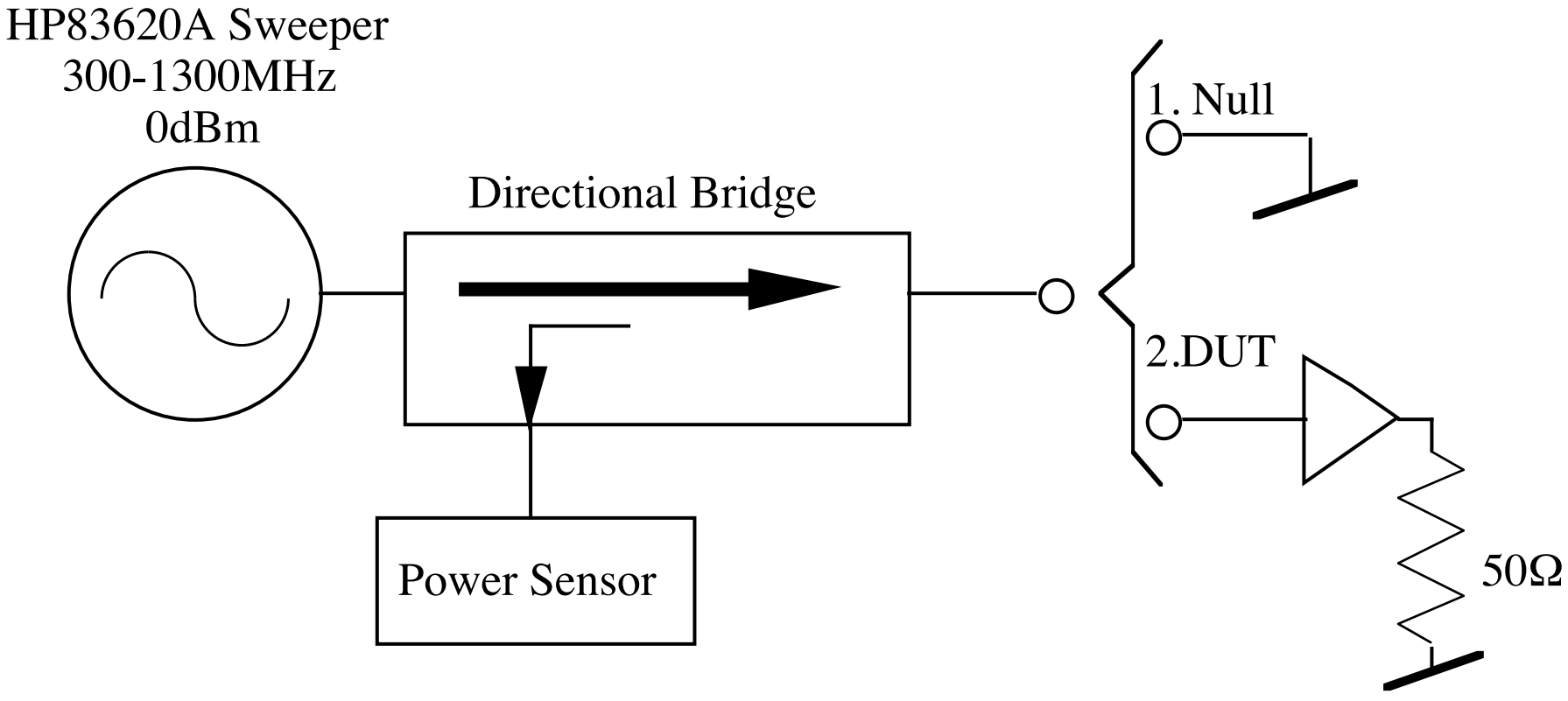

Figure 3.9 Apparatus for measurement of the reflection coefficient of the cryogenic amplifier at room temperature

The signal from the sweeper is incident on the load via the directional bridge. Power reflected off the load is directed onto a power sensor. A null measurement is first done with the bridge terminated in a short so all the incident RF power is reflected. Next, the short is replaced with the device under test. The difference between the reflected power with the amplifier connected and the reflected power with the short connected is the power reflection coefficient. Figure 3.10 is the power reflection coefficient as a function of frequency. The vertical axis is the reflection coefficient in dB with the cursor indicating 0dB and 5dB per division. The horizontal axis is frequency between 300 and 1300MHz at 100MHz per division.

As expected from a balanced amplifier, the power reflection coefficient is very good, better than -18dB over a bandwidth of at least 700MHz. The reflection coefficient is not expected to have a strong temperature dependence and any serious worsening of the reflection coefficient would manifest itself through poor gain and high noise temperature. As we shall see from the results of cryogenic measurements later in this section, at low temperatures the gain of the amplifier increases and its noise decreases, indicating that the good match at the amplifier input persists at cryogenic temperatures.

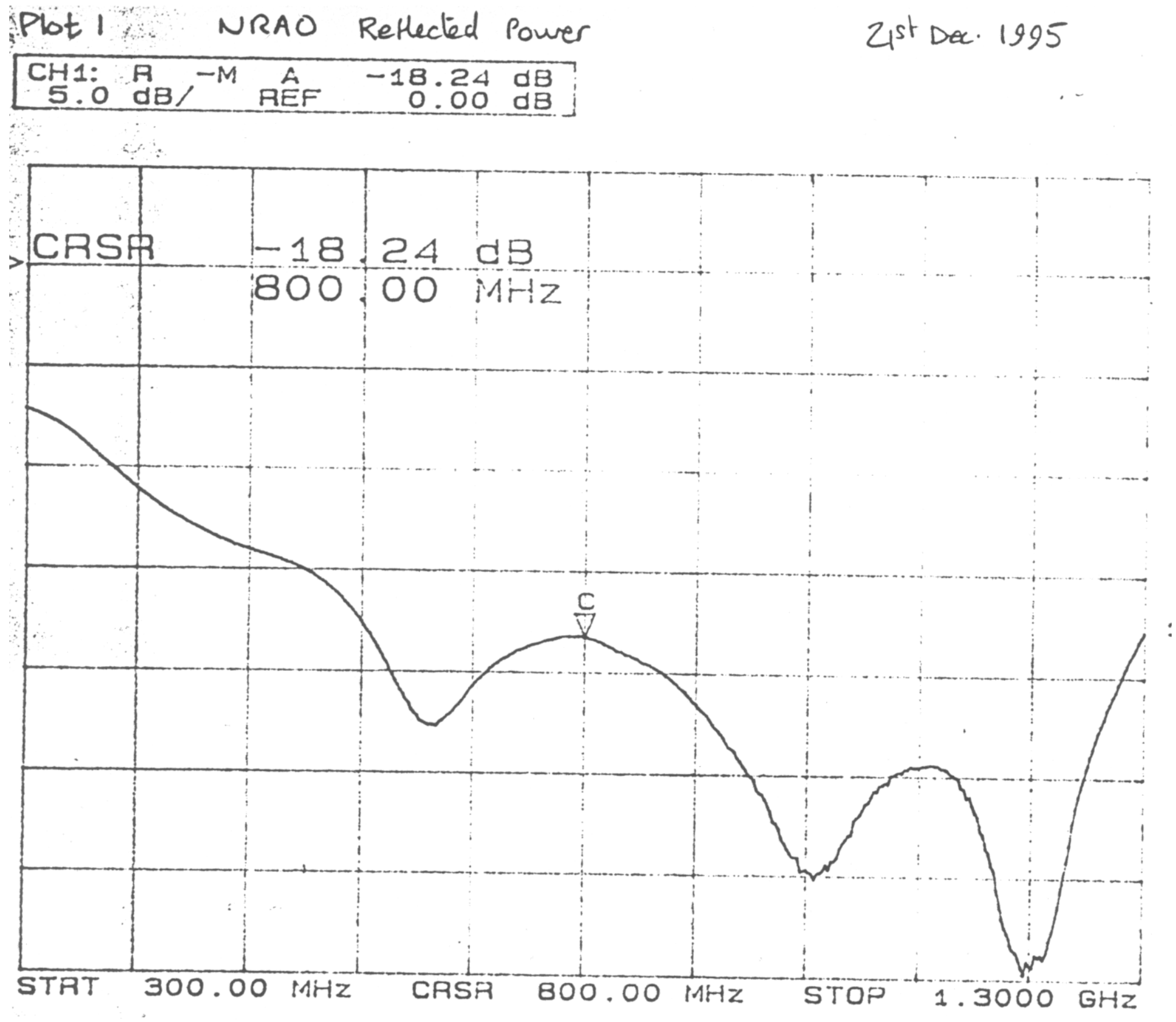

Figure 3.10 Power reflection coefficient at the input of the cryogenic amplifier at room temperature. The horizontal axis is frequency. The horizontal scale is 100MHz per division from 300MHz on the left hand side to 1300MHz on the right. The vertical axis is the power reflection coefficient in dB. The 0dB level (all power reflected) is the 2nd grid line from the top marked with an arrow on the left. The scale is 5dB per division. The power reflection coefficient is better than -18dB everywhere between 600 and 1300 MHz.

### 3.4.4 The 1st Cryogenic Amplifier Coupled to the Resonant Cavity

Consider: What happens to thermal noise from the two internal 50Ω terminations shown in Figure 3.8? The (bottom right) internal termination is not a significant source of noise since there is negligible power transfer from the terminator to the amplifier output, and any contribution it might make is divided by the power gain of the amplifier. The (top left) terminator emits the same noise power into the input of the each HEMT via the left hand $90^0$ hybrid, but by the same argument used for signals injected at the input port, the amplified noise from the two HEMT outputs combines in-phase only at the (bottom right) terminator and is absorbed. However, there is a path where thermal noise from the (top left) terminator is emitted through the input port. If the load is not a 50Ω impedance then a fraction of this power reflects off the load and appears back at the amplifier as a signal. Since the cavity impedance is rapidly varying with frequency near resonances it is important to understand the contribution of the internal terminations to the system noise with a cavity connected to the amplifier input.

The following experiment is an indicator of how noise sources in the amplifier contribute to the amplifier noise with a cavity load. The cryogenic amplifier is critically coupled to the resonant cavity at room temperature. The power spectrum at the amplifier output is measured using an HP8569B spectrum analyzer, first with the amplifier at room temperature then with the amplifier cooled in liquid nitrogen. Figure 3.11 shows the two power spectra.

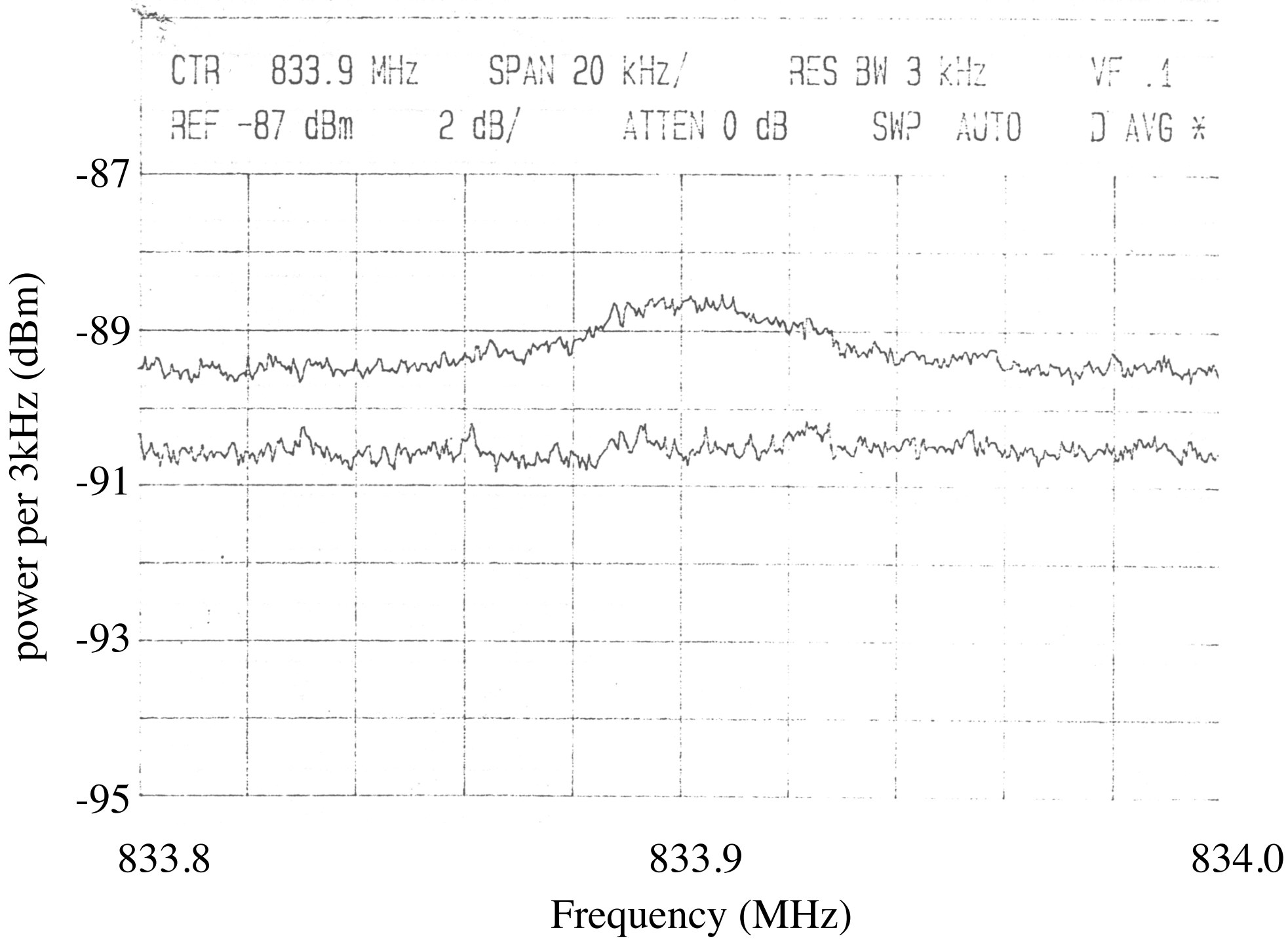

Figure 3.11 Power spectra measured at the balanced amplifier output with the amplifier input critically coupled to the resonant cavity. The cavity resonance was centered at 833.9MHz. On the vertical axis is power per 3kHz bandwidth, on the horizontal axis is frequency. The upper power spectrum was taken with the amplifier cooled in liquid nitrogen, the lower power spectrum was taken with the amplifier at room temperature.

Consider the upper trace, taken when the cavity was warmer than the amplifier. On resonance, the cavity acts as a load perfectly matched to the amplifier input. Therefore all power from the cryogenic amplifier internal termination that is emitted at the input port is absorbed in the cavity and not reflected. The noise temperature at the input is the physical temperature of the cavity. Off resonance the cavity is poorly matched and power from the amplifier internal termination emitted towards the cavity will be reflected back towards the amplifier input. The noise temperature at the amplifier input is now the temperature of the amplifier. Since the amplifier is cooler than the cavity, the power spectrum is peaked at the cavity resonance where the amplifier physical temperature dominates the noise spectrum. The lower trace was taken with the cavity and amplifier at the same temperature, so the noise power spectrum is flat.

This qualitative explanation of the shape of noise power spectra is based on a simplified model of the amplifier and is not suitable for making quantitative predictions of the behavior of the cavity-amplifier combination. A more sophisticated equivalent circuit approach used for analysis of the raw data is described in Appendix 4.

### 3.4.5 The Gain of the Cryogenic Amplifiers at Cryogenic Temperatures and in High Magnetic Field

It is important to check that the good gain, noise and bandwidth of the cryogenic amplifiers remain in the experiment's high magnetic field. An 'in situ' measurement of the gain of the 1st cryogenic amplifier was made, and the results compared with gain data supplied by NRAO. I measured the gain of the cryogenic amplifier (DUT) mounted on a cold test-stand just above the resonant cavity. Figure 3.11 is a schematic of the test stand apparatus.

A swept RF signal was applied through the directional coupler at the amplifier input. The ratio of the power at the output of the amplifier chain to the power from the swept source was first measured with the cryogenic amplifier at room temperature using a scalar network analyzer. The result was divided by the cryogenic amplifier's power gain at *room* temperature which had been previously determined using the network analyzer (see Appendix 4). The resulting curve is the loss in all the long cables and the directional coupler between the sweeper and network analyzer. I will refer to this curve as the connector response. The insert was then lowered into the magnet where the physical temperature of the cryogenic amplifier is typically 3K and the magnetic field at the amplifier is typically 4T. The ratio of the power at the amplifier chain output to the power from the swept source with the cryogenic amplifier cold was divided by the room temperature connector response, and the result is the gain of the cryogenic amplifier at a physical temperature of 3K.

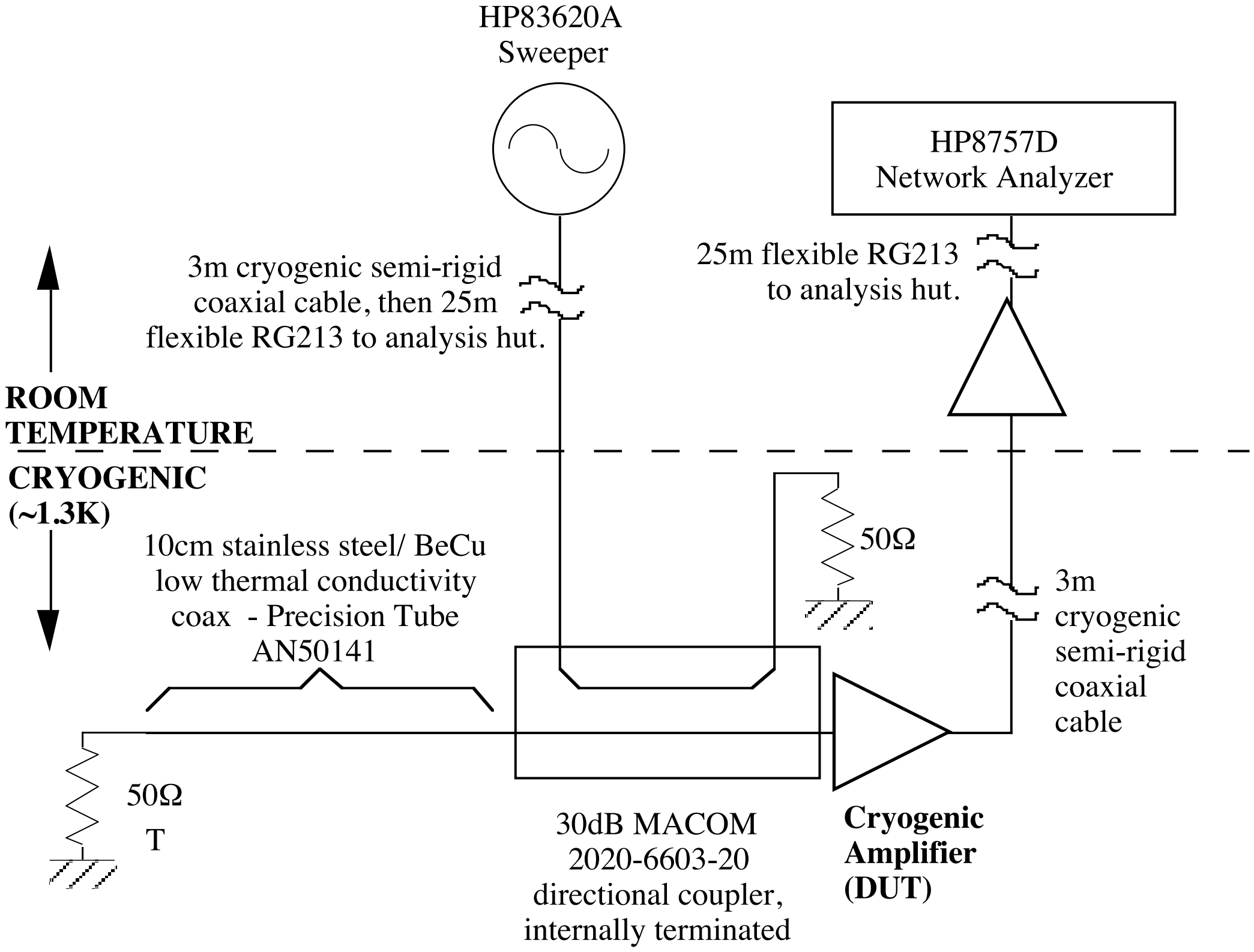

Figure 3.12 Schematic of the cryogenic test stand for the gain measurement. The cryogenic amplifier is thermally heat sunk to the cavity at 1.3K and is in the same position in which it is used when connected to the resonant cavity. The magnetic field at the amplifier is 4T.

This method of measuring the cryogenic gain is independent of power losses in the passive components of the chain, assuming that those losses are temperature independent. Previous cryogenic tests of cables and the directional coupler indicated that temperature dependent losses in these components were negligible. Figure 3.13 is the results of this measurement. My results are in good agreement with specifications from NRAO.

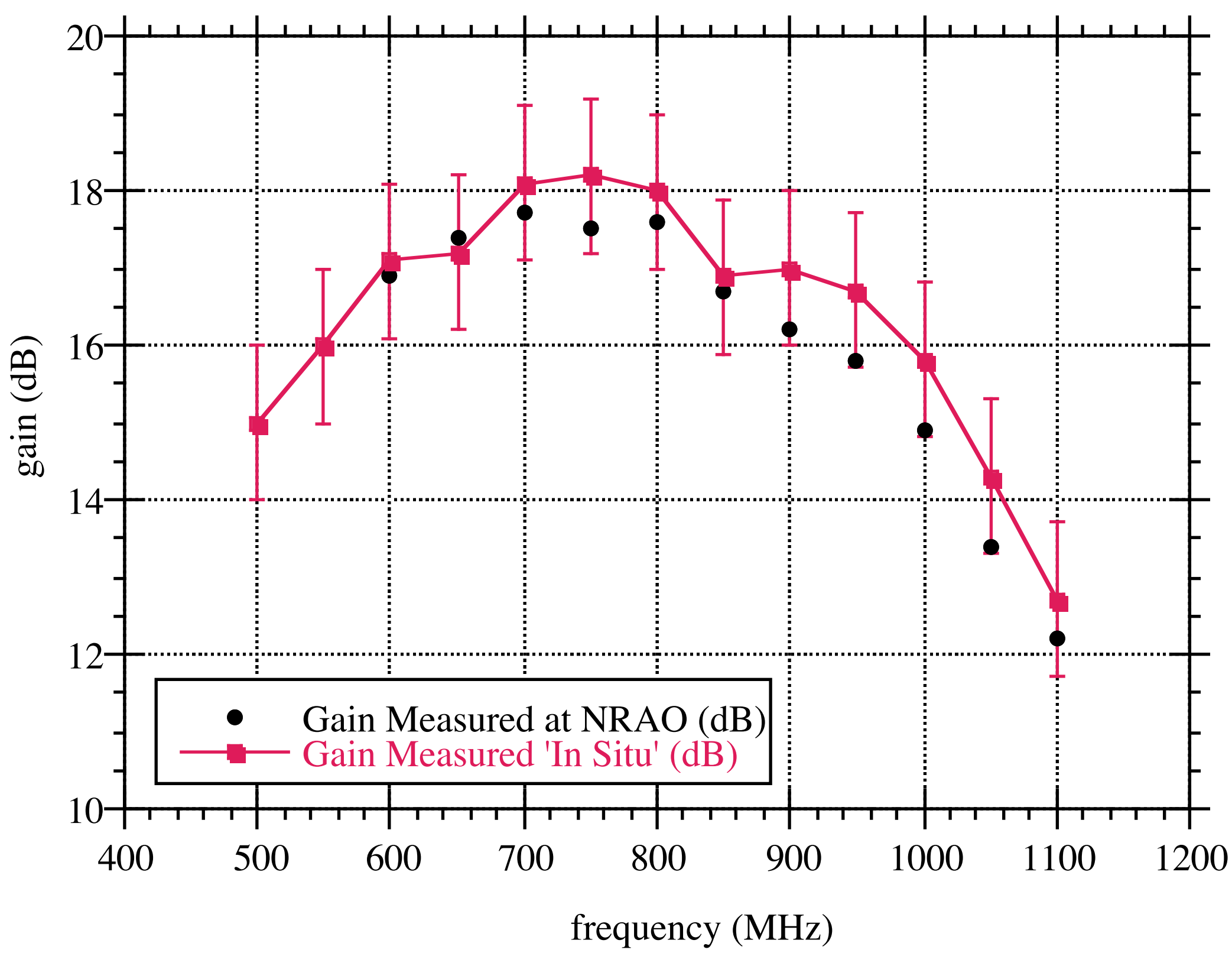

Figure 3.13 The gain of the 1st-stage cryogenic amplifier measured 'in situ' The error bars
are my estimate of how much cable losses change between room and cryogenic temperatures and in flexing the cables whilst lowering the insert into the magnet bore. There is good agreement with the specifications supplied by NRAO.

### 3.4.6 'In Situ' Measurement of the Cryogenic Amplifier Noise Temperature

The noise temperature of the 1st cryogenic amplifier was measured with the amplifier installed on the axion receiver. I argued in section 3.4.1 that the 1st cryogenic amplifier is the only component in the receiver chain whose noise temperature affects the sensitivity of the search. This measurement is a check that the amplifier noise temperature in the high magnetic field of the detector remains low. The amplifier noise temperature is referenced to the temperature of the Johnson noise in the cavity. The cavity temperature is varied, and by looking at the output power of the cryogenic amplifier as a function of the cavity temperature, the amplifier noise temperature can be deduced.

Refer to Figure 3.1. The cryogenic amplifier is critically coupled to the cavity using the reflection method described in Section 3.3.3. For the remainder of the measurement, the directional coupler is not used. On-resonance, the impedance of the cavity viewed through the critically coupled probe is 50Ω. Hence the cavity may be used as a source of thermal noise at a known temperature. The cavity temperature is initially 1.3K. Before warming the cavity the liquid helium in the reservoir below the cavity is boiled off with a 100W heater. Care is taken not to allow the heater to remain on after the helium had evaporated. Nevertheless, the cavity typically warms up to about 3K even after heater is turned off. A small amount of helium liquid is then released into the cavity space, forming a vapor of a few torr. This ensures that there is good thermal contact between the tuning rod and the rest of the cavity, since otherwise the rate of heat transfer from the cavity to the tuning rods is low and the rods may not reach thermal equilibrium.

Heaters on the top and bottom of the cavity are then turned on and the cavity is warmed up over 1 hour to about 10K. As it warms, the data acquisition system is run. The DAQ software first injects a swept signal through the weak port and measures the transmission to the receiver chain using a network analyzer, then takes a 90 second power spectrum of a 30KHz bandwidth about the cavity resonance, as discussed in Section 3.1. The swept signal is used to correct for the gain of the amplifier which may change by as much as 1dB as the cavity is warmed up, since the amplifier and cavity are in thermal contact. It is known from previous studies that the noise temperatures of HEMT amplifiers do not change much below a physical temperature of 10K. A plot of the power per 125Hz bin in the center of the power spectra (on-resonance) corrected for the varying gain of the amplifier as a function of the physical temperature of the cavity yields a straight line. Figure 3.14 is this power vs. frequency plot for our in situ noise temperature measurement at a cavity resonant frequency of 700MHz.

The power $P_{OUT}$ at the output of the axion receiver is given by:

$$P_{OUT} = k_B T_C G + k_B T_N G \qquad (3.4)$$

where $T_N$ is the amplifier noise temperature, $T_C$ is the cavity temperature, and G is the power gain of the axion receiver. Therefore the noise temperature is the absolute value of the intercept of a straight line through the data points with the temperature axis. The inferred value of the noise temperature is 4.6K. This value is consistent with the noise temperature specifications for the amplifier supplied by NRAO. Figure 3.15 is generated from the NRAO specifications of noise temperature vs. frequency. The data point from the 'in situ' measurement described above is included on the plot.

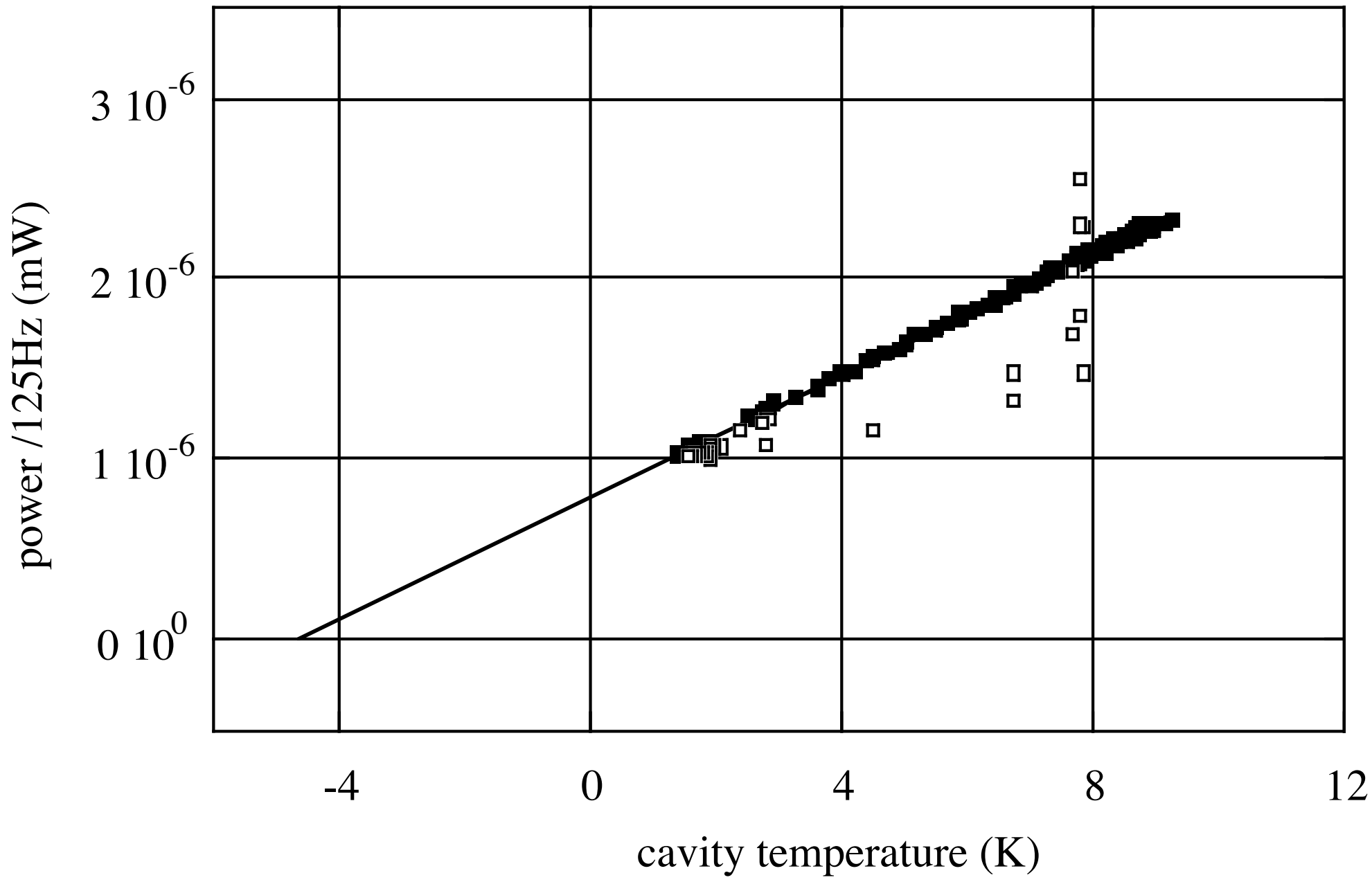

Figure 3.14 Power per 125Hz bin at the cavity resonant frequency vs. physical temperature of the cavity. The power measurement was made using the axion receiver and SRS760 FFT spectrum analyzer. The straight line is a least squares fit to the filled points. The unfilled points that were not included in the fit were excluded as they were taken after the cavity was rapidly cooled back to ~2K, or they were taken after this rapid cooling phase when the cavity temperature is slowly returning to 1.3K. For these points the temperature sensors do not accurately measure the cavity temperature.

Unfortunately it is difficult to make noise temperature measurements of the cryogenic amplifier 'in situ' to an accuracy of better than 10%. This is due to difficulty in controlling the exact temperature of the resonant cavity as it is warmed up. The value measured 'in situ' is consistent with the value quoted by NRAO, and gives me confidence that the cryogenic amplifier is working correctly in the magnetic field.

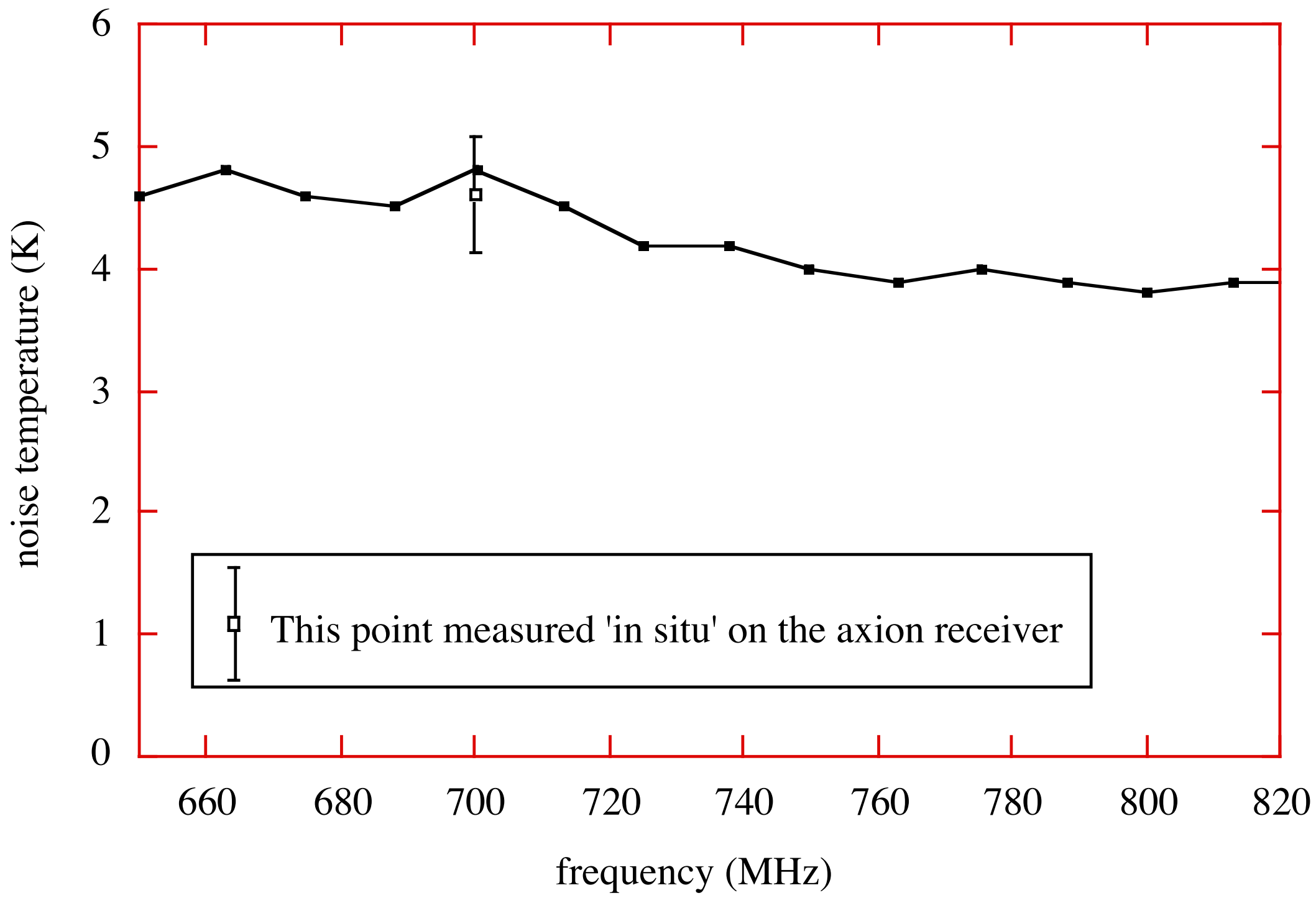

Figure 3.15 The measured value of the noise temperature of the axion experiment first stage cryogenic amplifier and noise temperature data from the manufacturers data sheet. The 10% error bars are on our measurement are an estimate of the uncertainty in the noise temperature determined from the plot in Figure 3.13

I discovered during commissioning that if the first cryogenic amplifier is oriented so that the magnetic field threads the HEMT junctions in the normal direction, the amplifier noise temperature rises to around 8K. A simple model explains this effect: The Lorentz force on the electrons in the HEMT junctions distorts the electron paths so that their trajectory across the gate region is longer. In the simplest version of the model, the HEMT amplifier noise temperature is proportional to the length of the electron path through the gate. This model predicts an increase in noise temperature consistent with that seen in my experiment [16]. After observing this effect the amplifier was re-oriented into its original position on the test stand, as for the in situ measurement of noise temperature marked on Figure 3.15. This amplifier position and orientation was maintained during acquisition of all the production data.

## 3.5 The Room Temperature Electronics

### 3.5.1 The RF Components

The output of the cryogenic amplifier is further amplified using a MITEQ low noise room temperature post-amplifier mounted in an RF shielded enclosure on top of the cryostat (see Figure 3.1). I will refer to this amplifier as the post amp. The gain of the post amp is about 35dB in the frequency range 300MHz-1GHz. Its noise temperature is 90K. The gain of the previous cryogenic amplifiers is 34dB, hence the noise temperature contribution of this amplifier to the overall noise referenced to the receiver input is less than 0.03K and can be neglected. The enclosure also connects the signal lines to type N hardware suitable for mating with the flexible RG213 cables that connect the RF electronics in the cryostat to the analysis hut. Overall there are 3 RF channels in use during data taking, one for the cryogenic amplifier output, one for the weakly coupled port and one for the third port of the directional coupler.

There is a total gain of the total gain is 69dB, or a factor of $8\times10^6$ between the cavity and the output of the post amp. Hence the noise power from the cavity and amplifier at the post-amp output in a 125Hz bandwidth is $8.3\times10^{-14}$W. The equivalent temperature of this noise power is $4.8\times10^7$K. Hence broadband thermal noise in the receiver electronics after the post amp is insignificant compared to amplified noise from the cavity, and shielding downstream of the post amp from room temperature broadband noise is not critical. Occasionally a narrowband signal may leak into our receiver chain and give rise to a spurious power excess. Our experience during running has been that the number of these background power excesses has been very small, of the order of 5 events of this type in a 100MHz bandwidth have been large enough to penetrate our shielding. This background is easily eliminated by detecting it outside the axion detector using the room temperature section of the axion receiver with a stub antenna connected to its input.

The final RF component in the receiver chain is an image reject mixer (IRM) used to shift RF power from frequencies near the $TM_{010}$ resonance to a fixed IF frequency range centered around 10.7 MHz. The image rejection is necessary to avoid mixing off-band RF power into the IF. To understand how an IRM works, consider first an ordinary mixer shown in Figure 3.16.

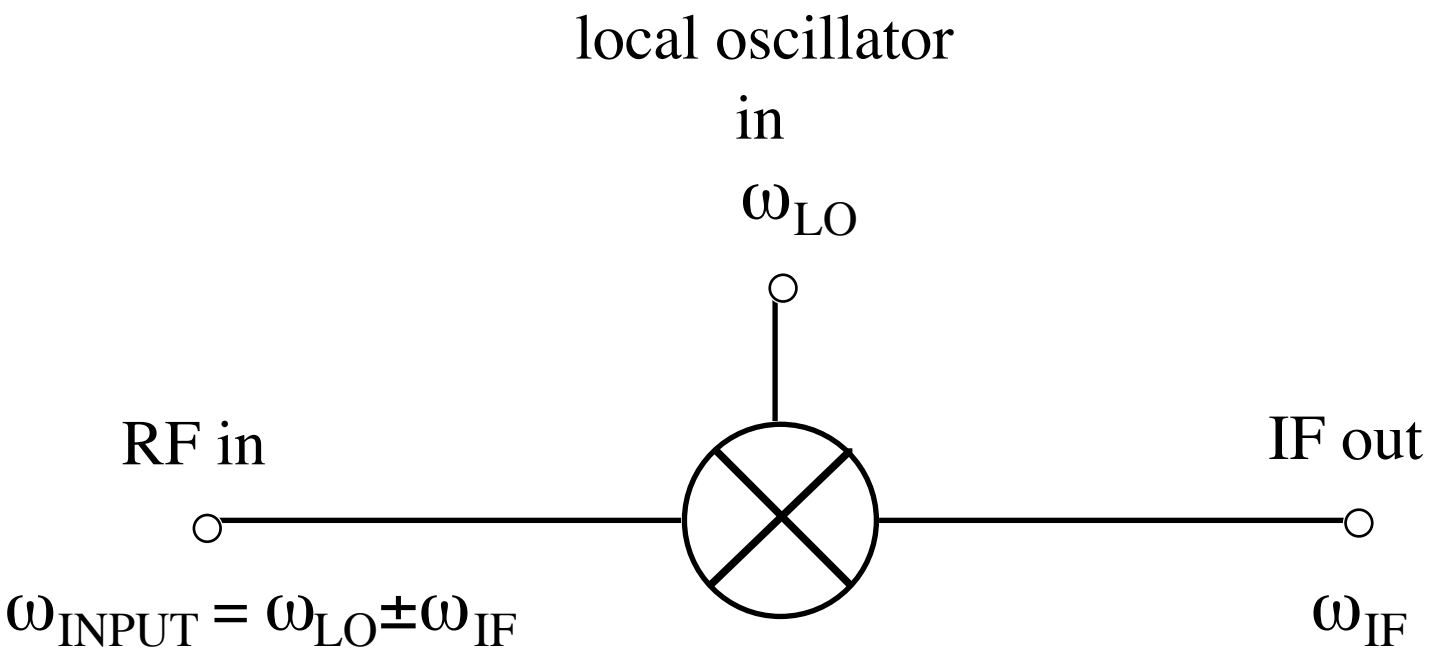

Figure 3.16 A simple mixer

A signal at frequency $\omega_{INPUT}$ is multiplied by the local oscillator signal which is a sine wave at frequency $\omega_{LO}$. The product of the two sine waves yields a component at the sum frequency $\omega_{INPUT}+\omega_{LO}$ which can be ignored and a second component at the difference frequency $\omega_{INPUT}-\omega_{LO}$ which is the IF frequency $\omega_{IF}$. The problem with this type of mixer is that power mixed down to the 'negative frequency' $-\omega_{IF}$ also appears at the IF frequency $\omega_{IF}$. This power originates from the RF image frequency $-\omega_{IF}+\omega_{LO}$. In our experiment $\omega_{INPUT}$ is the cavity resonant frequency for which a typical value is $2\pi(700\text{MHz})$ and $\omega_{IF}$ is $2\pi(10.7\text{MHz})$, so the image frequency is $-\omega_{IF}+\omega_{LO}$ or $\omega_{INPUT} - 2\omega_{IF} \sim 2\pi(682\text{MHz})$. With the cavity resonance at 700MHz and its width is typically 7KHz, the image frequency contains no signal, but equal noise power per bandwidth to the cavity frequency. Thus the effect of the image is to double the noise power in the IF electronics.

The image reject mixer solves this problem by splitting the input signal into two and introducing a $\pi/4$ phase lag into one of the components. Figure 3.17 is a schematic of an image reject mixer. I have written the phases of signals originating from the cavity frequency and the image frequency as the two signals propagate through the receiver chain. The upper expression is the phase for the cavity frequency that we want at the output, the lower expression is the phase for the image frequency that we want surpressed at the output. The left hand hybrid splits each signal into two components, adding a $\pi/4$ phase shift to the components sent to the upper arm. The signals in each arm are then mixed using two ordinary mixers fed by local oscillator signals of equal amplitude and phase from a common RF source.

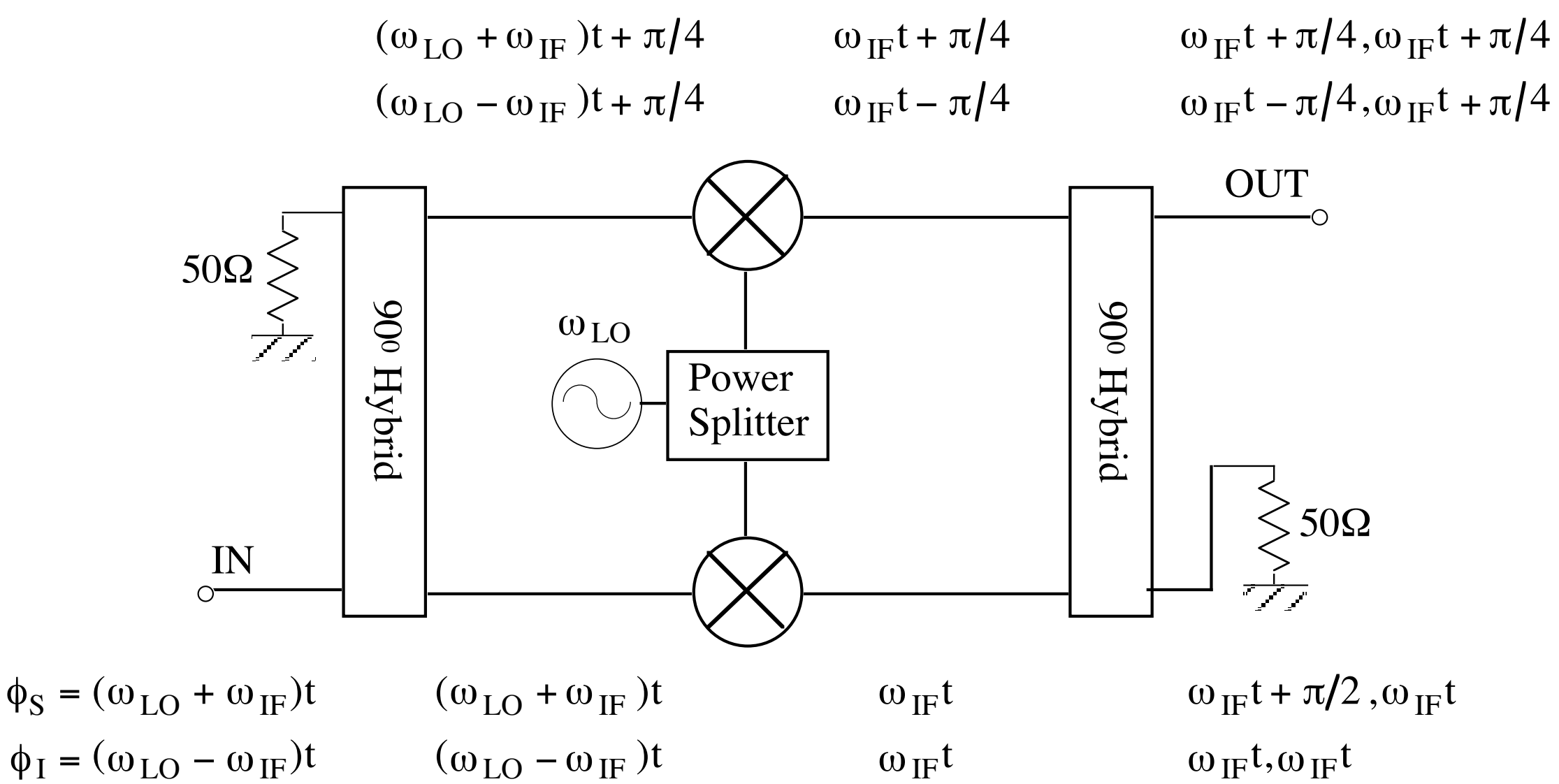

Figure 3.17 An image reject mixer

Notice that when the image component is mixed down to the IF frequency the sign of the phase shift introduced by the 1st hybrid is flipped. After the mix down to IF, the two branches are re-combined in a second hybrid. At the output, the two contributions to the IF signal originating from the RF frequency traveling through the upper and lower signal path between the two hybrids have both received an overall phase change of $\pi/4$ with the same sign. Hence they combine constructively. But the phase changes for the same two components originating from the image frequency are opposite in sign, hence there is an overall $\pi/2$ phase difference and the contribution from the image frequency vanishes at the output. At the lower output arm of the right hand hybrid the opposite is true and the image frequency signals combine constructively whilst the cavity frequency signals are rejected. Hence this port is terminated.

The image reject mixer employed in our first run was a MITEQ IRM045-070-10.7. It requires a local oscillator signal of +15dBm and has an insertion loss of 6dB. The image rejection defined as the ratio of image power to RF frequency power, is better than 20dB. During normal running the local oscillator frequency is always adjusted to 10.7MHz below the measured cavity resonant frequency so that the cavity resonance is centered at 10.7MHz in the IF.

## 3.5.2 The Intermediate and Audio Frequency Electronics

The output of the image reject mixer is fed into a sequence of IF components before being mixed down to audio frequencies. Figure 3.18 is a schematic of these components.

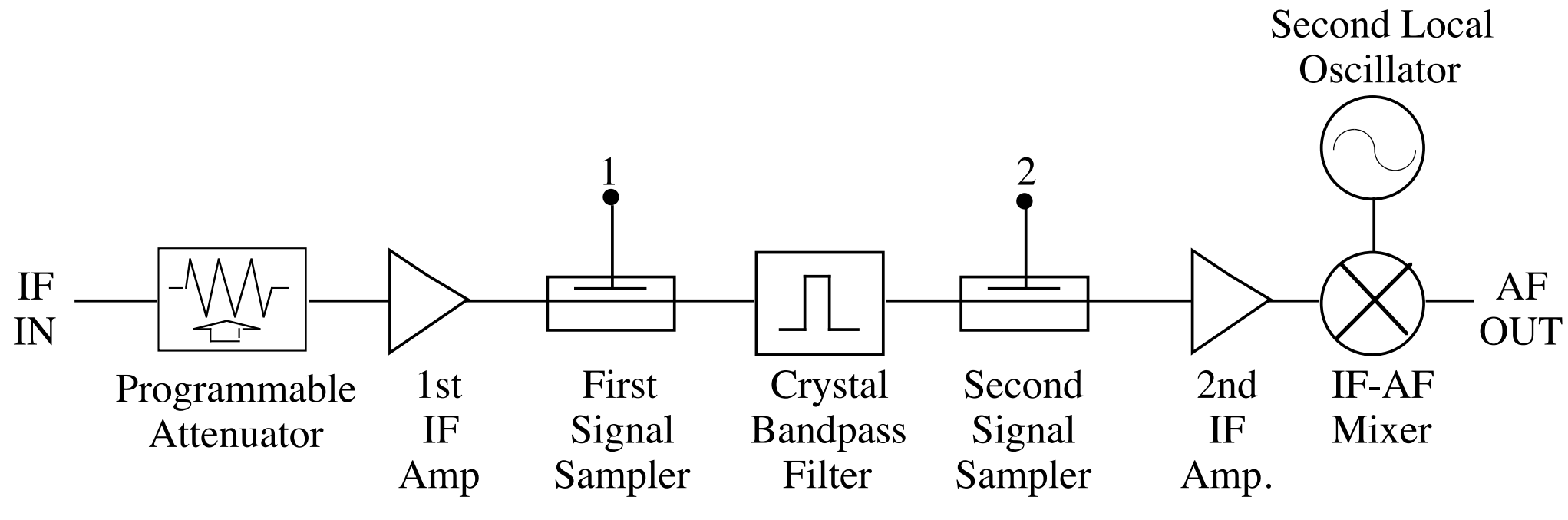

Figure 3.18 Schematic of the intermediate (IF) and audio frequency (AF) components in the receiver electronics

The receiver is 'double heterodyne', which means it employs two mixing stages in cascade.

The first mixing stage was discussed in the previous section. After the first mixer is a programmable attenuator which can be set between 0 and 63dB in 1dB steps. This attenuator is used to avoid saturating the receiver when testing with the 'cryogenic' components at room temperature. Next the IF signal is amplified by about 20dB and passes through a signal sampler used for monitoring the signal power level. During normal running the sampling ports (labeled 1 and 2) of the signal samplers are terminated with 50Ω.

The next receiver component is a bandpass filter, used to surpress noise outside a 30kHz bandwidth about 10.7MHz. The purpose of this filter is to remove noise in the far sidebands that otherwise be imaged in to the audio frequency (AF) bandwidth. With the filter removing this sideband noise, the second mixer does not need to be an image reject type. This filter has a 30kHz bandwidth. The maximum attenuation of signals within the passband is 3dB. The filter is enclosed in a temperature controlled insulated ovenized box and maintained at a temperature of 40K±0.5K. The filter is an 8 pole type and there is ripple in the attenuation across its passband. The ripple can be inferred from attaching a broadband noise source to the input of the axion receiver and averaging many power spectra at the receiver output. I will refer to this linear average of many power spectra with a noise source at the receiver input as the crystal filter passband response. Accurate measurement of the crystal filter passband response is important since it multiplies the final power spectra that form our raw data. Application of the crystal filter passband response to data analysis is discussed in Chapter 4.

The apparatus used to measure the crystal filter passband response is shown in Figure 3.19.

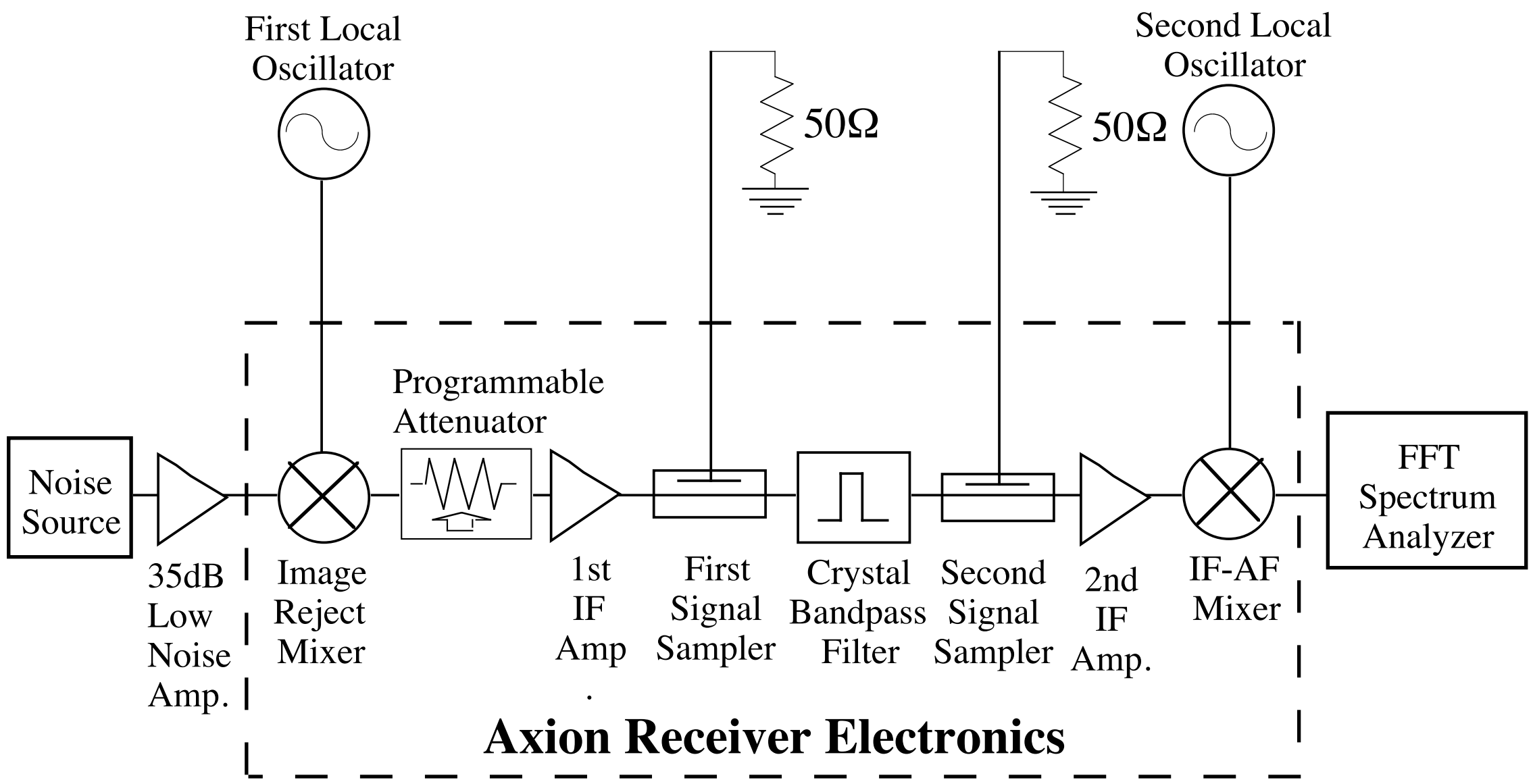

Figure 3.19. Apparatus for the measurement of the crystal filter passband response

Thermal noise generated using a Noise/Com 346B broadband $10^4$K noise source is amplified by 35dB injected at the input of the MITEQ IRM045-070-10.7 image reject mixer. The mixer output is connected the IF section of the axion receiver electronics. An SRS 760 FFT spectrum analyzer is used to measure $10^6$ 400-bin power spectra is taken covering a 50kHz bandwidth about 10.7MHz, the center frequency of the filter passband The total integration time is 2.2 hours. Figure 3.20 is a measurement of the crystal filter ripple.

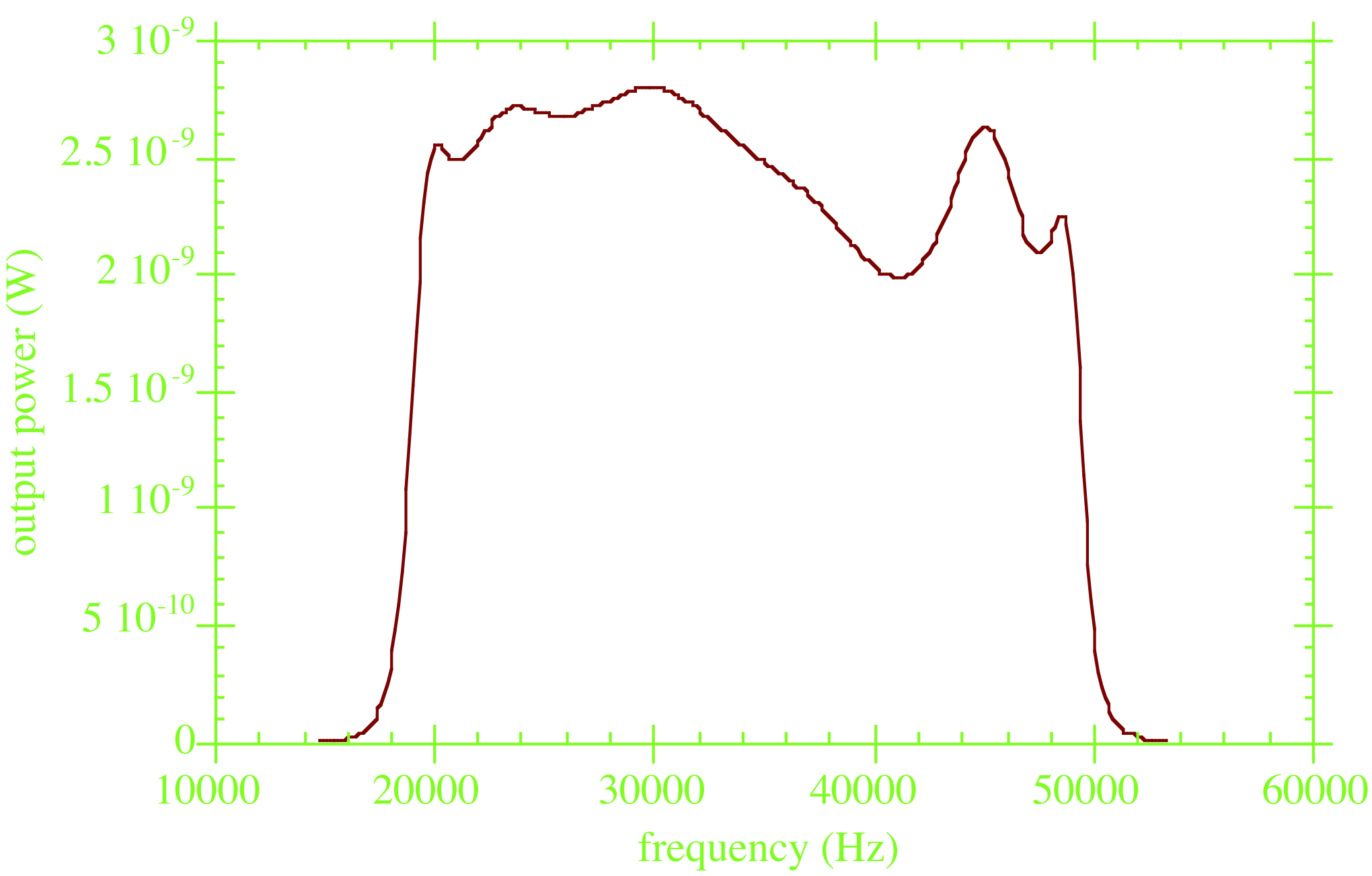

Figure 3.20 the power out of the crystal passband filter with a noise source at the input. On the vertical axis is power / 125Hz at the output of the receiver electronics; on the horizontal axis is frequency after the AF mixing stage

The search for axion-like signals in the power spectra produced by our receiver is sensitive to even very small changes in this response with time. Consider: a typical integration time for the acquisition of a single averaged FFT power spectrum on our experiment is 80s. With a 125Hz bin width this means we are averaging 10,000 spectra. Hence the rms fluctuations in Gaussian noise in the power spectrum are expected to be $1/\sqrt{10000}=0.01$ of the raw noise power, by the radiometer equation. This means that the crystal filter response in the passband only needs to change by a significant fraction of 1% over the run time of the experiment to have a noticeable systematic effect on the shape of the spectra. Changes of the order of 0.2% in the crystal response over the first 1.5 years of data taking are in fact seen. These changes will be discussed in detail in Chapter 4.

After the crystal filter is a second signal sampler and another 20dB IF amplifier. The IF signal is mixed down again, to audio frequencies (AF) in the range 10-60kHz. The output of the second mixer is fed to a commercial FFT spectrum analyzer, and the resulting power spectrum is stored as part of what I will call our 'raw data'.

### 3.5.3 Signal and Noise Power in the Receiver Chain

Table 3.1 shows the approximate power at the output of the major components of the electronics. Power over two different bandwidths is important. Firstly, power over the full bandwidth of the component in question is given, to check that the power output of a component over its full bandwidth is not so large that the next component in the receiver chain is saturated. Secondly power over a 125Hz bandwidth, the width of the frequency bins in the FFT spectrum analyzer, is shown. The starting point is the power over a 125Hz bandwidth incident on the first cryogenic amplifier, which is $10^{-20}$W=-170dBm At each point in the receiver chain the total broadband noise power over the bandwidth of the next component is insufficient to cause saturation. Notice that the crystal filter significantly reduces the broadband power by reducing the bandwidth from order of 1GHz to 35kHz.

| Component | Gain (dB) | Power per 125Hz at Output (dBm) | Total Output Power (dBm) |
|---|---|---|---|
| Cavity | - | -170* | -101* |
| Cryogenic Amplifiers | 34 | -136 | -67 |
| Room Temperature 'Post Amplifier' | 35 | -101 | -32 |
| Flexible Cable to Analysis Hut | -6 | -107 | -38 |
| Image Reject Mixer | -7 | -114 | -45 |
| 1st IF Amplifier | 30 | -84 | -15 |
| Crystal Filter | -3 | -87 | -60 |
| Second IF Amplifier | 30 | -57 | -30 |
| IF - AF Mixer | -7 | -64 | -37 |

*Assumes that the 1st cryogenic amplifier has a bandwidth of 1GHz and that the sum of the cavity and amplifier noise temperatures is 6K.

Table 3.1 Power levels at different points in the receiver electronics chain

### 3.5.4 The FFT Spectrum Analyser and a Typical Power Spectrum.

The FFT spectrum analyzer is a Stanford Research Systems SRS760. It has a maximum sampling rate of 256kHz. Each FFT spectrum is acquired over an 80s integration time and consists of 400 125Hz bins spanning the frequency range 10-60kHz. Uniform windowing is employed. Figure 3.21 shows a typical trace (recall that a trace is defined as the linear average of 10,000 power spectra taken at fixed cavity tuning settings $f_0$) taken during production running.

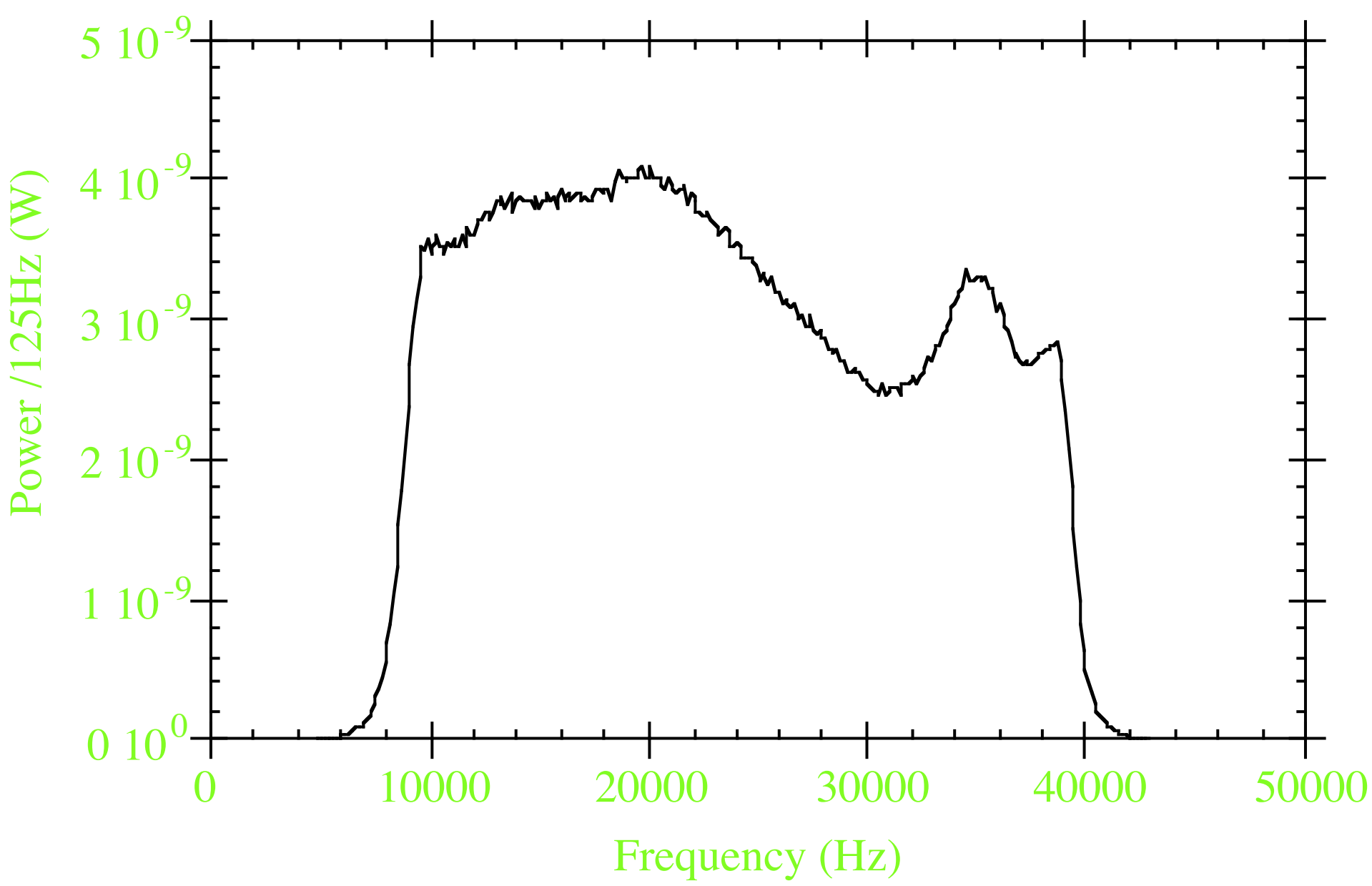

Figure 3.21 A typical trace from the raw data

The most obvious feature of this typical trace is the rapid falloff in power outside the frequency range 20-50kHz, caused by attenuation outside the passband of the IF crystal filter. Notice also that in the crystal filter passband attenuation plot of Figure 3.19 there is a broad hump centered at 24kHz, but in the trace of Figure 3.20 no hump is visible. This difference in shape is due to the coupling of the receiver electronics to the resonant cavity. The cavity is at 1.3K during normal running and the amplifier is at a physical temperature of ~3K, therefore the Johnson noise power on-resonance (where the cavity noise source dominates) is less than the Johnson noise power off-resonance (where internal noise sources in the amplifier dominate). The difference in noise powers on- and off- resonance is responisble for the decreased on-resonance power. Understanding the shapes of traces from the raw data is important for data analysis since the signal from axions is predicted to be power excess over 6 bins of a trace above the noise background. A model for the noise background is presented in Chaper 4 and Appendix 4.

# 3.6 Magnet and Cryogenics

## 3.6.1 The Magnet

The magnetic field was supplied by a superconducting solenoidal magnet manufactured by Wang NMR of Livermore, CA. The magnet consists of niobium-titanium windings immersed in a liquid $^4$He cryostat. The helium consumption of the magnet is typically 500$\ell$ every 9 days. During normal running the field at the center of the solenoid was 7.6T, falling off to about 70% of this value at the center of the end plates of the cavity. Figure 3.22 is a plot of the magnetic field strength vs. distance along the solenoid axis.

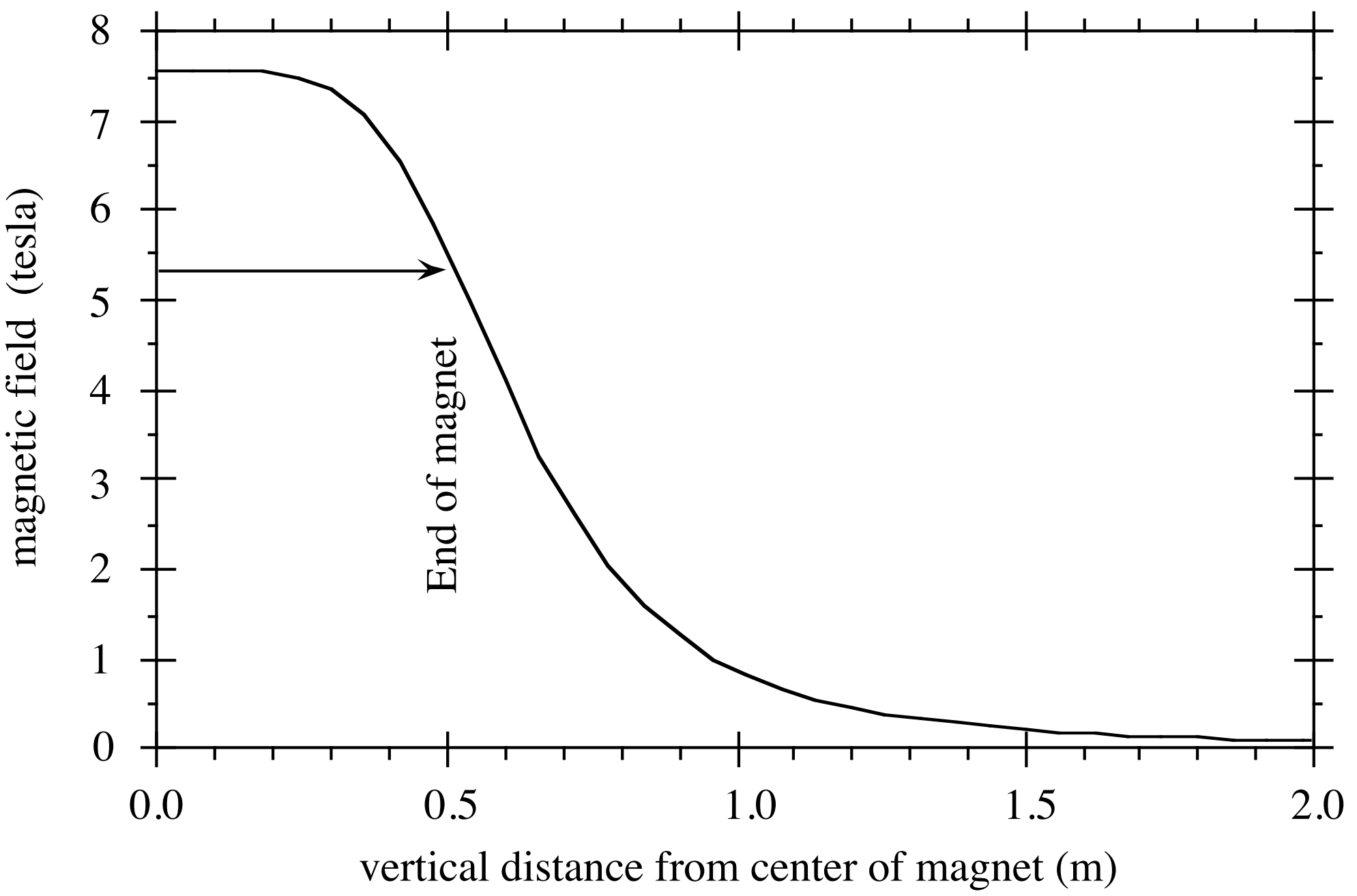

Figure 3.22 The on-axis magnetic field as a function of displacement from the cavity center

In Section 2.3 I assumed that the B field was constant throughout the cavity when deriving a formula for the axion-photon conversion power. This assumption is not valid in our experiment. However, the result is unchanged provided we employ a form factor defined as follows:

$$f_{nlm} = \frac{\left( \int_V dV \mathbf{E}(\mathbf{x},t) \cdot \mathbf{B}(\mathbf{x}) \right)^2}{B_0^2 V \int_V dV \varepsilon_r E^2} \qquad (3.5)$$

where $B_0$ is the B field at the center of the solenoid. This formula was used in the computation of the form factor displayed in Figure 3.7.

### 3.6.2 The Cryogenic Apparatus

Figure 3.23 is a schematic of the insert and the magnet showing the cryogenic system

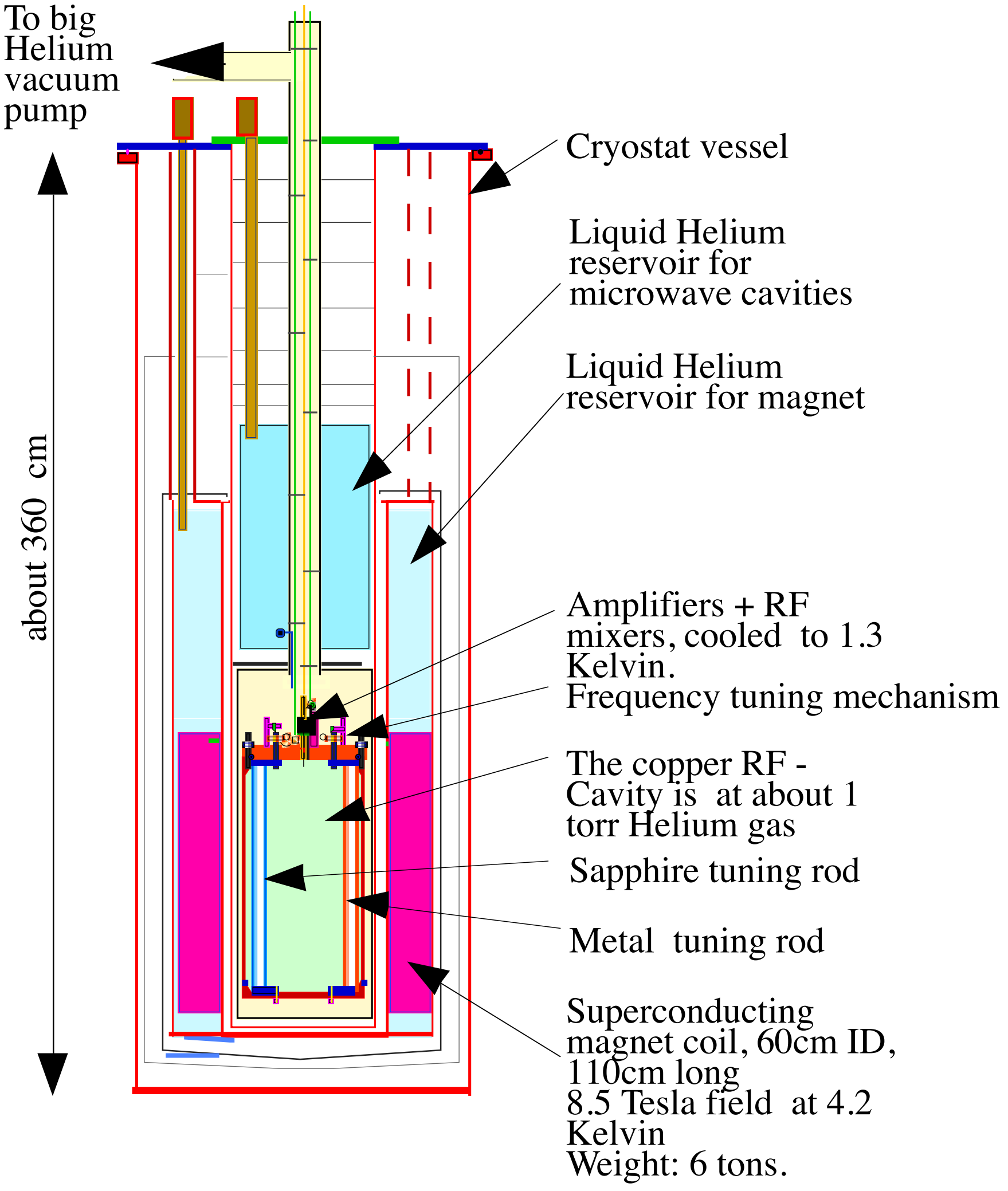

Figure 3.23 The cavity is contained in a cylindrical insert which can be withdrawn from the magnet bore without warming the magnet. This diagram shows the insert installed in the magnet bore .

The cavity is surrounded by a stainless steel vessel. Liquid helium is released into the bottom of this vessel from the liquid helium reservoir directly above the cavity space through a thin capillary line and a motor-controlled Joule-Thomson (JT) valve. During normal operation the liquid helium forms a lake a couple of inches deep below the bottom of the cavity. A liquid helium level gauge is attached to the bottom of the cavity to measure the depth of the helium lake. A feedback loop in the data acquisition software reads the depth of this lake and opens the JT valve to release more liquid into the vacuum space and maintain the lake at a constant depth. A column 4 inches in diameter rises from the top of the vessel enclosing the resonant cavity through the center of the liquid helium reservoir, past some heat exchange baffles to a tube protruding from the top of the magnet. Through this tube a large Roots blower pumps on the helium lake below the cavity, cooling the space to a physical temperature of 1.3K. The pressure of the helium gas in the insert is typically 1 torr. During scanning in mode crossing regions, the depth of the liquid helium lake is increased so that the entire cavity is immersed in liquid. The liquid helium reservoir supplying the insert space requires a refill approximately once a fortnight. Between the insert and the inner wall of the cylindrical magnet dewar is an insulation vacuum space. During normal running the vacuum in this space is better than $10^{-6}$torr. Heat losses from the insert are slight because it is surrounded on three sides by the magnet dewar at 4.2 and on the 4th side by several layers of baffles kept cold by the passage of helium vapor up the insert column.

# 4. Analysis

## 4.1 Introduction to Data Analysis

### 4.1.1 The Axion Search Strategy

The aim of my analysis is to search for signals that could be the signature of axions in the raw data from the axion detector. The search is conducted in two channels. The first search channel is optimized for axion signals that are significantly narrower than 1 frequency bin or 125Hz. The statistic used in the 1-bin search is power excess in single 125Hz frequency bins. The motivation for this channel is the possibility that some or all of the axions may have a very small velocity dispersion; the signature for such axions in our detector would be power excess in a single bin. The second search channel is optimized for the more likely case of axion signals that are 6 frequency bins (750Hz) wide. The statistic used for this 6-bin search is power excess in the sum of 6 adjacent frequency bins. The motivation for the 6-bin channel is the thermalized axion dark matter model discussed in Section 2.2.3. The 1-bin channel is more speculative and was chosen partly because of the simplicity of analysis.

### 4.1.2 The Single Bin Search Channel

The 1 bin search proceeds as follows: First, data is acquired covering the entire 701-800MHz frequency range. This data set consists of $2.5 \times 10^5$ raw traces and covers the entire frequency range with sufficient integration time to achieve a target sensitivity at all frequencies (see Section 4.3.3) . I will refer to this data set as the run 1 raw data. The run 1 raw data is used to generate a combined data set consisting of a continuous power spectrum covering the entire 701-800MHz frequency range at a bin width of 125Hz. Also included in the combined data is an estimate of the signal-to-noise ratio for axion signals in each frequency bin. A cut is made in the combined data on power excess in single 125Hz bins. Frequencies passing the cut are added to the 'single bin candidate list'. A second set of raw data is now taken, which I will refer to as run R. Enough power spectra are taken in run R to reproduce the sensitivity in run 1 at each frequency in the single bin candidate list. The run R raw data is then used to generate a second combined data set. The same cut is made on the run R combined data as was made on the run 1 combined data. Any frequencies that pass the cut in both run 1 and run R are examined further and several procedures are employed to identify any signals observed at these frequencies with known backgrounds. If all frequencies surviving both cuts can be identified with known sources of background, the cuts made in runs 1 and R are used to compute an exclusion limit on the axion photon coupling. The exclusion limit is computed analytically from the Gaussian statistics of the single bin combined data.

### 4.1.3 The Six Bin Search Channel

The 6 bin search proceeds as follows: The run 1 combined data described in the previous paragraph is searched for 6 bin power excess. A cut is made in the combined data on power excess in the sum of 6 adjacent frequency bins. Frequencies passing this cut are added to the 'six bin candidate list'. A second set of raw data is taken in run R at each of the frequencies on this list. For the six bin candidate search, the raw data from run R is combined with the raw data from the original run 1, yielding a power spectrum with increased SNR at frequencies on the six bin candidate list. I will refer to this combined data as the run (1+R) combined data. A second cut is made, again on power excess in the sum of 6 adjacent bins. Frequencies passing the 6 bin power excess cuts in both the run 1 and the run (1+R) combined data sets are added to the 'six bin persistent list'. A third set of raw data which I refer to as run P is taken in which 280,000 power spectra are taken at each frequency on the six bin persistent list. The run P raw data is combined with the run 1 and the run R raw data. A third and final cut is made on the overall combined data, once again using power excess in the sum of 6 adjacent frequency bins as the cut statistic. Any frequencies passing all 3 cuts are examined further and several procedures are used to identify any signals at these frequencies with known backgrounds. Assuming that there are no surviving frequencies, Monte Carlo simulation techniques are used to determine the sensitivity and confidence level of the 6 bin search channel.

### 4.1.4 Layout of this Analysis Chapter

The rest of this chapter is divided into 4 sections. In section 4.2 I discuss the algorithm for combining raw traces to give a combined data set in which the cuts for the two search channels are implemented. In Section 4.3 I discuss the run 1 combined data, which is used both in the 1 bin and in the 6 bin searches. In Section 4.4, I discuss the 1 bin search and the determination of our sensitivity to narrow axion signals. In Section 4.5 I discuss the 6 bin search and the determination of our sensitivity to the more likely thermal axion signals.

## 4.2 The Data Combining Algorithm

### 4.2.1 What is Power Excess ?

Both search channels described in Section 4.1 use power excess as the search statistic. What is a power excess? Consider: The search is for a narrow peak above the background noise of the cavity. The first step in my analysis is to isolate the power spectrum of the noise background, which is a slowly varying function of frequency, from the fluctuations about this smooth background level. It is in the fluctuations about the smooth background that we search for the narrow-peak signature of axions. If the fluctuations are due to nothing but Johnson noise in the cavity they should average to zero after a sufficiently long integration time by the radiometer equation. If there is axion to photon conversion at some frequency, the fluctuation at that frequency does not average to zero but instead averages to a positive value, a narrow peak above the smooth background noise level. This is what is meant by power excess.

### 4.2.2 Preliminary Treatment of Raw Traces

Refer to Figures 3.20, a typical trace from raw data. The first stage in the processing of a raw trace is to crop off frequency bins outside the passband of the filter. There are a total of 400 bins in each trace of which the first 100 and the last 125 bins are discarded. The reason for discarding more bins at the higher frequency end can be seen in the raw trace. There is a narrow pole in crystal filter response at the right hand end of the filter passband. This pole has a width of about 10 bins, too close to the axion signal width, so the region of each power spectrum around the pole is discarded. I will refer to the section of each trace remaining after the end bins are discarded as the cropped trace.

Now refer to Figure 3.19, the crystal filter passband response. Comparing Figure 3.19 with the trace in Figure 3.20, it is easy to see that much of the slowly varying structure in the power spectrum is due to the pole structure of the filter. Therefore the next step in data analysis is to divide the cropped trace by the corresponding bins in the crystal filter passband response. The result of performing this operation on the trace in Figure 3.19 is shown in Figure 4.1:

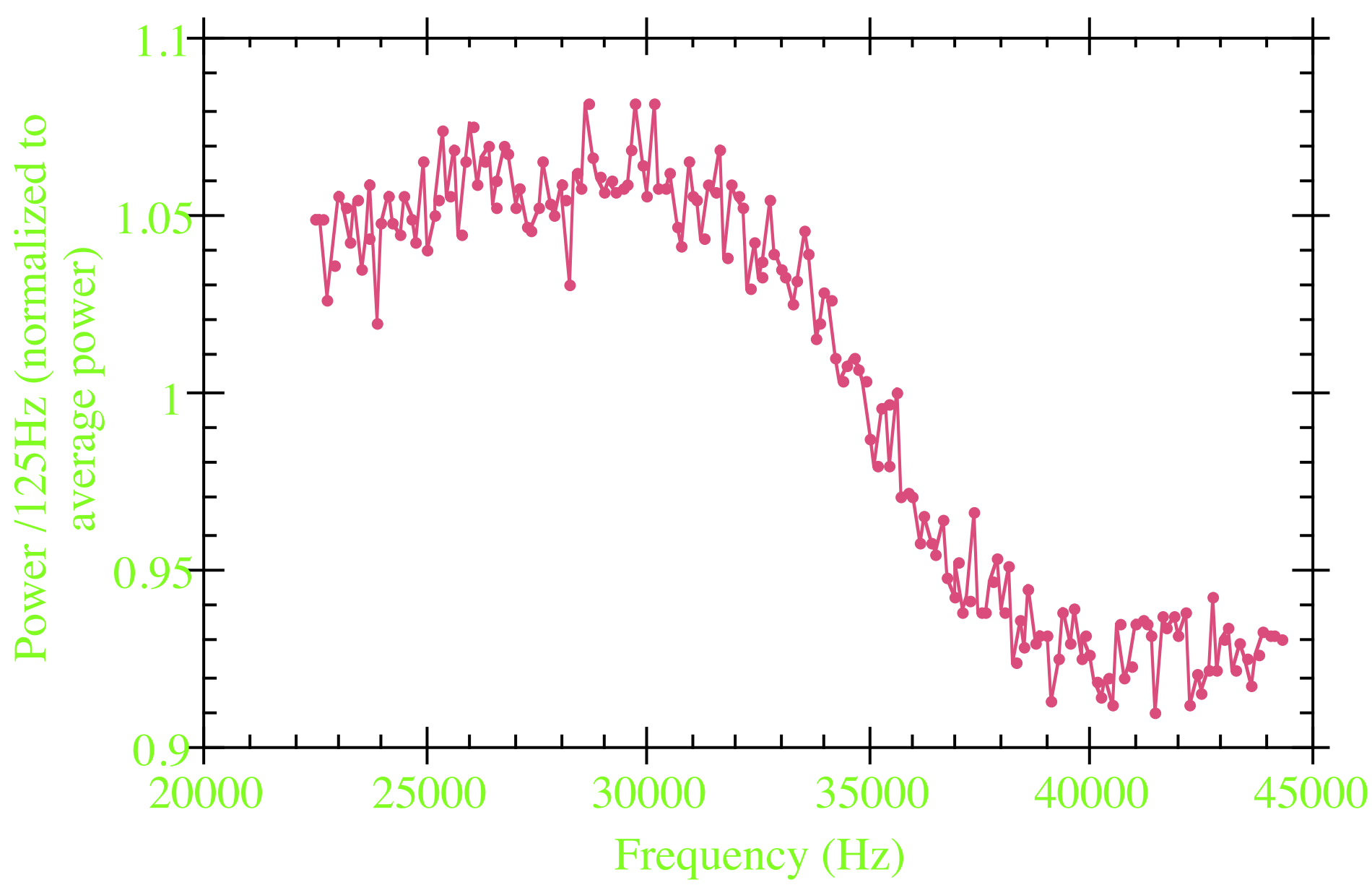

Figure 4.1 A raw trace after the 1st 100 and last 125 frequency bins have been removed, and after it has been divided by the crystal filter passband response. On the vertical axis is power in units of the average power over the 175 bins in trace. The 100th bin is power downconverted from the measured cavity resonant frequency $f_0$.

I will refer to the traces normalized with the receiver passband attenuation as receiver corrected traces. Notice that there is additional smooth structure in the receiver corrected traces. Furthermore, this structure is not symmetric about the center of the trace, which corresponds to the resonant frequency of the cavity.

### 4.2.3 Origin of Asymmetry in the Receiver Corrected Traces

If all the structure in the raw traces were attributable to the frequency dependent effects in the receiver components downstream of the RF amplifiers, the receiver corrected traces would consist of Gaussian noise on a background shape symmetric about the cavity resonant frequency. It is clear from figure 4.1 that in reality the power spectra are not symmetric. Furthermore, receiver corrected traces from data taken with the cavity $TM_{010}$ mode at different frequencies exhibit different structure. Figure 4.2 shows four receiver corrected traces taken with 4 different values of $f_0$.

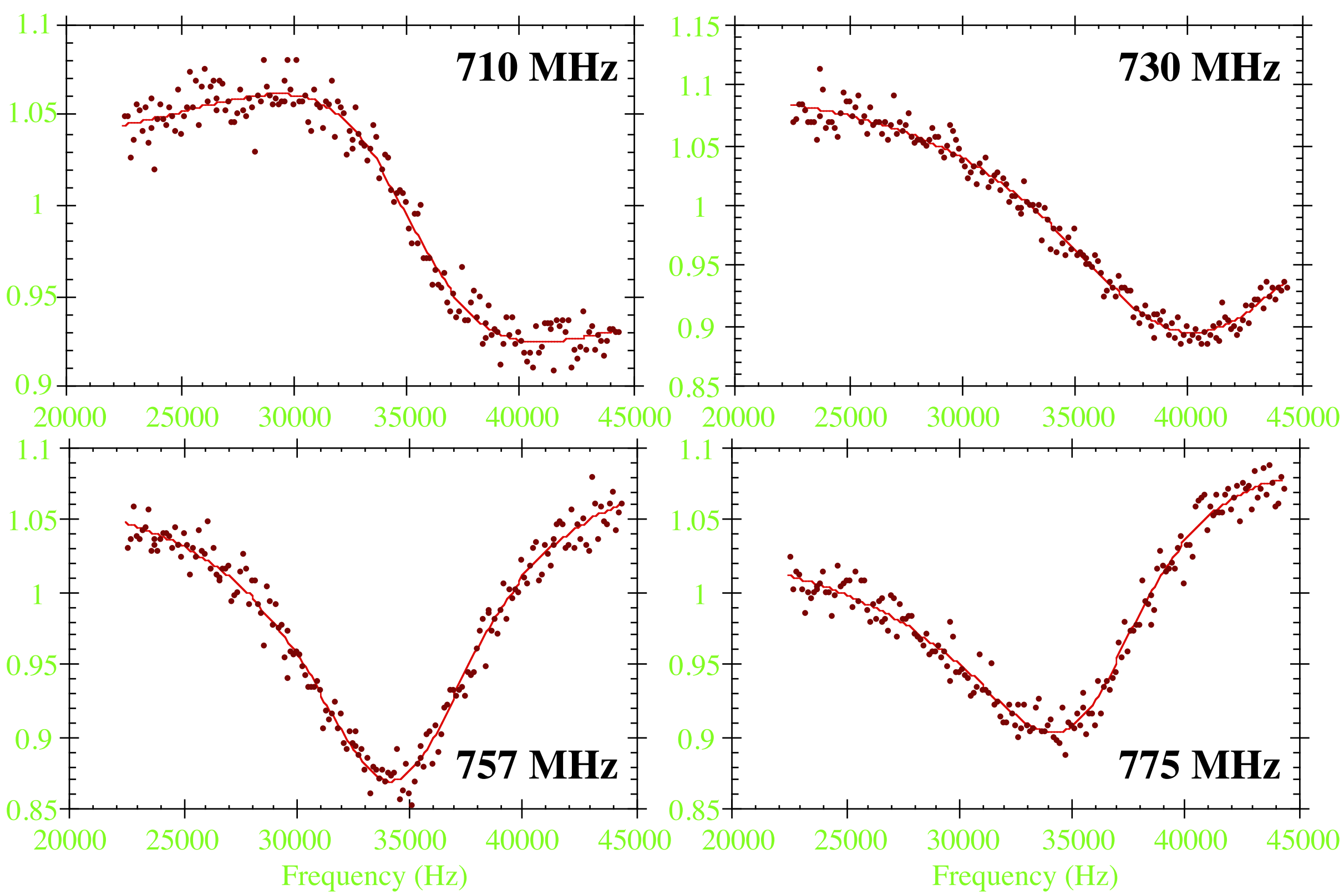

Figure 4.2 Four receiver corrected traces at different cavity resonant frequencies. The collection of points is the power spectrum, the smooth lines are the fit to the power spectrum to be discussed in Section 4.2.4.

Not only are the receiver corrected traces asymmetric, the asymmetry is frequency dependent. This implies it is related to the coupling between the resonant cavity and the 1st cryogenic amplifier. This coupling is quite complicated because there are noise sources both in the cavity and in the amplifier, and a non-negligible length of transmission line between the two. In order to understand the asymmetric structure I must develop an equivalent circuit model of the amplifier coupled to the cavity.

### 4.2.4 Equivalent Circuit Model for the Resonant Cavity Coupled to the First Cryogenic Amplifier

My equivalent circuit for the resonant cavity coupled to the amplifier shown in Figure 4.3.

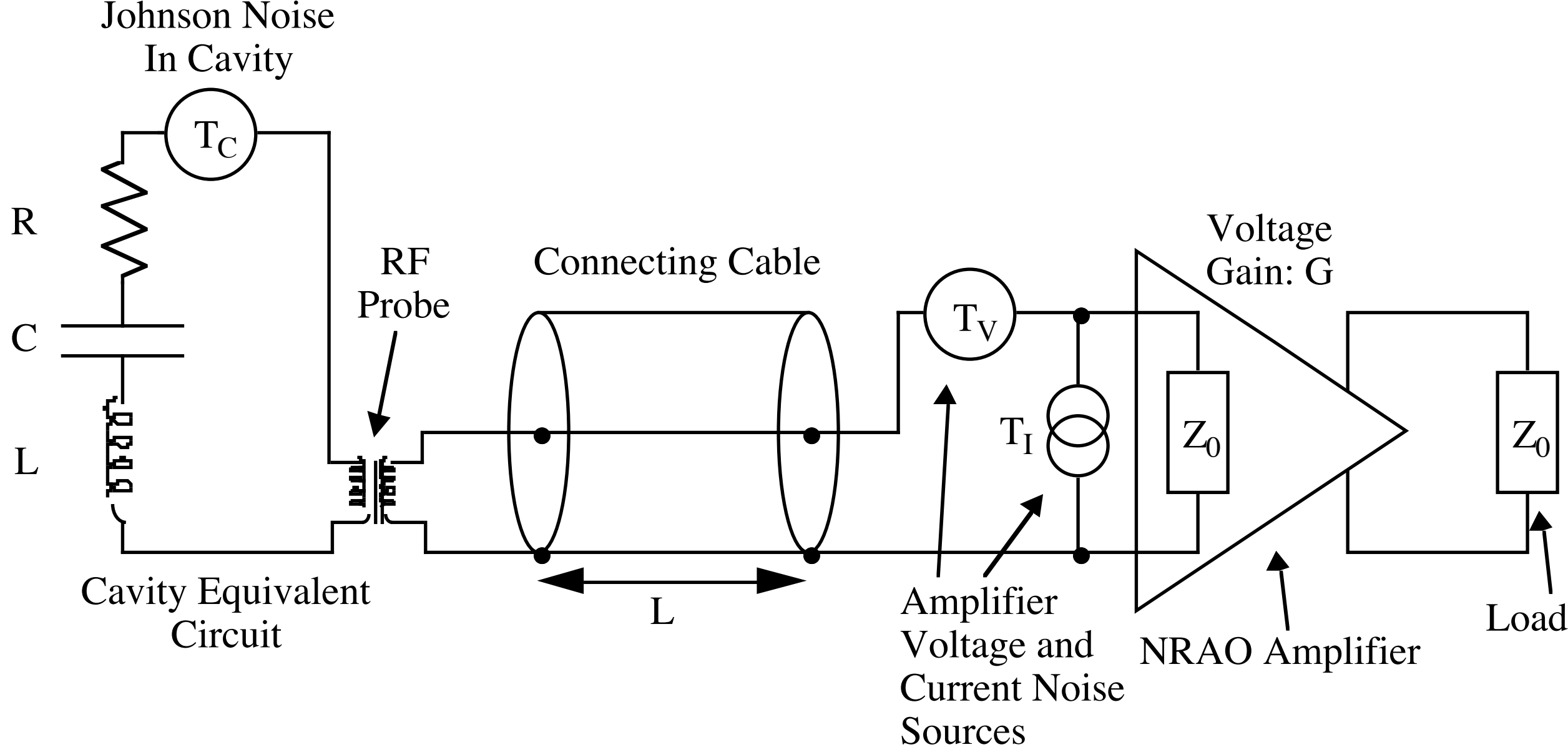

Figure 4.3 The equivalent circuit for the resonant cavity coupled to the 1st cryogenic amplifier

It is important to realize that this equivalent circuit is only an approximation. I have assumed that the amplifier, which is actually a sophisticated balanced design, can be represented as a perfectly matched 50Ω load together with a current and voltage noise sources. I calculate the power spectrum at the amplifier output based on this equivalent circuit in Appendix 4. I give only the result of the calculation here. Define P as the power at the amplifier output and Δ is displacement from the center frequency of the power spectrum in units of 125Hz. Then the power spectrum from the equivalent circuit is:

$$P(\Delta) = \frac{a_1 + 8a_3\left(\frac{\Delta - a_5}{a_2}\right)^2 + 4a_4\left(\frac{\Delta - a_5}{a_2}\right)}{1 + 4\left(\frac{\Delta - a_5}{a_2}\right)^2} \tag{4.1}$$

where the coefficients $a_1$ - $a_4$ are:

$$\begin{aligned} a_1 &= T_C + T_I + T_V \\ a_2 &= \frac{f_0}{Q} \\ a_3 &= (T_I + T_V + (T_I - T_V)\cos 2kL) \\ a_4 &= (T_I - T_V)\sin 2kL \end{aligned} \tag{4.2}$$

where k is the wavenumber corresponding to frequency $f=f_0+125\Delta$, and the rest of the parameters are defined in the equivalent circuit of Figure 4.3. The last coefficient $a_5$ is not an equivalent circuit parameter. Remember that the receiver downconverts power from the measured cavity resonant frequency to a band centered at 35kHz. Therefore in theory the 100th bin of the receiver corrected spectrum should be the cavity resonant frequency. In reality, the cavity resonant frequency not perfectly measured; there is uncertainty at the ±500Hz level. So $a_5$ is the bin displacement of the cavity resonant frequency from the 100th bin in the receiver corrected trace.

### 4.2.5 Tests of the Equivalent Circuit Model on Run 1 Raw Data

Unfortunately, many of the physical parameters used in the equivalent circuit of Figure 4.3 are not measurable. However, Equation 4.1 can still be used as a fit function with 5 free parameters to fit the raw traces. The smooth lines overlaid on the traces of Figure 4.2 are the result of fitting the 4 traces to the fit function 4.1. By eye, the fit function does a good job of finding the smooth background level.

There are two more tests to assess the performance of the fit on data over the entire range of cavity frequencies. The first test is to fit each raw trace in the data set to my fit function. The fit parameters that result are plotted as a function of frequency. I then see if the frequency dependence of the fit parameters agrees with the frequency dependence I would expect from the definitions of the fit parameters in (4.2). Figure 4.4 shows fit parameters $a_2$ - $a_4$ plotted as a function of frequency with the fit applied to a representative sample of the traces from the run 1 raw data.

Is the variation of the fit parameters with frequency shown in Figure 4.8 consistent with the definitions in Equation 4.2? The simplest fit parameter is $a_2$ which is the width of the cavity resonance. Referring to the upper plot of figure 4.8, I have plotted the value of $a_2$ along with the resonance width measured by the network analyzer. The agreement is good.

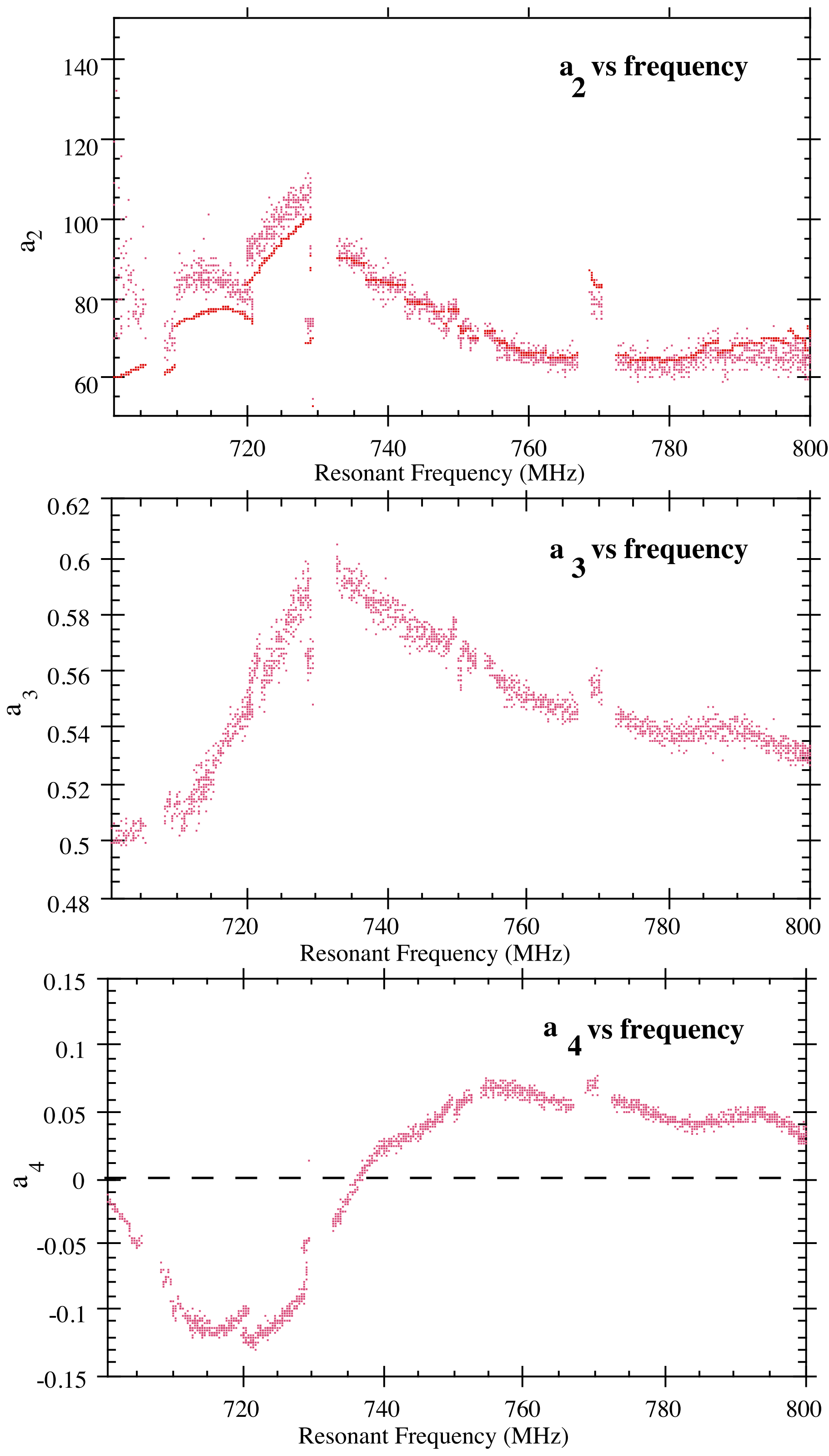

Figure 4.4 Fit Parameters $a_2$, $a_3$, and $a_4$ vs. resonant frequency for each trace in run 1.

Now consider the third plot in Figure 4.4, of $a_4$ vs. frequency. From Equation 4.2, this fit parameter is a numerical coefficient times the sine of twice the electrical length of the cable connecting cavity to the amplifier. If the noise temperatures $T_V$ and $T_I$ were constant with frequency I would expect this fit parameter to oscillate about zero as the frequency, and hence the electrical length of the cable, are increased. The plot of this fit parameter is clearly not a sine wave, but the plot does oscillate about zero, and I do not expect a sine wave anyway since the current and voltage noise in the amplifier are in reality frequency dependent. Furthermore, from Equation 4.2 we might expect $a_3$ vs. frequency to be a constant plus an oscillating term $\pi/2$ out of phase with the plot of $a_4$. Indeed, at the lower frequencies you can see that the zeros of $a_4$ at 700 and 735MHz seem to correspond to a minimum and maximum of $a_3$ which is consistent with the $a_3$ being $\pi/2$ out of phase with $a_4$. I conclude that, though my equivalent circuit model is an approximation intended only as a sensible guess at the origin of the shapes of power spectra encountered, there are encouraging signs that the behavior of the fit parameters with frequency are consistent with their definitions in my equivalent circuit model.

### 4.2.6 Fit Parameters and the Radiometer Equation

The purpose of doing the 5 parameter fit to each receiver corrected trace is to draw a line which represents our best estimate of the average power per 125Hz bin for each bin of each receiver normalized trace. The bin to bin statistical fluctuations about this average power level should be Gaussian distributed and the rms size of those fluctuations should be consistent with the radiometer equation. I shall now check that this is the case for real data.

Recall that each trace is divided by the corresponding fit function, and the resulting trace consists of fluctuations about one, since on average the ratio of the power in a 125Hz bin to the average power in the same bin will be unity. I subtract one from each bin and am left with Gaussian distributed fluctuations about zero. Figure 4.9 is the receiver normalized trace from the top left corner of Figure 4.2 normalized again using the 5 parameter fit.

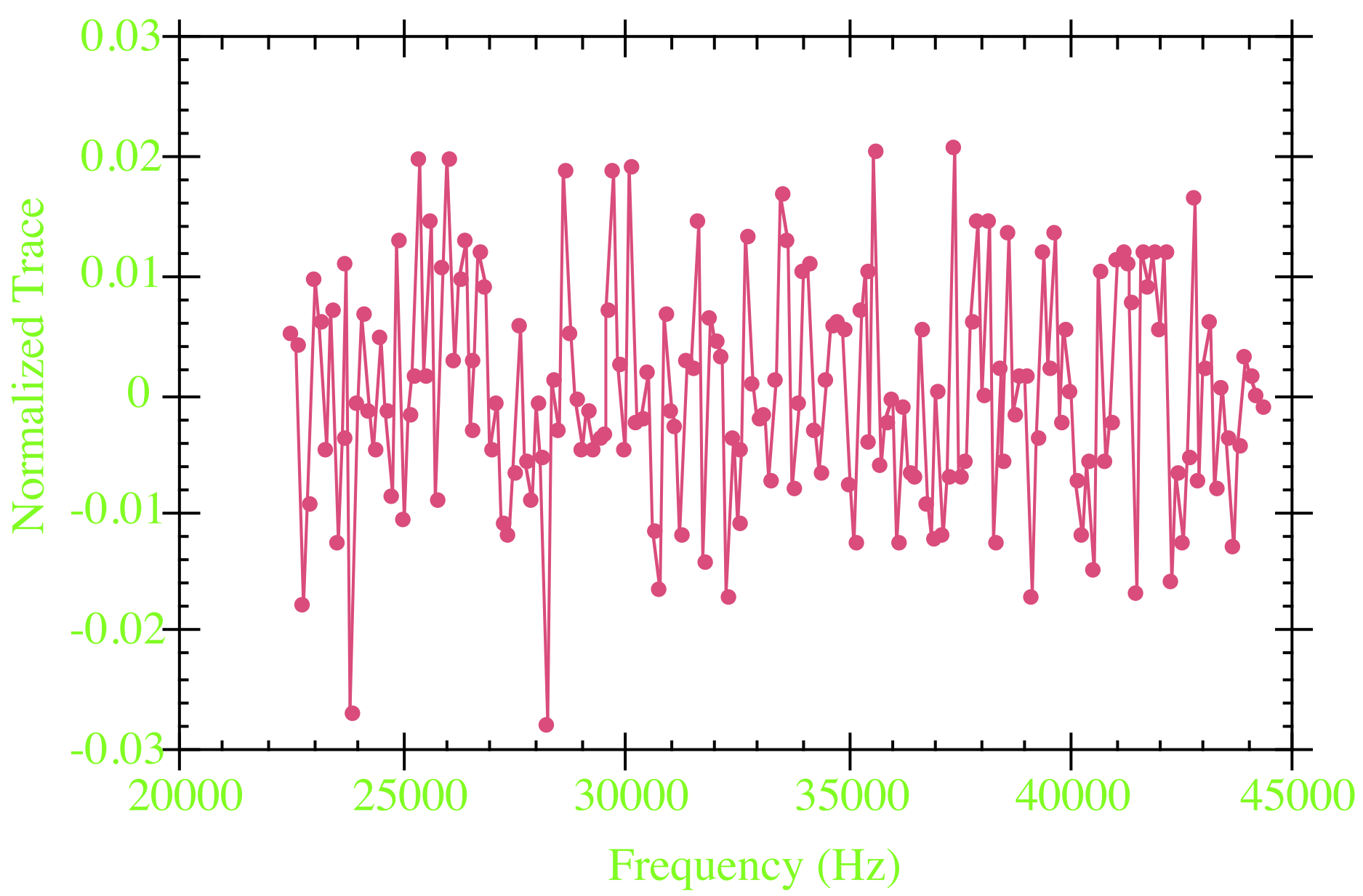

Figure 4.5 A single receiver normalized trace after normalization to the 5 parameter fit function.

From now on I will refer to the fluctuations about zero in the normalized traces as 'deltas'. Let me emphasize at this stage that these deltas are dimensionless, since after processing each is the ratio a power fluctuation to an average power, and both quantities are in Watts. An important quantity for our analysis is the ensemble rms size of these fluctuations for a single trace, $\sigma$.

$$\sigma = \sqrt{\sum_{i}^{n_b} \frac{\delta_i^2}{n_b}} \qquad (4.3)$$

where $n_b$ is the number of bins and $\delta_i$ is the delta from the ith bin. There happens to be a simple relationship between $\sigma$ and N, the number of spectra used in the linear average taken by the FFT to form the trace. This relationship comes from the radiometer equation.

$$\frac{P_S}{\sigma_{P_N}} = \frac{P_S}{P_N}\sqrt{N} \qquad (4.4)$$

where $P_s$ is the power level of a signal, $P_n$ is the average noise power and $\sigma_{P_N}$ is rms size of the noise power fluctuations. Canceling the signal power, I get:

$$\frac{\sigma_{P_N}}{P_N} = \frac{1}{\sqrt{N}} \qquad (4.5)$$

$$\text{or,} \quad \sigma = \frac{1}{\sqrt{N}} \qquad (4.6)$$

The last formula, a familiar result in statistics, provides a simple relationship between the rms spread of points in our normalized trace and the number of spectra averaged together to form the trace. In reality, $1/\sqrt{N}$ is the expectation value of the rms spread for an average of N spectra; because I only have a finite number of bins (175) in each trace I expect $\sigma$ to be distributed about $1/\sqrt{N}$. Let u be the sum of the squares of the deltas. Then the probability distribution for u is:

$$p(u = \sum_{i=1}^{n} \delta_i^2) = \frac{u^{\frac{n}{2}-1} e^{-\frac{u}{2\sigma^2}}}{\sigma^n 2^{\frac{n}{2}} \Gamma(\frac{n}{2})} \qquad (4.7)$$

where the rms spread in the underlying distribution is $\sigma$ given by Equation 4.6. From the effective chi squared distribution I deduce the probability distribution for the number of averages $N_{eff}$ inferred from the rms deviation of 175 deltas. The distribution is:

$$p(N_{eff}) = p(u)\left|\frac{du}{dN_{eff}}\right| = (\frac{n}{2})^{\frac{n}{2}} \frac{N_{eff}^{-(1+\frac{n}{2})} e^{\frac{-n}{2N_{eff}\sigma^2}}}{\sigma^n \Gamma(\frac{n}{2})} \qquad (4.8)$$

Figure 4.6 is a histogram of $N_{eff}$ calculated using Equation 4.6 from the measured rms spread for all the normalized traces in the 1st run data. There are approximately 250,000 points used for the histogram, and the points represent the heights of the histogram bins.

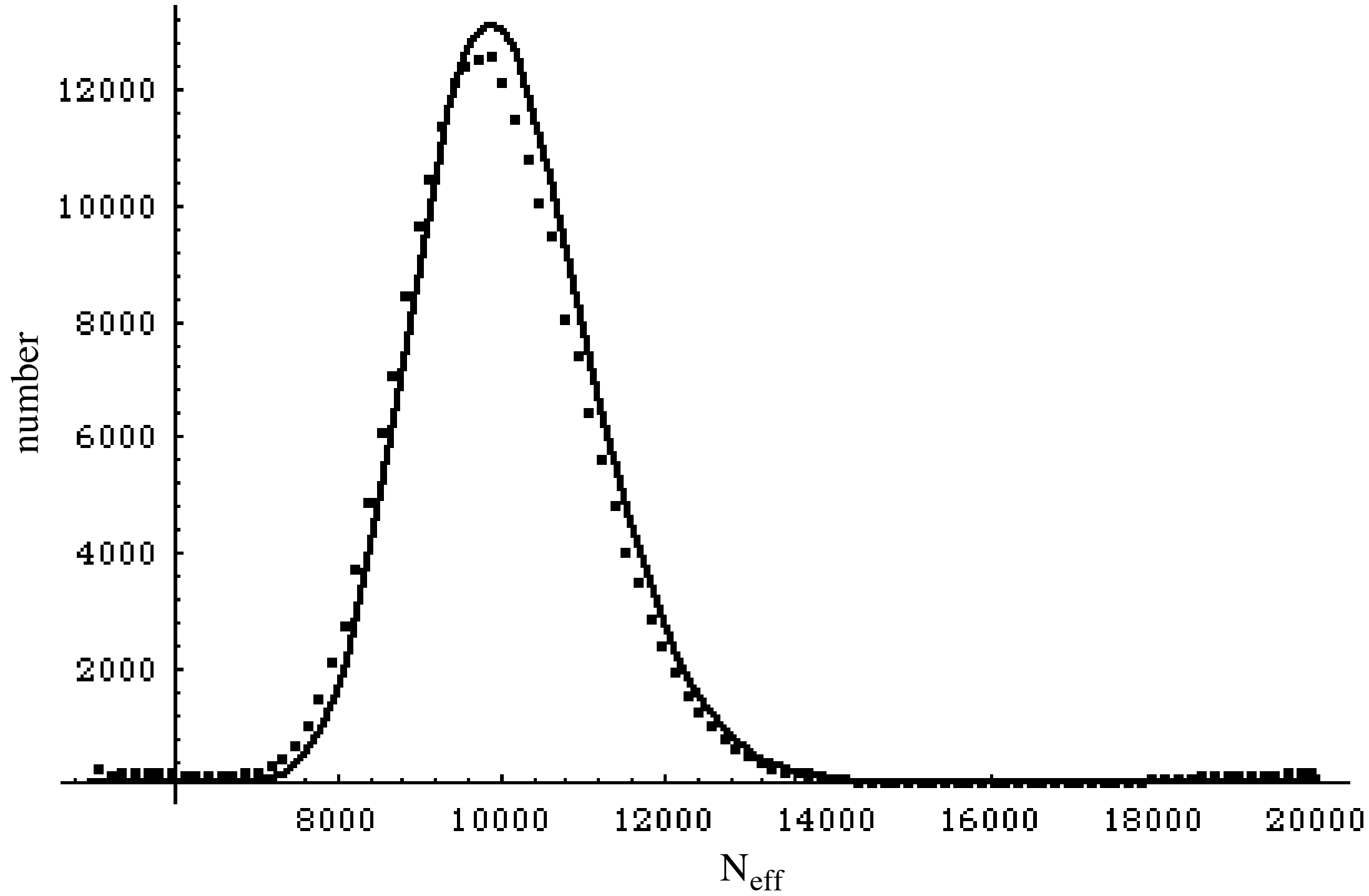

Figure 4.6 Histogram of $N_{eff}$ inferred by from the measured rms spread in each receiver and fit normalized trace in the run 1 data. The points are the heights of histogram bins and the smooth line is the theoretical probability distribution from Equation (4.8).

The measured distribution is centered on just under 10,000 which is the number of spectra averaged for each trace in the first run data. The smooth curve is equation 4.8 multiplied by the number of data points in the histogram; the agreement between the theoretical and measured distributions is good. I therefore have confidence that I do not have a large contribution from spurious non-Poisson or non-stationary noise. The remainder of the analysis, which depends on stationary Poisson noise, is validated.

### 4.2.7 Signal Power in Normalized Traces

Consider the same normalized trace shown in Figure 4.5, except that I have inserted an artificial candidate peak at 32.5kHz. This is shown in Figure 4.7:

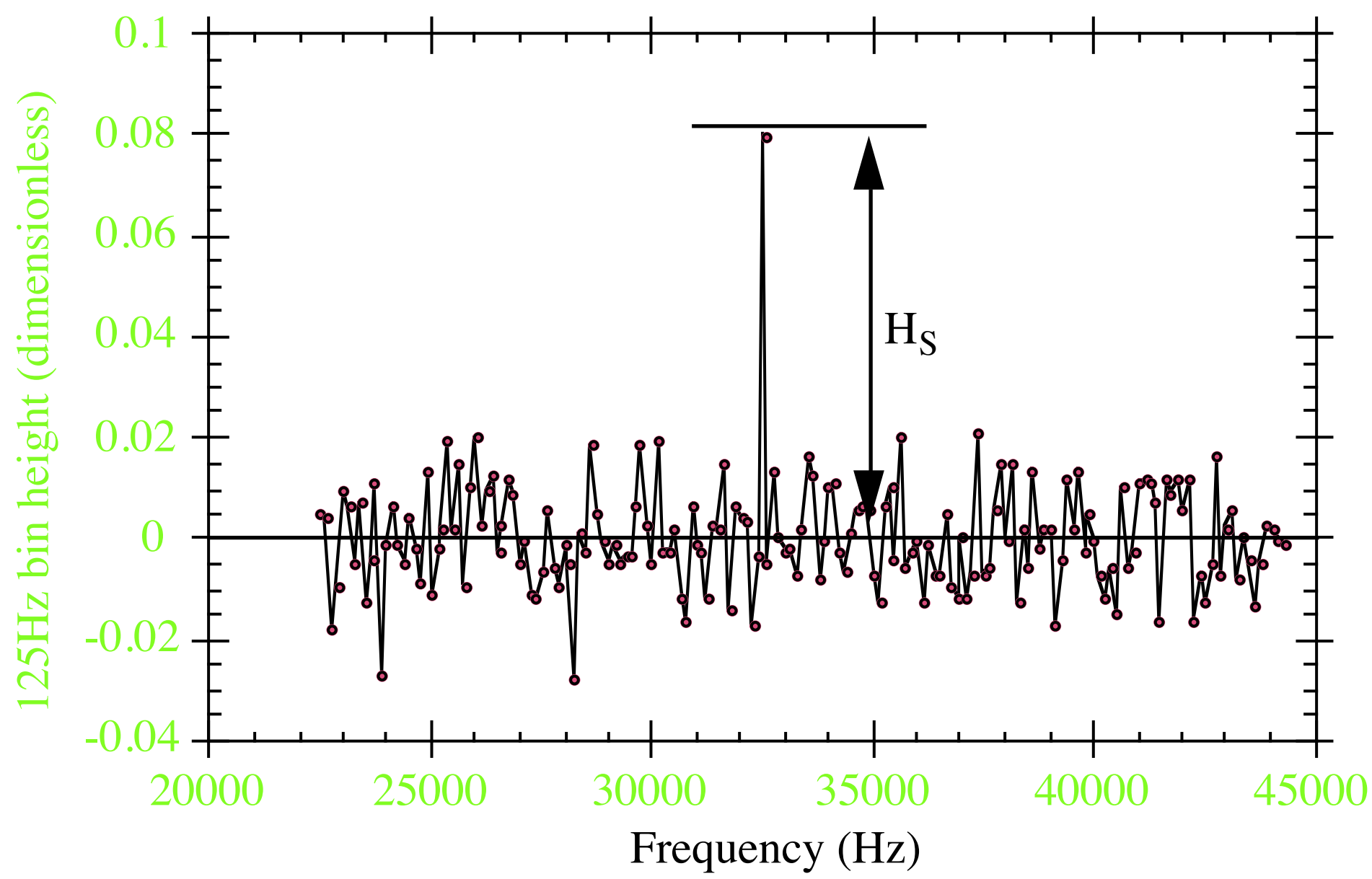

Figure 4.7 A single trace from the raw data with an artificial single bin peak inserted

What power in Watts results in a peak of height $H_S$ on this dimensionless scale? I use the definition of $\sigma$:

$$\sigma = \frac{\sigma P_N}{\overline{P_N}} \qquad (4.9)$$

The average power $\overline{P_N}$ can be expressed in terms of the receiver noise temperature $T_N$, the bin width at the FFT B, and Bolzmann's constant $k_B$.

$$\overline{P_N} = k_B T_N B \qquad (4.10)$$

$$\text{hence } \sigma P_N = \sigma k_B T_N B \qquad (4.11)$$

$$\text{and the signal power is given by } P_S = \frac{H_S}{\sigma} \sigma k_B T_N B \qquad (4.12)$$

$$\text{or } P_S = H_S k_B T_N B \qquad (4.13)$$

How does the system noise temperature affect the sensitivity? Consider: if an axion gives rise to power $P_S$ at some frequency, the height $H_S$ of the signal seen in the normalized trace is inversely proportional to the system noise temperature.

I explore the issue of the effect of system noise temperature by considering two otherwise equivalent traces at two different system noise temperatures. In both cases, a fake axion peak at the same conversion power is overlaid on the middle bin. The height of the peak relative to the Gaussian noise fluctuations is shown in Figure 4.8.

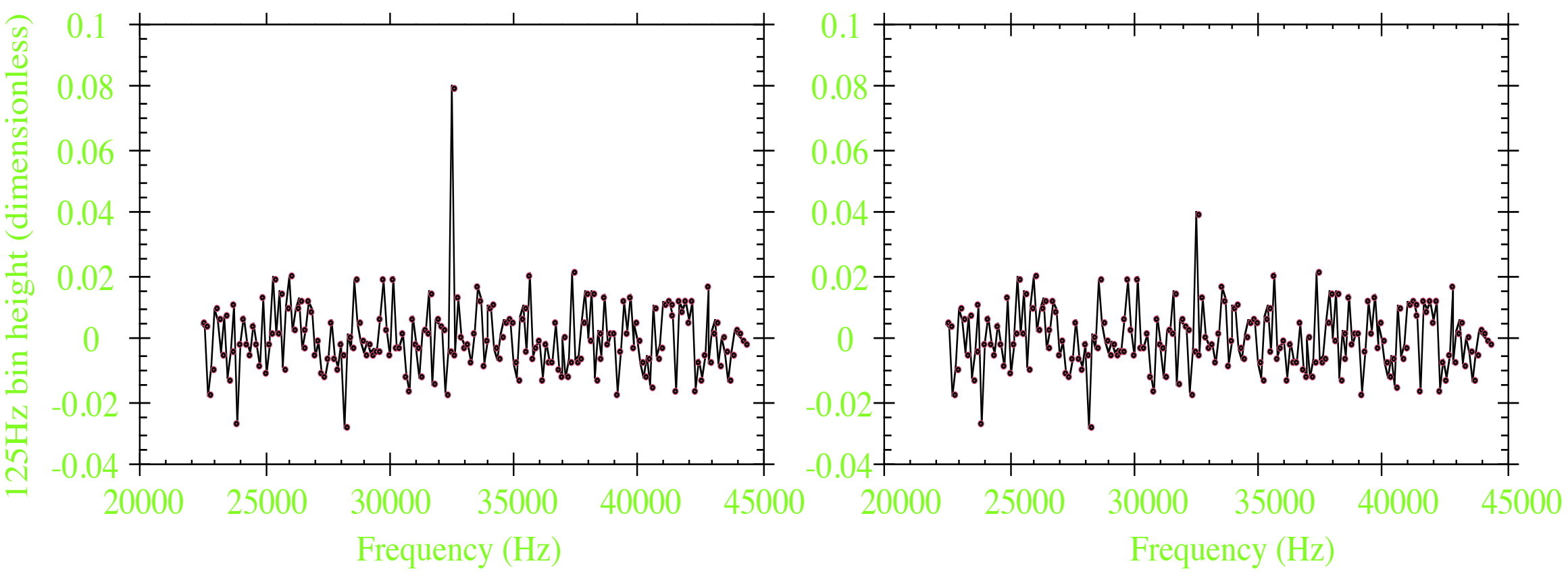

Figure 4.8 An artificial peak of a fixed power inserted into a normalized trace for 2 values of the noise temperature differing by a factor of 2. The rms spread in the noise background is the same in the two cases, but the height of the peak is halved if the noise temperature is doubled.

As doubling the noise temperature doubles both the average noise power and the rms noise power fluctuations, the rms of the deltas, $\sigma$, is the same in the two cases. However, the normalized signal height is degraded by a factor of two with a doubling of $T_N$. This is because the rms noise power fluctuation doubles, but the signal power remains the same.

It seems somehow more direct to have traces where a fixed signal power leads to a peak height independent of $T_N$ with the rms noise fluctuation proportional to the system noise temperature. How do I achieve this? Just multiply the deltas by $k_B T_N B$. After this operation the traces from Figure 4.8 are transformed into those in Figure 4.9.

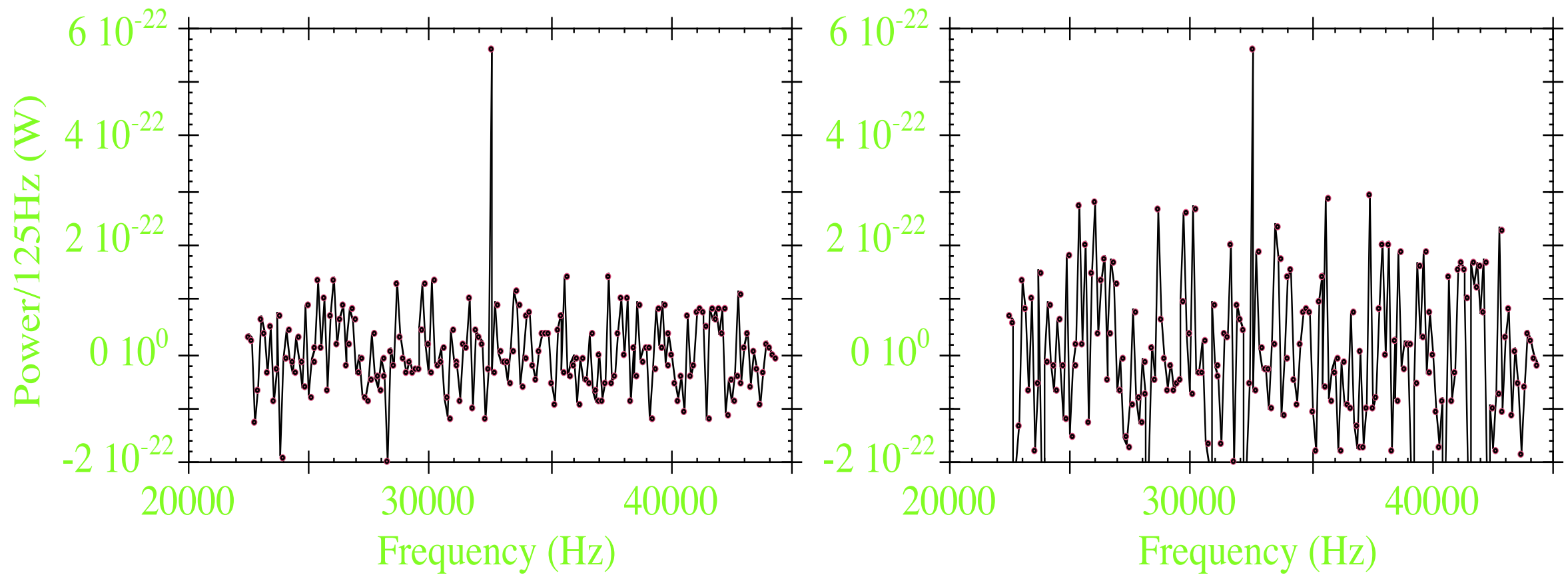

Figure 4.9 Receiver and fit corrected traces after multiplication by $k_B T_N B$. The size of the noise fluctuations now reflects the noise temperature and the height of the artificially injected peak is the same in both cases.

Note that the fluctuations are now in units of Watts. Particularly note that the deltas are now power (in Watts) referenced to the input of the first cryogenic amplifier. If the rms of the fluctuations in noise power at the input of the first stage cryogenic amplifier is $10^{-22}$W, then in the processed data stream, the rms fluctuation in the deltas is $10^{-22}$W. This correspondence permits a direct and simple estimation of our axion power sensitivity.

## 4.2.8 Signal to Noise Ratio in a Single Trace

Using results from Section 4.2.7 I now calculate, as an exercise, the signal to noise ratio for a hypothetical KSVZ axion in a single trace in a raw data file. A single trace is the linear average of $10^4$ power spectra, so $\sigma$ will typically be about $10^{-2}$. Assume the noise temperature of the amplifier is 4.6K, and the cavity physical temperature is 1.3K, the total system noise temperature is 5.9K. Hence the rms of the background noise fluctuations $\sigma^w$ is:

$$\sigma^w = k_B T_N B\sigma = 1.4 \times 10^{-23} \times 5.9 \times 125 \times 0.01 = 10^{-22}\,\mathrm{W} \qquad (4.14)$$

The power in the cavity due to a KSVZ axion is calculated from Equation 2.28. Assuming V=220l, B=7.6T, $f_{nlm}$=0.5, $g_\gamma$=0.97, $\rho_a = 0.45\,\mathrm{GeV/cc}$, $f_a$=750MHz, $Q_C$=70,000 I obtain a power level of $7.5 \times 10^{-22}$W from axion to photon conversion. At critical coupling half of this power, $3.3 \times 10^{-22}$W, enters the amplifier. So the ratio of the power entering the amplifier to the rms noise per 125Hz is 3.3. This signal to noise ratio is not high enough for an axion search in a single trace. However, in real data taking many traces overlap with every cavity frequency. In the next section I discuss a method for combining traces. The signal to noise ratio in the combined data set will be much higher.

### 4.2.9 Factors Affecting the Relative Weights of Data from Different Raw Traces

As mentioned in Section 3.1, the change in $TM_{010}$ mode frequency between adjacent traces is typically 1/15th of the width of the receiver passband. As a result, there will be many traces that contain a data point in the same 125Hz bin of frequency space. I must decide on a scheme by which these data points may be combined in a manner which maximizes our sensitivity. I wish to discuss four factors which determine the weight given to a data point in the combined data. Firstly, the system noise temperature $T_N$ when the trace was taken. Secondly, the number of spectra N averaged to form the trace. Thirdly, the difference between the frequency of the data bin and the center frequency of the $TM_{010}$ mode. And finally the power from axion to photon conversion which may vary from trace to trace due to changes in the magnetic field, cavity mode form factor, etc.

As discussed in section 4.8, the initially dimensionless deltas are multiplied by $kT_NB$ so they are the power fluctuations (in watts) referenced to the input of the first cryogenic amplifier. Hence the rms deviation for the deltas is proportional to the system noise temperature. From the radiometer equation, the rms noise fluctuations are also proportional to $N^{-1/2}$.Thus the rms noise fluctuations in watts, $\sigma^W$, can be used to derive both N and $T_N$. Recall that after processing, the height of an axion induced signal at constant power is independent of the noise temperature. Hence the signal to noise ratio is inversely proportional to $T_N$ and directly proportional to $\sqrt{N}$ , as expected

Next, I discuss the effect of the position of the $TM_{010}$ mode on the weighting of a bin from a trace. I define h as the ratio of the signal height in a bin from a-$\gamma$ conversion, to the signal height seen if the bin is at the center of the $TM_{010}$ mode resonance. Thus if the bin is at resonance, h=1. If the bin is far off resonance, h=0. Between these two extremes, h as a function of frequency is Lorentzian. Recall that in the experiment, the first local oscillator is set to center the cavity resonance in the middle of the trace. If n is the number of 125Hz wide bins to the center of the resonance then h is given by:

$$h(n) = \frac{1}{1 + \frac{4(125)^2 n^2}{\Gamma^2}} \qquad (4.15)$$

The factor of 4 comes from the definition of Γ as the Lorentzian full width at half height.

Finally, the signal to noise ratio is proportional to the power P from axion-photon conversion. This power is dependent on the physical parameters of the detector (see Equation 2.28)

### 4.2.10 The Scheme for Combining Raw Traces

In this section I will state without proof the weighting strategy used in combining our raw traces. Proof that this weighting scheme leads to a maximal signal to noise ratio in the weighted combined data follows in Section 4.2.11. In section 4.2.12 I write down some results that are useful when using the combined data sets.

Suppose I have two traces that overlap. This means that the two traces have frequency bins that were downconverted from the same frequency in the resonant cavity. Let the bin from the 1st (2nd) trace have power excess $\delta_1$ ($\delta_2$), Lorentzian height $h_1$ ($h_2$) rms fluctuations $\sigma_1^w$ ($\sigma_2^w$) and theoretical axion signal power from Equation 2.28 of $P_1$($P_2$). Then the weighted sum of the power excesses from two data sets $\delta_{ws}$ is:

$$\delta_{ws} = \frac{h_1 P_1 \delta_1}{{\sigma_1^w}^2} + \frac{h_2 P_2 \delta_2}{{\sigma_2^w}^2} \qquad (4.16)$$

The signal to noise ratio S in each trace is given by S=hP/σ for that trace. The signal to noise ratio for the weighted sum data is the sum in quadrature of the signal to noise ratios in the two raw traces.

$$S_{WS} = \sqrt{\left(\frac{h_1 P_1}{\sigma_1}\right)^2 + \left(\frac{h_2 P_2}{\sigma_2}\right)^2} = \sqrt{S_1^2 + S_2^2} \qquad (4.17)$$

These results also apply when combining any number of raw traces

### 4.2.11 Proof that the Scheme for Combining Raw Traces Maximizes the Signal to Noise Ratio in the Combined Data.

I now find a weighted average for combining contributions to a single bin from two traces that the signal to noise ratio (SNR) in the combined data. The result turns out to be the scheme quoted without proof above. Let the relevant bin in trace 1 have $h=h_1$ and let the trace have rms fluctuations $\sigma^W=\sigma_1^W$ .and axion conversion power $P_1$. Similarly, for the relevant bin from the second trace, $h=h_2$ , $\sigma^W=\sigma_2^W$ and the conversion power is $P_2$. For the weighted average, I add delta from trace 1 to delta from trace 2 multiplied by c, a weighting factor. I wish to maximize the ratio of the total signal power $P_{tot}$ to the rms noise $\sigma_{tot}^w$ in the combined data. $P_{tot}$ is proportional to the weighted sum of the signal heights in the two spectra. If a is the proportionality constant then $P_{tot}$ is:

$$P_{TOT} = a(h_1P_1 + ch_2P_2) \quad (4.18)$$

$\sigma_{TOT}^W$ is given by:

$$\sigma_{tot}^w = \sqrt{\sigma_1^{w2} + c^2\sigma_2^{w2}} \quad (4.19)$$

To maximize sensitivity in the combined data point, I maximize $f(c) = P_{TOT} / \sigma_{TOT}^W$ with respect to c. Figure 4.10 is a family of f(c) curves, for different σ and $P_{TOT}$.

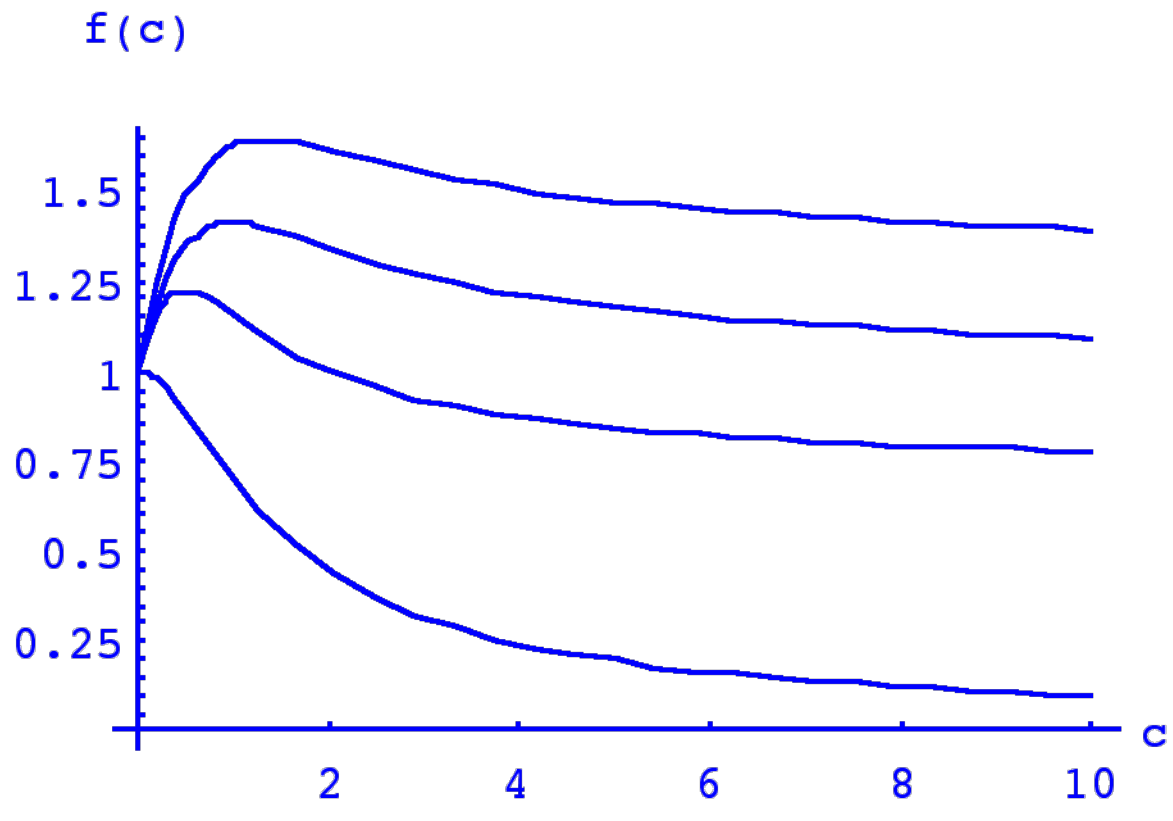

Figure 4.10. f(c) for different values of $h_1$,$h_2$,$P_1$ and $P_2$

$$\text{The maxima are at } c = \frac{h_2P_2\sigma_1^{w2}}{h_1P_1\sigma_2^{w2}} \quad (4.20)$$

Note that this is equivalent to the more familiar procedure of multiplying delta from the first trace $\delta_1$ by $h_1P_1/\sigma_1^{w^2}$ and $\delta_2$ by $h_2P_2/\sigma_2^{w^2}$. Does this weighting scheme extend to the case of contributions from an arbitrary number of traces?

I will prove by induction that the extension of the above scheme to an arbitrary number of traces is valid. First assume that the method is optimal for the weighted average of deltas from k traces, where the bin from the $i^{th}$ trace is parameterised by $h_i$, $\sigma_i$ and $P_i$. Then the weight $w_i$ by which the ith raw trace is multiplied to form the optimal weighted sum is:

$$w_i = \frac{h_i P_i}{\sigma_i^{w^2}} \quad (4.21)$$

In the weighted sum of k traces the signal height $h_k^{total}P_k^{total}$ is:

$$h_k^{total}P_k^{total} = h_1P_1 \times \frac{h_1P_1}{\sigma_1^{w2}} + h_2P_2 \times \frac{h_2P_2}{\sigma_2^{w^2}} + \ldots + h_kP_k \times \frac{h_kP_k}{\sigma_k^{w2}} = \sum_i^k \frac{h_i^2 P_i^{\,2}}{\sigma_i^{w^2}} \quad (4.22)$$

Similarly the rms noise level $\sigma_k^{w,total}$,

$$\sigma_{k1}^{w,total} = \sqrt{\sigma_1^{w^2} \times \left(\frac{h_1P_1}{\sigma_1^{w^2}}\right)^2 + \sigma_2^{w2} \times \left(\frac{h_2P_2}{\sigma_2^{w^2}}\right)^2 + \ldots + \sigma_k^{w^2} \times \left(\frac{h_kP_k}{\sigma_k^{w^2}}\right)^2} = \sqrt{\sum_i^k \frac{h_i^2P_i^{\,2}}{\sigma_i^{w^2}}} \quad (4.23)$$

I now treat the delta from the weighted average as a single data point, and combine it with a delta from a k+1th trace, parameterised by $h_{k+1}$ and $\sigma_{k+1}$. The weighting factor for delta from the combined traces is $h_k^{total}P_k^{total}/\sigma_k^{w,total^2} = 1$. The weighting factor for the $k+1^{th}$ delta is $h_{k+1}P_{k+1}/\sigma_{k+1}^{w\;\;2}$ The overall signal height $h_{k+1}^{total}$ and $\sigma_{k+1}^{w,total}$ are

$$h_{k+1}^{total}P_{k+1}^{total} = \left(\sum_i^k \frac{h_i^2P_i^{\,2}}{\sigma_i^{w^2}}\right) \times 1 + h_{k+1}P_{k+1} \times \frac{h_{k+1}P_{k+1}}{\sigma_{k+1}^{w\;\;2}} = \sum_i^{k+1} \frac{h_i^2P_i^{\,2}}{\sigma_i^{w^2}} \quad (4.24)$$

Similarly,

$$\sigma_{k+1}^{w,total} = \sqrt{\left(\sqrt{\sum_i^k \frac{h_i^2P_i^{\,2}}{\sigma_i^{w^2}}}\right)^2 \times 1^2 + \sigma_{k+1}^{w\;\;2} \times \left(\frac{h_{k+1}P_{k+1}}{\sigma_{k+1}^{w\;\;2}}\right)^2} = \sqrt{\sum_i^{k+1} \frac{h_i^2P_i^{\,2}}{\sigma_i^{w^2}}} \quad (4.25)$$

These are of the same form as $h^{total}$ and $\sigma^{w,total}$ for the weighted average of k deltas. Hence I have shown that, given that my proposed method for forming a weighted sum of deltas from k traces, the signal height and rms noise level in the weighted sum of k+1 traces has the same algebraic form as the weighted sum of k traces. So if the scheme is optimal for k traces, it is also optimal for k+1 traces. Finally, I know it is optimal for k=1, hence my weighting scheme optimizes sensitivity in a weighted average of contributions from an arbitrary number of traces. This concludes the proof.

## 4.11 Some Useful Quantities from the Combined Data

In the previous section I devised a weighting scheme for the deltas such that the weighted sum maximizes our sensitivity to axion signals. The weighted sum of the deltas, $\delta_{WS}$ in a single bin of the combined data, is given by:

$$\delta_{WS} = \sum_i \frac{h_i P_i \delta_i}{\sigma_i^{w^2}} \qquad (4.26)$$

where the index i refers to the ith spectra contributing to the single bin in the combined data stream. The standard deviation for the weighted sum of the deltas, $\sigma_{WS}$ is:

$$\sigma_{WS} = \sqrt{\sum_i \frac{h_i^2 P_i^2}{\sigma_i^{w^2}}} \qquad (4.27)$$

The data array in which we will search for peaks is the deltas in units of the standard deviation, obtained by dividing Equation (4.26) by Equation (4.27):

$$\frac{\delta_{WS}}{\sigma_{WS}} = \frac{\sum_i \frac{h_i P_i \delta_i}{\sigma_i^{w^2}}}{\sqrt{\sum_i \frac{h_i^2 P_i^2}{\sigma_i^{w^2}}}} \qquad (4.28)$$

Notice that neither $\delta_{WS}$ nor $\sigma_{WS}$ is in units of watts; I destroyed the normalization by multiplying each delta contributing to a bin in the combined data stream by a weighting factor, $w_i$, that had dimensions of 1/watts (see Equation 4.21). To restore the correct normalization to the weighted sum of the deltas, divide by the sum of the weighting factors, given by:

$$\sum_i w_i = \sum_i \frac{h_i P_i}{\sigma_i^{w^2}} \qquad (4.29)$$

Using the sum of the $w_i$ s to normalize the weighted sum of the deltas, I get the following expression for fluctuation about the mean power in watts, $\delta_j(W)$, in the jth bin of the combined data.

$$\delta_j(W) = \frac{\sum_i \frac{h_i P_i \delta_i}{\sigma_i^{w^2}}}{\sum_i \frac{h_i P_i}{\sigma_i^{w^2}}} \qquad (4.30)$$

Similarly, by dividing the RHS of Equation 4.27 by the sum of the weights, I obtain the standard deviation for in watts, $\sigma_j(W)$, in the jth bin of the combined data.

$$\sigma_j(W) = \frac{\sqrt{\sum_i \frac{h_i^2 P_i^{\,2}}{\sigma_i^{w^2}}}}{\sum_i \frac{h_i P_i}{\sigma_i^{w^2}}} \qquad (4.31)$$

Notice that if I divide 4.30 by 4.31 I recover 4.28, delta in units of sigma, as expected. The theoretical signal power in watts $P_S^W$ in the combined data is obtained by dividing the right hand side of Equation 4.24 by the sum of the weighting factors, Equation 4.29:

$$P_S^W = \frac{\sum_i \frac{h_i^2 P_i^{\,2}}{\sigma_i^{w^2}}}{\sum_i \frac{h_i P_i}{\sigma_i^{w^2}}} \qquad (4.32)$$

Finally the signal to noise ratio SNR in the combined data is obtained by dividing the signal power in watts (Equation 4.32) by the rms noise level in watts (equation 4.31). The result is:

$$\mathrm{SNR} = \sqrt{\sum_i \left( \frac{h_i P_i}{\sigma_i^w} \right)^2} \qquad (4.33)$$

The SNR in the optimally combined data is obtained by adding in quadrature the SNRs from the contributing bins in the raw traces. This reproduces the intuitive result stated without proof in Section 4.2.10.

### 4.2.13 The Data Combining Algorithm In Practice

The algorithm for combining the raw traces described in Sections 4.2.10 - 4.2.12 is implemented in the programming language C. Figure 4.12.1 is a block diagram of the procedure. In practice, the weighting function $w_i$ defined in Equation 4.21 is computationally unwieldy because it contains quantities in watts which have magnitudes of order $10^{-22}$. In manipulating these numbers in C software there is a danger on some machines that quantities too small to be represented as a floating point number would be generated. To avoid this possibility, the theoretical axion power P was replaced with $P/P_0$, where $P_0=3.5.10^{-22}$W, and the rms noise level $\sigma^w$ and raw trace bin contents $\delta$ were replaced with $\sigma^w/k_B$ and $\delta/k_B$ respectively. These changes have no effect on the relative weights of different raw traces in the combined data. However, extra numerical factors need to be inserted in order that the quantities discussed in section 4.11 can be derived from the combined data files as they were written in the actual combined data files. In figure 4.12 I have written two of the most important quantities with the numerical factors correctly inserted

The combined data is written to disk in 50MHz blocks, each block beginning at a multiple of 50MHz. The format of the combined data files is as follows. Firstly, a header consisting of 20 4-byte floating point numbers. For the duration of the first data-taking run, the first floating point number stores the start frequency of the combined data file in MHz; the remaining 19 header variables are unused. Immediately following the header are the four arrays listed in Figure 4.12.1. Each array consisted of 400,000 4-byte floating point numbers. The total file size is 6.4MBytes for the 4 arrays + 80Bytes for the header. This data structure allows for the addition of more arrays without the necessity of re-writing pre-existing combined data reading software.

Crop off the 1st 100 and last 125 bins of each raw spectrum. Divide spectrum by IF transferred power. Fit the IF normalized spectrum to a five parameter equivalent circuit model Divide the spectrum by the fit function and subtract 1.

For each trace normalized read/measure:

1) $\Gamma$, the width of the TM010 resonance.
2) $T_N$, the system noise temperature.
3) N, the number of averaged spectra per trace.
4) $\sigma(\delta)$, the rms noise level in the normalized trace.

Multiply the fluctuations $\sigma(\delta)$ in each raw trace by $k_B B$.

Make 4 combined data streams for each 125Hz bin accross the frequency range scanned

1. Number of Raw Spectra Contributing
2. $\sum_i \frac{h_i P_i \delta_i}{\sigma_i^{w2}}$
3. $\sum_i \frac{h_i^2 P_i^2}{\sigma_i^{w2}}$
4. $\sum_i \frac{h_i P_i}{\sigma_i^{w2}}$

data array for peak search

$$\frac{\sum_i \frac{h_i P_i \delta_i}{\sigma_i^{w2}}}{\sqrt{\sum_i \frac{h_i^2 P_i^2}{\sigma_i^{w2}}}}$$

Signal-to-Noise Ratio (SNR) in jth bin of combined data

$$\mathrm{SNR} = \frac{P_0}{k_B}\sqrt{\sum_i \frac{h_i^2 P_i^2}{\sigma_i^{w2}}} = 25.3\sqrt{\sum_i \frac{h_i^2 P_i^2}{\sigma_i^{w2}}}$$

deviation from mean power (W)

$$k_B \frac{\sum_i \frac{h_i P_i \delta_i}{\sigma_i^{w2}}}{\sum_i \frac{h_i P_i}{\sigma_i^{w2}}}$$

rms noise level (W/125Hz)

$$k_B \frac{\sqrt{\sum_i \frac{h_i^2 P_i^2}{\sigma_i^{w2}}}}{\sum_i \frac{h_i P_i}{\sigma_i^{w2}}}$$

power deposited in amplifier by KSVZ axions (halo density $=7.5.10^{-25}\mathrm{gcm}^{-3}$)

$$3.5 \times 10^{-22}\,\mathrm{W}\frac{\sum_i \left(\frac{h_i P_i}{\sigma_i^{w}}\right)^2}{\sum_i \frac{h_i P_i}{\sigma_i^{w2}}}$$

Figure 4.11 A block diagram of the data combining algorithm as implemented in software

## 4.3 The Run 1 Combined Data

The run 1 combined data was generated using the algorithm described in Section 4.2. Peak searches in the run 1 combined data are critical to sensitivity. This is because the number of the first cut in the 1 bin and 6 bin search channels reduce the number of frequencies being searched from $\sim 10^6$ per 100MHz search range to $\sim 10^3$. In the run 1 channel, for example, the number of frequencies passing the run 1 combined data cut was 652 out of the 792,000 125Hz bins between 701 and 800MHz. The 1-bin and 6-bin cuts on the run 1 combined data set are the most important factor in determining the overall sensitivity of the axion search.

### 4.3.1 Cuts on the Raw Data

Between April 8th 1996 and October 25th 1997, production data was taken in the frequency range 701-800MHz, corresponding to a range of axion masses of 2.90-3.31μeV. The raw data consists of $4.2.10^5$ raw traces. Of these, 6058 (1.4%) were eliminated due to disturbances of the axion detector. Three different cuts were used.

The first cut was on traces taken with the pressure in the resonant cavity over 0.8 torr. These high pressures almost always occurred as sudden pressure rises, and the result was rapidly variations in the cavity resonant frequency as the pressure of the helium vapor in the cavity changed and altered its effective volume.

The second cut discarded data which occurred when the change in cavity resonant frequency between successive spectra was more than 7kHz. The data that failed this cut was heavily correlated with the data that failed the first cut, as sudden changes in cavity resonant frequency were often caused by sudden changes in pressure in the cavity. Data taken during rapid frequency changes was likely to correspond to a poorly measured cavity Q and resonant frequency which make meaningful analysis of the power spectra impossible.

The third cut on the raw data was a cut on all traces written when the cavity temperature was over 2K. Such a high cavity temperature is a sign of a failure in the vacuum system or that the detector is warming up because the liquid helium supply has failed.

## 4.3.2 The First Run Data

Figure 4.12 is a plot of resonant frequency vs. time during run 1. This run began on April 8th 1996 and ended on January 24th 1997.

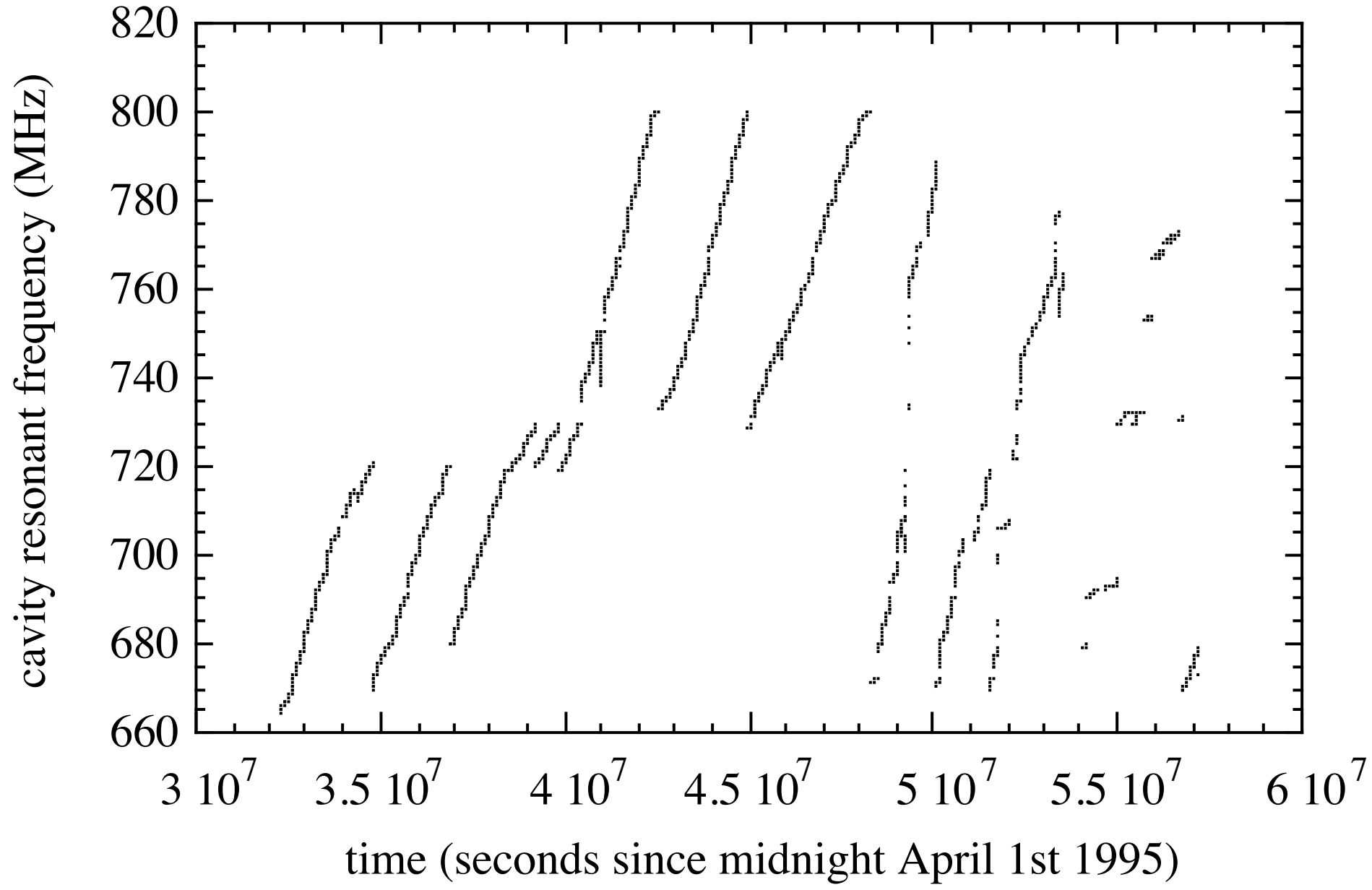

Figure 4.12 Cavity resonant frequency as a function of time during run 1.

Notice that rather than making a single sweep of the frequency range, we made several sweeps which were well separated in time at each frequency. This allows us to reject background from transient RF peaks; if a peak is visible at some frequency, but fails to reoccur at the same frequency weeks or months later, it is unlikely to be an axion.

Also, in this plot you can see that run 1 went through 3 distinct stages. First, 3 sweeps of the entire frequency span were made between April 8th 1996 and October 11th 1996. Secondly, a fourth less continuous sweep of parts of the span was made between October 11th 1996 and December 16th 1996 . Thirdly, the gaps left in the 1st 4 sweeps due to mode crossings were filled in with the cavity flooded with liquid $^4$He as discussed in Section 3.3.4 between December 16th 1996 and January 24th 1997.

### 4.3.3 Signal to Noise Ratio in the Run 1 Combined Data

Run 1 consists of $2.5.10^5$ traces covering the frequency range 701-800MHz. Figure 4.13

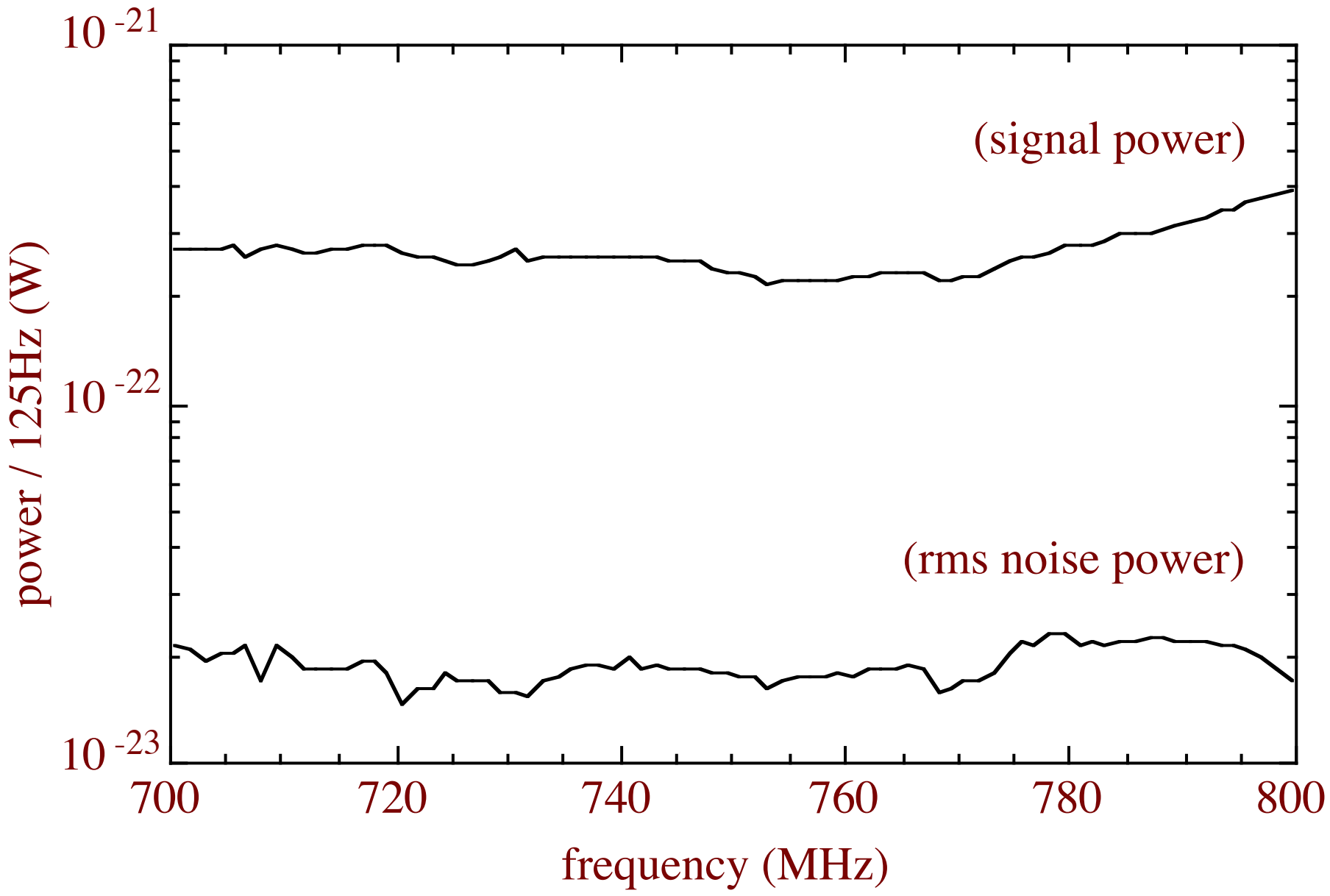

Figure 4.13 Signal and noise power vs. frequency. The noise power is per 125Hz bin.

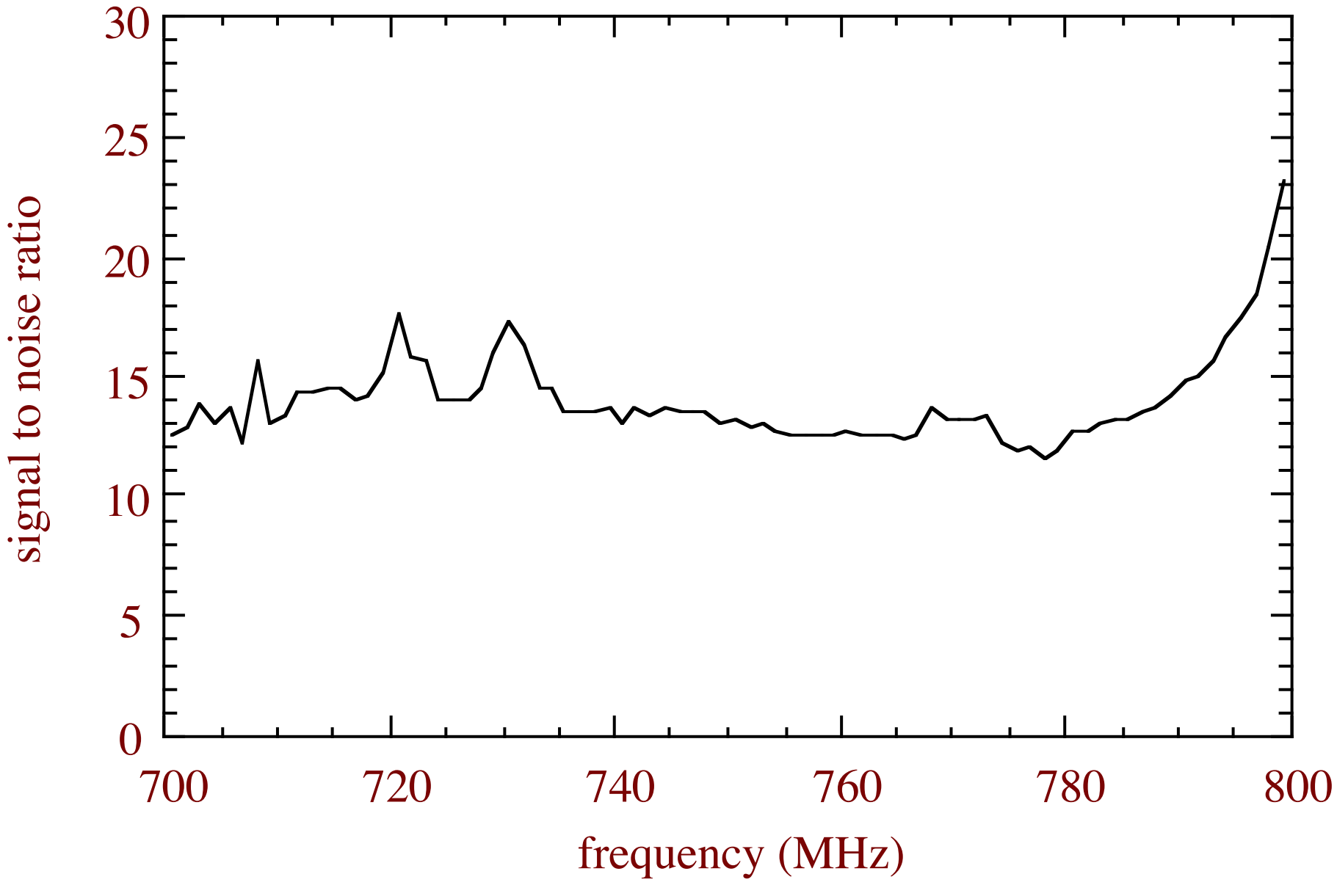

Figure 4.14 Signal to noise ratio vs. frequency

is a plot of expected power from axions deposited in the RF electronics and rms noise level
vs. frequency, generated using the combined data and the results of Section 4.2.12. Figure 4.14 is the signal to noise ratio vs. frequency. I assume a number density of axions of 0.45GeV/cc and the KSVZ axion model. To generate each of these figures, every 100th bin was read, so there were 80 bins read per MHz. These 80 bins were averaged to yield a single value per MHz for each of the quantities displayed.

This signal to noise ratio assumes that all the signal power appears in a single 125Hz bin. In the case where the signal power is spread over 6 bins, the rms noise level increases by a factor of $\sqrt{6}$. This means that the ratio of the KSVZ signal power to the rms noise fluctuations in a 750Hz (6 bin) bandwidth is at least 4.7 everywhere in our frequency range. However, there are additional reductions in the signal to noise ratio in the 6 bin channel which will be discussed in Section 4.5.

### 4.3.4 The Statistics of the Run 1 Combined Data

Figure 4.15 is a histogram of the dispersion of the combined data from run 1, where for each combined data frequency the power excess in units of the rms deviation of power excess is calculated using Equation 4.28. The points are the heights of histogram bins; the smooth line is the theoretical Gaussian of unit sigma for 792,000 data points.

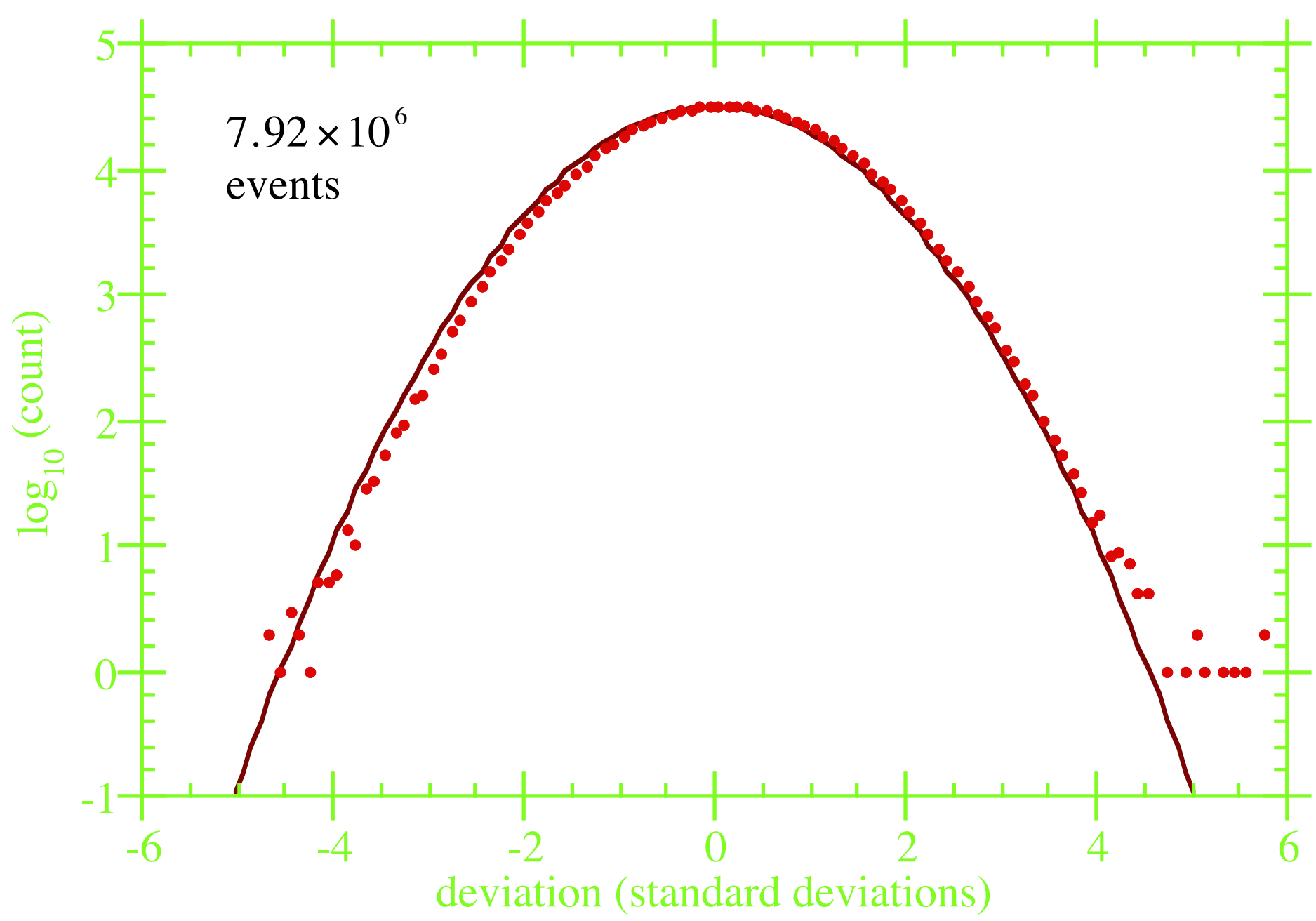

Figure 4.15 Dispersion of the run 1 combined data. The points are the heights of histogram bins, the smooth curve is the theoretical Gaussian.

There is an deviation from Gaussian distribution in this histogram. Specifically, although the width of the histogram looks about right, the mean of the histogram not zero. Clearly, for some reason, the weighted sum made in each bin of the combined data is of variables (deviations from mean power in the raw spectra) that are not perfectly Gaussian distributed. This means that in our analysis we have not done a perfect job of removing the non-Gaussian components of the power spectra discussed in Sections 4.2.2 - 4.2.5.

We should not be surprised at failing to entirely remove the non-Gaussian components from the power spectra. Our knowledge of the underlying physical parameters governing the shape of the power spectra is limited. For instance, there are many noise sources in the HEMT amplifier, yet our equivalent circuit assumes that there are only two; furthermore it completely ignores the effects of the internal $90^0$ hybrids and the directional coupler between the cavity and the amplifier input. Introducing more fit parameters is undesirable since real signals appearing in spectra are unacceptably surpressed. A second example of our imperfect knowledge of the contributions to power spectra shapes is the crystal filter. We have assumed that the transfer function of the IF electronics is well known, and can be divided out of each trace as the first step in data analysis (see section 4.2). The crystal filter was kept in a temperature controlled box and its temperature was regulated to better than $\pm 0.5^0$C, yet it is possible, for example, that some aging process in the crystal causes the transfer function to change over time. Alternatively, one of the other components in the receiver chain could have some time and frequency dependent gain modulation. Our normalization curve (see Figure 3.19) would then be inaccurate.

### 4.3.5 The Origin of the Non-Gaussian Component in the Normalized Power Spectra

In Figure 4.16 there are 25 curves plotted sequentially in time. Each curve is the average of 1000 raw traces that have been normalized for crystal filter transfer function and then normalized to a 5 parameter fit to the resulting spectrum as discussed in Sections 4.2.2 - 4.2.5. The scale on the vertical axis is in units of average power per 125Hz bin, and I have plotted every 5th bin of each trace average along the frequency axis.

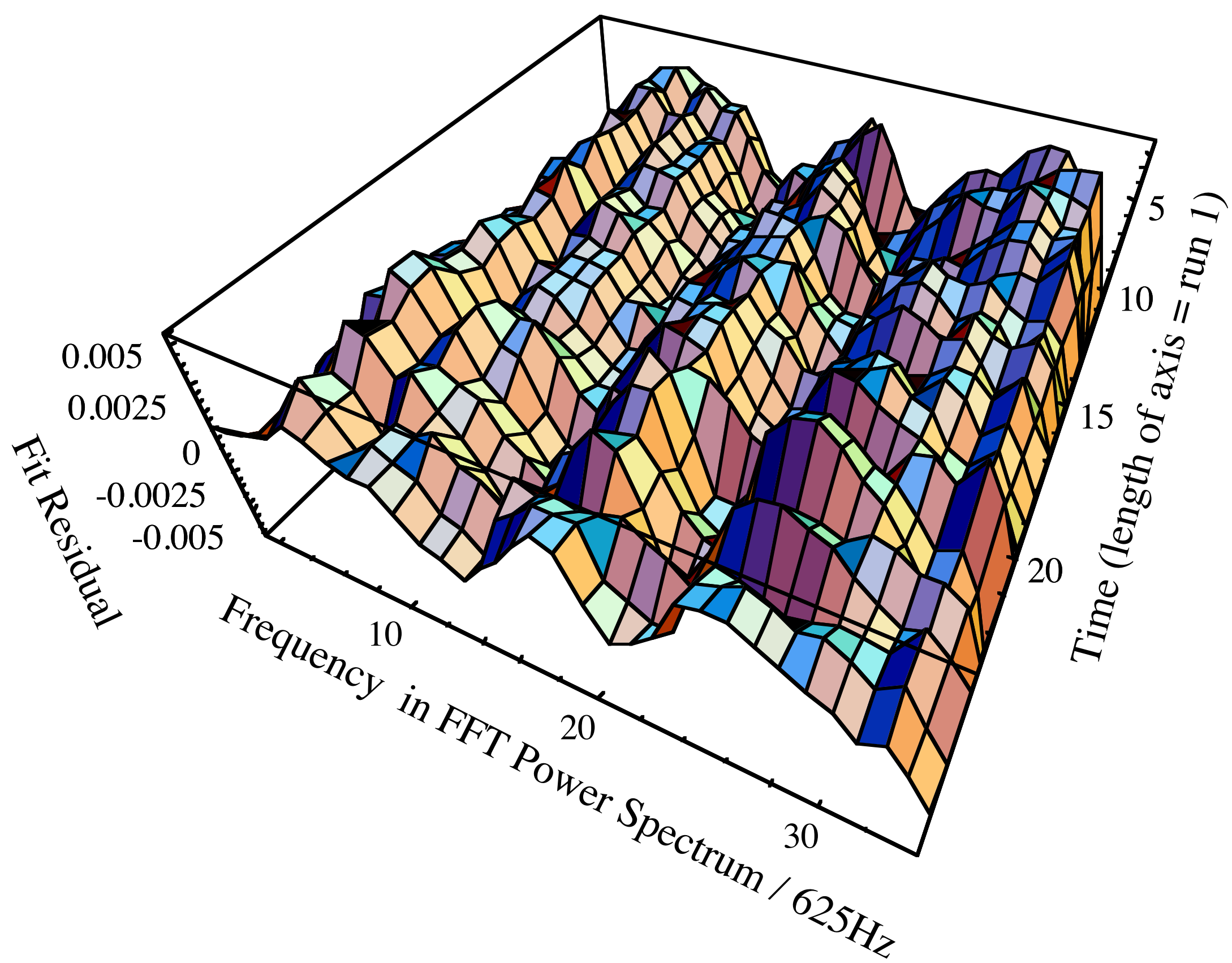

Figure 4.16 Averages of 1000 raw traces ordered sequentially in time, taken throughout run 1. A non-Gaussian component is visible in each average, that appears largely time-independent. On the vertical scale, 0.01 = 1 standard deviation a single raw trace.

Firstly, notice that the non-Gaussian 'residual' component of the power spectrum is clearly visible. If there were no non-Gaussian component, each curve would consist of Gaussian fluctuations about zero with a much smaller amplitude. I will hereafter refer to the non-Gaussian component of a crystal filter normalized trace as the fit residual.

Secondly, notice that the shape of the fit residual is fairly constant in time. As can be seen in Figure (4.2), the shapes of the traces varied considerably throughout the run. It is unlikely that an imperfection in the fit would produce such a time-independent non

Gaussian residual. However, one would expect that an error in the crystal filter transfer function that was very slowly varying in time might lead to a time-independent residual.

Thirdly, notice that the are about 5 ripples common to most of the averages of corrected traces plotted in Figure (4.16). By eye, the bandwidth of the ridges is typically 2-3kHz. If the residuals originated in the RF section of the electronics, you would not expect structure narrower than the cavity full width, which is typically about 7kHz. However, the crystal filter or some other component in the IF section of the receiver electronics could easily generate the residuals seen above.

I conclude that the non-Gaussian component of the power spectra originated in the IF electronics. As a further test of this hypothesis, I average together all the traces taken in run 1 in Figure 4.17. The vertical axis is normalized to the mean power per 125Hz bin. Each trace is the linear average of $10^4$ power spectra, so by the radiometer equation, the rms scatter in a single trace is 0.01 of the mean power.

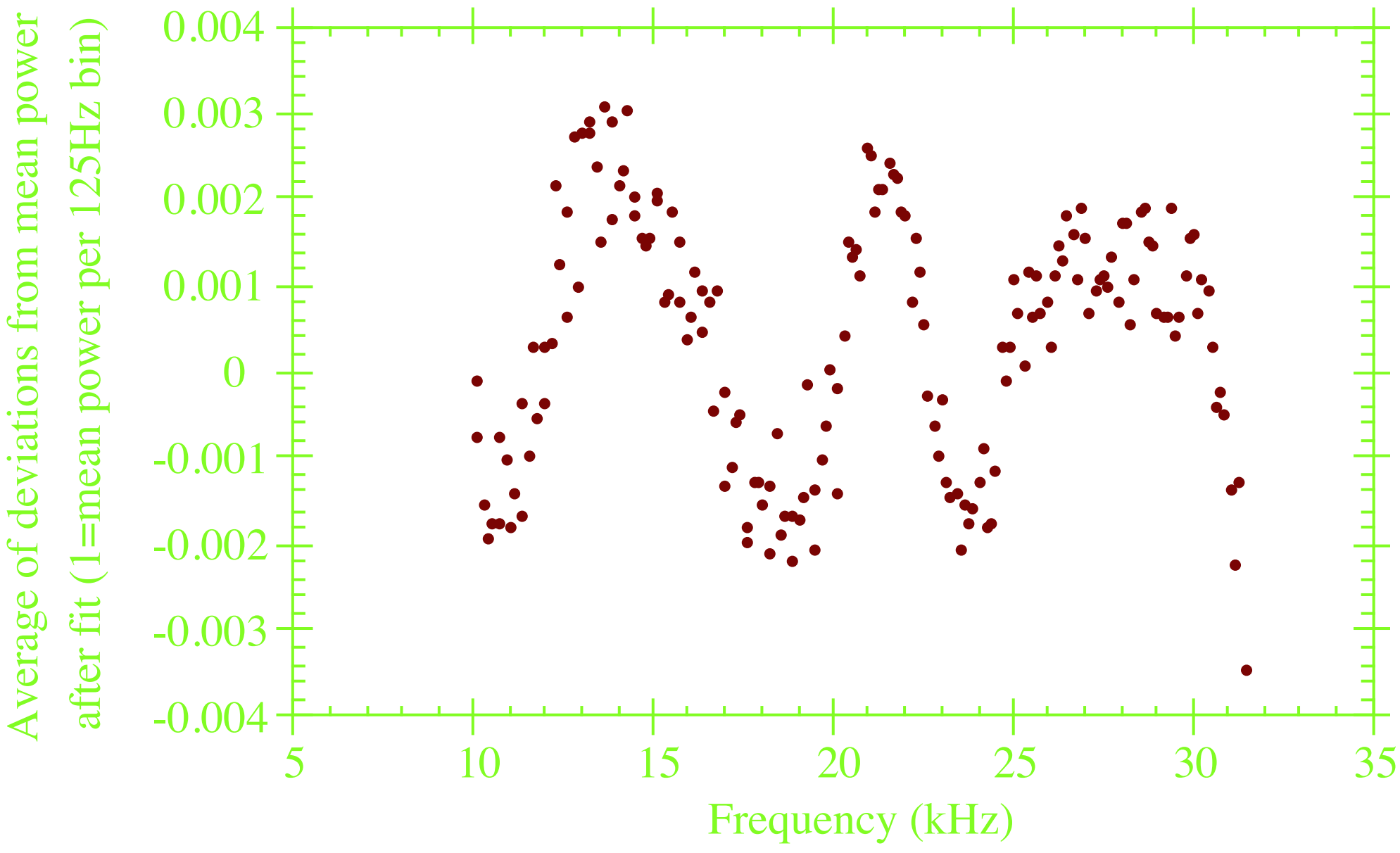

Figure 4.17 The average of all the receiver and fit normalized raw traces in the run 1 combined data. The vertical axis is power in units of the mean power in a trace, hence the rms of the Gaussian noise fluctuations in a single trace is 0.01 on this scale.

Recall that the step in frequency between acquisition of successive traces is about 2kHz. Therefore, into each bin of the combined data go contributions from raw trace bins at all positions across the IF bandwidth. The size of the systematic offset from zero is therefore randomized, depending on the exact positions on the IF bandwidth of the bins from raw traces that contribute to the bin of combined data under consideration. Since the number of raw traces contributing to each combined data bin is ~50, I argue that the systematic contribution to each combined data bin is approximately Gaussian. Each raw trace contributes an offset that could be anywhere in the interval $(-0.2\sigma,+0.2\sigma)$ where $\sigma$ is the rms Gaussian fluctuations per raw trace. I model the contribution of the IF systematic as a Gaussian error of rms $\sigma_{IFS} = 0.2\sigma/\sqrt{2} = 0.14\sigma$.

Recall from Section 4.2.2 that each raw trace is divided by the response of the IF electronics to a noise source at its input. I propose that the systematic deviation in the run 1 data stems from this normalization being inaccurate. I now estimate the contribution of the error produced by this effect to the overall statistics of the combined data. The contribution of the systematic shift from a raw trace will be modeled as Gaussian. Using a standard formula for error propagation, the overall rms deviation in the combined data introduced by dividing the trace by the normalization is:

$$\sigma_{TOTAL} = \sqrt{\sigma^2 + \sigma_{IFS}{}^2} = \sigma\sqrt{1+0.14^2} = 1.01 \qquad (4.34)$$

My model predicts the result of the systematic error in the crystal filter passband response is an 1% increase in the rms scatter in the combined data. This small excess in the spread of the data points in run 1 is consistent with observation. In particular, look again at Figure 4.6. The discrepancy between the histogram of $N_{eff}$ and the theoretical histogram implied by the radiometer equation is small, but an offset in the modal effective number of averages between theory and observations of a single bin, or $\Delta N_{eff}$~200 can be seen. If $\sigma_{TOTAL}$ in the real data were 1.01 of the value (10,000) used to generate the theoretical histogram, the resultant mean effective number of averages would be $1/(1.01/\sqrt{10,000})^2 = 9,800$. This is consistent with the position of the histogram peak in Figure 4.6.

Finally, note also that the increase in rms spread in the combined data is not visible in Figure 4.15, because the *measured* standard deviation from each raw trace is used to compute the standard deviation in the combined data, as discussed in Section 4.2.6. The increased spread in each raw spectrum is therefore normalized out in the data combining procedure and is therefore invisible in the combined data dispersion plot. However, we can see an offset of the mean of the dispersion from zero. This offset occurs because the central bins of a raw trace are weighted more heavily than the edge bins when the trace is multiplied by the cavity Lorentzian. Since from Figure 4.17 the systematic offset is positive at the center of the trace, the systematic tends to shift the mean of the combined data in a positive direction.

### 4.3.6 Time Evolution of the Crystal Filter Passband Response

We have seen that the IF systematic appears to be fairly constant over the timescale of run 1, at least after dividing by the crystal filter passband response (CFPR) and by the fit function. On what timescale does the IF systematic change? The IF systematic is suspected to be due to evolution of the CFPR. Figure 4.18 is the ratios of some measurements of the CFPR, measured as described in Section 3.5.2. The 3 ratios were of CFPR measurements separated in time by the periods indicated on the plots.

The upper plot in Figure 4.18 is the ratio of a CFPR measured on 18th July 1996 (5 months after run 1 was completed) and an CFPR measured during the commissioning run on 26th October 1995, before run 1 began. The CFPR measured during commissioning was used in the initial normalization of all the run 1 traces. There is as much as 2% discrepancy between the two measurements, yet it is improbable that there was any evolution of the CFPR during run 1 because such evolution would have been evident in Figure 4.16, the average of the receiver and fit normalized traces as a function of time. There are 2 possibilities that are consistent with no discernible CFPR evolution in run 1. The first is that there was, in fact, evolution, but due to some error in measuring the CFPR during commissioning, the evolution is swamped by a larger, time independent systematic error. This systematic error is mostly removed by the fit procedure, but a residual time independent component gives rise to the structure dominating figure 4.16. The second possibility is that the CFPR changed discontinuously before run 1 and again sometime between the end of run 1 and the second CFPR measurement on April 26th 1997. In either case, the systematic fortunately causes only a minimal distortion of the otherwise Gaussian statistics of the combined data, and has no observable effect on the sensitivity of the experiment as discussed in Section 4.3.5.

## 4.18 Simulations of the Run 1 Combined Data Set

We now have a hypothesis that the fit residuals are due to variation in the transfer function of the IF section of the electronics. Furthermore, the residuals appear to have the same shape throughout run 1. To test this hypothesis, I run a simulation. The raw material for this simulation is the fit residuals averaged over every raw trace in the run 1 data. (Figure 4.17).

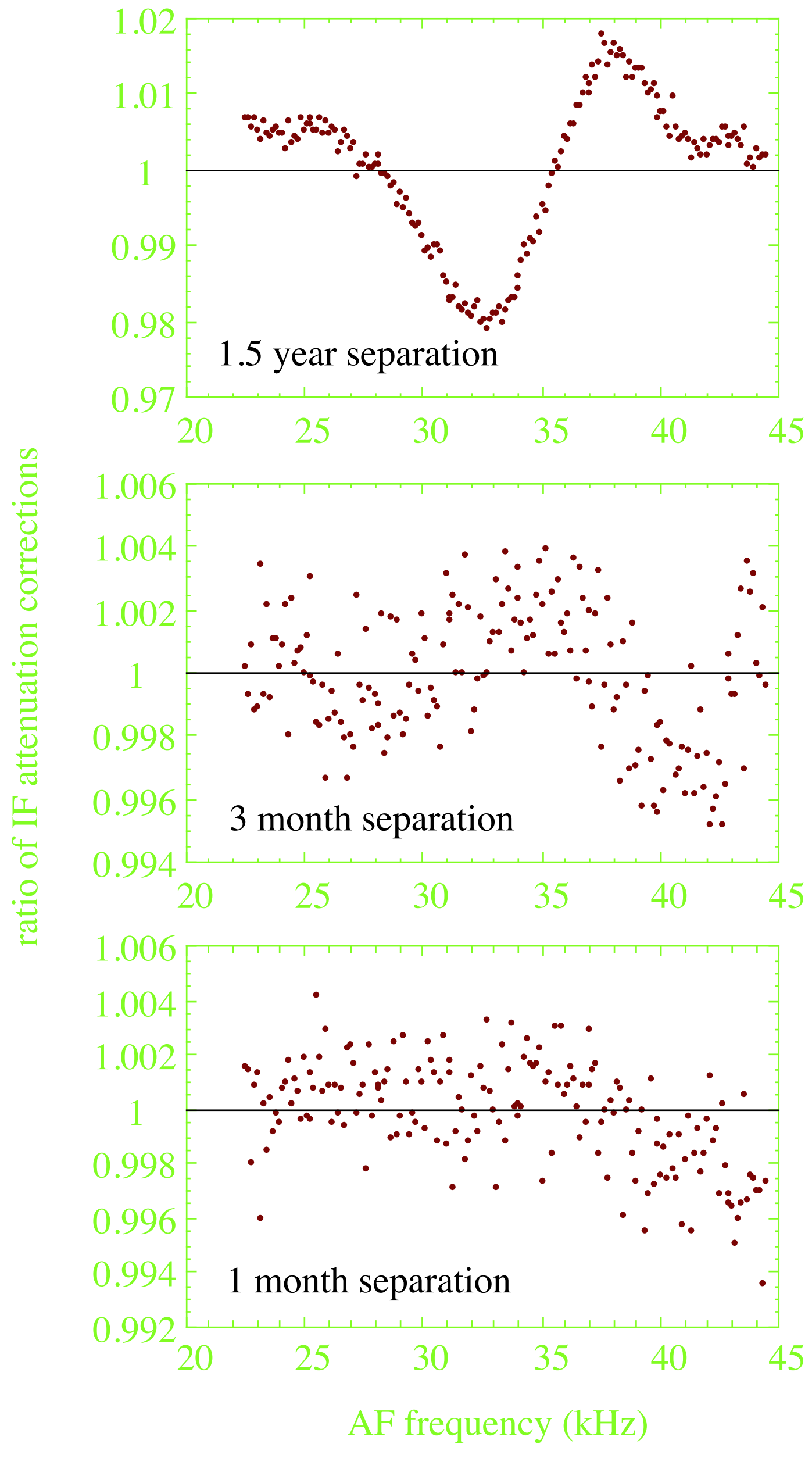

Figure 4.18 Ratios of measurements of the crystal filter passband response separated by the time intervals indicated on the plots. Each passband response is normalized to its average value before the ratio is calculated. On the horizontal axis is frequency at the FFT spectrum analyzer. Note the changes of scale on the vertical axis.

In my simulation, I take each raw trace in the run 1 data. I normalize the trace with the 'known' transfer function of the IF receiver and fit the normalized trace to the 5 parameter cavity-amplifier fit. I keep the best fit curve through the data and discard the real trace. I then build an artificial trace to replace the real one as follows. I start with the average fit residuals from Figure 4.17. I add to them computer generated Gaussian distributed noise of rms 0.01 (as would be expected from a trace which is an average of 10,000 power spectra). I then multiply this curve by the best fit curve to the real trace. The resulting 'fake' trace is then analyzed exactly like a real trace from the data. It is fitted and normalized with the 5 parameter fit, and the resulting normalized trace is added to a fake combined traces exactly as the real trace would have been in the true analysis.

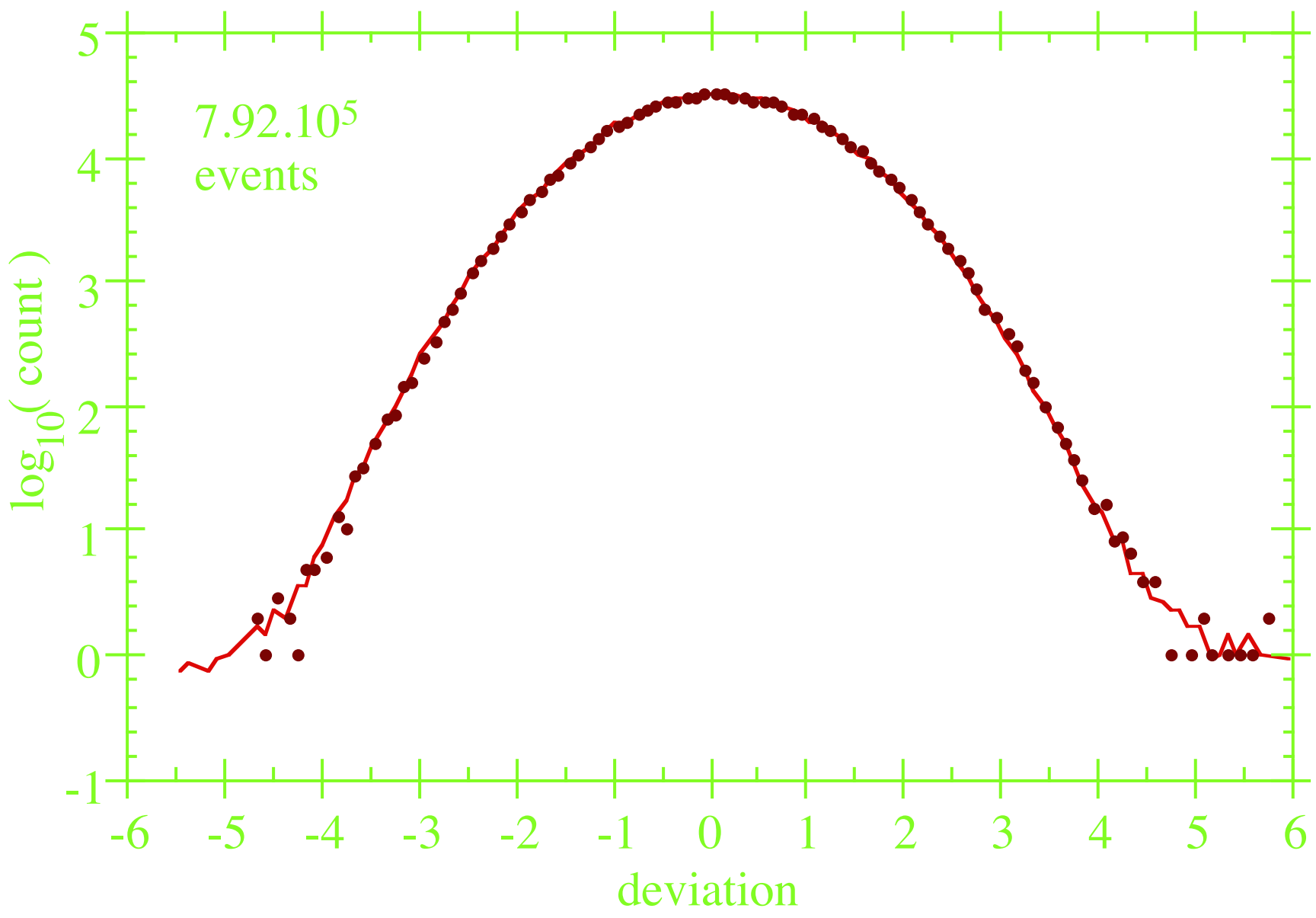

Figure 4.19 The dispersion in the run 1 data with the results of the simulation overlaid. the simulation overlaid. The points are the heights of the histogram bins for the dispersion in the true run 1 data set. The continuous line is the result of the Monte Carlo.

This procedure is repeated for every raw trace until a complete set of artificial combined traces has been produced. To improve the statistics of the resulting data set, the procedure is repeated 10 times, so the artificial set of combined traces resulting from the simulation has roughly 10 times the statistics of the real combined data. In Figure 4.19 the histogram of the real combine data (points represent heights of histogram bins) are overlaid on the results of the simulation adjusted for the different normalizations of the real and simulated data. The agreement between the simulated data histogram and the real data is excellent.

## 4.4 The Single Bin Peak Search

The single bin peak search is intended to be simple and intuitive. It relies on the underlying Gaussian nature of the noise fluctuations. Because the distribution of thermal noise is Gaussian, the sensitivity to axions in this channel can be calculated analytically without recourse to Monte Carlo simulations. I will discuss the single bin search in the run 1 data, followed by the single bin rescan. Finally I will list the candidates that pass the cut both in run 1 and in the rescan and for each candidate I will I explain how the candidate frequency was eliminated as an axion candidate. The exclusion plot for the axion 1 bin search channel is given in the Chapter 5.

### 4.4.1 Single Bin Peak Search in the Run 1 Data

The starting point is the array of power excess in units of standard deviation in each of the 792000 125Hz-wide frequency bins between 701 and 800MHz. To search for single bin power excess, a cut is made on the run 1 combined data at a fixed number of standard deviations above the mean power. These frequencies are then re-scanned to see if the power excess is persistent. In the single bin channel the same cut on power excess will be applied in the rescan as was applied in run 1.

To get a flavor for the important issues, consider a hypothetical search where a cut is made at $2\sigma$ above the mean power. This will clearly be a more sensitive search than one where a cut is made at $4\sigma$, but if the cut is set too low then the number of frequencies on the rescan list will get too large. For instance, the run 1 data set consists of a 99MHz bandwidth or 792,000 125Hz frequency bins. From Gaussian statistics the expected fraction of these bins passing a 2σ cut is 2%. Therefore $1.6 \times 10^4$ bins are expected to pass the cut and require further coverage. This would take a prohibitively long time. If a 4σ cut is used the expected number of frequency bins passing the cut is 13. Clearly a compromise is required.

In the single bin search channel I use a cut at 3.2σ. As discussed in Section 4.3.5, the effect of the crystal filter passband response systematic on the single bin data distribution is to shift the mean of the Gaussian distribution upwards by 0.1σ. Therefore to make a 3.2σ cut on the Gaussian data centered at +0.1σ, the cut must be made at +3.3σ in the combined data set. This cut generated 538 single bin candidates. A useful formula giving the expected number of rescans $n_r$ in a sample of N bins from a Gaussian distribution of unit width when examining every event above p sigma is:

$$n_r = \frac{N}{2}\left(1 - \mathrm{erf}\left(\frac{p}{\sqrt{2}}\right)\right) \qquad (4.35)$$

Due to the IF systematic discussed in sections (4.16-4.17), there is a deviation in our run 1 data from purely Gaussian statistics. Setting a 3.2σ cut in the 1-bin search channel lead to 538 candidates. The expected number of bins above the cut, using equation (4.35) is 544. This expectation is consistent with the number of frequencies in the 1-bin rescan list.

### 4.4.2 The Single Bin Candidate Rescan

Each of the 538 candidates in the 1 bin channel was rescanned, with enough traces acquired at each frequency to give the same or higher signal to noise ratio as was achieved in run 1. The data from the rescan was combined, again applying the combining algorithm described in Section 4.2.10. The same 3.2σ cut was applied in the rescan data. Of the 538 candidates from the run 1 search, 8 survived the rescan cut. Table 4.1 is a table of these candidates.

### 4.4.3 Investigation of Persistent 1 Bin Candidates

The 4th column of Table 4.1 contains one or more code letter(s) summarizing how the persistent candidate was eliminated as a possible axion signal. The codes are as follows:

A: Elimination by detection with an antenna. Refer to Figure 3.1. The cable connecting
the output of the MITEQ AMF-3f-003010-10-10P post amplifier to the input of the MIT axion receiver is disconnected, and a stub antenna is connected to the receiver input in its place. A trace is taken to see if the candidate corresponds to a background signal in the laboratory. If it does, this is an strong indication, but not proof, that the peak in the detector is due to RF leakage from outside the detector into the cavity or electronics.

T: To establish that a peak is due to an external RF background, the two main sources of leakage into the detector, the weakly coupled port and the line connected to the directional coupler weak port, are terminated at the RF feedthroughs on top of the cryostat. If the peak disappears then the candidate is discarded. If the peak is diminished in size so that it does not pass the 3.2σ cut, then it is also discarded.

| Frequency (MHz) | Run 1 Peak Height | Run R Peak Height | Comment and Identification Method |
|---|---|---|---|
| 729.910750 | 3.5655 | 17.958 | Radio Peak, T,A |
| 748.944125 | 3.7367 | 3.4041 | NR |
| 750.014000 | 3.5827 | 3.5450 | NR |
| 771.218500 | 28.893 | 3.6456 | NR |
| 771.218625 | 28.893 | 3.6456 | NR |
| 771.234375 | 23.576 | 5.3310 | NR |
| 771.265750 | 27.365 | 9.5139 | NR |
| 780.299625 | 3.5043 | 3.3814 | NR |

Table 4.1 Persistent candidates in the 1 bin search channel. The 1st column gives the candidate frequency. The 2nd and 3rd columns give the heights of the candidate signals in run 1 and the rescan. The final column gives the results of investigation of the persistent candidate by manual scanning

NR: Non-recurring peak. On line software is used to make an average of at least 100 traces at the frequency of a persistent candidate. The size of the Gaussian noise fluctuations is measured, and the noise temperature of the apparatus is used to determine the power in Watts equal to a one sigma noise fluctuation. The power at the persistent candidate bin is determined in Watts. The number of traces taken is sufficiently large that unless this peak is non statistical, the peak height will be a tiny fraction of the power corresponding to the 3.3σ cut applied in run 1.. If the probability, calculated using Gaussian statistics, that the bin height measured is a peak of power 3.2 sigma in run 1 plus a large negative statistical fluctuation is less than 0.5%, then the peak is rejected as a candidate.

As can be seen from Table 4.1, all of the candidates were rejected using the above 3 techniques. The exclusion limit resulting from the single bin search channel is calculated in Chapter 5.

# 4.5 The Six Bin Peak Search

## 4.5.1 Losses in Signal Power in the 6 Bin Channel

Figure 2.6 in Section 2.2.3 shows a well-motivated prediction for the axion line shape and width. The six bin (750Hz) is well matched to a search for axions in this scenario. In the six bin channel the problem of power leakage into adjacent bins encountered in the narrow axion case is not significant (This problem is discussed in Appendix 5). However there are other complications in the 6-bin channel which I will now consider.

Firstly, because the lineshape is roughly Maxwellian, we do not expect all axions to convert into photons within a 750Hz bandwidth. Assuming the linewidth in Figure 2.6, 20% of the power is in the long tail of the Maxwellian. This power loss degrades the SNR by 20%.

Secondly, the Maxwellian peak will enhance the power in several adjacent bins in any raw trace where it appears. The fit procedure described in section 4.2.4 will tend to deviate from the shape it would have in the absence of the excess power from axions and obscure some percentage of the signal. For KSVZ-strength signals in a single raw trace, the degradation of the signal strength due to this effect is typically between 10% and 15%.

The third effect concerns the statistic we use to search for 6-bin power excess. Conceptually, the simplest procedure is to split n 125Hz-wide bins into n/6 750Hz-wide bins, each of which is the sum of 6 underlying 125Hz wide bins. Unfortunately, any axion signal will be split between 2 adjacent 750Hz bins in this case. Depending on the position of the peak with respect to the bins chosen, the maximum power excess in a 750Hz bin is between 1/2 and 1 of the signal power. Using this statistic the expected signal strength is degraded, typically by ~25%.

## 4.5.2 Use of Overlapping Frequency Bins

A more sensitive procedure is to make overlapping 6 bin segments. The ith 6 bin segment is the sum of the ith to the (i+5)th 125Hz bins, the (i+1)th 6 bin segment is the sum of the (i+1)th to the (i+6)th 6 bin segment, and so on. Hence a set of n 125Hz bins yields (n-5) 6 bin segments. This algorithm ensures that the bulk of the power in the axion line will be deposited in at least one of the 6-bin segments. The price of this gain in SNR over the previous method is that there are now ~n 750Hz wide bins compared to the n/6 that I get using the above non-overlapping method. Adjacent bins in the array are highly correlated, so that a Maxwellian peak will lead to power excesses in several neighboring 6 bin sets.

To get around this problem, the single bin data array is first used to construct overlapping 6 bin sets. I will refer to this procedure as 'coadding' the data. Start by setting a cut at a high value ($6\sqrt{6}\sigma$ sigma is used in our peak search). If any bin passes the cut then the frequency of that bin is added to the rescan list. Then this bin and the surrounding 16 bins are set to a low value (say -10), so that none of the bins correlated with the one first passing the cut will appear on the rescan list. Step the cut level down by a small amount (I use 0.1 sigma) and repeat the process some target cut level or number of rescan frequencies is achieved. I will refer to this technique as a 'running sigma cut'.

Now back to the problem of having ~n (correlated) bins. Optimistically one might guess that if the same cut level was set in the n/6 uncorrelated 6-bin segments generated using a simple 6 bin cut, and in the n correlated 6-bin segments generated using a running sigma cut, one might get the same number of candidates in the two cases because the running sigma cut prevents 2 bins correlated with each other from both passing the cut. This guess turns out to be wrong. In simulation studies a search on n correlated bins with a running sigma cut consistently yielded 2-3 times as many candidates as a search on n/6 uncorrelated bins at the same cut level.

In spite of this effect, the correlated bin search is adopted in our 6 bin search channel. The signal power loss due to splitting of a 6 bin wide signal between non-overlapping bins in the uncorrelated bin search unacceptably reduces sensitivity for the uncorrelated 6 bin search method. Unfortunately, the correlated bin search method is non-statistical, and the sensitivity in peak searches using this method must be determined using a simulation.

### 4.5.3 Simulation of the 6 Bin Peak Search in the Run 1 Data.

In order to determine the cut level required in the run 1 data, a Monte Carlo simulation is used. Briefly, it works by creating a set of combined traces with artificially injected KSVZ axion-like signals injected at a random set of frequencies throughout the combined traces (see Appendix 5). So the combined traces closely resemble the actual run 1 combined data set, but they contain axion-like signals at known frequencies. The array of power excesses from the combined traces coadded to make correlated 6 bin data, and a running sigma cut peak search is implemented.

Figure 4.20 is a plot of the fraction of the injected peaks located in the peak search on the Monte Carlo generated combined traces as a function of the cut level.

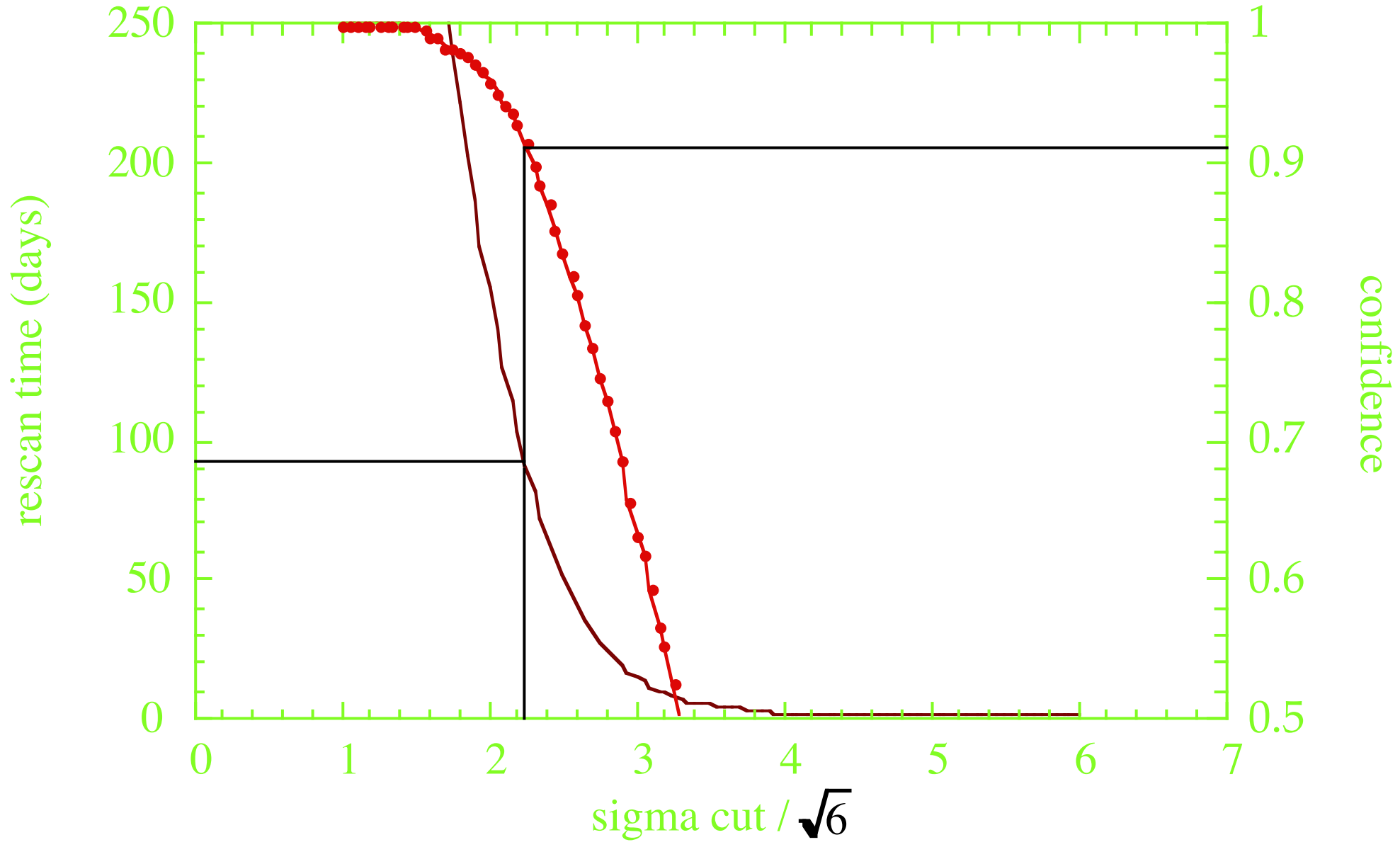

Figure 4.20 Results of Monte Carlo simulation of the run 1 combined Data. The line with points overlaid is the fraction of the injected signals that passed the cut on the x axis. The fraction of peaks detected is the confidence that an axion will pass the cut, read from the right hand scale the right hand scale. The scale on the left hand side is the estimated time to rescan the bins passing the cut that are not due to an injected signal

Based on the results of this simulation, a cut was made in the 6 bin peak search on the true run 1 data at the $2.25\sqrt{6}\sigma$ cut level. 6535 frequencies passed this cut. These 6535 candidates were re-scanned in the second round of data taking.

### 4.5.4 Re-scanning the 6 Bin Candidates

The $2.25\sqrt{6}\sigma$ cut applied to the 6-bin combined data from Run 1 yielded 6535 candidates. Each candidate was subsequently rescanned in a run which I will hereafter refer to as run R.

The same procedure is employed for the 6- and 1-bin candidates. The candidate list is used to generate a script file which is read by the data acquisition (DAQ) software. The DAQ software employs a feedback loop where the tuning rods are moved and the cavity resonant frequency measured as described in Section 3.3.2, to tune the cavity resonance to within ±1kHz of the frequency to be rescanned. A set of traces, each the linear average of 10,000 power spectra, is taken. The total number of traces is adjusted so that the predicted signal to noise ratio (SNR) is at least that obtained at the same frequency in run 1. As for run 1 the raw traces are combined using the algorithm described in Section 4.2.10, yielding a data set which I will refer to as the run R combined data.

The aim of run R in the six bin search channel is to improve the SNR at the rescan frequencies selected by the run 1 cut. The maximal SNR will be obtained by combining run 1 and run R data near each rescan frequency. As when combining raw traces I must combine data from runs 1 and R in a manner which optimizes the SNR in the overall data stream. I will refer to data resulting from combining the data from run 1 with the data from run R as the 1+R combined data. Before discussing the data combining algorithm and the statistical properties of the 1+R combined data I will describe a preliminary processing step performed on the run R data.

### 4.5.5 Systematic Correction in the Run R Data

The combined data from run R differs from than in run 1 in one important aspect. In section (4.17) I discussed the effect of the crystal filter systematic on the run 1 data and I showed that the effect of the systematic on the combined data is small. This is because the magnitude and sign of the systematic contribution from different traces changes as the cavity frequency is stepped in ~2kHz increments past each combined data frequency and many different bins of the systematic curve shown in figure (4.17.2) add each bin of the run 1 combined data.

In run R, the cavity resonant frequency is held at a fixed value while a single candidate frequency is being re-scanned. Hence the same systematic shift is applied with every new trace that is added to the combined data. If the systematic shift is $\delta_{SYS}$ and the rms spread in Gaussian fluctuations in a single trace is $\sigma$, then the ratio $R_S$ of the amplitude of the systematic to the rms spread in the Gaussian noise in the sum of n traces weighted equally is:

$$R_S = \frac{\delta_{SYS}\sqrt{n}}{\sigma} \qquad (4.36)$$

Any systematic error in the combined data will grow as the square root of the number of traces combined. Left uncorrected, this systematic would cause large deviations from Gaussian statistics in the combined data. Furthermore, during run 1, the systematic shift observed in raw spectra appeared to be the same for all traces. During run 2, since many traces were taken with the cavity frequency at the same value, much smaller changes in the systematic (which may occur, for example, if the systematic were due to aging of the IF filter crystal) may have a significant effect on the statistics of the combined data.

To correct the run R combined data for the IF systematic, I exploit the fact that in the run R combined data, the rescan frequency is always very close to the center of the raw traces. Hence, traces taken within a time scale of weeks of each other all contribute the same systematic to the combined data. I take the set of run R combined data points near to rescan frequencies. This consists of 6535 data points. In combining the run R data, a 5th array is added to the combined data set (see Section 4.2.13 for a summary of the other four arrays). The extra array contains the average of the times (in seconds since midnight on April 1st 1995) when the traces contributing to each combined data bin were taken. The combined data power excesses at the rescan frequencies are sorted are placed in order of increasing average time. Each point is divided by $\sqrt{n}$ , and the points were grouped into sets of 300. A array consisting of the linear average of each group of points forms a relatively smooth line through the set of time ordered power excesses. An array consisting of a single element for each rescan frequency is generated, where each array element is the value on the smooth curve for the average time associated with the rescan frequency multiplied by $\sqrt{n}$. The correction array is used to subtract the IF systematic from the vicinity of each rescan frequency in the combined data.

If equation (4.24.1) is a good approximation and if the raw traces contributing to each candidate point are taken within a short enough time span so that the evolution of the systematic is negligible, then the statistics of the corrected combined data should again be Gaussian. Figure 4.21 is the dispersion at candidate frequencies in the combined data from run R. The smooth line is a theoretical Gaussian. It is in good agreement with the points which are are heights of the histogram bins.

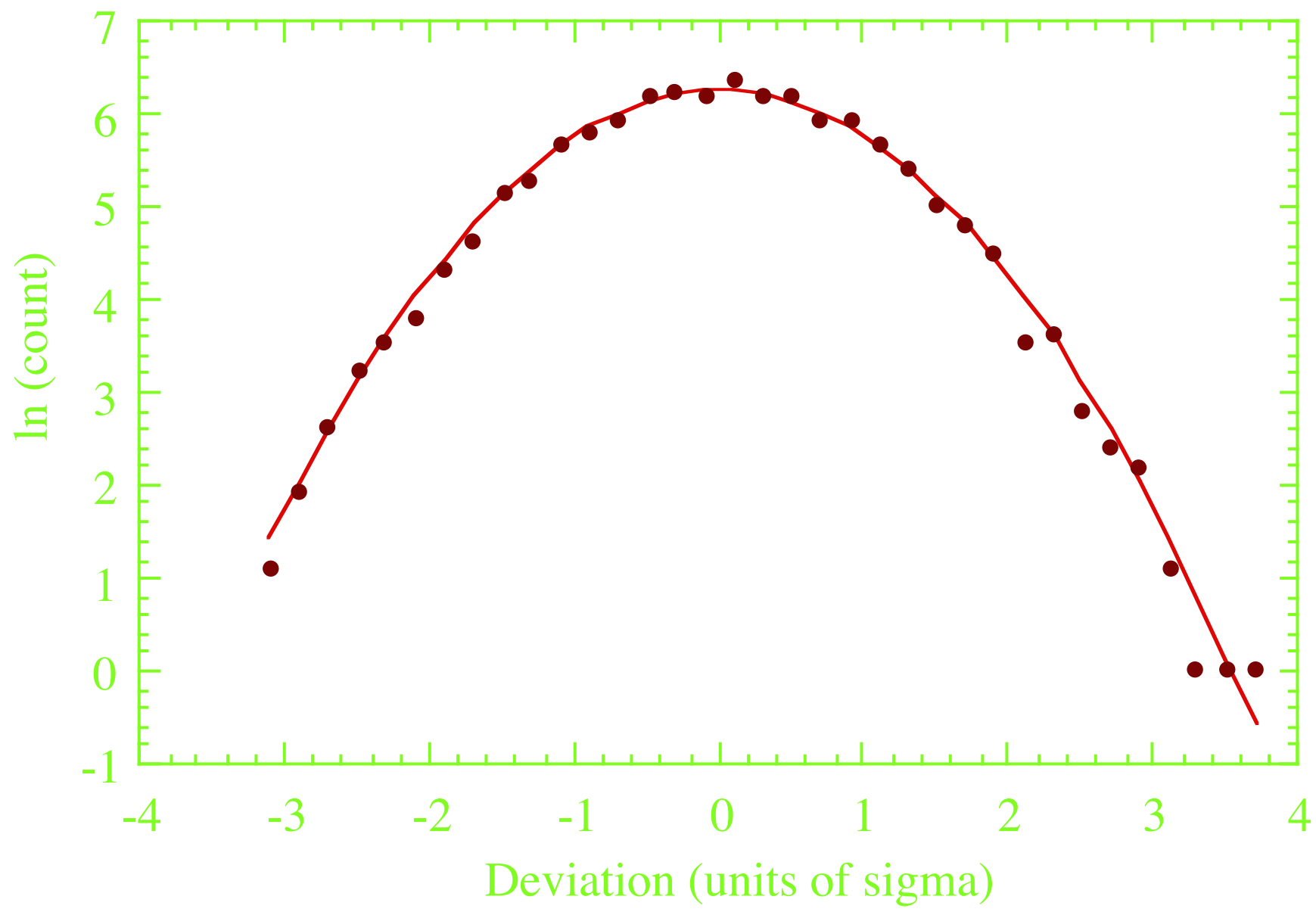

Figure 4.21 dispersion in the run (1+R) combined data

### 4.5.6 Combining the Run 1 with the Run R Data

The run 1 combined data and the run R combined data near each rescan frequency are combined using a weighting scheme similar to that described in Sections 4.2.10 - 4.2.12. 41 125Hz frequency bins surrounding each rescan frequency are taken for each of the two data sets. Matching bins from the 2 data sets are combined as follows: if $\delta_1$ ($\delta_R$) is the contents of one of the 41 run 1 (run R) bins then the weighted sum $\delta_{1+R}$ of bin from run 1 with the corresponding bin from run R is:

$$\delta_{1+R} = \frac{\delta_1 S_1 + \delta_R S_R}{\sqrt{S_1^{\ 2} + S_R^{\ 2}}} \qquad (4.37)$$

As with the weighting scheme described in Section 4.2.10, the SNR in the combined data set is the sum in quadrature of the SNRs in the two components.

This combining algorithm yielded a single combined data set which I will refer to as the run (1+R) combined data. This data was only used for the 6 bin search channel.

### 4.5.7 Cut on the 1+R 6-Bin Combined Data

In the single bin channel there is an obvious statistic with which to search for peaks, the heights of single bins in the combined data. A narrow axion line does not move from one 125Hz bin to another between run 1 and run R, so the sum of the heights same frequency bin from the two runs is the most powerful search statistic. The 6-bin channel is more complex. Consider: the axion peak is many 125Hz bins wide, and each bin also contains Gaussian noise. A peak may pass the run 1 cut due to a power excess in the ith bin; if in the rescan the ith bin happens to contain a large negative fluctuation from noise, but the jth bin (where $|j-i|<$axion linewidth) passes the cut, then by re-examining only the same frequency bin where we first saw the power excess we would miss the event. It is most likely that the same bin will pass the cut in the two data sets (i=j), with the probability of two bins separated by (i-j) passing cuts in runs 1 and R decreasing as i-j gets larger.

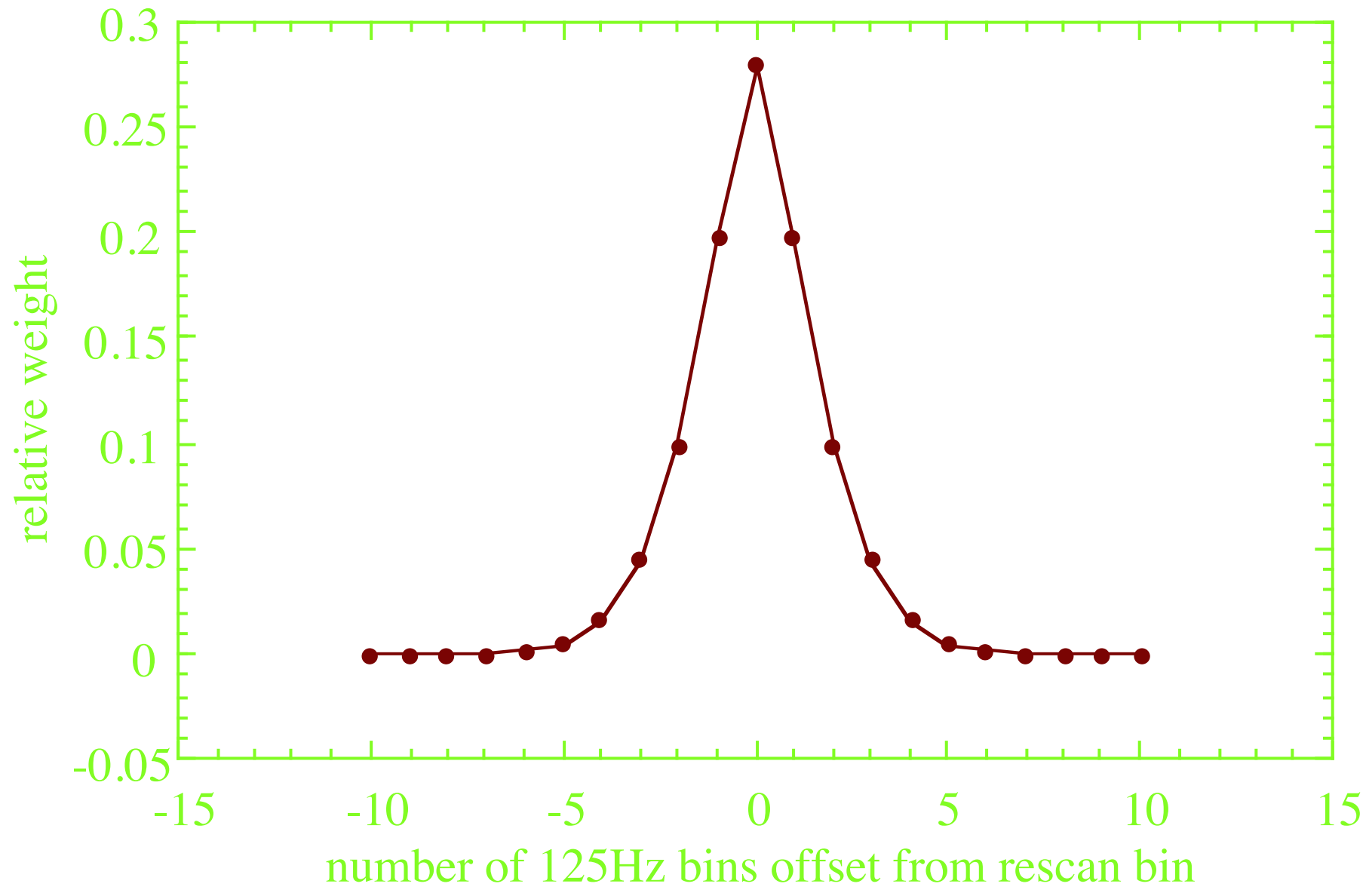

Figure 4.22 Weighting function for the run (1+R) cut

The candidate search in the run (1+R) combined data employs the following procedure: First the arrays of 41 bins about each rescan frequency are coadded, generating overlapping 6-bin sets as defined in Section 4.5.2. The 41 element array is then multiplied by the function shown in Figure 4.22, with the peak of the function centered on the rescan frequency.

The weighting function is generated in a simulation where 1000 fake axion signals are injected into raw data as described in Appendix 5. A peak search (see Section 4.5.2) is carried out with a $2.25\sqrt{6}\sigma$ cut and a distribution is generated of the difference between the frequency where each injected signal is located and its injection frequency. This distribution is then convolved with a similar distribution using the same $2.25\sqrt{6}\sigma$ cut, and the convolution of the two distributions is normalized to unit area, yielding the distribution function.

I shall refer to the product of the overlapping six bin sets near each rescan frequency and the weighting function as the weighted (1+R) distribution. The cut implemented is to select the top 10% of rescan candidates in the weighted (1+R) distribution. 10% of candidates in the weighted (1+R) distribution appeared above a $2.06\sqrt{6}\sigma$. This cut reduces the number of 6-bin candidates to 654. These candidates are scanned for a 3rd time in 'run P'.

### 4.5.8 6-Bin Search Sensitivity in the Run (1+R) Data

The question I wish to address to determine our sensitivity in the 6-bin channel is: At what axion power level is there a 90% probability that the resultant signal in the run (1+R) data will pass the cuts in the 6-bin search channel. The sensitivity of the experiment when the axion signal has to pass both the $2.25\sqrt{6}\sigma$ cut in the run 1 data and the $2.06\sqrt{6}\sigma$ cut in the run (1+R) data is determined using a simulation.

The simulation is somewhat more complex than before. Fake combined traces are created corresponding to the real run 1 and the real run R data. Axion-like signals are injected as before (see Appendix 5), but this time the frequency list is a subset of the frequencies that passed the run 1 cut. This is because in the run R data only the frequencies on the run 1 candidate list have a good signal to noise ratio. The peak search procedure mirrors that used for the real data set as closely as possible. First a $2.25\sqrt{6}\sigma$ cut is made on the fake combined traces corresponding to the run 1 data. A list is made of the frequencies that pass the cut and are correlated with the frequencies at which artificial peaks were injected. Then the fake combined traces corresponding to runs 1 and R are combined as described in Section 4.5.7. A $2.06\sqrt{6}\sigma$ cut is made on these fake combined traces. Finally, a list is made of the frequencies of injected signals that passed both cuts.

This simulation is repeated at different levels of power in the injected signals. There are ~1500 injected signals per simulation. The full frequency range 701-800MHz is split up into 20 bins. The output of the simulation is the fraction of peaks injected in each of the 20 bins that passed both cuts as a function of the injected signal power. Inverting this curve, the signal power detectable at any confidence level as a function of frequency is obtained.

The Monte Carlo was implemented in C and ran on the M.I.T. Abacus computer farm. Figure 4.23 is a plot of the power level that passed both cuts in the Monte Carlo simulation at 93% confidence as a function of frequency.

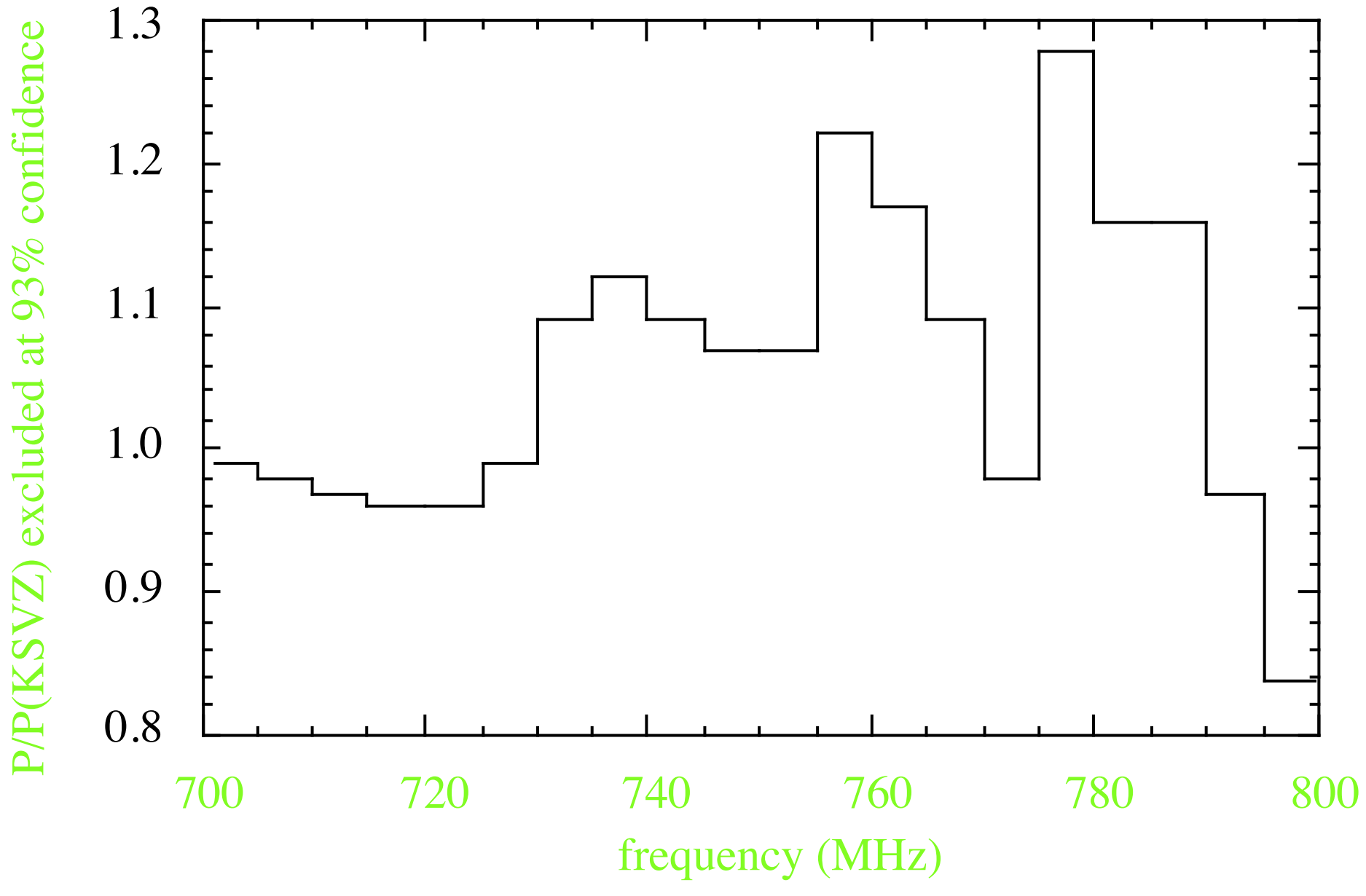

Figure 4.23 Power excluded at 93% confidence after applying run 1

and run (1+R) cuts, from Monte Carlo simulation.

The power level is in units of the power expected from the KSVZ axion model using Equation 2.28. I assume a halo density of 0.45GeV/cc. Referring back to Figure 4.14, notice that the regions where we are sensitive to the weakest signals correspond to the regions of highest signal to noise ratio in the run 1 combined data. This gives us some confidence in the results of the simulation.

### 4.5.9 The Run P Combined Data

The 654 6-bin candidate frequencies passing the $2.06\sqrt{6}\sigma$ cut in run R were scanned again between August 22nd and October 21st 1997. I refer to this scan as run P. Figure 4.24 is a plot of the cavity frequency vs. time during this run.

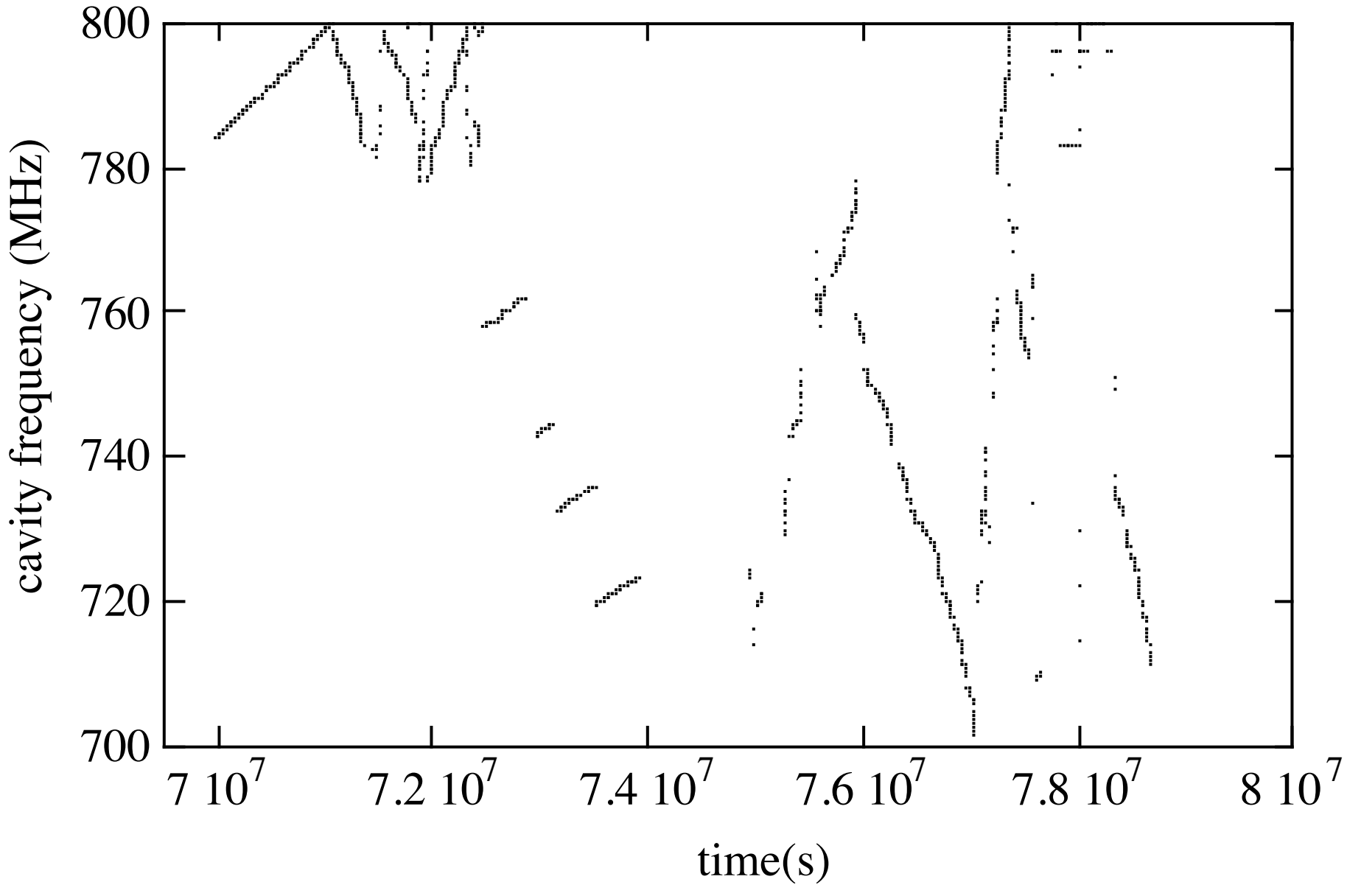

Figure 4.24 Cavity frequency vs. time during scanning of 6-bin persistent candidate frequencies

Three distinct phases can be seen in the run P data taking. First persistent candidates between 778MHz and 800MHz are scanned. Next, persistent candidates near to mode crossings were scanned with the cavity filled with liquid helium. Finally, the rest of the run P candidates were scanned. 28 traces of 10,000 power spectra each were acquired at each candidate frequency.

To remove the crystal filter passband response systematic from the run P data, the crystal filter passband response was re-measured every month, and the updated measurement used to normalize traces taken within ± 2 weeks measurement. Evidence that there was no observable systematic shift on a one month time scale during the run P rescan is given in the 3rd plot of Figure 4.18, which is the ratio of the two crystal filter passband response measurements taken 1 month apart during run P. The important bins in the for rescan data taking are the center bins which always coincide with the rescan frequency. The drift in these bins is negligible.

The run P combined data set is made using the algorithm described in Sections 4.2.10 - 4.2.12. Raw data files acquired between July 11th and 17th 1997 are normalized using an crystal filter passband response measured on July 18th 1997; the remainder of the run P, acquired between August 22nd and September 22nd 1997 is normalized using a 2nd crystal filter passband response measured on 29th August. The run P is combined with the run 1 and run R data using an algorithm similar to that used to make the run (1+R) combined data. 41 bin sections around each persistent 6-bin candidate frequency are made in each of the 3 data sets. The section from the run R data is corrected for the IF systematic (see Section 4.5.5), then the optimal sum of the 3 data sets is made in a manner analogous to that employed in making the run (1+R) combined data. Let $\delta_1$ be the deviation from mean power in one of the 41 bins near a rescan frequency in the run 1 combined data and $\delta_R$ ($\delta_P$) be the corresponding power excess in the run R (P) combined data. Denoting the signal to noise ratios at that frequency by $S_1$, $S_R$ and $S_P$ respectively, then the weighted sum $\delta_{1+R+P}$ of the three data streams is:

$$\delta_{1+R+P} = \frac{\delta_1 S_1 + \delta_R S_R + \delta_P S_P}{\sqrt{S_1^2 + S_R^2 + S_P^2}} \qquad (4.38)$$

and the signal to noise ratio of the combined data is the sum in quadrature of the signal to noise ratio of the components.

### 4.5.10 Cut on the 6-bin Run (1+R+P) Data

The 41 bin regions about each persistent candidate frequency in the run (1+R+P) data were coadded and a weighted sum of these bins was made with the weighting function shown in Figure 4.25. Also shown is the weighting function used for the run (1+R) data.

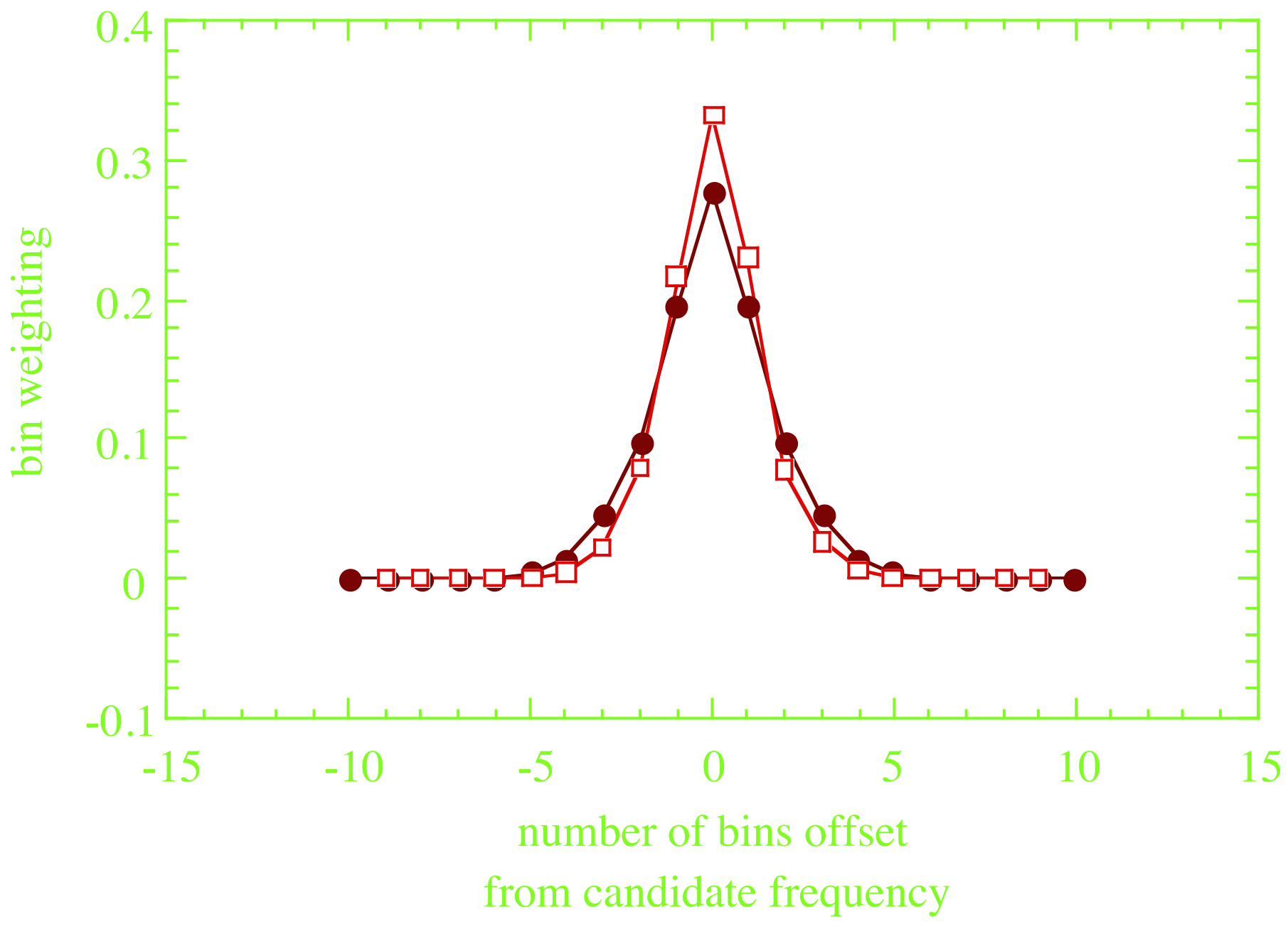

Figure 4.25 The weighting function used in the weighted sum of neighboring bins in the run (1+R+P) combined data (square data points), with the weighing function used for the run (1+R) data (round data points)

The weighting function used for the run (1+R+P) data is narrower than the one used to make the run (1+R) data. This is because the SNR in the (1+R+P) data is higher, hence it is less likely that fluctuations due to the background Gaussian noise will cause a bin far in frequency from the bin with the maximum signal power to pass the cut.

Figure 4.26 is the weighted sum for each rescan frequency in the run (1+R+P) data. Also shown is the cut applied to determine which frequencies to re-examine 'by hand'.

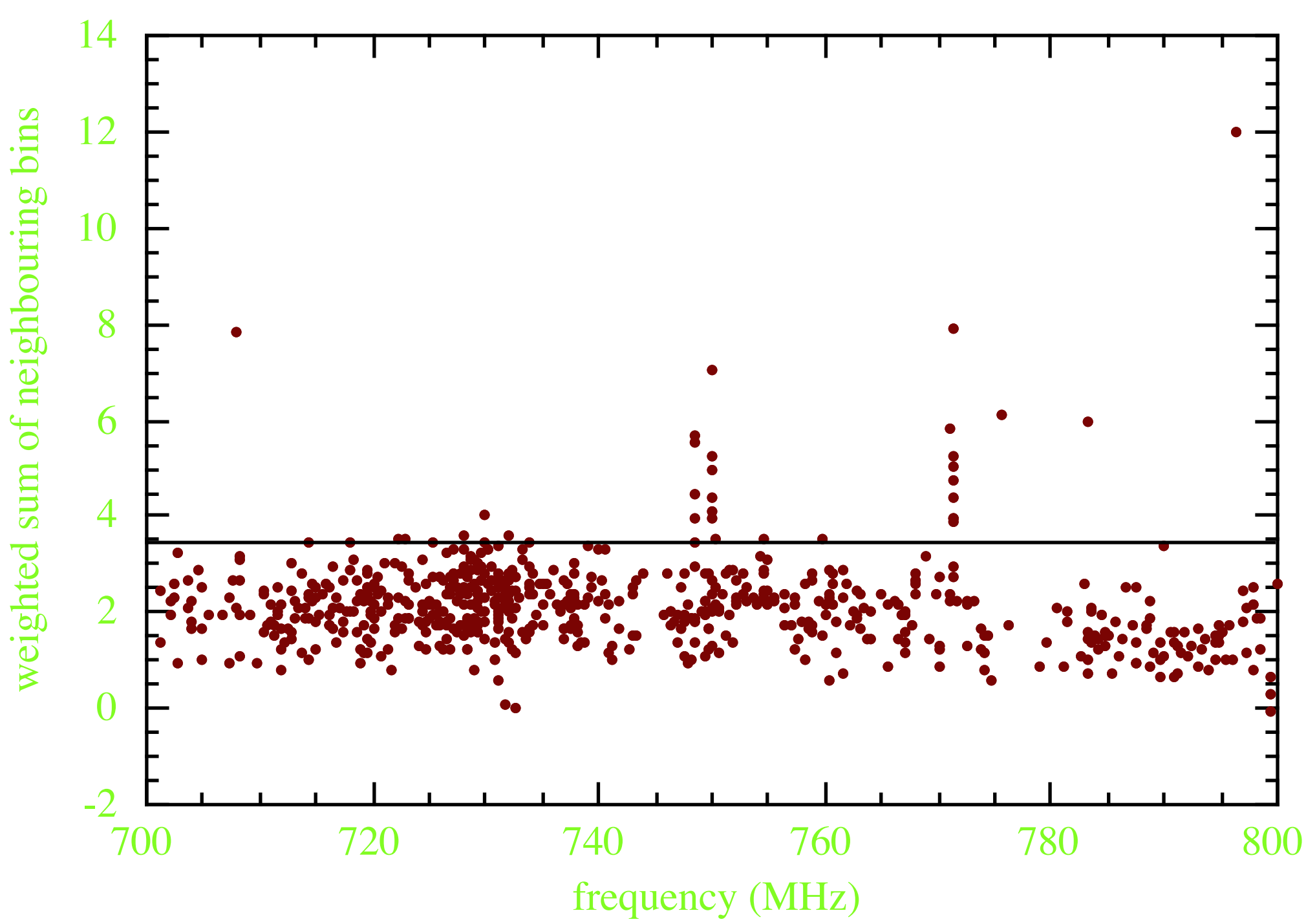

Figure 4.26 weighted sum of bins surrounding
persistent candidate frequencies.

### 4.5.11 Effect of the Run P Cut on Sensitivity

The effect of the run P cut on sensitivity is determined using a simulation similar to that applied to the run (1+R) combined data. Axion signals were added to a set of fake raw traces mirroring the real run 1, run R and run P data (see Appendix 5). This time, the power of axion signal added was frequency dependent; the signal power injected as a frequency f was set equal to the signal power detectable at 93% confidence in runs 1 and R, multiplied by a scale factor. The simulation was run for a set of different values of the scale factor. Figure 4.27 is a plot of the fraction of signals passing the cuts in the run 1, run (1+R) and the run (1+R+P) combined data as a function of the scale factor.

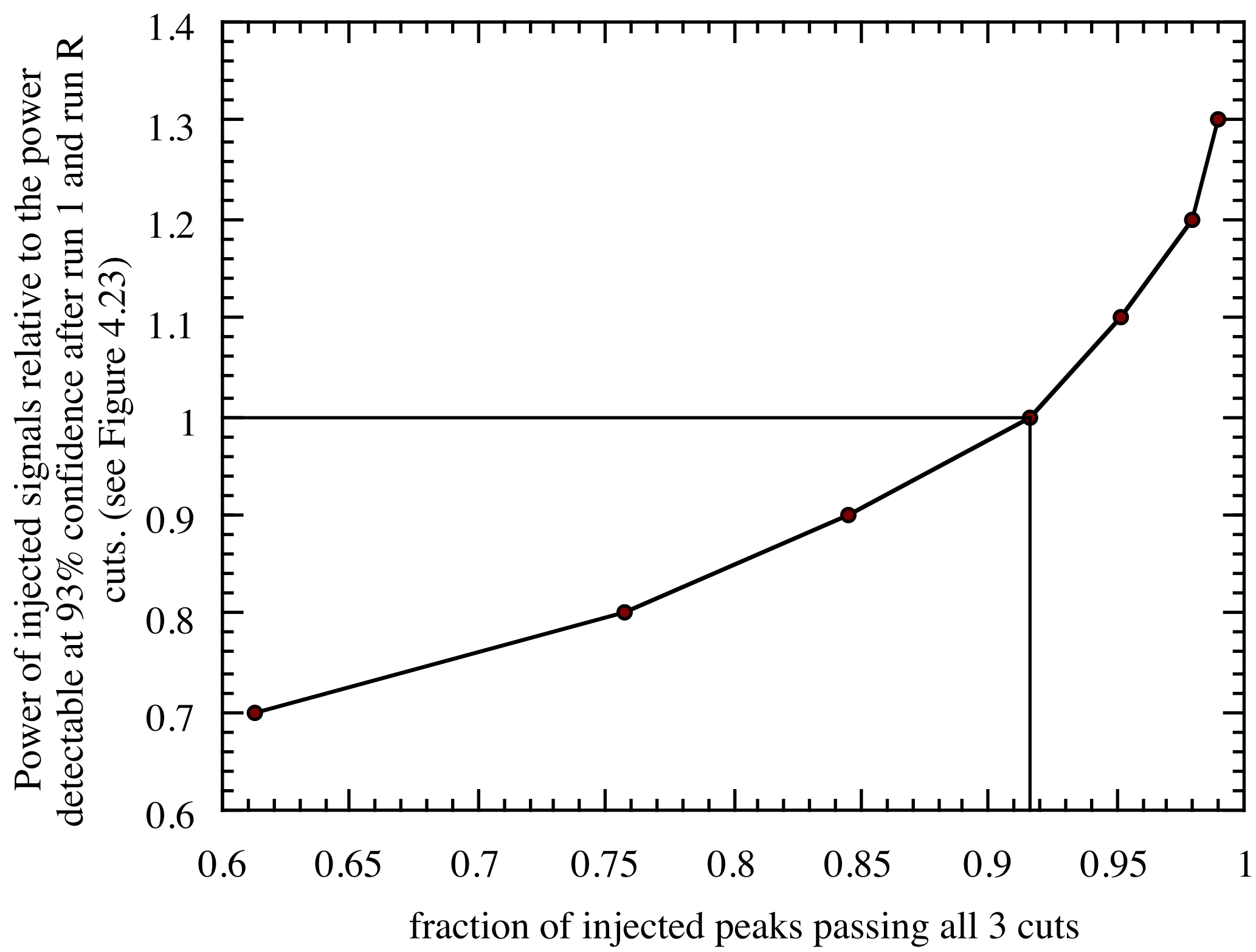

Figure 4.27 On the horizontal axis is the fraction of injected peaks passing all 3 cuts in the 6 bin search channel. On the vertical axis is the injected power in units of the power which passed the (1+R) combined data cuts at 93% confidence.

I conclude from this plot that the 6 bin search channel is sensitive to axions at the 90% confidence level for axion power as a function of frequency shown in Figure 4.23.

### 4.5.12 Elimination 'By Hand' of Surviving 6 Bin Candidates

Table 4.2 shows all the 6-bin candidate frequencies that survive the cuts in runs 1, R, and P, and all the 1-bin candidate frequencies that survive the cuts in runs 1 and R. The 2nd, 3rd, and 4th columns in the table are the peak heights the 3 data sets. The 5th column is a code to describe what method to eliminate that candidate. The different methods for candidate elimination are explained below.

| Frequency (MHz) | Run 1 Peak Height | Run R Peak Height | Run P Peak Height | Method of Elimination |
|---|---|---|---|---|
| 707.792250 | 2.2633 | 7.0965 | 7.8610 | RP,T |
| 714.362250* | 2.2589 | 2.2014 | 3.4795* | NR |
| 718.078125* | 2.2712 | 2.6417 | 3.4418* | NR |
| 722.143875 | 3.2541 | 3.1852 | 3.5557 | NR |
| 729.489750* | 2.4225 | 2.1543 | 2.9880* | NR |
| 729.910750 | 3.3062 | 2.6648 | 4.0381 | RP,T |
| 731.953750 | 2.6281 | 2.5627 | 3.6342 | NR |
| 740.692500* | 2.7180 | 2.0656 | 3.2812* | NR |
| 748.426625 | 6.4740 | 6.4740 | 5.5094 | NR |
| 748.434250 | 6.9001 | 5.3010 | 5.6607 | NR |
| 750.427125 | 6.2973 | 4.2480 | 3.5190 | RP,T |
| 754.498125 | 3.4347 | 3.4347 | 3.5136 | NR |
| 759.703000 | 3.2476 | 2.7648 | 3.5429 | NR |
| 775.749500 | 8.6955 | 4.8389 | 6.1293 | RP,T |
| 771.249375 | 15.559 | 9.1932 | 5.2231 | RP,T |
| 783.249375 | 8.5138 | 5.6073 | 5.9795 | RP,T |
| 790.000000* | 2.4320 | 2.2307 | 3.3705* | NR |
| 796.315000 | 3.1884 | 2.8133 | 12.017 | NR |
| 799.976250* | 5.4750 | 3.1005 | 2.5656* | RP,T |

Table 3.2 A list of candidate frequencies surviving cuts in runs 1, R, and P. The 2nd column is the power excess in the sum of 6 neighboring bins at the candidate frequency in run 1. The 3rd and 4th columns are the weighted sum of sets of 6 neighboring
bins around the candidate frequency for runs R and P, respectively. The last column gives the method by which the candidate frequency was eliminated in manual scanning. NR means that during manual scanning a power excess statistically consistent with a KSVZ axion signal was not seen. RP,T means that the signal at the candidate frequency was identified with a radio peak detectable with an antenna in the lab and that the
peak was subsequently eliminated in the axion detector by terminating the weak and directional-coupler RF ports. These terms are explained in detail below. Candidates marked with an asterisk were below the run P cut, but were examined manually anyway. All heights are divided by $\sqrt{6}$, the enhancement in the rms noise from coadding 6 bins.

For manual scanning of the few remaining candidates, on-line trace combining software is used. Many raw traces can be acquired at a candidate frequency and the average of these traces computed in real time. Overlapping sets of 6 bins are coadded for the six bin peak search.

RP, T Refer to Figure 3.1. The axion receiver is disconnected from the long cable connecting it to the room temperature MITEQ post amplifier. A stub antenna is attached and one or more traces are acquired at the candidate frequency. RP means that a radio peak was detected in the lab at the same frequency as the candidate. This is an indicator that the cause of the candidate may be leakage of this radio signal into the axion detector. For such candidates, the weakly coupled port and the line to the cryogenic directional coupler weak port (see Figure 3.1) are terminated on top of the cryostat. If the peak disappears it was eliminated as a candidate.

NR During manual scanning no peak was seen at this frequency. To establish that there is no peak at high confidence, the coadded power spectrum is calibrated by calculating the power level corresponding to 1 sigma rms in the Gaussian noise background from the amplifier and cavity noise temperature. Figure 4.23 is used to determine the power in the signal being searched for. The actual height of the candidate frequency bin is measured. If the probability is less than 0.5% that the bin height seen is produced by a signal of the power level from Figure 4.23, then the candidate is eliminated.

None of the 6 bin persistent candidates in Table 4.2 were left after manual scanning. I now proceed to derive limits for the 1 bin and 6 bin axion search channels.

# 5. Results

## 5.1 General Strategy for Generating Exclusion Limits

The result of this experiment is that no persistent narrow peaks were observed in the frequency range 701-800MHz, corresponding to an axion mass range of $2.9-3.3\mu eV$. The minimum power signal power $P_{exc90}$ excluded at 90% confidence is different for the two search channels, and within a single search channel it is frequency dependent. To turn this minimum excluded power into a limit, I start with the fact that the power of the axion to photon conversion signal is proportional to the coupling constant $g_{a\gamma\gamma}$ squared and the halo density. It is customary to assume a fixed halo density and give an exclusion limit on the coupling constant. I assume a local halo density of 0.45GeV/cc. Using Equation 2.22 I can write:

$$\frac{g_{a\gamma\gamma}{}^2(\mathrm{KSVZ})}{g_{a\gamma\gamma}{}^2(\mathrm{excl90})} = \frac{P_{\mathrm{KSVZ}}}{P_{\mathrm{excl90}}} \qquad (5.1)$$

Here $P_{KSVZ}$ is the signal power we would see in our cavity due to a KSVZ axion, $g_{a\gamma\gamma}(\mathrm{KSVZ})$ is the axion photon coupling constant in the KSVZ model, $P_{exc90}$ is the signal power our experiment is sensitive to at 90% confidence and $g_{a\gamma\gamma}(\mathrm{excl90})$ is the axion photon coupling that our experiment can rule out with 90% confidence. In the KSVZ axion model the axion mass is related to $g_{a\gamma\gamma}(\mathrm{KSVZ})$ by:

$$m_a = 0.62\mathrm{eV}\frac{\pi\left[g_{a\gamma\gamma}(\mathrm{KSVZ})\right](\mathrm{GeV}^{-1})}{0.97(10^{-7}\mathrm{GeV}^{-1})\alpha} \qquad (5.2)$$

Using (5.1.1) and (5.1.2) the upper limit on the axion-photon coupling as a function of axion mass is:

$$g_{a\gamma\gamma}{}^2(\mathrm{excl}) = \left(10^{-7}\mathrm{GeV}^{-1}\right)^2\left(\frac{m_a}{0.62\mathrm{eV}}\right)^2\left(\frac{0.97\alpha}{\pi}\right)^2\left(\frac{P_{\mathrm{excl90}}}{P_{\mathrm{KSVZ}}}\right) \qquad (5.3)$$

To deduce an upper limit on $g_{a\gamma\gamma}{}^2$ assuming a different value for the halo density $\rho_{NEW}$, multiply the upper limit for $g_{a\gamma\gamma}{}^2$ in our limit plots by the ratio ($7.5\times10^{-25}\,\mathrm{gcm}^{-3}/\rho_{NEW}$).

## 5.2 The Exclusion Limit in the Single Bin Peak Channel

In the 1-bin peak search we assume that the axion signal power is concentrated in a narrow spike with width significantly less than the 125Hz-wide bins in our FFT traces. We take the 1-bin data to Gaussian-distributed.

As I discuss in Appendix 5, although one might expect all the power from the a peak much narrower than the 125Hz frequency bin width to be seen in a single bin of the combined data, in reality this is only true if the axion signal is at the exact center of one of the frequency bins. Let $\xi$ represent the position of the delta function spike in a frequency bin where $\xi=0$ if the spike is at the bin center and $\xi=1$ if the spike is at bin edge. The result from Appendix 5 is that the fraction of the axion power in the frequency bin encompassing the signal is given by

$$f(\xi) = \frac{4}{\pi^2 \xi^2} \sin^2\left(\frac{\pi\xi}{2}\right) \qquad (5.4)$$

Using this result, knowing the SNR for KSVZ axions and assuming that the 1-bin data is Gaussian distributed, I can derive an exclusion limit. In both run 1 and run R we I a cut on the 1-bin combined data at $3.3\sigma$. If the axion signal is a spike of power P in a single frequency bin offset from the center of the bin by $\xi$, then the signal power deposited in that bin is $Pf(\xi)$. The probability of this signal passing the cut is given by:

$$\begin{aligned} cl(P,\xi) &= \frac{1}{2\pi} \int_{-\infty}^{Pf(\xi)-3.3} dx \exp\left(\frac{-x^2}{2}\right) \\ &= \frac{1}{2}\left(\frac{1+\operatorname{erf}(Pf(\xi)-3.3)}{\sqrt{2}}\right) \end{aligned} \qquad (5.5)$$

To obtain the confidence level, take the average of $cl(P,\xi)$ from the center of a bin to the edge. Combining (5.4) and (5.5) and averaging over $\xi$ I get:

$$\overline{cl}(P) = \frac{1}{2} \int_{\xi=0}^{1} d\xi \left(1 + \operatorname{erf}\left(\frac{1}{\sqrt{2}}\left(P \operatorname{sinc}^2\left(\frac{\pi\xi}{2}\right) - 3.3\right)\right)\right) \qquad (5.6)$$

Figure 5.1 is a plot of $\overline{\text{cl}}(P)$ vs. P generated by doing the integral numerically. The lower plot is a magnified section of the upper plot.

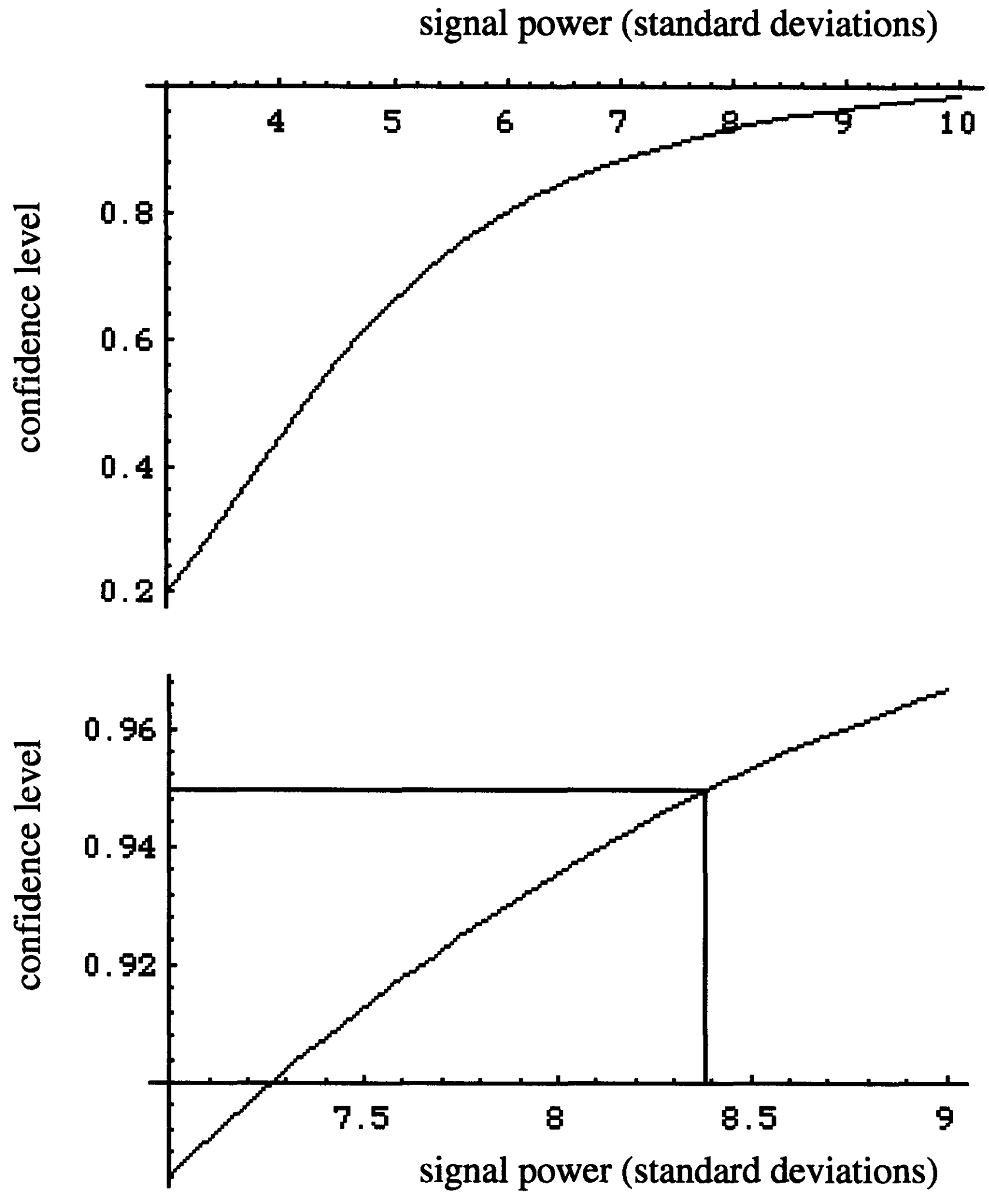

Figure 5.1 Confidence level vs. signal power in the single bin channel, assuming that the axion linewidth is small compared to 125Hz, with a cut of 3.3σ applied in to both the run 1 and the run R data

A 3.3σ cut was applied in both run 1 and run R. I require that the overall confidence of detecting a signal at power P to be 90%. Therefore in each of the 2 runs, the probability of a signal passing the cut must be $\sqrt{0.9} = 0.949$. From the above plot, a signal of power P=8.38σ will pass the 3.3σ cut at 94.9% confidence. Hence in our 1-bin analysis we are sensitive to 8.38σ axion signals at 90% confidence. More powerful signals are detected at higher confidence.

To turn this result into a limit I use the SNR as a function of frequency in run 1 (the SNR as a function of frequency in run R always equaled or exceeded that in run 1, so this method is conservative). Using Equation (5.1) where the ratio of the KSVZ axion power to the axion power detectable at 90% confidence is SNR/8.38 I obtain $g_{a\gamma\gamma}^2$ excluded at 90% confidence as a function of frequency. The exclusion region is shown in Figure 5.2.

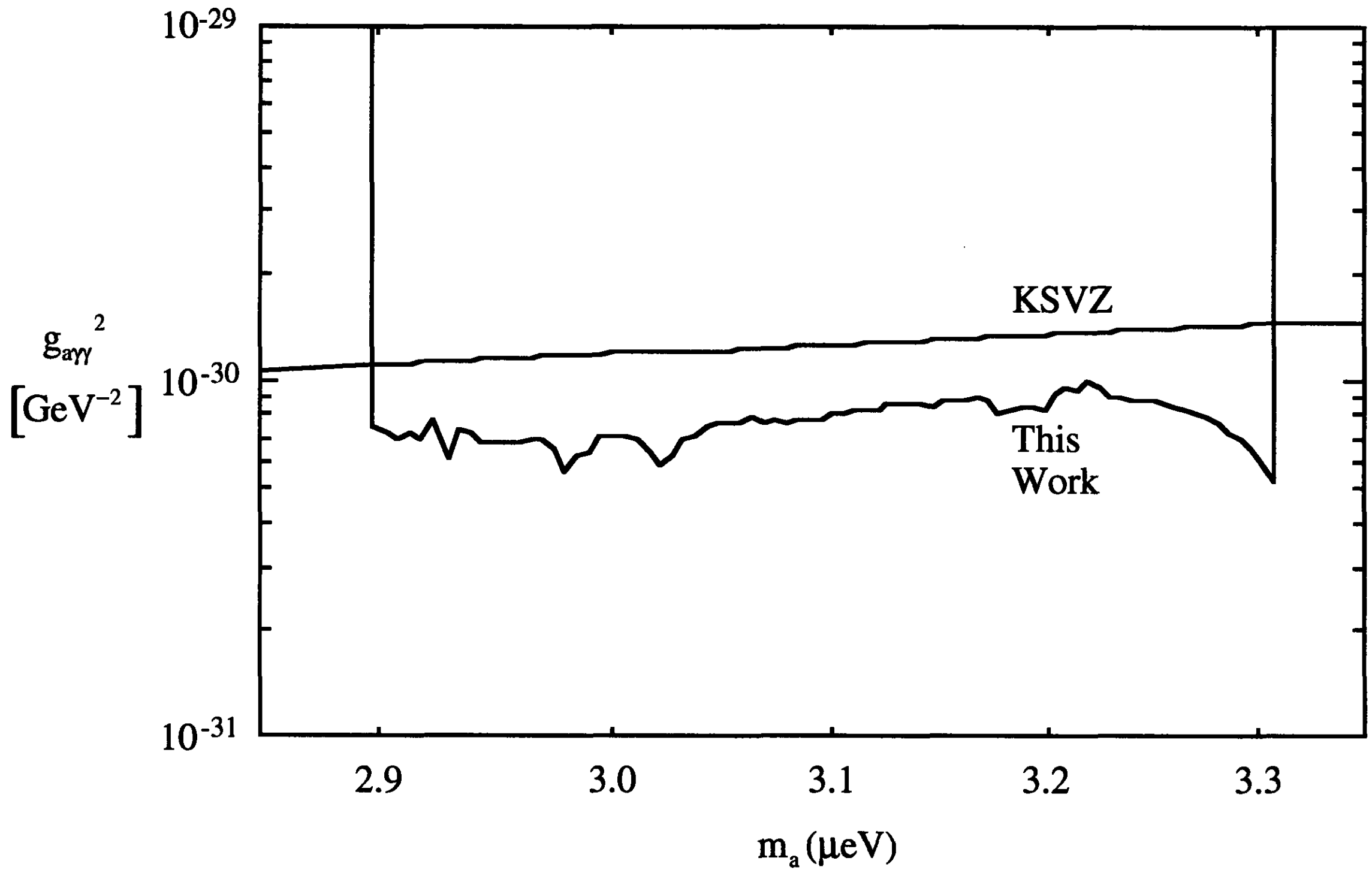

Figure 5.2, axion-photon couplings excluded at 90% confidence
by this work assuming axion line is narrow compared to 125Hz

Also shown is $g_{a\gamma\gamma}^2$ vs. axion mass for the KSVZ axion model. As expected, the experiment is sensitive at the sub-KSVZ level for the whole of the mass range we covered if the axion signal is significantly narrower than 125Hz.

## 5.3 The Exclusion Limit in the Six Bin Channel

The sensitivity to axions in the 6-bin channel is deduced from the results of the Monte Carlo simulation described in Section 4.5.8. Summarizing the simulation method, I divide the frequency range in which we have raw data into 20 segments. I make a combined data set starting with the real traces, but replacing some of the traces with computer generated ones containing artificially generated Maxwellian peaks overlaid with Gaussian noise at a pre-determined set of frequencies. In this way a set of combined data containing axion-like peaks is built up. We first make peak-injected run 1 data, then apply the run 1 cut to it. Next we analyze the run R data in the same way, injecting peaks only at the frequencies that passed the run 1 cut. Finally we apply the run R cut to this combined data and list the subset of the original injection frequencies in each of the 20 frequency bins that passed both cuts. In this way we obtain the efficiency of the peak search for a known injected signal power as a function of frequency. Finally, we run the simulation multiple times with different injected powers, and make a plot of fraction of peaks passing both cuts as a function of injected signal power for each frequency bin. The power detectable at 90% confidence in each bin is read off from these plots.

Figure 4.23 is the results of our simulation on runs 1 and R at 93% confidence. Further degradation in efficiency from the run P cut at 3.5 sigma reduces the confidence level to 90%. The results from Figure 4.23 can be converted directly into an exclusion plot using Equation 5.3. Figure 5.3a shows the region of axion masses and couplings to photons excluded at 90% confidence in our experiment. On the vertical axis is the square of the coupling constant divided by the square of the mass. Figure 5.3b is an exclusion plot of a larger range of axion masses showing results from the two previous pilot experiments [12,13]. Also shown are the KSVZ and DFSZ model predictions. The experiment has achieved an improvement in sensitivity of two orders of magnitude, and is the first axion search experiment to probe a region of parameter space occupied by realistic axion models.

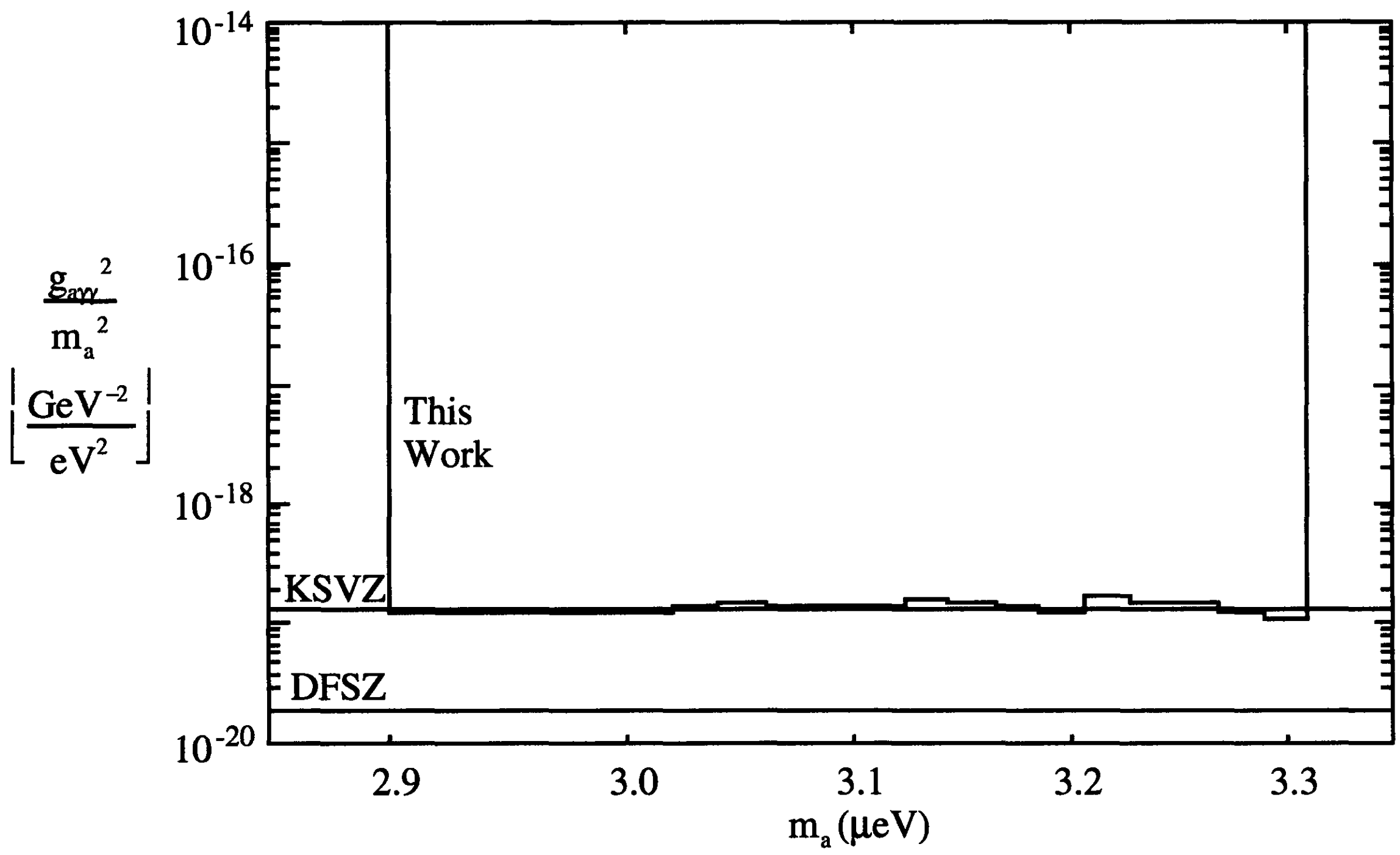

Figure 5.3a Exclusion limit for this experiment at 90% confidence in the 6 bin search channel. The vertical axis is in units of (coupling constant / axion mass)$^2$

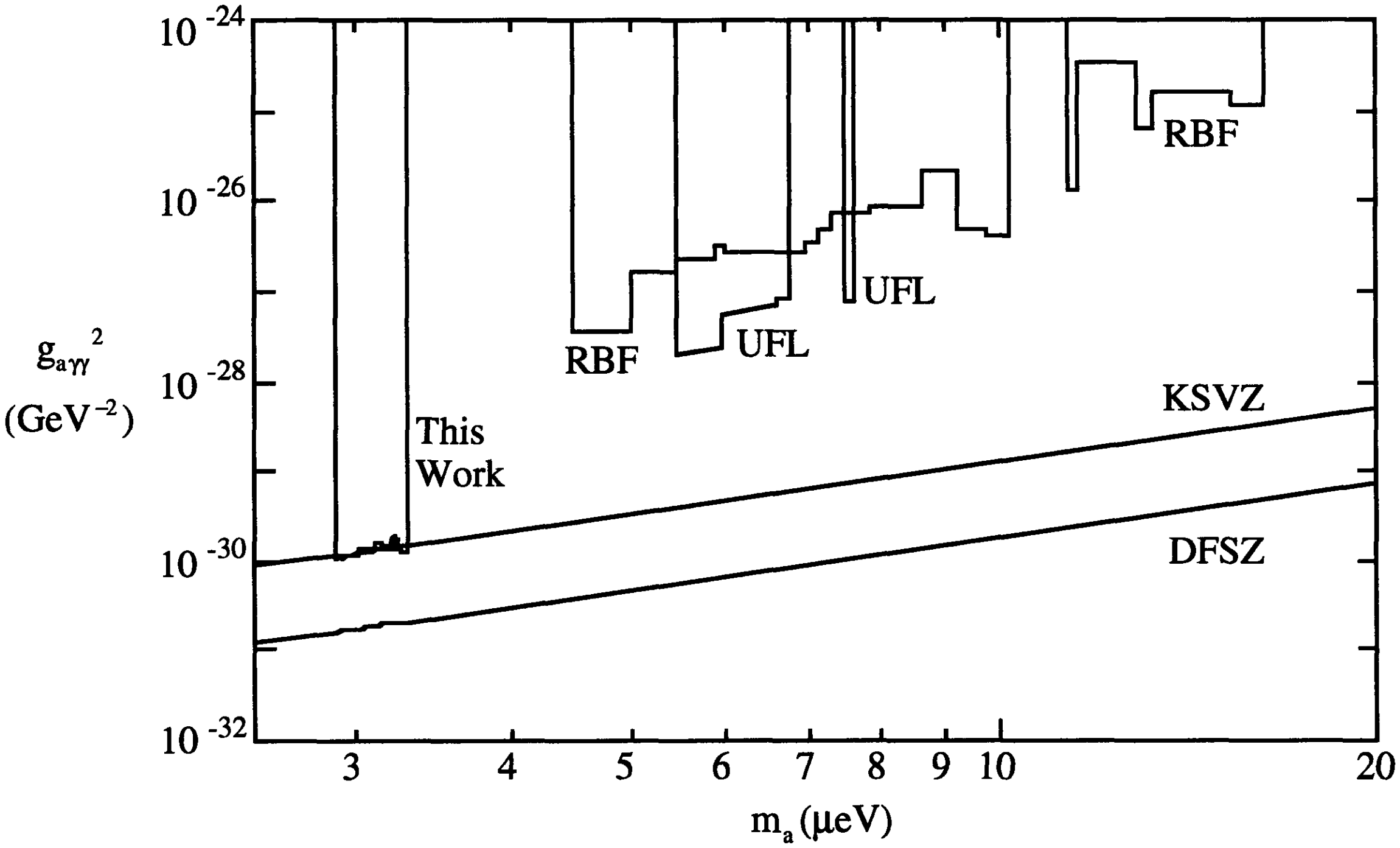

Figure 5.3b Exclusion limit on the coupling constant at 90% confidence. Also shown are the results of previous pilot experiments.

# 6. Conclusions

I have described the axion search experiment and the analysis of the first production data. The result of this analysis is an exclusion limit on axions at masses and couplings consistent with theoretical predictions and at number densities comparable with predictions from astrophysics. In its first production run, our experiment was sensitive to realistic axion halo cold dark matter for the first time - it has been a great success.

The experiment continues to take data. The mass range covered in the first run was 2.9 - 3.3 μeV covered in about 1.5 years of data taking. Furthermore the experiment is capable of covering the full mass range without major modification, and could cover the bulk of that mass range at a significantly higher rate with modest modifications based on existing technology. In addition, the experiment has spawned several interesting research and development projects.

The 'invisible' axion may not be invisible for long...

# 6. Conclusions

I have described the axion search experiment and the analysis of the first production data. The result of this analysis is an exclusion limit on axions at masses and couplings consistent with theoretical predictions and at number densities comparable with predictions from astrophysics. In its first production run, our experiment was sensitive to realistic axion halo cold dark matter for the first time - it has been a great success.

The experiment continues to take data. The mass range covered in the first run was 2.9 - 3.3 μeV covered in about 1.5 years of data taking. Furthermore the experiment is capable of covering the full mass range without major modification, and could cover the bulk of that mass range at a significantly higher rate with modest modifications based on existing technology. In addition, the experiment has spawned several interesting research and development projects.

The 'invisible' axion may not be invisible for long...

## Appendix 1: Axion Field Theory in MKS Units

In order to derive an expression for the power deposited in our resonant cavity due to axion-photon conversion, I first need to write the relevant terms in the QCD-axion Lagrangian in MKS units. Using the symbols $\lambda$ and $\kappa$ to denote quantites that we will deduce using dimensional analysis, the Lagrange density in natural units ($\hbar$ = c = 1) is:

$$\mathcal{L}=\frac{1}{2}(\partial_\mu a)^2+\frac{1}{2}\mu m^2 a^2-\kappa g_{a\gamma\gamma}a\mathbf{E}.\mathbf{B} \qquad \text{(A1.1)}$$

In the following I will use square brackets [] to denote 'dimensions of' and M, L, T and Q to denote mass, length, time and charge respectively. I start by working out the dimensions of the lagrange density. The action is the integral of the lagrange density, $S=\int ã d^4x$, and the dimensions of the axion are $[S]=[\hbar]=ML^2T^{-1}$. Hence $[\mathcal{L}]=[S]/[d^4x]=ML^{-2}T^{-1}$. Using this with the 1st 'kinetic' term in the Lagrange density (A1.1) I can deduce the dimensions of the axion field: $[\mathcal{L}]=[(\partial_\mu a)^2]\Rightarrow[a^2]=[ã]L^2=MT^{-1}$. Knowing $[a^2]$ I can deduce $[\mu]$ from the second term in (A1.1): $[\mathcal{L}]=[a^2]M^2[\mu]\Rightarrow[\mu]=M^{-2}L^{-2}=[c^2/\hbar^2]$. The third term contains the product of two quantities, $\kappa$ and $g_{a\gamma\gamma}$ neither of which has pre-determined dimensions. For convenience I will try to find a quantity $\kappa$ such that $[\kappa][\mathbf{E.B}]=[\mathcal{L}]$ and $[g_{a\gamma\gamma}\ a]=1$. Using $[\mathbf{E}]=MLT^{-2}Q^{-1}$ and $[\mathbf{B}]=MT^{-1}Q^{-1}$ I get $M^2LT^{-3}Q^{-2}[\kappa]=[ã]=ML^{-2}T^{-1}\Rightarrow[\kappa]=M^{-1}L^{-3}T^2Q^2$. Thus $\kappa$ has the dimensions of $\varepsilon_0$, the permittivity of free space. It is natural that this quantity should appear in a Lagrangian term containing the **E** and **B** fields. I can now write the Lagrange density in MKS units:

$$\mathcal{L}=\frac{1}{2}(\partial_\mu a)^2+\frac{1}{2}\frac{m^2c^2}{\hbar^2}a^2-\varepsilon_0 g_{a\gamma\gamma}a\mathbf{E}.\mathbf{B} \qquad \text{(A1.2)}$$

Because $[g_{a\gamma\gamma}\ a]=1$ I can deduce $[g_{a\gamma\gamma}{}^2]=M^{-1}T$. Another important quantity in axion physics is the mass density of axions $\rho_a$, related to the mean square axion field by:

$$\langle a^2\rangle=\eta\frac{\rho_a}{m^2} \qquad \text{(A1.3)}$$

Again I have inserted an unknown constant $\eta$ to be deduced by dimensional analysis. Using [a] calculated from the lagrangian we get $[\eta]=MT^{-1}.M^{-1}L^3.M^2=M^2L^3T^{-1}=[\hbar^2/c]$. Hence in MKS units (A1.3) becomes:

$$\langle a^2 \rangle = \frac{\rho_a \hbar^2}{m^2 c} \qquad \text{(A1.4)}$$

To construct a dimensionless quantity out of the axion mass, density and coupling to photons, exploit the fact that $[g_{a\gamma\gamma}\, a]=1$ to deduce that:

$$\left[\frac{g_{a\gamma\gamma}{}^2 \hbar^2 \rho_a}{m^2 c}\right] = 1 \qquad \text{(A1.5)}$$

This dimensionless quantity is useful for calculating the power for axion to photon conversion. We would like a power in watts but we prefer to write the axion mass and axion-photon coupling in natural units. Using (A1.5) we can write the formula as the product of dimensionless quantity in parentheses and a string of cavity parameters in MKS units. The quantity in parentheses can be evaluated in natural units by setting $(\hbar = c = 1)$.

# Appendix 2: Thermal Noise and Nyquist's Theorem

## A2.1 Nyquist's Theorem

The average mean square size of the voltage fluctuations $\langle V^2 \rangle$ due to thermal noise across a resistor at temperature T is given by an important theorem first derived by Nyquist. I will derive Nyquist's theorem by considering the circuit in Figure A2.1:

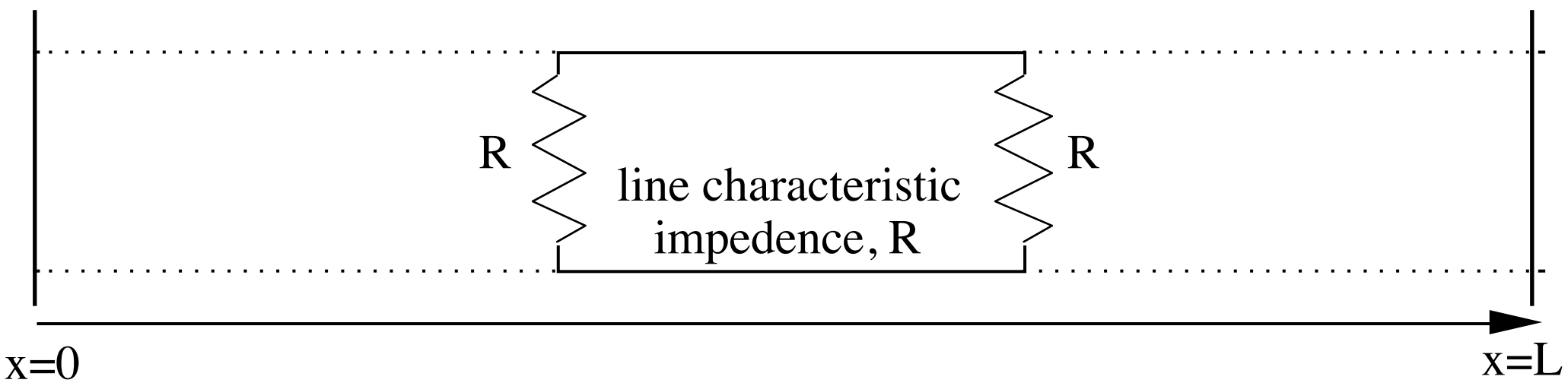

Figure A2.1 A transmission line of real characteristic impedance R perfectly matched at both ends.

A transmission line of characteristic impedance R is terminated at each end by a resistor also of impedance R. Thus traveling waves propagating down the line in either direction will be totally absorbed at the terminating impedance. The apparatus is actually a 1-dimensional resonant cavity and the terminating resistors are the walls of a black body enclosing the resonant cavity. Let the terminating resistors be in thermal equilibrium at temperature T. Inside the transmission line is black body radiation also at temperature T in thermodynamic equilibrium with the terminating resistors.

Since the transmission line is perfectly matched at both ends it supports no standing waves, and the solutions of the wave equation on the line are traveling waves with any frequency moving in either direction at speed c. The rate of energy transfer from one terminating resistor to the other can be calculated by considering population of the modes of oscillation of the line which represent traveling waves in one direction only. Unfortunately, since the line is a continuum there is a non-denumerable infinity of such modes. I can get around this problem by pretending that the transmission line is a small section of a much longer line, of length L. At the ends of this line I impose periodic boundary conditions i.e.,

$$\psi(x=0)=\psi(x=L) \qquad \text{(A2.1)}$$

where $\psi(x)$ is the amplitude of the wave. Periodic boundary conditions reduce the number of modes to a denumerable infinity, with wavenumbers given by:

$$k_p = \frac{2\pi p}{L} \text{ where } p \in \{1,2,...\} \qquad \text{(A2.2)}$$

Negative p give the wavenumbers of traveling waves propagating to the left, which I ignore here since I am interested in energy transfer in one direction only. Hence the number of right-propagating modes n in a bandwidth B per unit length of line is:

$$n = \frac{B}{c} \qquad \text{(A2.3)}$$

Notice that this result is independent of L. In the classical limit where $h\nu << k_B T$, each mode carries an average of $k_B T$ in energy. So the average energy density $\overline{U}$ in the line due to radiation in bandwidth B propagating to the right is:

$$\overline{U} = \frac{k_B TB}{c} \qquad \text{(A2.4)}$$

The average energy $\bar{\varepsilon}$ deposited in the right hand resistor in time t is the average energy in the right-traveling modes in the length of line ct to the left of the resistor:

$$\bar{\varepsilon} = \overline{U}ct = k_B TBt \qquad \text{(A2.5)}$$

So the average power $\overline{P}$ deposited in the right hand resistor due to black body radiation in the line at temperature T is:

$$\overline{P} = k_B TB \qquad \text{(A2.6)}$$

The sole source of the radiation in the line that is absorbed in the right hand resistor is the left hand resistor. So what would be the size of voltage fluctuations in series with the left hand resistor necessary to cause the power given by Equation A2.6 to be absorbed in the right hand resistor? P can be related to the current around the loop made by the transmission line and the two resistors:

$$P = \langle I^2 \rangle R = \frac{\langle V^2 \rangle}{4R} \qquad \text{(A2.7)}$$

where V=2RI by Ohm's law around the loop. Hence the rms. voltage fluctuation across a resistor in thermodynamic equilibrium at temperature T due to its own internally generated Johnson noise is:

$$\langle V^2 \rangle = 4k_B TRB \qquad \text{(A2.8)}$$

Equations A2.6 and A2.8 are both statements of Nyquist's theorem. Note that real circuits the source and load resistors are often at different physical temperatures. Results A2.6 and A2.8 are still correct in this case. The reason that the above derivation of Nyquist's theorem is still valid is that the left hand resistor is in thermal equilibrium with the right-propagating modes in the cable and the right-hand resistor is in thermal equilibrium with the left propagating modes. Nowhere in the above derivation do I require that the left- and right- propagating modes are in equilibrium with each other, hence if the left hand resistor is at temperature $T_L$ then the power it emits is $kT_LB$, regardless of the temperature of the right hand resistor. There is a net transfer of energy from the left handed resistor to the right handed resistor when they are at different temperatures. However, if both the resistors are immersed in large heat baths then the physical temperatures of the two resistors remain fixed and different.

Note that Equation A2.8 for the mean square voltage fluctuations across a resistor at temperature T means the voltage fluctuations that give rise to the power emitted by that resistor given in Equation A2.6. The *total* voltage fluctuations across a resistor in the circuit of Figure A2.1 is the sum of fluctuations due to its own internally generated thermal noise and those due to thermal noise in other resistor. Where $T_L$ ($T_R$) is the temperature of the left (right) hand resistor, the total mean square voltage fluctuation across the resistor $<V^2>_{TOT}$ is:

$$\langle V^2 \rangle_{TOT} = 4k_B (T_L + T_R) RB \qquad \text{(A2.9)}$$

Equation A2.9 has important consequences for our experiment. Consider: a source (the cavity) is connected to a load resistor (the 1st cryogenic amplifier input impedance). The amplifier can be considered as a voltage multiplier. Its output is the voltage across the internal termination multiplied by the amplifier gain G. Assume that the amplifier and cavity are perfectly matched. Figure A2.2 shows the equivalent circuit.

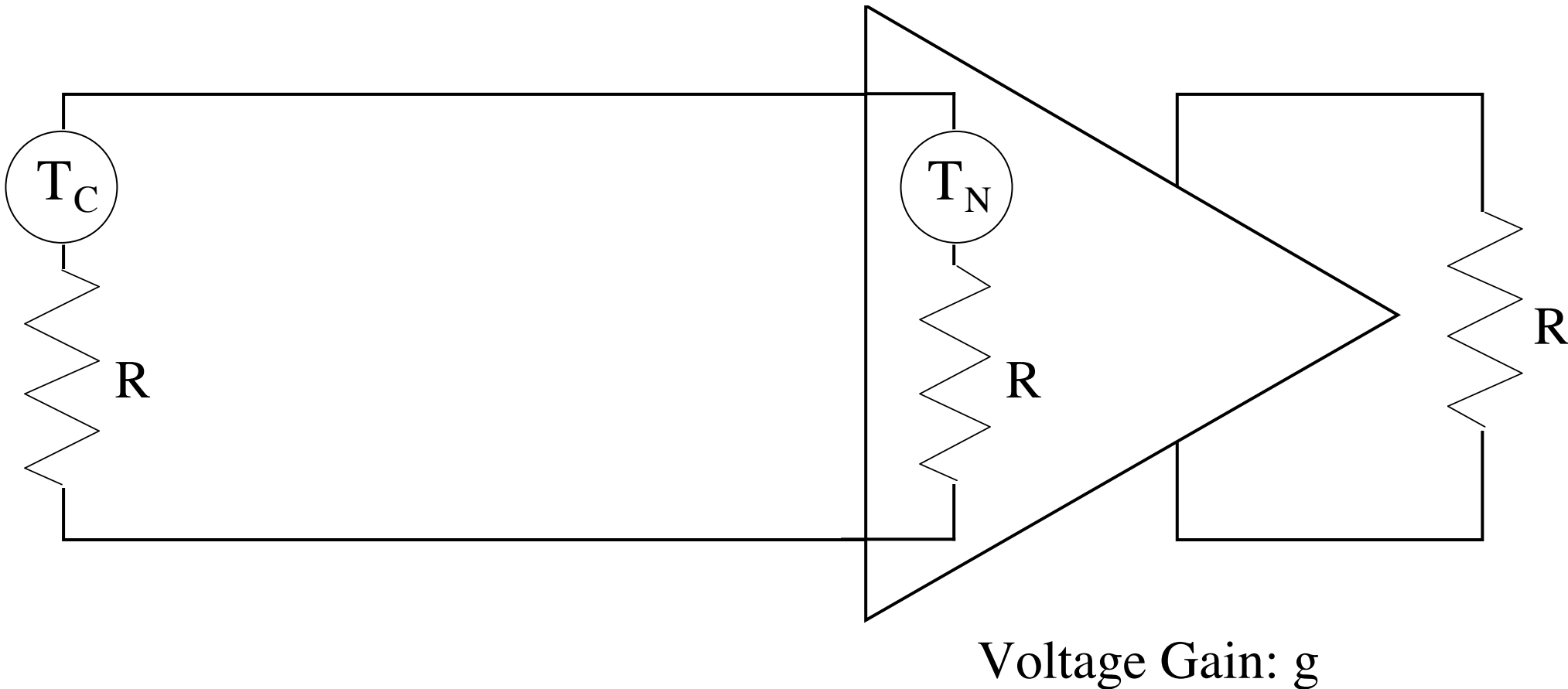

Figure A2.2 Resistor coupled to the input of a noisy amplifier.

Using Equation A2.9, the power $P_{NTOT}$ at the load resistor connected to the amplifier output due to both noise sources in the input circuit is:

$$P_{NTOT} = k_B\left(T_C + T_N\right)Bg^2 \qquad \text{(A2.10)}$$

So the temperature associated with the Johnson noise in the input circuit is $T_C$+$T_N$.

## A2.2 Fluctuations about the Average Energy Density

Nyquist's theorem furnishes an expression for the average energy deposited in a load resistor in time t due to a source resistor at temperature T to which it is perfectly matched. However, if I was to measure the actual energy ε deposited in the resistor in many different time intervals each of length t, I would get a set of values for ε distributed about the average value $\bar{\varepsilon}$ with some rms. spread $\sigma_\varepsilon$. To derive an expression for $\sigma_\varepsilon$ I will use some formalism from statistical mechanics first written down by Gibbs in 1902 [17].

Consider a system in thermal contact with a heat reservoir at temperature T. The probability distribution for the energy of the system is:

$$p(\varepsilon) = \int d\Omega\, e^{\frac{\alpha - e(p_i, q_i)}{\Theta}} \delta(e(p_i, q_i) - \varepsilon) \qquad \text{(A2.11)}$$

where the integral is over the phase space $(p_i,q_i)$ accessible to the system and $\Theta=kT$. $\alpha$ is a function of $\Theta$ but is independent of the phase space position $(p_i,q_i)$. The normalization is:

$$\int d\varepsilon\, p(\varepsilon) = \int d\Omega\, e^{\frac{\alpha-\varepsilon}{\Theta}} = 1 \qquad \text{(A2.12)}$$

The average value of a function $f(\varepsilon)$ of the energy of the system is given by:

$$\overline{f(\varepsilon)} = \int d\varepsilon\, f(\varepsilon)\, p(\varepsilon) = \int d\Omega\, f(\varepsilon)\, e^{\frac{\alpha-\varepsilon(p_i,q_i)}{\Theta}} \qquad \text{(A2.13)}$$

So the average energy $\bar{\varepsilon}$ is:

$$\bar{\varepsilon} = \int d\Omega\, \varepsilon\, e^{\frac{\alpha-\varepsilon}{\Theta}} \qquad \text{(A2.14)}$$

Next we evaluate $\alpha$. Differentiating Equation A2.12 with respect to $\Theta$ I get:

$$\int d\Omega \left( \frac{\varepsilon-\alpha}{\Theta^2} + \frac{1}{\Theta}\frac{d\alpha}{d\Theta} \right) e^{\frac{\alpha-\varepsilon}{\Theta}} = 0 \qquad \text{(A2.15)}$$

or,

$$\int d\Omega\, \varepsilon\, e^{\frac{\alpha-\varepsilon}{\Theta}} - \left( \alpha - \Theta\frac{d\alpha}{d\Theta} \right) \int d\Omega\, e^{\frac{\alpha-\varepsilon}{\Theta}} = 0 \qquad \text{(A2.16)}$$

Using the normalization condition of Equation A2.12 and the formula for the average value of a function of energy from Equation A2.13, I obtain the following differential equation for $\alpha$:

$$\alpha - \Theta\frac{d\alpha}{d\Theta} = \bar{\varepsilon} \qquad \text{(A2.17)}$$

I want an expression for the rms. deviation from the mean energy, $\sigma_\varepsilon \equiv \sqrt{\overline{\varepsilon^2} - \bar{\varepsilon}^2}$. Differentiating Equation A2.14 with respect to $\Theta$ I get:

$$\frac{d\bar{\varepsilon}}{d\Theta} = \int d\Omega\varepsilon\left[\frac{\varepsilon - \alpha}{\Theta^2} + \frac{1}{\Theta}\frac{d\alpha}{d\Theta}\right]e^{\frac{\alpha-\varepsilon}{\Theta}}$$

$$= \int d\Omega\frac{\varepsilon}{\Theta^2}\left[\varepsilon - \left(\alpha - \Theta\frac{d\alpha}{d\Theta}\right)\right]e^{\frac{\alpha-\varepsilon}{\Theta}} \qquad \text{(A2.18)}$$

$$= \int d\Omega\frac{\varepsilon}{\Theta^2}\left[\varepsilon - \bar{\varepsilon}\right]e^{\frac{\alpha-\varepsilon}{\Theta}}$$

where for the last line I have used Equation A2.17. Rearranging A2.18 I get:

$$\Theta^2\frac{d\bar{\varepsilon}}{d\Theta} = \int d\Omega\varepsilon^2 e^{\frac{\alpha-\varepsilon}{\Theta}} - \bar{\varepsilon}\int d\Omega\varepsilon e^{\frac{\alpha-\varepsilon}{\Theta}}$$

$$= \overline{\varepsilon^2} - \bar{\varepsilon}^2 \qquad \text{(A2.19)}$$

$$= \sigma_\varepsilon^{\ 2}$$

or,

$$\sigma_\varepsilon^{\ 2} = kT^2\frac{d\bar{\varepsilon}}{dT} \qquad \text{(A2.20)}$$

Equation A2.20 is a very useful and general result. Given a system which is in thermodynamic equilibrium at temperature T having some energy with the probability distribution of Equation A2.11 proposed by Gibbs, I can deduce the mean square size of the fluctuations in energy of the system about its mean value. No assumptions are made about the physical nature of the system.

## A2.3 The Radiometer Equation

Now I are ready to reconsider the terminated line discussed in the previous derivation of Nyquist's theorem. I take as my system in thermodynamic equilibrium at temperature T the right-propagating modes in bandwidth B in a length of line (ct) immediately to the left of the right hand resistor (Figure A2.1). From Equation A2.5 the average energy stored in these modes is:

$$\bar{\varepsilon} = k_B TBt \qquad \text{(A2.21)}$$

The probability distribution for the energy stored in modes of a resonant cavity with black walls is the Gibbs distribution - this is how one derives the classical average energy $k_B T$ per mode. Hence the mean square deviation from the mean energy dissipated in the resistor is given by Equation A2.20:

$$\sigma_\varepsilon^{\ 2} = k_B^{\ 2} T^2 Bt \qquad \text{(A2.22)}$$

Combining Equations A2.21 and A2.22 I get:

$$\frac{\sigma_\varepsilon}{\bar{\varepsilon}} = \frac{1}{\sqrt{Bt}} \qquad \text{(A2.23)}$$

Finally, the rms. deviation $\sigma_\varepsilon$ and average energy $\bar{\varepsilon}$ in this particular set of modes are equal to the energy absorbed by the terminating resistor in time t. Hence Equation A2.23 can be rewritten in terms of the average power $\overline{P_N} = \bar{\varepsilon}/t$ and rms. deviation from the average power $\sigma_P = \sigma_\varepsilon / t$ deposited in the load resistor:

$$\frac{\sigma_P}{\bar{P}} = \frac{1}{\sqrt{Bt}} \qquad \text{(A2.24)}$$

Equation A2.24 is the radiometer equation. To cast it in a more familiar form, suppose that I measure the energy simultaneously in the same bandwidth about many different frequencies, as in a spectrum analyzer. Suppose at one of these frequencies there is excess power $P_S$. The signal to noise ratio SNR is the ratio of $P_S$ to the rms. of the noise fluctuations. Using Equation A2.24, the signal to noise ratio is:

$$\mathrm{SNR} = \frac{P_S}{\sigma_P} = \frac{P_S}{\overline{P_N}} \sqrt{Bt} \qquad \text{(A2.25)}$$

where $\overline{P_N}$ is the average noise power.

# Appendix 3: Equivalent Circuit Model

## A3.1 Equivalent Circuit of a Resonant Cavity Mode

An oscillation mode of a resonant cavity can be represented by the series RLC circuit shown in Figure A3.1.

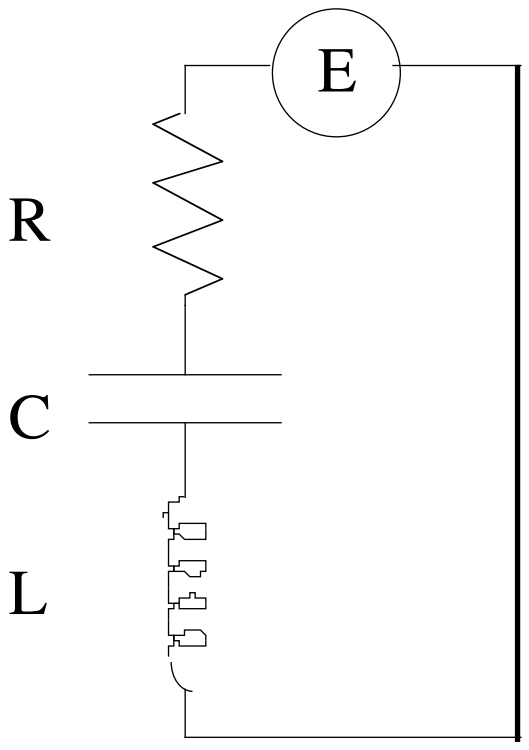

Figure A3.1 Equivalent circuit of a resonant cavity with Johnson noise.

The voltage source generates a sine wave of angular frequency ω. The impedance of the RLC network at this frequency is

$$Z = R + i(\omega L - \frac{1}{\omega C}) \qquad (A3.1)$$

The complex part of the impedance vanishes at $\omega_0$ where

$$\omega_0 = \frac{1}{\sqrt{LC}} \qquad (A3.2)$$

So we can rewrite (1) as

$$Z = R + iL(\frac{\omega^2 - \omega_0^{\ 2}}{\omega}) \qquad (A3.3)$$

Let $\omega - \omega_0 = \Delta$. We assume that $\Delta << \omega_0$ and write the impedance to lowest order in Δ.

$$Z = R + 2iL\Delta \qquad (A3.4)$$

The power dissipated in the circuit is given by

$$P = \Re e\{\frac{|E|^2}{Z}\} = E_0^{\ 2}\Re e\{\frac{1}{R + 2iL\Delta}\} = \frac{E_0^{\ 2}}{R}\frac{1}{1 + 4\Delta^2(\frac{L}{R})^2} \qquad \text{(A3.5)}$$

So power dissipated as a function of frequency is a Lorentzian centered on the resonant frequency. Rewriting the impedance and power dissipation in terms of the full width $\Gamma$=R/L of the Lorentzian:

$$Z = R(1 + \frac{2i\Delta}{\Gamma}) \qquad P = \frac{E_0^{\ 2}}{R}(\frac{1}{1 + \frac{4\Delta^2}{\Gamma^2}}) \qquad \text{(A3.6)}$$

We define the quality factor of the resonance, $Q = \omega_0/\Gamma$. The assumption $\Delta << \omega_0$ can be rewritten $\Delta/\Gamma << Q$. For modes in our resonant cavity Q~$10^5$, so the assumption will always be valid within a few full widths of the resonant frequency. In the following analysis I will assume that our cavity resonance is perfectly Lorentzian.

## A3.2 Coupling the Cavity to an External Circuit

In our apparatus the resonant cavity is coupled to an RF preamplifier by an electric field probe. Consider: how is the insertion depth of the probe related to the amount of power dissipated at the preamplifier input? To understand this problem, let us first consider the problem of a signal source connected to a load of resistance $R_L$. (Figure A3.2). The source is represented by the Thévenin equivalent circuit of a voltage source in series with the source resistance $R_S$.

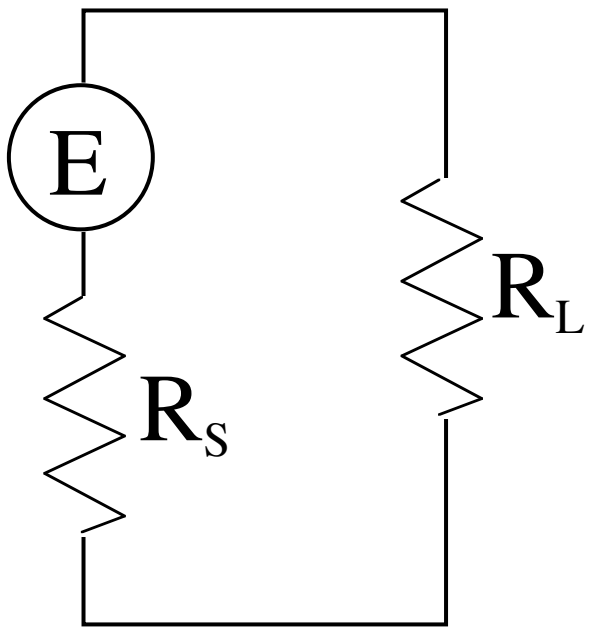

Figure A3.2 Equivalent circuit of a cavity coupled to an external load.

The current I around the circuit is:

$$I = \frac{E}{R_S + R_L} \qquad \text{(A3.7)}$$

So the power $P_L$ dissipated in the load resistance is

$$P_L = I^2 R_L = \frac{E^2 R_L}{(R_S + R_L)^2} \qquad \text{(A3.8)}$$

The total power $P_T$ dissipated in the circuit is

$$P_T = I^2 (R_S + R_L) = \frac{E^2}{(R_S + R_L)} \qquad \text{(A3.9)}$$

From (A3.8), the power dissipated in the load is maximized for $R_S = R_L$. As $R_L/R_S$ tends to zero, the total power dissipated in the circuit, $P_T$ is maximized, but no power is dissipated in the load resistor. As $R_L/R_S$ tends to infinity, the total power $P_T$ tends to zero, so again no power is dissipated in the load. Figure 3 is a plot of $P_L$ as a function of $R_L/R_S$.

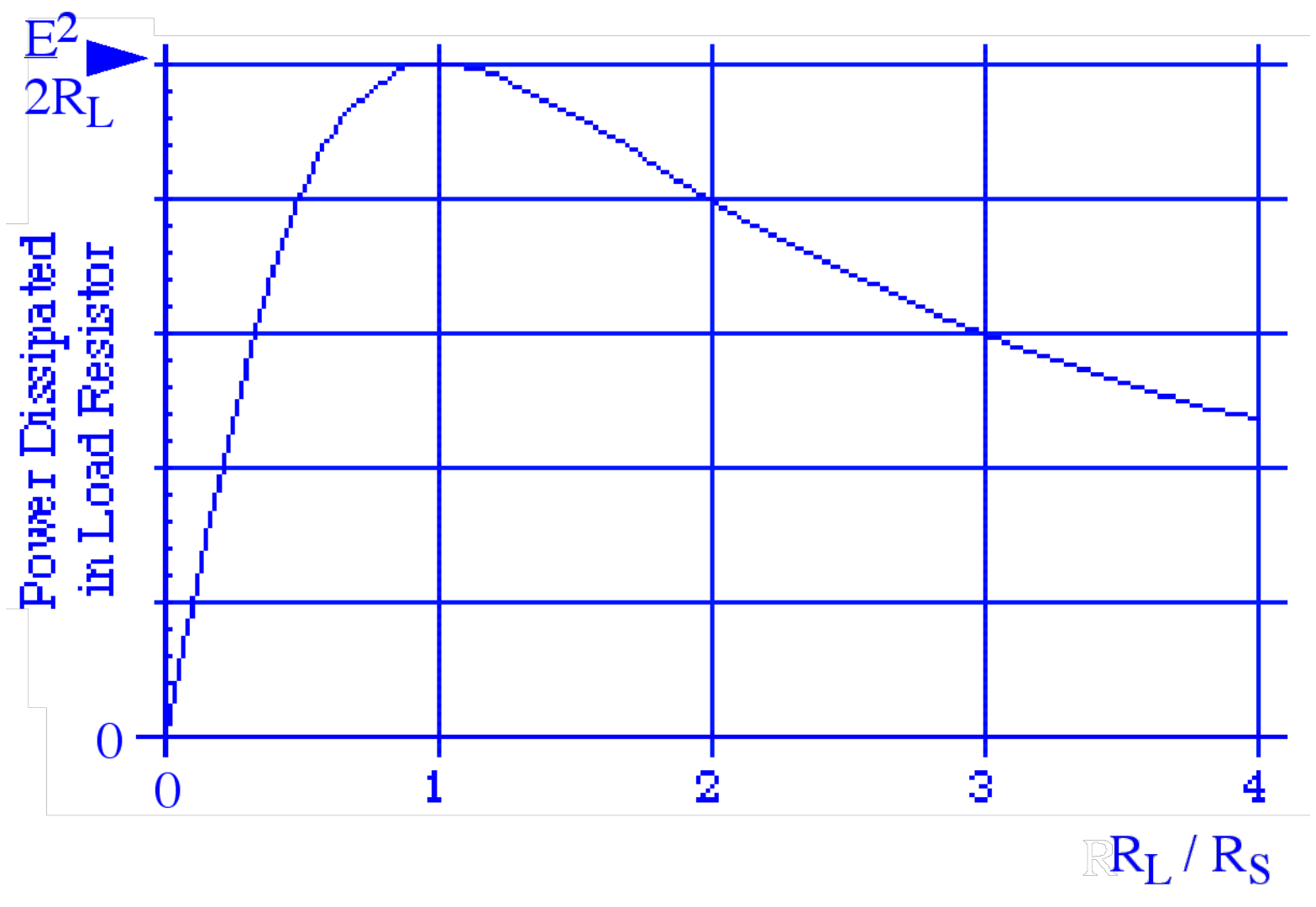

Figure A3.3 Power dissipated in load resistor vs $R_L$ / $R_S$.

In our experiment, the $TM_{010}$ mode of our resonant cavity is coupled by an electric field probe to the input of our first stage amplifier. Figure 4 is an equivalent circuit for the apparatus, where for now I neglect the effects of noise sources in the amplifier and the transmission line between the electric field probe and the amplifier. We represent Johnson noise in the cavity mode with a voltage source. The probe is represented by a transformer of turns ratio n:1. The turns ratio changes as the insertion depth of the electric field probe is varied. The secondary winding of the transformer is connected to an external load which represents the input impedance of the first stage amplifier. As discussed previously, the HEMT amplifier used in our search is very well matched to the 50Ω transmission line, and I assume throughout this analysis that it has a 50Ω input impedance.

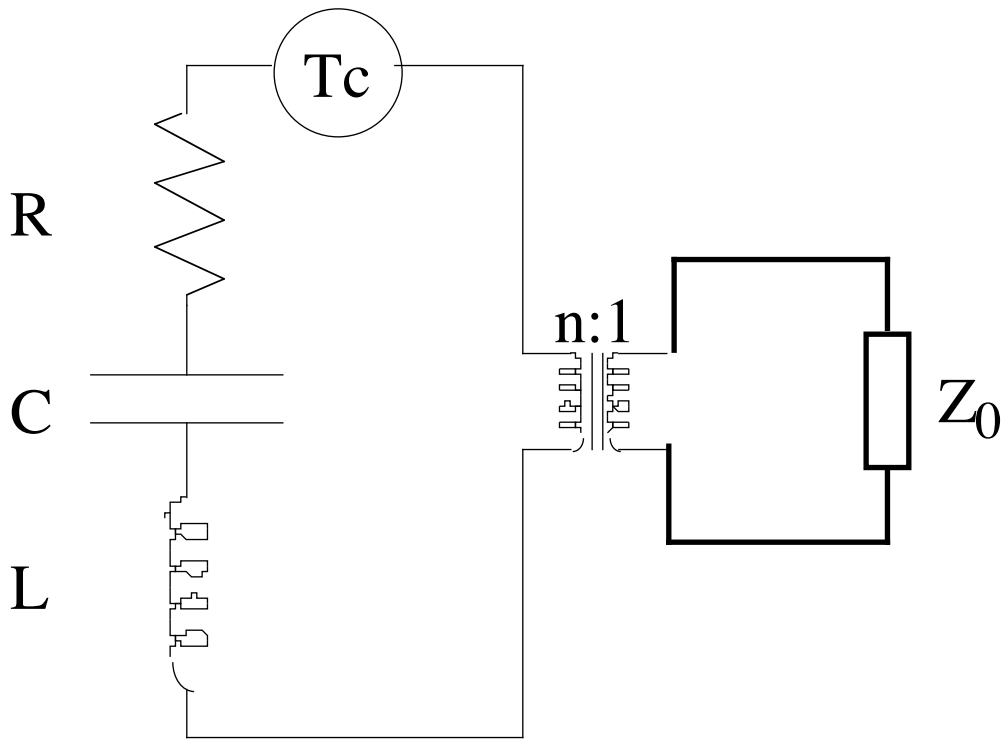

Figure A3.4 Equivalent circuit of a noisy cavity with variable coupling to an external load.

I define the effective impedance $Z_{eff}$ of the cavity as the ratio of the voltage across to the current through the secondary winding of the transformer. $Z_{eff}$ is given by:

$$Z_{eff} = \frac{R}{n^2}\left(1 + \frac{2iL\Delta}{R}\right) \qquad \text{(A3.10)}$$

At the resonant frequency $\omega_0$ the imaginary part of $Z_{eff}$ vanishes. Hence by my previous argument, maximum power is dissipated in the external impedance $Z_0$ where $Z_{eff} = Z_0$. Hence the optimal turns ratio is $n^2 = R/Z_0$ at the resonant frequency $\omega_0$. I define the coupling $\beta \equiv n^2 Z_0/R$. Maximum power is transferred for $\beta=1$. We say that the cavity mode is *critically coupled* to the external circuit.

During our first run, the probe insertion depth was adjusted to keep the cavity mode critically coupled to the electronics. I will take $\beta=1$ for the remainder of this circuit analysis. When $\beta$=1, the resistance in the cavity equivalent circuit doubles due to the resistance of the external load. This means that the width of the cavity resonance at critical coupling is double that of the resonance with no external load attached. I will make extensive use of $\Gamma_C=2\Gamma$, the resonance width at critical coupling.

## A3.3 Resonant Cavity Coupled to a Noisy Amplifier

I now consider the case where the first stage amplifier is noisy. For now, I continue to ignore the effect of the cable connecting the cavity to the amplifier input. The amplifier noise is represented by voltage and current noise sources at the amplifier input, as shown in figure 5. I assume that the voltage and current noise are uncorrelated.

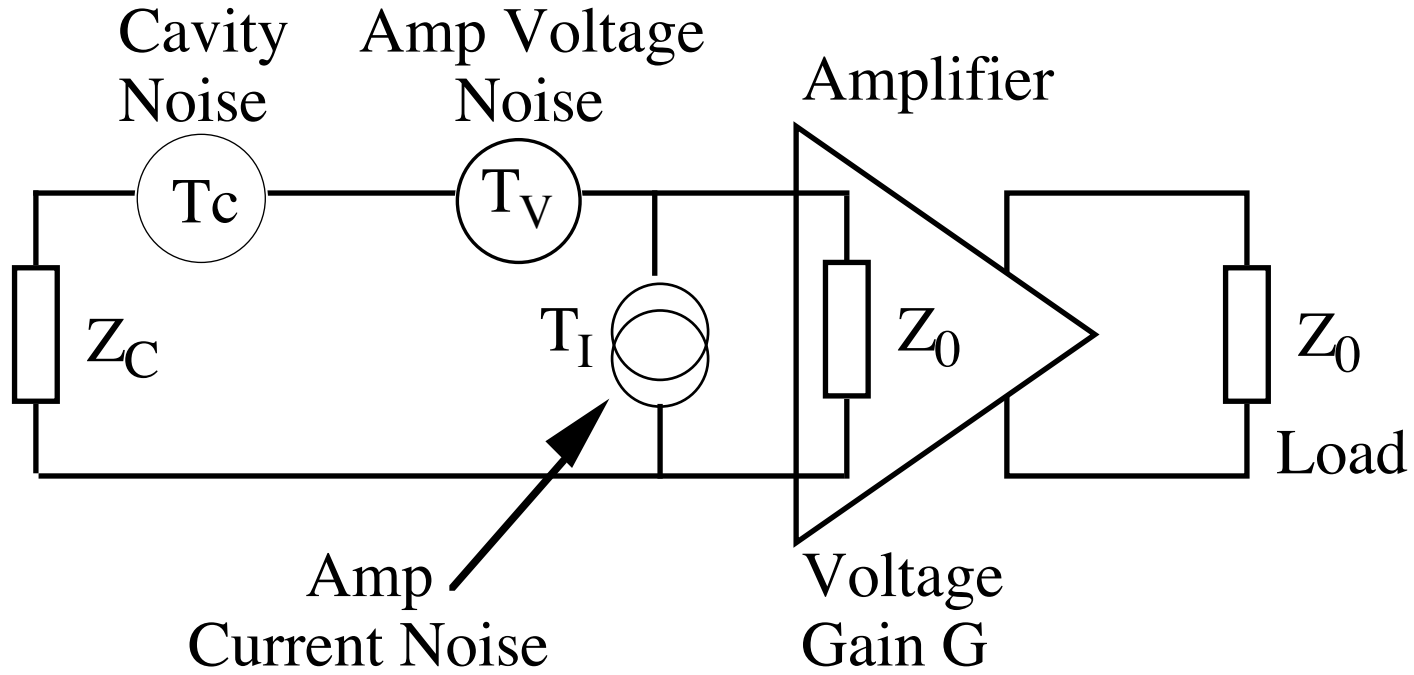

Figure A3.5 Equivalent circuit of cavity and amplifier input assuming a short connecting cable

The impedance $Z_C$ is the effective impedance of the cavity, $Z_{eff}$ that would be measured through the field probe at critical coupling. I use equation (A3.6) to write $Z_C$ in terms of the resonance full width at critical coupling. Let $E_N$ be the voltage across the amplifier voltage noise source, $E_C$ be the voltage across the cavity noise source and $I_N$ be the current through the amplifier current noise source. The rms of the current and voltage fluctuations are related to the noise temperatures of these noise sources by Nyquist's theorem:

$$\frac{\left\langle |E_C|^2 \right\rangle}{Z_0} = 4k_B T_C B \ , \ \frac{\left\langle |E_N|^2 \right\rangle}{Z_0} = 4k_B T_V B \ \text{and} \ \left\langle |I_N|^2 \right\rangle Z_0 = 4k_B T_I B \qquad \text{(A3.12)}$$

Where B is the bandwidth. Using Ohms law, I write the instantaneous voltage across the amplifier input impedance due to the three noise sources.

$$V_{in} = \frac{(E_C + E_N)Z_0 + I_N Z_0 Z_C}{(Z_C + Z_0)} \qquad \text{(A3.13)}$$

Hence the power $P_{LOAD}$ at the load is given by:

$$P_{LOAD} = |G|^2 \frac{\left(\frac{\langle |E_C|^2 \rangle + \langle |E_N|^2 \rangle}{Z_0} + \langle |I_N|^2 \rangle \frac{|Z_C|^2}{Z_0}\right)}{\left|\frac{Z_C}{Z_0} + 1\right|^2} \qquad \text{(A3.14)}$$

Using Nyquist's theorem (12) to rewrite $P_{LOAD}$ in terms of noise temperatures,

$$P_{LOAD} = \frac{4k_B B|G|^2(T_C + T_V + T_I |Z_C/Z_0|^2)}{\left|\frac{Z_C}{Z_0} + 1\right|^2} \qquad \text{(A3.15)}$$

## A3.4 Traveling Wave Approach to Cavity - Amplifier Coupling

A different approach to analysis of our equivalent circuit consists essentially of solving the wave equation on the coaxial transmission line connecting the cavity to the amplifier input. At each end of the cable are loads where we can impose boundary conditions. The noise at frequency ω from the various noise sources (Figure A3.5) is represented by traveling waves propagating down the transmission line away from the noise source in both directions (towards the amplifier and towards the cavity), with equal amplitudes. At the impedances terminating the ends of the transmission line, the reflection coefficients are determined from the line characteristic impedance and the impedances of the terminations. With arbitrary impedances at the two ends of the line, multiple reflections of the plane waves must be considered. However, because the input impedance of our amplifier is essentially 50Ω, the same as that of the transmission line, none of the power incident on the amplifier is reflected. So we have a relatively simple picture. Each noise source emits traveling waves in both directions. The wave traveling towards the amplifier is totally absorbed at its input. The wave traveling towards the cavity is partially reflected, and the reflected signal is incident on the amplifier where it too is totally absorbed. The two amplitudes are added coherently to determine the amplitude of the signal at the amplifier input.

A crucial quantity for this analysis is the reflection coefficient r at the cavity termination. Consider a plane wave traveling down a transmission line of characteristic impedance towards the cavity. The incident and reflected voltage and current are given by:

$$\text{Incident wave:} \qquad V_{incident} = V_i e^{i(\omega t + kx)} \qquad I_{incident} = I_I e^{i(\omega t + kx)} \qquad \text{(A3.16)}$$

$$\text{Reflected Wave:} \qquad V_{reflected} = V_r e^{i(\omega t - kx)} \qquad I_{reflected} = I_r e^{i(\omega t - kx)} \qquad \text{(A3.17)}$$

The boundary conditions at the cavity impedance are $V_i + V_r = V_c$ and $I_i - I_r = I_c$, where $V_c$ and $I_c$ are the voltage across and current through the cavity impedance. Using the two boundary conditions we obtain:

$$\frac{Z_C}{Z_0} = \frac{V_i + V_r}{Z_0 I_i - Z_0 I_r} = \frac{1+r}{1-r} \qquad \text{or} \qquad r = \frac{Z_C - Z_0}{Z_C + Z_0} \qquad \text{(A3.18)}$$

We may rewrite the expression (A3.10) for the cavity impedance in terms of the width $\Gamma_c$ of the resonance at critical coupling:

$$Z_C = Z_0 \left(1 - \frac{4i\Delta}{\Gamma_C}\right) \qquad \text{(A3.19)}$$

Using this result, the cavity reflection coefficient (18) can also be re-written:

$$r = \frac{1}{1 - \frac{i\Gamma_C}{2\Delta}} \qquad \text{(A3.20)}$$

Another useful relationship is obtained by combining (A3.18), (A3.19) and (A3.20)

$$Z_C + Z_0 = \frac{Z_C - Z_0}{r} = -4 Z_0^{\,2}\left(\frac{1}{2} + \frac{i\Delta}{\Gamma_C}\right) \qquad \text{(A3.21)}$$

or,

$$\left|1 + \frac{Z_C}{Z_0}\right|^2 = 4\left(1 + \frac{4\Delta^2}{\Gamma_C^{\,2}}\right) \qquad \text{(A3.22)}$$

We will now re-derive (A3.15) using the traveling wave approach. First, let us consider the voltage noise. According to our model, the amplitude at the amplifier input is given by adding the amplitude of the signals incident on the amplifier directly from the noise sources with no reflections and the signals which are reflected off the cavity before reaching the amplifier. We first calculate the voltage at the amplifier input $E_{amp}$ due to the two voltage noise sources in the equivalent circuit:

$$E_{amp} = \frac{E_N + E_C}{2}(1 - r) = \frac{E_N + E_C}{2}\left(\frac{1}{1 + \frac{2i\Delta}{\Gamma_C}}\right) \qquad \text{(A3.23)}$$

The '-'sign before the reflection coefficient accounts for the $180^0$ phase difference between the voltages of the traveling waves emitted in opposite directions by each voltage noise source. Therefore the power $P_{VL}$ at the load resistor due to the two voltage noise sources is:

$$P_{VL} = \frac{\left(\left\langle |E_N|^2 \right\rangle + \left\langle |E_C|^2 \right\rangle\right) G^2}{4Z_0}\left(\frac{1}{1 + \frac{4\Delta^2}{\Gamma_C{}^2}}\right) = \frac{k_B(T_C + T_V)BG^2}{4Z_0{}^2(1 + \frac{4\Delta^2}{\Gamma_C{}^2})} = \frac{k_B(T_C + T_V)BG^2}{\left|1 + \frac{Z_C}{Z_0}\right|^2} \qquad \text{(A3.24)}$$

Next let us consider the effect of the current noise source at the amplifier input. Adding the current incident directly on the amplifier to that which first reflects off the cavity input, we obtain the current $I_{amp}$ at the amplifier input due to the current noise source.

$$I_{amp} = \frac{I_N}{2}(1 + r) \qquad \text{(A3.25)}$$

The resulting voltage $V_I$ at the amplifier input given by

$$V_I = \frac{I_N Z_0}{2}\frac{(2 - \frac{i\Gamma_C}{2\Delta})}{(1 - \frac{i\Gamma_C}{2\Delta})} \qquad \text{(A3.26)}$$

So the power $P_{IL}$ at the load resistor as a result of the current noise source is:

$$P_{IL} = \frac{G^2 \left\langle |V_I|^2 \right\rangle}{Z_0} = \frac{G^2 \left\langle |I_N|^2 \right\rangle Z_0}{4} \frac{(1 + \frac{16\Delta^2}{{\Gamma_C}^2})}{(1 + \frac{4\Delta^2}{{\Gamma_C}^2})} = \frac{k_B T_I B G^2 \left| \frac{Z_C}{Z_0} \right|^2}{\left| 1 + \frac{Z_C}{Z_0} \right|^2} \tag{A3.27}$$

The total noise power in the load resistor is the sum of the power due to voltage noise (A3.24) and the power due to current noise (A3.27). Notice that adding these two results gives exactly the expression calculated in (A3.15) using the a simple Ohm's law approach.

## A3.5 Incorporating a Long Connecting Cable

What do we gain by using a traveling wave model? The answer is that in the traveling wave picture it is easier to introduce circuit elements that cannot be represented in an equivalent circuit by a simple impedance network. An example of such a circuit element is the length of transmission line connecting the cavity to the amplifier input. Consider: looking towards the cavity down the cable from the amplifier input one 'sees' a highly frequency dependent, complex impedance. Looking the other way down the cable from the cavity towards the amplifier input, one 'sees' simply a 50Ω real impedance, independent of the cable length, because the amplifier and the cable are perfectly matched. It turns out that one can only approximately represent a length of coaxial cable by an equivalent circuit containing a finite number of impedances. This is not surprising; the cable is a continuum and hence has an infinite number of oscillation modes, so it is naturally impossible to model it exactly using a finite number of elements.

Figure A3.6 is the equivalent circuit including a length L of transmission line connecting the cavity and RF probe to the amplifier input.

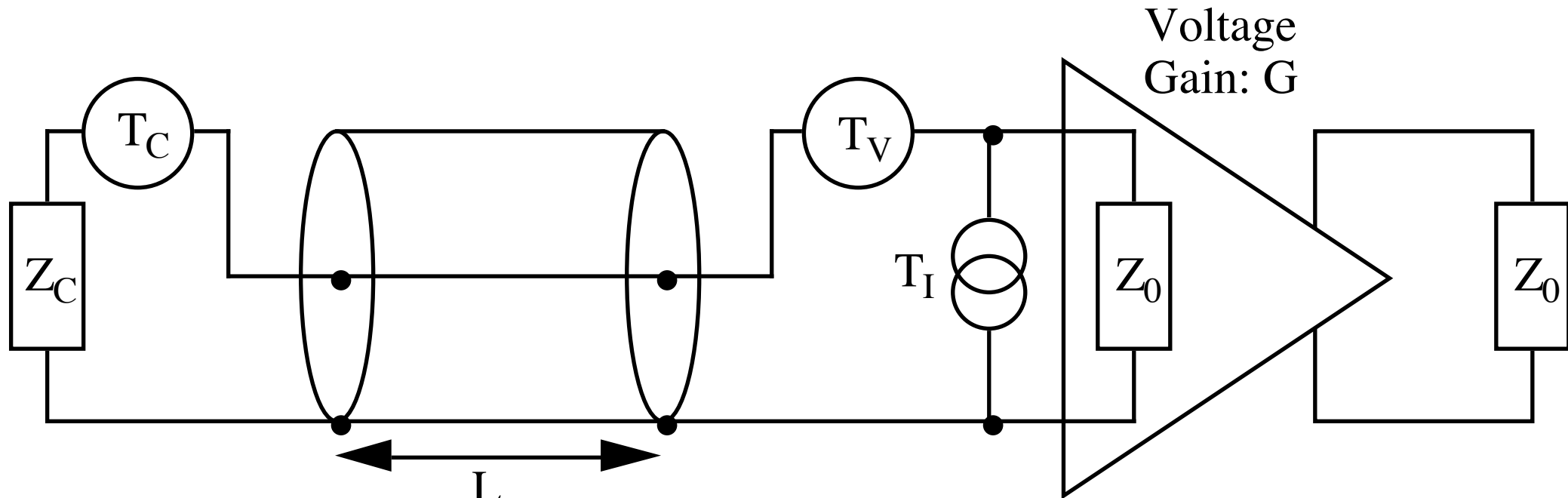

Figure A3.6 An equivalent circuit model of a cavity coupled to a balanced amplifier.

The noise power $P_{CAV}$ at the load resistor due to the cavity noise source is the same as it was when we neglected the transmission line, because the impedance looking from the cavity towards the amplifier is 50Ω independent of the cable length L. We just extract the cavity noise contribution from the formula for voltage noise (24) calculated for L=0.

$$P_{CAV} = \frac{k_B T_C B G^2}{4{Z_0}^2(1 + \frac{4\Delta^2}{{\Gamma_C}^2})} \qquad \text{(A3.28)}$$

The power at the load resistor due to the amplifier voltage noise source $P_V$ and that due to the current noise source, $P_I$ are both dependent on the cable length. This is because the impedance looking from the amplifier noise sources back down the cable towards the cavity is dependent on the cable length. I assume that losses in the cable can be neglected. The effect of the transmission line is to introduce a phase lag between waves emitted from a noise source towards the amplifier and waves which travel down the cable, partially reflect off the cavity and then travel back to the amplifier. The voltage $V^V_{IN}$ at the amplifier input due to the amplifier voltage noise source is

$$V^V_{IN} = \frac{E_N}{2}(1 - re^{-2ikL}) \qquad \text{(A3.29)}$$

Notice the phase factor due to the length of transmission line. This voltage results in a power $P_V$ in the load resistor given by:

$$P_V = \frac{\langle |E_N|^2 \rangle G^2}{4Z_0}\left|1 - re^{-2ikL}\right|^2$$

$$= \frac{k_B T_V B G^2}{(1 + \frac{{\Gamma_C}^2}{4\Delta^2})}\left(2 + \frac{{\Gamma_C}^2}{4\Delta^2} - (1 - \frac{i\Gamma_C}{2\Delta})e^{2ikL} - (1 + \frac{i\Gamma_C}{2\Delta})e^{-2ikL}\right)$$

$$= \frac{k_B T_V B G^2}{(1 + \frac{4\Delta^2}{{\Gamma_C}^2})}\left(1 - \frac{4\Delta}{\Gamma_C}\sin 2kL + \frac{8\Delta^2}{{\Gamma_C}^2}(1 - \cos 2kL)\right) \qquad \text{(A3.30)}$$

The voltage $V^I_{IN}$ at the amplifier input due to the amplifier current noise source is:

$$V^I_{IN} = \frac{I_V Z_0}{2}(1 + re^{-2ikL}) \qquad \text{(A3.31)}$$

resulting in a power $P_I$ in the load resistor given by:

$$P_I = \frac{\left\langle |I_V|^2 \right\rangle Z_0 G^2}{4}\left|1 + re^{-2ikL}\right|^2$$

$$= \frac{k_B T_I B G^2}{(1 + \frac{\Gamma_C{}^2}{4\Delta^2})}(2 + \frac{\Gamma_C{}^2}{4\Delta^2} + (1 - \frac{i\Gamma_C}{2\Delta})e^{2ikL} + (1 + \frac{i\Gamma_C}{2\Delta})e^{-2ikL})$$

$$= \frac{k_B T_I B G^2}{(1 + \frac{4\Delta^2}{\Gamma_C{}^2})}(1 + \frac{4\Delta}{\Gamma_C}\sin 2kL + \frac{8\Delta^2}{\Gamma_C{}^2}(1 + \cos 2kL)) \qquad \text{(A3.32)}$$

Combining (A3.28),(A3.30) and (A3.32), the total noise power $P_{OUT}$ at the load resistor is:

$$P_{OUT} = \frac{k_B B G^2}{(1 + \frac{4\Delta^2}{\Gamma_C{}^2})}((T_C + T_V + T_I) + \frac{4\Delta}{\Gamma_C}(T_I - T_V)\sin 2kL + \frac{8\Delta^2}{\Gamma_C{}^2}((T_I + T_V) + (T_I - T_V)\cos 2kL)) \qquad \text{(A3.33)}$$

Notice that this expression includes a term linear in $\Delta$, so the power spectrum of the noise is not in general symmetric about the resonant frequency.

## Appendix 4: FFT Response to a Narrow Axion Signal

Naively one might guess that the power spectrum of a signal consisting of a single frequency is a single bin spike with no leakage into surrounding bins. This guess turns out to be wrong. In a real experiment a power spectrum consists of a finite number N of equally spaced measurements of the voltage across a load. Let $h_k$ be the result of the kth measurement, and $\Delta$ be the period between successive measurements. I refer to the set of measurements $\{h_k, k=0,1,...,N-1\}$ as the time series. The Fourier transform of the time series yields a series of N complex numbers, where the nth element $H_n$ is:

$$H_n = \sum_{k=0}^{N-1} h_k e^{\frac{2\pi ikn}{N}} \qquad \text{(A4.1)}$$

By the Nyquist sampling theorem, the maximum frequency measurable in the Fourier transform is $f_m=1/2\Delta$. I will consider the range of the Fourier transform to be n=0 to n=N-1; the Fourier transform is periodic with period N.

Consider the case where the input signal is a single frequency f where $|f|<f_m$. Adopting the complex representation the time series $s_k(f)$ for a real sinusoidal signal can be written:

$$s_k(f) = \cos\left(\frac{2\pi fk}{2f_m}\right) = \frac{1}{2}\left(\exp\left(-\frac{2\pi ifk}{2f_m}\right) + \exp\left(+\frac{2\pi ifk}{2f_m}\right)\right) \qquad \text{(A4.2)}$$

The exponential representation separates the components at frequencies +f and -f. For simplicity I take only the component at frequency +f. The time series $h_k$ is:

$$h_k(f) = \exp\left(-\frac{2\pi ifk}{2f_m}\right) \qquad \text{(A4.3)}$$

The Fourier transform of $h_k$ is:

$$H_n(f) = \sum_{k=0}^{N-1} \exp\left(\frac{2\pi i}{N}\left(n - \frac{Nf}{2f_m}\right)\right)$$

$$= \frac{1 - \exp\left(2\pi i\left(n - \frac{Nf}{2f_m}\right)\right)}{1 - \exp\left(\frac{2\pi i}{N}\left(n - \frac{Nf}{2f_m}\right)\right)} \qquad \text{(A4.4)}$$

The two sided power spectral density (PSD) is:

$$P_n(f) = \left|H_n(f)\right|^2 = \frac{\sin^2\left(\pi\left(n - \frac{Nf}{2f_m}\right)\right)}{\sin^2\left(\frac{\pi}{N}\left(n - \frac{Nf}{2f_m}\right)\right)} \qquad \text{(A4.5)}$$

This equation looks exactly like that for the intensity of light reflected by a diffraction grating with narrow rulings illuminated by a monochromatic source. The N time samples correspond to N rulings on a grating and the total time duration of the string of time samples is analogous to the total width of the grating. The two sided PSD consists of N frequency bins corresponding to n=0,1,2,...,(N-1), each of bandwidth $2f_m/N$. Each bin is centered on a multiple of $2f_m/N$. Notice that if the frequency f is equal to a bin central frequency, $P_n(f)$ is zero for all n except one, where it takes the value $N^2$.

I consider the case when f is within the bandwidth of one of these frequency bins, but is not necessarily at the bin central frequency. If f is within the nth bin, then:

$$f = n - \frac{f_m\xi}{N} \quad \text{and} \quad -1 \le \xi \le +1 \qquad \text{(A4.6)}$$

Where $\xi$=0 the signal frequency is exactly at the center of the bin. Where $\xi=\pm1$ the signal frequency is right on the boundary between adjacent bins. The one sided PSD becomes:

$$P_n(\xi) = \frac{\sin^2\left(\frac{\pi\xi}{2}\right)}{\sin^2\left(\frac{\pi\xi}{2N}\right)} \qquad \text{(A4.7)}$$

For large N, the fraction f(ξ) of the signal power in the frequency bin enclosing the signal is given by:

$$f(\xi) = \frac{4}{\pi^2 \xi^2} \sin^2\left(\frac{\pi\xi}{2}\right) \qquad \text{(A4.8)}$$

Figure A4.1 is a plot of f(ξ). The smooth line is the theory from equation A4.8. The points are the results of taking a power spectrum on a finite number of samples of a sinusoidal wave. The Fourier transform was taken using a Macintosh 7600/132, the Fourier transform package used was the standard routine included with the LabView software package, version 4. The agreement is good.

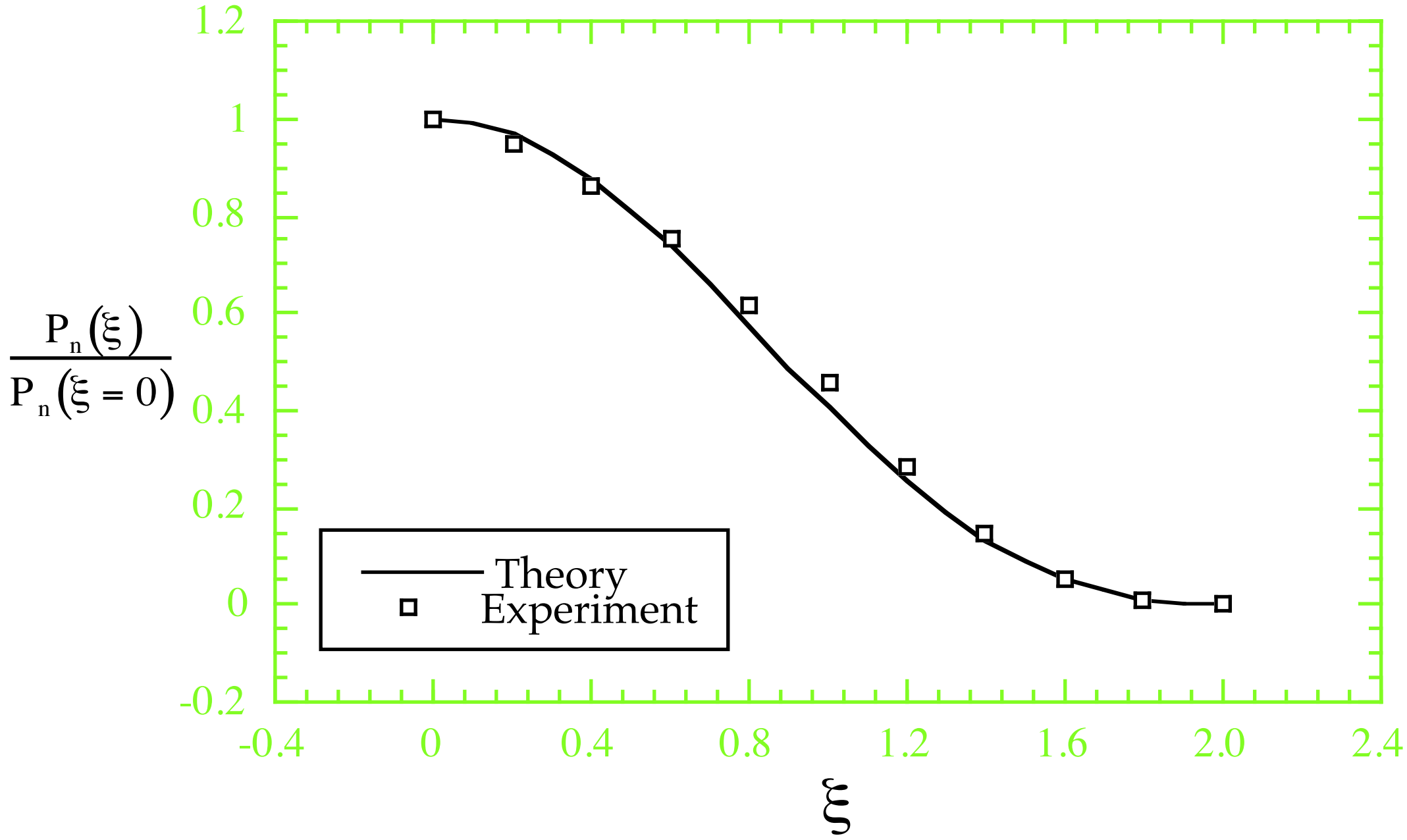

Figure A4.1 Theoretical and experimentally measured leakage in the power spectrum of a single frequency. When ξ=0 the frequency of the sinusoidal wave is exactly the center frequency of one of the FFT bins. When ξ=1/2, the frequency of the sinusoidal wave is on the boundary between two bins. The square points are the result of taking the power spectrum of an actual sinusoidal signal using a commercial Fourier transform package. The smooth line is the result of equation A4.8.

## Appendix 5: Simulation of Raw Traces

The simulations used to estimate the sensitivity of the axion detector to thermal KSVZ axion signals in the 6-bin search channel were largely based on a method for generating simulated raw traces containing signals at power levels comparable to those expected from axion conversion. This appendix is a brief outline of how this Monte Carlo simulation works.

As mentioned in the main body of the text, the simulation programs generate a list of frequencies at random where the simulated signals will be injected into the artificial traces. I will describe the method for generation of a artificial trace used to generate a simulated set of run 1 data.

I start with a real trace from the raw data that overlaps with a frequency where I wish to inject a signal in a simulated data set. I normalize that trace with the crystal filter response and perform a 5 parameter fit on the receiver corrected trace. So far, this is the same procedure as used in the analysis of the production data. For the purpose of the Monte Carlo, I now discard the real trace, but keep the 5 fit parameters.

The next stage is to start building up the simulated trace containing the signal. For the run 1 data, the baseline on which I overlay the Maxwellian shaped signal is the fit residual curve discussed in Section 4.3.5. The simulated Maxwellian and the background fit residual curve are then multiplied by the cavity Lorentzian to reduce the signal power to the correct level (see Equation 4.15). Figure A5.1 is a plot of the crystal filter response with a simulated axion signal overlaid on top of it near the center.

For reasons discussed at length in Section 4.3, the crystal filter response does not contribute significantly to the run 1 sensitivity. Therefore it is often sufficient to overlay the simulated Maxwellian on a flat background and not on the crystal filter residuals even in simulations of run 1. For simulation of runs R and P, I assume that any crystal filter normalization errors have been successfully eliminated by the methods discussed in Sections 4.5.5 and 4.5.9, and always overlay the simulated Maxwellian peak on a flat background.

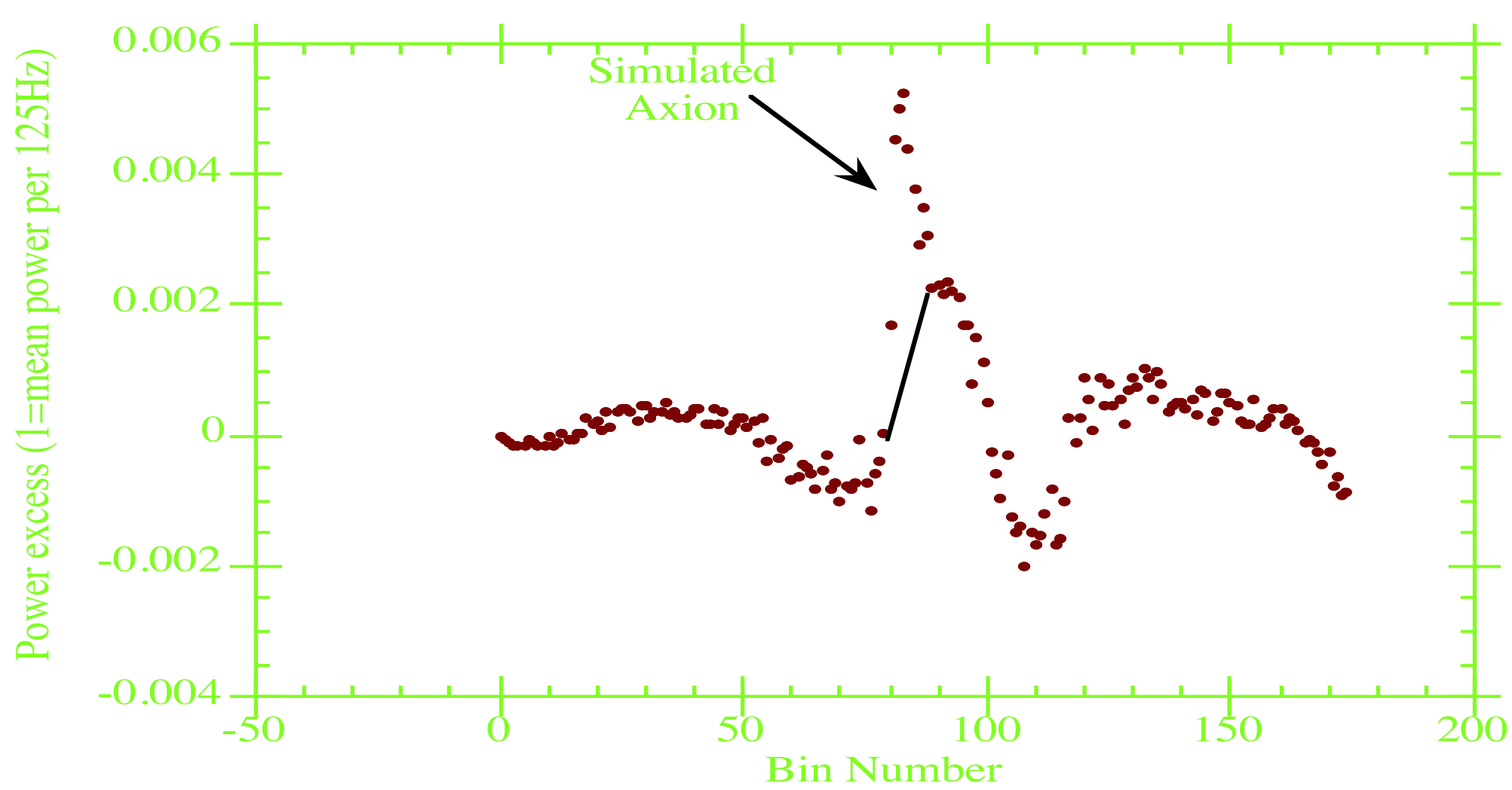

Figure A5.1 A fake Maxwellian on top of the run 1 crystal filter residual, and weighted with the Lorentzian profile of the cavity resonance. The straight line shows the baseline under the peak.

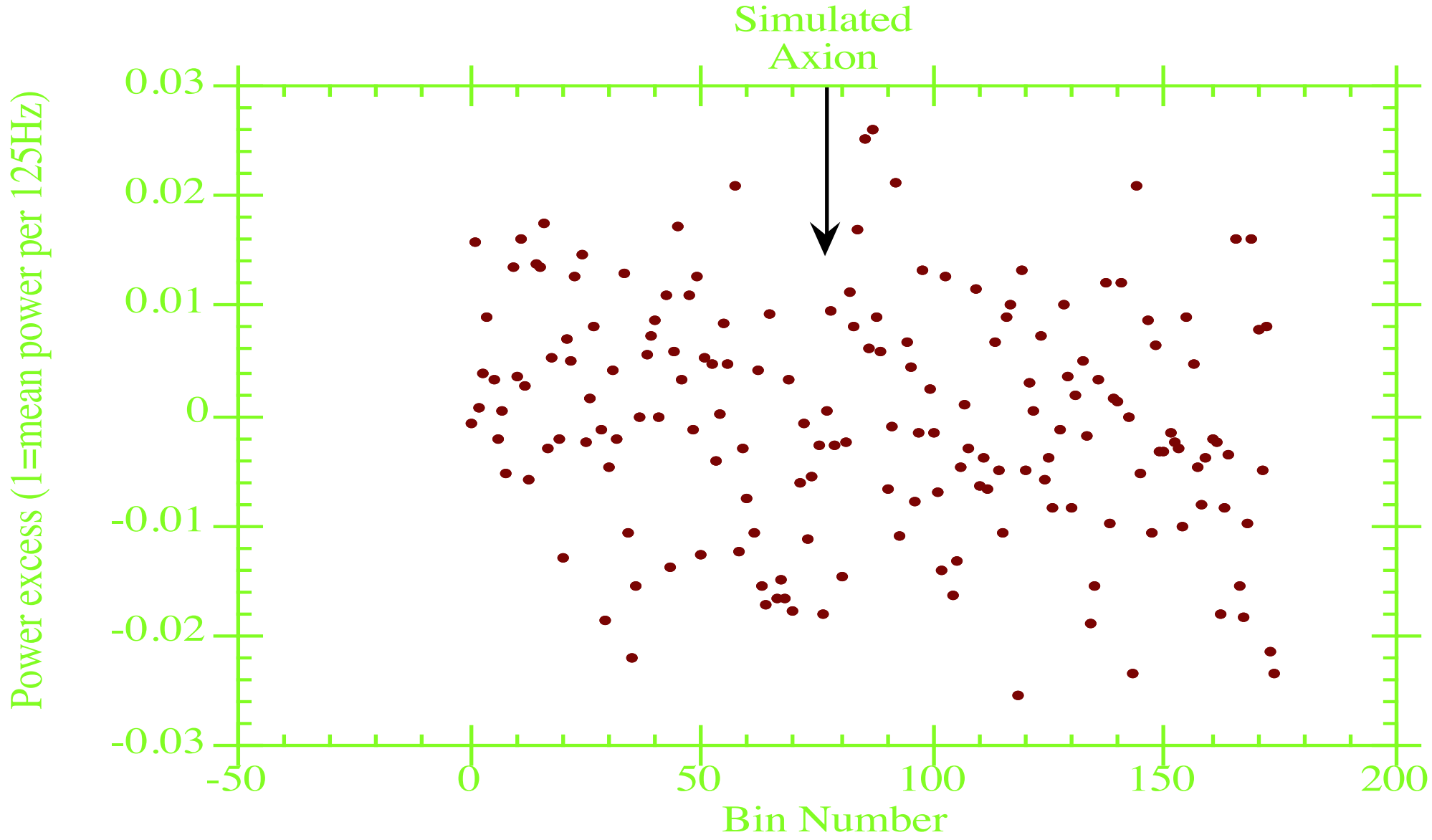

Figure A5.2 The same fake Maxwellian with computer generated Gaussian noise of rms 0.01 overlaid on top of it.

As expected in a single trace, the height of the simulated KSVZ signal is small compared to the overlaid Gaussian noise. Finally, each bin of the simulated spectrum is incremented by 1, and the result multiplied by the 5 parameter fit to the original spectrum . The result is shown in Figure A5.3.

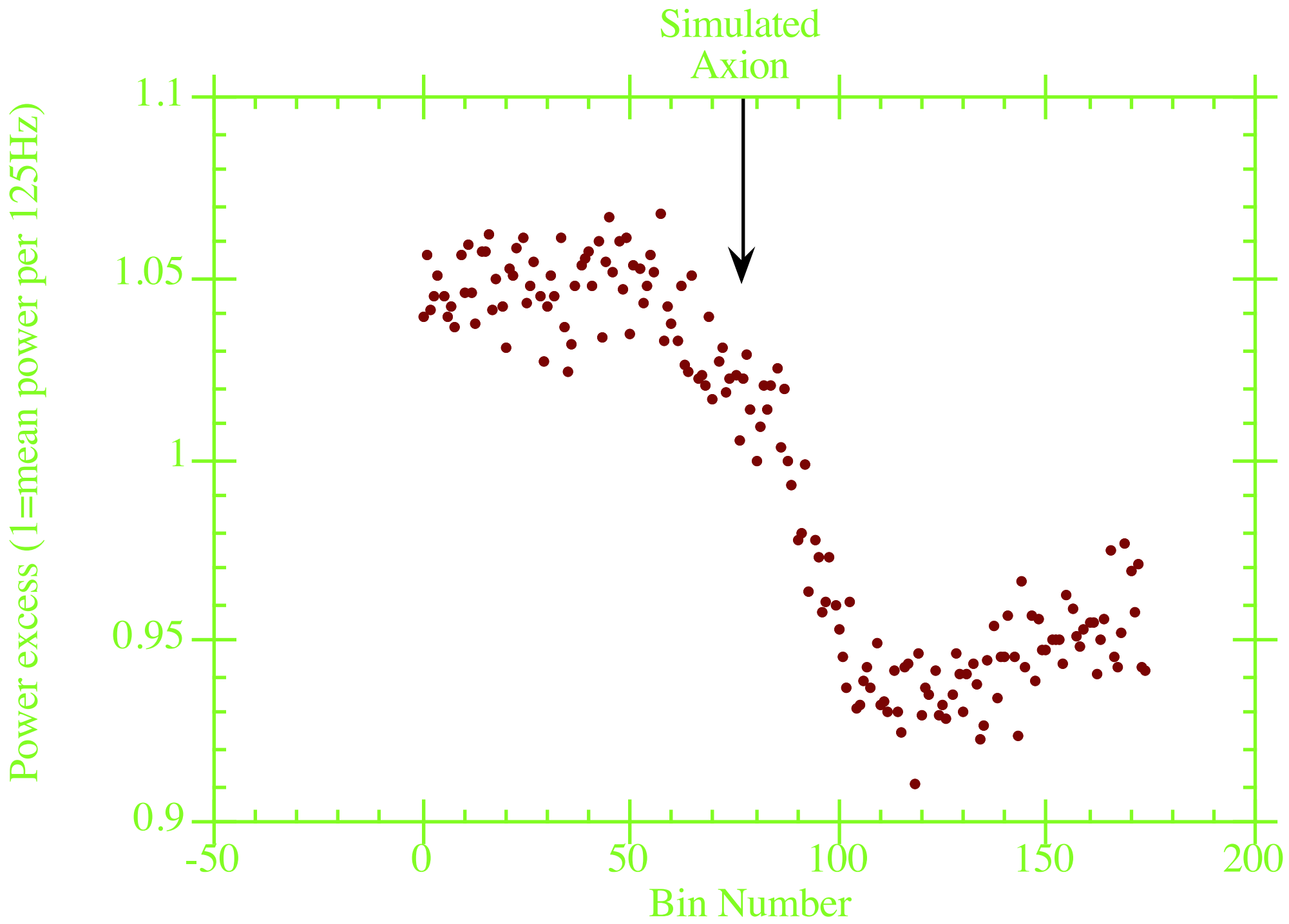

Figure A5.3 A simulated receiver corrected trace with an artificial KSVZ axion - like peak injected

This process is repeated for every raw trace in the real data set that overlaps with one of the randomly selected injection frequencies. The fake traces are treated in the rest of the simulation exactly like the real production data. In particular, they are re-fitted using the same 5 parameter fit to establish what fraction of the injected signal is lost due to distortion of the fit function in the vicinity of the injected peak. By performing peak searches on the simulated combined data sets, I can deduce the confidence with which the experiment would detect KSVZ axions at different power levels.